\documentclass[preprint,12pt]{elsarticle}
\usepackage{graphicx,psfrag,color}
\usepackage{mathbbol,amsmath,amsfonts,amssymb,bm,dsfont}
\usepackage{wasysym}
\usepackage{amsthm}

\usepackage{graphicx,psfrag,color}
\usepackage{dcolumn}
\usepackage{bm}
\usepackage{mathbbol, appendix}
\usepackage{multirow}
\usepackage{booktabs}
\usepackage{enumitem}
\usepackage{mathrsfs}
\usepackage{tikz}   
\usepackage{mathtools}  
\usetikzlibrary{arrows.meta} 
\usepackage{pgfplots}
\usepackage{xxcolor}
\pgfplotsset{compat=1.10}
\pgfdeclareradialshading[tikz@ball]{ball}{\pgfqpoint{0bp}{0bp}}{%
 color(0bp)=(tikz@ball!0!white);
 color(7bp)=(tikz@ball!0!white);
 color(15bp)=(tikz@ball!70!black);
 color(20bp)=(black!70);
 color(30bp)=(black!70)}
\makeatother
\usepackage{pgfplots}
\usepackage{xxcolor}
\pgfplotsset{compat=1.10}

\pgfdeclareradialshading[tikz@ball]{ball}{\pgfqpoint{0bp}{0bp}}{%
 color(0bp)=(tikz@ball!0!white);
 color(7bp)=(tikz@ball!0!white);
 color(15bp)=(tikz@ball!70!black);
 color(20bp)=(black!70);
 color(30bp)=(black!70)}
\makeatother

\journal{Nuclear Physics B}

\makeindex
\begin{document}
\begin{frontmatter}

\title{Macroscopic Length Correlations in Non-Equilibrium Systems and Their Possible Realizations}

\author[]{Z. Nussinov\footnote{zohar@wustl.edu}}
\address{Department of Physics, Washington University, St.
Louis, MO 63160, USA.}

\begin{abstract}
We consider general systems that start from and/or end in thermodynamic equilibrium while experiencing a finite rate of change of their energy density or other intensive quantities $q$ at intermediate times. We demonstrate that at these times, during which $q$ varies at a finite rate, the associated covariance, the connected pair correlator $G_{ij} = \langle q_{i} q_{j} \rangle - \langle q_{i} \rangle \langle q_{j} \rangle$, between any two (far separated) sites $i$ and $j$ in a macroscopic system may, on average, become finite. Once the global mean $q$ no longer changes, the average of $G_{ij}$ over all site pairs $i$ and $j$ may tend to zero. However, when the equilibration times are significant (e.g., as in a glass that is not in true thermodynamic equilibrium yet in which the energy density (or temperature) reaches a final steady state value), these long range correlations may persist also long after $q$ ceases to change. We explore viable experimental implications of our findings and speculate on their potential realization in glasses (where a prediction of a theory based on the effect that we describe here suggests a universal collapse of the viscosity that agrees with all published viscosity measurements over sixteen decades) and non-Fermi liquids. We discuss effective equilibrium in driven systems and derive uncertainty relation based inequalities that connect the heat capacity to the dynamics in general open thermal systems. These rigorous thermalization inequalities suggest the shortest possible fluctuation times scales in open equilibrated systems at a temperature $T$ are typically ``Planckian'' (i.e., ${\cal{O}}(\hbar/(k_{B} T))$). We briefly comment on parallels between quantum measurements, unitary quantum evolution, and thermalization and on how Gaussian distributions may generically emerge.
\end{abstract}

\end{frontmatter}

\section{Introduction}
In theories with local interactions, the connected correlations between two different sites $i$ and $j$ often decay with their spatial separation $|i-j|$. Indeed, connected correlations decay exponentially with distance in systems with finite correlation lengths. In massless (or critical) theories, this exponential decay is typically replaced by an algebraic drop. The detailed understanding of these decays was achieved via numerous investigations that primarily focused on venerable equilibrium and other systems with fixed control parameters, e.g., \cite{Ruelle,GJ,matt,matt',jens,jens1,jens',sergey,jens2,ali1,ali2} including long range correlations at high temperatures in disparate systems associated with generalized screening lengths \cite{hightus}. 
Pioneering studies examined work-free energy relations in irreversible systems \cite{chris',chris,campisi,Talkner}. We wish to build on these notions and ask what occurs in a general (quantum or classical) non-relativistic system, when an intensive parameter such as the average energy density (set, in all but the phase coexistence region where latent heat appears, by the temperature) or external field is varied so that, during transient times, the system is forcefully kept {\it out of thermal equilibrium}. We will illustrate that, under these circumstances, extensive fluctuations will generally appear. These large fluctuations will imply the existence of connected two point correlation functions that will, on average, {\it remain finite for all spatial separations}. If the system returns to equilibrium, these long range correlations may be lost. In focusing on driven non-equilibrium systems, the quantum facets of our work complement investigations on nontrivial aspects of the interplay between entanglement and thermalization that have witnessed a flurry of activity in recent years in, e.g., studies of operator scrambling \cite{op-scram1,op-scram2} and entanglement growth \cite{ent-growth}. Earlier celebrated analysis also suggested fundamental quantum mechanical ``chaos'' bounds in thermal systems \cite{juan-martin}. In the current work, we will largely focus on the more precise quantum descriptions. Much of our analysis can be replicated for the classical limit of these systems. 

Although our considerations are general, we may couch these for theories residing on $d-$dimensional hypercubic lattices of $N=L^{d}$ sites; the average energy density $\epsilon \equiv E/N$ with $E$ the total energy. In theories with bounded local interactions, we may express (in a variety of ways) the Hamiltonian $H$ as a sum of $N'= {\cal{O}}(N)$ terms ($\{{\cal{H}}_{i}\}_{i=1}^{N'}$) that are each of finite range and bounded operator norm, 
\begin{eqnarray}
\label{HsH'}
H= \sum_{i=1}^{N'} {\cal{H}}_{i}.
\end{eqnarray} 
Our principal interest lies in the thermodynamic ($N \gg 1$) limit. Since our focus is on general non-equilibrium systems, the (general time dependent) Schrodinger picture probability density matrix $\rho(t)$ need not be equal to the any of the standard density matrices describing equilibrium systems. Our analysis will be largely quantum; the Ehrenfest equations typically reproduce the classical equations of motion. Aspects of classical dynamics may also be directly investigated along lines similar to those that we will largely pursue for the quantum systems. With a Liouville operator replacing the Schrodinger picture Hamiltonian, the quantum dynamics may generally replicate the classical canonical equations of motion \cite{Liouville1,Liouville2}. In the individual Sections of this work, we note which results also hold for classical systems.  

\begin{figure*}
	\centering
	\includegraphics[width=1.7  \columnwidth, height=.4 \textheight, keepaspectratio]{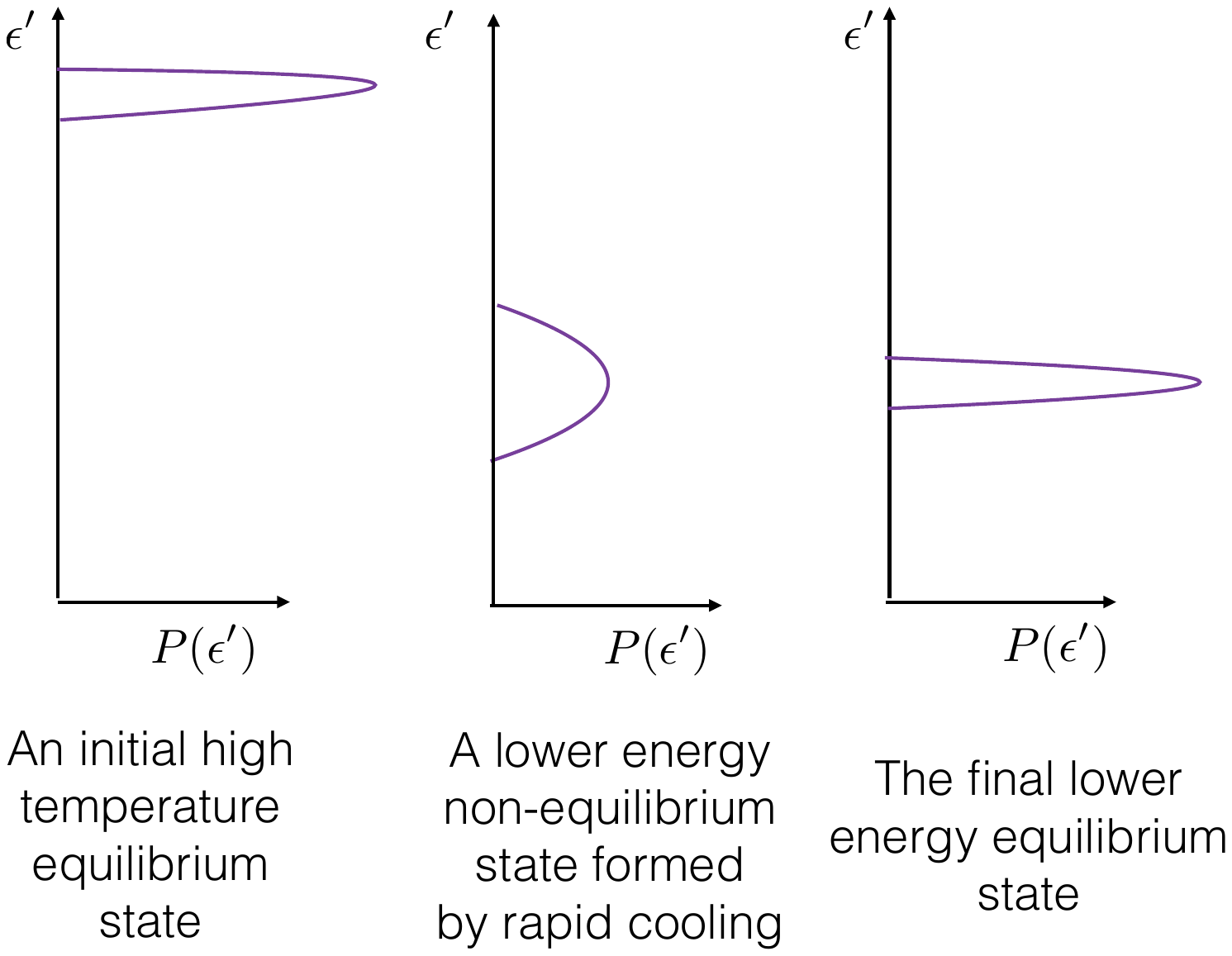}
	\caption{A schematic of the probability distribution $P(\epsilon')$ of the energy density (Eq. (\ref{PER})) for a rapid cooling process. Left: An initially equilibrated system at high temperatures where the energy density is sharply defined (in the thermodynamic limit, the distribution is a delta-function). Center: The system is rapidly cooled to a final state such that its energy density drops down at a finite rate as a function of time. During this cooling process, as it is being driven, $P(\epsilon')$ obtains a finite standard deviation (even for macroscopic systems). The demonstration of such a generic widening of the distribution is a principal objective of this paper. A finite standard deviation of $P(\epsilon')$ implies correlations that extend over length scales comparable to the system size. Right: After the cooling ceases, (if and) when the formerly driven system re-equilibrates, the distribution $P(\epsilon')$ becomes a delta-function once again (yet now at the lower temperature (smaller average energy density $\epsilon$) to which the system was cooled). Similar broadening may occur for general intensive quantities $q$.}
	\label{Idea.}
\end{figure*}

\section{Sketch of main result}
\label{sec:sketch}

In a nutshell, in order to establish the existence of long range correlations we will show the following:
\newline

\fbox{\begin{minipage}{30em}
 $\bullet$ If the expectation value of the Hamiltonian $H$ of the original (undriven) system varies in the time evolved (driven) state such that $\frac{d \epsilon}{dt} \equiv \frac{d}{dt} Tr\Big(\rho(t) \frac{H}{N}\Big) \neq 0$ then the energy density fluctuations  $\sigma_{\epsilon} \equiv \sigma_{\frac{H}{N}}$ as computed with $\rho(t)$ will, generally, also be finite. Similar results apply to all other intensive quantities.
 \end{minipage}}
\newline

As we will explain in Section \ref{intuition} and thereafter, starting from an equilibrated system, there is a minimal time $t_{\min}$ associated with the onset of a finite $\frac{d \epsilon}{dt}$ and standard deviation $\sigma_{\epsilon}$. Once the driving ceases and $d \epsilon/dt=0$, the time scale required for the system to re-equilibrate and return to its true equilibrium state with $\sigma_{\epsilon}=0$ may depend on system details (see, e.g., Section \ref{2b2b} for an approach to glasses in which the latter return time scale may be very large).  
 
While we will largely employ the more general quantum formalism, our central result holds for both quantum and classical systems. 
The central function that we will focus on to further quantify these fluctuations 
is the probability density of global energy density,
\begin{eqnarray}
\label{PER}
P(\epsilon') \equiv Tr \Big[\rho(t)~ \delta(\epsilon'- \frac{H}{N}) \Big].
\end{eqnarray}
To avoid cumbersome notation, in Eq. (\ref{PER}) and what follows, the time dependence of $P(\epsilon')$ is not made explicit; the reader should bear in mind that, throughout the current work, $P(\epsilon')$ is time dependent. In equilibrium, the energy density (similar to all other intensive thermodynamic variables) is sharply defined; regardless of the specific equilibrium ensemble employed, the distribution of 
Eq. (\ref{PER}) is a Dirac delta-function, $P(\epsilon') = \delta(\epsilon'-\epsilon)$. This is schematically illustrated in the left and righthand sides of Figure \ref{Idea.}. As we highlighted above, the chief goal of the current article is to demonstrate that when a system that was initially in equilibrium is driven at intermediate times (by, e.g., rapid cooling) such that its energy density varies at a finite rate as a function of time, the distribution $P(\epsilon')$ will need not remain a delta-function. A caricature of this feature is provided in the central panel of Figure \ref{Idea.} \cite{me}. Because the final state displays a broad distribution of energy densities, our result implies that the ``work'' per site, in the context of its quantum mechanical definitions as energy differences between final and initial states \cite{chris',chris,campisi,Talkner,natana} is not necessarily sharp (even in the $N \to \infty$ limit). Since the variance of $P(\epsilon')$ is a sum of pair correlators $G_{ij} \equiv \langle {\cal{H}}_{i} {\cal{H}}_{j} \rangle - \langle {\cal{H}}_{i} \rangle \langle {\cal{H}}_{j} \rangle$, this latter finite width of $P(\epsilon')$ of the system when it is driven implies (as we will explain in depth) that the correlations $G_{ij}$ extend over macroscopic length scales that are of the order of the system size. (Here, $\langle \cdot \rangle$ denotes the average as computed with  $\rho(t)$.)

Whenever the formerly driven system re-equilibrates, $P(\epsilon')$ becomes a delta-function once again (right panel of Figure \ref{Idea.}). We will investigate driving implemented by either one of two possibilities:\newline

(1) Endowing the Hamiltonian with a non-adiabatic transient time dependence leading to a deviation from $H$ only during a short time interval during which the system is driven (Sections (\ref{sec:product}, \ref{sec:dual},\ref{sec:Magnus},\ref{section:short}, and \ref{sec:effective})). In this case, between an initial and a final time,
the Schrodinger picture Hamiltonian differs from $H$, i.e.,  
$H(t_{i}=0 < t' < t_{f}) \neq H$.  \newline

(2) Including a coupling to an external bath yet allow for no explicit time dependence in the fundamental terms forming the Hamiltonian (this approach is invoked in Section \ref{intuition} (in particular, in its second half describing Eq. (\ref{usfe}), Section \ref{2-Ham}), and \ref{LR_explain}, \ref{app:triv1}, and \ref{semi-class-prob})). By comparison to procedure (1) above, this approach is more faithful to the real physical system in which the form of all fundamental interactions is time independent. \newline

\begin{figure*}
	\centering
	\includegraphics[width=.5  \columnwidth, height=.1 \textheight, keepaspectratio]{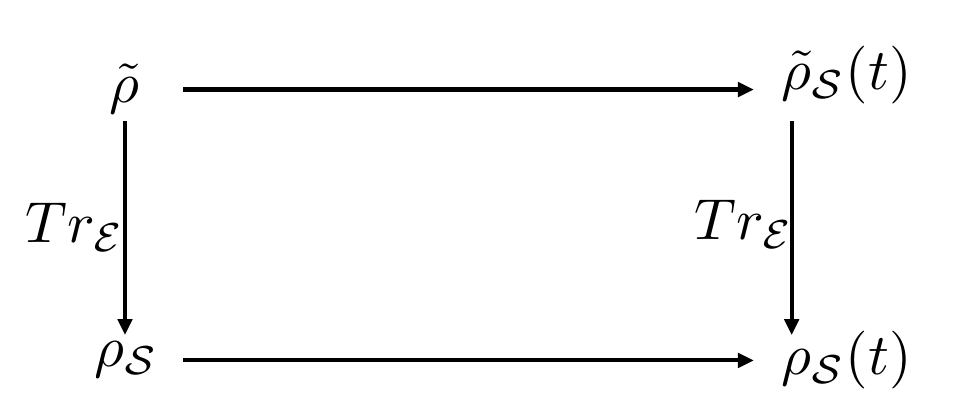}
	\caption{The evolution of the density matrix $\tilde{\rho} \to \tilde{\rho}(t) = \tilde{{\cal{U}}} \tilde{\rho} \tilde{{\cal{U}}}$ describing the system ${\cal{S}}$ and its environment ${\cal{E}}$ (when, together, they form a larger closed hybrid system ${\cal{I}} = {\cal{S}} \cup {\cal{E}}$) is unitary. After tracing over the environment (Eq. (\ref{usfe})), the resultant dynamical mapping $\rho_{\cal{S}}(t) = \Phi_{t}(\rho{\cal{S}}(0))$
	describing the reduced density matrix of the system alone is, generally, not a unitary transformation.
	In many situations of physical interest, the environment may, however, still be effectively captured by a modification of a system Hamiltonian. As we will explain in \ref{LR_explain}, causality constrains this effective system Hamiltonian. In systems with local interactions,
	the rate of energy density of the system cannot be made to change instantaneously from zero to a finite value. A minimal time (linear in the system size) must elapse before the environment (and effective interactions borne by the presence of the environment) can couple to a finite fraction of the system.} 
		\label{dyn_map.}
\end{figure*}

In procedure (1), the density matrix of the system evolves unitarily 
$\rho \rightarrow \rho(t) = {\cal{U}}(t) \rho {\cal{U}}^{\dagger}(t)$. 
In the more realistic approach (2), the evolution of the density matrix of the system $\rho_{S}(t)$ (now a reduced density matrix after a trace over the environment is performed) is described by a general (non-unitary \cite{non-unitary}) dynamic map $\rho_{\cal{S}}(t) = \Phi_{t}(\rho_{\cal{S}}(0))$; a cartoon is provided in Figure \ref{dyn_map.}. \newline

In procedure (2), we will examine the probability distribution $P(\epsilon')$ of Eq. (\ref{PER}) with the replacement of $\rho(t)$ by $\rho_{\cal{S}}(t)$.  \newline

The divide between these unitary and non-unitary evolutions with and without an external environment is a feature that is not always of great pertinence; indeed though many common non-dissipative physical systems are not truly closed they are described to an excellent approximation by the standard unitary evolution of the Schrodinger equation. Complementing the standard distinction between unitary and non-unitary evolutions, there is another issue that we will highlight in the current work. As we will elaborate in \ref{LR_explain}, there are physical constraints on the possible transient time variations of the effective Hamiltonian (that are captured by analysis including the effect of the environment). Notably, in a theory with interactions that are of finite range and strength, due to causality, the allowed changes in the transient time Hamiltonian that captures the effects of the environment cannot be made to instantaneously vary over arbitrarily large distances. That is, the environment cannot couple (nor decouple) to a finite fraction of a macroscopic system instantaneously. Keeping in mind this constraint on the form of the possible variations of the effective Hamiltonian of approach (1), we will often use these two descriptions interchangeably. Our inequalities will bound, from below, (a) the variance of the distribution $P(\epsilon')$ and (b) the magnitude of the pair correlator $G_{ij}$ for sites $i$ and $j$ that are separated by a distance that is of the order of the system size \cite{adiabatic}. 
A similar broadening of the distribution $P(q')$ (and ensuing lower bounds on the associated pair correlators) may arise for general intensive quantities 
$q \equiv \langle Q \rangle/N$ (that include the energy density $\epsilon$ only as a special case).

\section{Outline}
\label{sec:outline}

A large fraction of the current work (Sections \ref{sec:product} - \ref{sec:effective}) establishing the central result of Section \ref{sec:sketch} and related effects will be somewhat mathematical in spirit. The sections towards the end of this paper (Sections \ref{2b2b},\ref{sec:electron}, and \ref{adiabatic-sec}) will touch on possible measurable quantities. In these later sections, our discussion is more speculative.

We now briefly summarize the central contents of the various Sections. In Section \ref{intuition}, we explain why, in spite of its seemingly striking nature, our main finding of large variances (even in systems with local interactions) and the macroscopic range correlations that they imply is quite natural. By {\it macroscopic range}, we refer, in any macroscopic $N \gg 1$ site system, to correlations that span the entire system size. As we explain in Section \ref{intuition} (and in \ref{extensive}, \ref{LR_explain}, \ref{app:triv1}, and  \ref{semi-class-prob}), in various physical settings, finite rates of change of the energy (and other) densities and concomitant long range correlations may appear only at sufficiently long time after coupling the system to an external drive. Next, in Section \ref{sec:product}, we discuss special situations in which our results do not hold- those of product states with an evolution given by separable Hamiltonians. This will prompt us to explore systems that do not have a probability density that is of the simple local product form and to further discuss various aspects of entanglement. Notwithstanding their simplicity and appeal, product states do not generally describe systems above their ground state energy density. Similarly, the finite temperature probability densities of interacting classical systems do not have a product state form. In Section \ref{sec:dual}, we turn to more generic situations such as those appearing in rather natural dual models on lattices in an arbitrary number of spatial dimensions for which a class of finite energy density eigenstates can be exactly constructed. These theories principally include (1) general rotationally symmetric spin models (both quantum and classical) in an external magnetic field and (2) systems of itinerant hard-core bosons with attractive interactions. We investigate the effects of ``cooling/heating'' and ``doping'' protocols on these systems and illustrate that, regardless of the system size, after a finite amount of time, notable energy or carrier density fluctuations will appear. In Section \ref{sec:CSM}, we similarly solve simple models in which the external environment exhibits uniform global fluctuations. Armed with these proof of principle demonstrations, we examine in Section \ref{sec:Magnus} the anatomy of a Dyson type expansion to see how generic these behaviors may be. Straightforward calculations illustrate that although there exist fine tuned situations in which the variance of intensive quantities such as the energy density remain zero (e.g., the product states of Section \ref{sec:product}) in rapidly driven systems, such circumstances may be rare. General non-adiabatic evolutions that change the expectation values of various intensive quantities may, concomitantly, lead to substantial standard deviations. In Section \ref{2-Ham}, we go one step further and establish that under a rather mild set of constraints, macroscopic range connected fluctuations are all but inevitable. (Yet another proof of these long range correlations will be provided in \ref{app:triv1} and \ref{semi-class-prob}). In Section \ref{sec:patuach}, we derive bounds on the fastest fluctuation rates in open thermal system by linking a generalized variant of the quantum standard time-energy uncertainty relations to the heat capacity. In Section \ref{sec:lti}, we illustrate that local quantities in thermal translationally invariant systems are similarly bounded. Our new thermalization bounds suggest that, under typical circumstances, up to factors of order unity, the smallest fluctuation times for thermal systems cannot be shorter than ``Planckian'' times ${\cal{O}}(\hbar/(k_{B} T))$. We next illustrate (Section \ref{deviation}) how general expectation values in these systems relate to equilibrium averages. Our effect has broad experimental implications: common systems undergoing heating/cooling and/or other evolutions of their intensive quantities may exhibit long range correlations. In Section \ref{sec:effective}, we demonstrate that the non-equilibrium system displays an effective equilibrium relative to a time evolved Hamiltonian. The remainder of the paper, largely focusing on candidate experimental and {\it in silico} realizations of our effect, is more speculative than the detailed exact solutions and derivations presented in its earlier Sections. In Sections \ref{2b2b} and \ref{sec:electron}, we turn to two prototypical systems and ask whether our findings may rationalize experimental (and numerical) results. In particular, in Section \ref{2b2b}, we discuss glasses and show a universal collapse of the viscosity data that was inspired by considerations similar to those that we describe in the current work. In Section \ref{sec:electron}, we ask whether the broadened distributions that we find may lead to ``non-Fermi'' liquid type behavior in various electronic systems. In Section \ref{adiabatic-sec}, we discuss adiabatic quantum processes and demonstrate how these may maintain thermal equilibration. We further speculate on possible offshoots of this result that suggest certain similarities between quantum measurements and thermalization. We conclude in Section \ref{sec:conclusions} with a synopsis of our results.

Various details (including an alternate proof of our central result, typical order of magnitude estimates, and further analysis) have been relegated to the appendices. \ref{extensive} provides simple estimates of the minimal time scale $t_{\min}$ that must be exceeded in order to establish finite rate of variation of the energy density (and concomitant long range correlations amongst the local contributions  $\{{\cal{H}}_{i} \}$ to the Hamiltonian). In \ref{LR_explain}, we prove that in typical non-relativistic systems with local interactions (where the Lieb-Robinson bounds apply), a finite rate of change of the energy density (and,  
similarly, that of other intensive quantities) is only possible at sufficiently long times $t>t_{\min}$.  As we briefly noted above, \ref{app:triv1} and \ref{semi-class-prob} will provide a complementary proof of our central result. 
In \ref{app:triv1}, we demonstrate that a finite a rate of variation of the energy density implies long range connected correlations between the environment driving the system and the system itself. \ref{semi-class-prob} then employs ``classical'' probability arguments to illustrate that the latter long range correlation between different sites in the system and its surrounding environment may lead to correlations between the sites in the system bulk even if these sites are far separated. A lightning review of several earlier known long range correlations is provided in \ref{rev:long-range}. In \ref{Ising_example}, we show that using entangled states (similar to those analyzed in Section \ref{sec:dual}) reproduces the finite temperature correlators of an Ising chain. In \ref{ent-ent-ent}, we demonstrate that the entanglement entropy of symmetric entangled states is logarithmic in the system size; this latter calculation will further illustrate that the entangled spin states studied in Section \ref{sec:dual} display such macroscopic entanglement. These examples underscore that, even in closed systems, eigenstates of an energy density larger than that of the ground state can very naturally exhibit a macroscopic entanglement. In \ref{trivial-maxS}, \ref{sxsx}, and \ref{appendix:preparation}, we discuss aspects related to the spin model example of Section \ref{sec:dual} (and, by extension, to some of the models dual to this spin model that are further studied in Section \ref{sec:dual}). \ref{trivial-maxS} details what occurs when adding a general number of $S=1/2$ spins. We connect the result in the limit of a large number of spins to the Gaussian distribution resulting from random walks (in the limit of large spins, the addition of spins naturally relates to the addition of classical vectors). \ref{sxsx} and \ref{appendix:preparation} underscore the correlations in the initial state of this spin model system. \ref{lrra} explicitly introduces these correlations. In \ref{inevit.}, we explain why such correlations are inevitable in various cases. (The discussion in these appendices augment a more general result concerning correlations in the initial state of various driven systems that is described in the text following Eqs. (\ref{trivHeisenb_4_referee}, \ref{trHHH@}) regarding generally more complex correlations.) The central aim of Section \ref{sec:dual} was to provide the reader with a simple solvable spin model and its duals where a finite $\sigma_{\epsilon}$ and associated long range correlations between ${\cal{H}}_{i}$ appear hand in hand with a finite rate of change of the energy density. The exact solvability of the spin model of Section \ref{sec:dual} hints that the correlations that its initial simple correlations exhibit are not necessarily generic. In \ref{appendix:preparation}, we outline a gedanken experiment in which the initial state of Section \ref{sec:dual} may be realized. In \ref{app-boost}, we discuss several situations in which the variance of the energy density remains zero even when the energy density itself changes at a finite rate. Several aspects of the viscosity fit discussed in Section \ref{2b2b} are elaborated on in \ref{explain-deviation}. \ref{prompt} provides intuitive arguments for the appearance of long time Gaussian distributions. Such long time Gaussian distributions were (a) invoked in our derivation of the 16 decade viscosity collapse of supercooled liquids and glasses and also appear (b) in standard textbook systems that have equilibrated at long times at general temperatures $T$ where (with $C_v$ denoting the heat capacity at constant volume), the width of the Gaussian distribution is given by $\sigma_{\epsilon} = \sqrt{k_{B} T^{2} C_{v}}/N \sim {\cal{O}}( N^{-1/2})$. Lastly, in  \ref{Higher_entanglement_entropy_states}, we explain that, generally, the entanglement entropy may be higher than of the states studied in \ref{ent-ent-ent}. 

\section{Intuitive arguments}
\label{intuition} 
To make our more abstract discussions clear, we first try to motivate why our central claim might not, at all, be surprising and expand on the basic premise outlined in Section \ref{sec:sketch}. Consider a system that is, initially, in thermodynamic equilibrium with a sharp energy density $\epsilon$. For an initial closed equilibrium system (described by the microcanonical ensemble), the standard deviation of $\epsilon$ scales as $1/N$ while in open systems connected to a heat bath, the standard deviation of $\epsilon$ is ${\cal{O}}(1/\sqrt{N})$. In either of these two cases, the standard deviation of $\epsilon$ vanishes in the thermodynamic limit (similar results apply to any intensive thermodynamic variable), see, e.g., the right-hand panel of Figure \ref{Idea.}.  Now imagine cooling the system. As the system is cooled, its energy density $\epsilon$ drops. Various arguments hint that as $\epsilon$ drifts (or is ``translated'') downwards in value, its associated standard deviation also increases (see the central panel of Figure \ref{Idea.}). This is analogous to the increase in width of an initially localized ``wave packet'' with a non-trivial evolution (with the energy density itself playing the role of the packet location). This argument applies to {\it both quantum and classical systems} (with the classical probability distribution obeying a Liouville or Fokker-Planck type equations instead of the von Neumann equation obeyed by the quantum probability density matrices). Thus, on a rudimentary level, it might be hardly surprising that the energy density obtains a finite standard deviation when it continuously varies in time. A finite standard deviation of the energy density implies long range correlations of the local energy terms. This is so since the variance of the energy density 
\begin{eqnarray}
\label{central}
0 < \sigma_{\epsilon}^2 =& \frac{1}{N^{2}} \sum_{i,j} (\langle {\cal{H}}_{i} {\cal{H}}_{j} \rangle - 
\langle {\cal{H}}_{i} \rangle \langle {\cal{H}}_{j} \rangle) 
= \frac{1}{N^{2}} \sum_{i,j} G_{ij} \equiv \overline{G}.
\end{eqnarray} 
Thus, if $\sigma_{\epsilon}$ is finite then the average $\overline{G}$ of $G_{ij}$ over all separations $|i-j|$ will be non-vanishing. More broadly, similar considerations apply to intensive quantities of the form $q=\frac{1}{N} \sum_{i} q_{i}$ that must have a sharp value in thermodynamic equilibrium.  
Thus, generally, if $q$ broadens as some parameters are varied, there must be finite
connected correlations $(\langle q_{i} q_{j} \rangle - \langle q_{i} \rangle \langle q_{j} \rangle)$ even when $|i-j'|$ is the order of the linear dimension of a macroscopic system. Identical conclusions to the ones presented above may be drawn for systems that end in thermodynamic equilibrium (instead of starting from equilibrium) while experiencing a finite rate of change of their energy density at earlier times at which Eq. (\ref{central}) will hold.
This effect may appear for quantum as well as classical systems. Generally, there are ``classical'' and ``quantum'' contributions \cite{class-quant+} to the variance $\sigma_{\epsilon}^2$. 

Empirically, in cases of experimental relevance, as in, e.g., cooling or heating a material, if the rate of change of its temperature (or energy density) is finite then Eq. (\ref{central}) will hold. Although heat (and other) currents associated with various intensive quantities $q$ traverse material surfaces, experimentally, even for thermodynamically large systems, the rate of change of energy density $\epsilon$, and other intensive quantities $q$
can be readily made finite, i.e., $dq/dt = {\cal{O}}(1)$. This common experimentally relevant situation of finite heat or other rate of change $dq/dt$ in macroscopic finite size ($N \gg 1$) samples is the focus of our attention (see \ref{extensive}). We nonetheless remark that if the energy density (or other intensive parameter) exchange rate are dominated by contributions in Eq. (\ref{central}) with $i$ and $j$ close to the surface then $dq/dt = {\cal{O}}(1/L)$ and the average connected correlator associated with $q$ for arbitrarily far separated sites $i$ and $j$ will be bounded by $\overline{G_{q}} \ge {\cal{O}}(N^{-2/d})$ \cite{explain-similar}.  
As we will further emphasize in Section \ref{2-Ham} (as will formally follow therein, e.g., from the Heisenberg picture Eq. (\ref{HHtH}) or its Ehrenfest theorem analog in the Schrodinger picture), in order to achieve a finite rate of change of any intensive quantity (including that of the energy density $d\epsilon/dt$ (or, equivalently, of the measured temperature $dT/dt$)), {\it the coupling (and correlations) between the system and its surroundings must be extensive} and involve minimal time scales (see \ref{extensive}, \ref{LR_explain}, and \ref{app:triv1}). In reality, due to the surface flow of the heat current from the surrounding environment to the system during periods of heating or cooling, the local energy density in the system is generally spatially non--uniform and may depend on the distance to the surrounding external bath from which heat flows to the system. 

\begin{figure*}
	\centering
	\includegraphics[width=1.2 \columnwidth, height=.21355 \textheight, keepaspectratio]{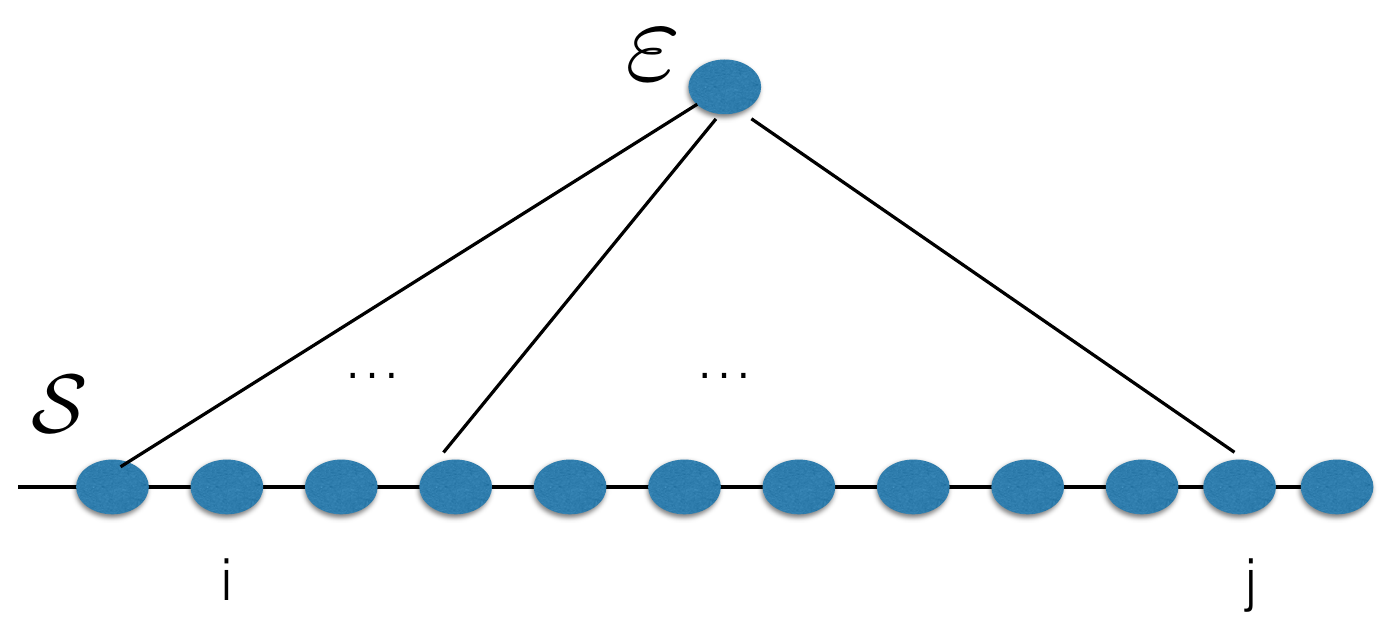}
	\caption{An intuitive representation of the effect. In order to drive the system ${\cal{S}}$ and vary its energy density at a finite rate, the environment ($\cal{E}$) must couple to a finite fraction of the number of sites in ${\cal{S}}$ (e.g., sites $i$ and $j$). The energy fluctuations at both $i$ and $j$
	are correlated with ${\cal{S}}$. This, consequently, allows for non-trivial correlations between the local energy fluctuations (those of ${\cal{H}}_{i}$ and ${\cal{H}}_{j}$ of Eq. (\ref{HsH'}))
	even when $i$ and $j$ are far apart. By virtue of both coupling to the ${\cal{E}}$, colloquially, any such pair of sites $i$ and $j$ are a graph distance of two (links) away  (``two degrees of separation'' apart) even though they may be very far away spatially. The above graph constitutes a simple example of a ``small worlds network'' where each node (site) is linked to all others by a small number of steps (in this case, two) \cite{2unitary}.  As we will explain in \ref{LR_explain}, in non-relativistic systems with local interactions, causality in the form of the Lieb-Robinson bounds \cite{Lieb_Robinson} mandates that a minimal time must elapse before an external drive may couple to sites in the bulk of the system ${\cal S}$. Physical estimates on lower bounds on minimal times are further briefly discussed in \ref{extensive}. The finite rate of the energy density implies that, on average, the correlation between the driving Hamiltonian (coupling the system to its environment) and each of the system degrees of freedom is of order unity and of uniform sign (\ref{app:triv1}). In Section \ref{sec:CSM}, we will examine simple models with an associated uniform coupling of the system degrees of freedom to a common environment.}
	\label{coupling.}
\end{figure*}

The physical origin of the long range correlations of Eq. (\ref{central}) in general systems (either quantum or classical) is symbolically depicted in Figure \ref{coupling.}. As noted above, in order to achieve a finite rate of cooling/heating in a system with bounded interaction strengths, a finite fraction of the fields/sites in the system must couple to the surrounding heat bath (see also \ref{app:triv1} for a simple brief demonstration of macroscopic length correlations between the surrounding environment and the system bulk in systems with time dependent $\tilde{H}$). If such a single bath/external drive couples to a finite fraction of all sites/fields in the system ${\cal{S}}$ so as to lower the average energy density then even fields that are spatially far apart become correlated by virtue of their non-local interaction with the common environment ${\cal{E}}$ (their shared bath or external drive). The full Hamiltonian $\tilde{H}$ describing the system ${\cal{S}}$ and its environment ${\cal{E}}$ (including the coupling between ${\cal{S}}$ and ${\cal{E}}$) provides the full time evolution ${\cal{\tilde{U}}}(t)$ for the initial density matrix $\tilde{\rho}$ on ${\cal{I}} = {\cal{S}} \cup {\cal{E}}$. We may trace or ``integrate'' over the bath/drive degrees of freedom in ${\cal{E}}$ (accounting for the driving (as well as dissipation) due to coupling to the environment) to arrive at the Schrodinger picture reduced density matrix $\rho_{\cal{S}}$ depending only on the degrees of freedom in ${\cal{S}}$. Thus, we consider
\begin{eqnarray}
\label{usfe}
&&\rho_{\cal{S}}(t) \equiv Tr_{\cal{E}} ({\cal{\tilde{U}}} (t)  \tilde{\rho} {\cal{\tilde{U}}}^{\dagger}(t)) , \nonumber
\\  &&{\cal{\tilde{U}}} (t) =  {\cal T}  \exp (-\frac{i}{\hbar} \int_{0}^{t} \tilde{H}(t') dt') 
\\   &&~~~~~\equiv   {\cal T} \exp (-\frac{i}{\hbar} \int_{0}^{t} (H_{\cal{S}} (t') + H_{\cal E}(t') + H_{{\cal S}-{\cal E}}(t'))  dt')  \nonumber. 
\end{eqnarray}
Here, ${\cal T}$ denotes time ordering, and three Hamiltonians (i) $H_{\cal S}$, (ii) $H_{\cal E}$, and (iii) $H_{{\cal S}-{\cal E}}$ describe, respectively, (i) the Hamiltonian involving only the degrees of freedom in ${\cal S}$, (ii) the Hamiltonian involving degrees of freedom in ${\cal E}$ alone, and (iii) the interaction between the system and its environment. $H_{{\cal S}-{\cal E}}$ may capture the coupling between different, far separated, fields (say at sites $i$ and $j$) in the system ${\cal S}$ to the same external drive/environment ${\cal E}$. Indeed, associated with the solvable systems of Section \ref{sec:dual}, initial long range correlations may be created by coupling all sites in the system to the same environment (a magnet of macrosopic magnetization (see \ref{appendix:preparation})).

The trace in the first line of Eq. (\ref{usfe}) over the external drive degrees of freedom ${\cal{E}}$ may generate a correlation between these two fields at $i$ and $j$ irrespective of their spatial separation (see Figure \ref{coupling.}). This correlation in $\rho_{\cal{S}}(t)$ between spatially distant fields may arise, rather universally, if in $H_{{\cal S}-{\cal E}}$ the latter two fields couple to the very same external drive or environment ${\cal{E}}$. For a uniform external drive, the coupling between all fields in ${\cal{S}}$ to those in ${\cal{E}}$ is of typical comparable strength. Thus, the resulting correlation in $\rho_{\cal{S}}(t)$ may be {\it non-local} even at short times $t$ (so long as at that these (or earlier) times, a finite fraction of the fields in ${\cal{S}}$ couple to the external drive/bath ${\cal{E}}$). A semi-classical motivation for this effect is sketched in \ref{semi-class-prob}. As alluded to in procedure (ii) of Section \ref{sec:sketch}, in real physical systems the form of the microscopic interactions is time independent
(corresponding to a time independent $\tilde{H}$ in Eq. (\ref{usfe}). 
 
 In relativistic theories, a strict minimal cutoff time $t_{\min}$ for a finite fraction of the fields in ${\cal{S}}$ to become coupled to an external drive/bath ${\cal{E}}$ is set by $t_{\min} = \ell_{\min}/c$. Here,  $\ell_{\min}$ is the minimal linear distance between the ``center of mass'' of ${\cal{S}}$ and the nearest point in ${\cal{E}}$ and $c$ is the speed of light for bona fide radiative coupling that changes the energy density $\epsilon$ (or temperature) of the system. Thus, since $\ell_{\min} = {\cal{O}}(L)$ for, e.g., radiative coupling to the environment, this minimal time $t_{\min} = {\cal{O}}(L/c)$ (as further discussed in \ref{extensive} while paying attention to absorption lengths). For generic spin models and other non-relativistic local theories, a similar bound on $t_{\min}$ on the time required for the environment to couple with a typical uniform strength or become entangled with a finite fraction of the sites in ${\cal{S}}$ is set by the effective (Lieb-Robinson (LR)) speed $v_{\sf LR}$ \cite{sergey,jens2,ali1,ali2,Lieb_Robinson} ($t_{\min} = t_{\sf LR}= {\cal{O}}(L/v_{\sf LR})$). In all cases (relativistic or non-relativistic) $t_{\min} = {\cal{O}}(L/v)$ with $v$ a finite relevant speed. Thus, no long-range correlations violating causality (either relativistic or non-relativistic Lieb-Robinson type) appear. Rather, our results concerning long-range correlations pertain to times $t>t_{\min}$. At such times, the relativistic or Lieb-Robinson light-cones (respectively given by $(ct)$ or ($v_{\sf LR} t$)) already {\it span most of the system ${\cal{S}}$}. Indeed, as seen from Eq. (\ref{usfe}), long range correlations may be generated from the coupling of the environment ${\cal{E}}$ to the bulk of ${\cal{S}}$. At sufficiently short times, no such coupling exists and, in tandem, the total energy of the system cannot change at a rate proportional to its volume (i.e., at these short times, the rate of change of the energy density vanishes, $d \epsilon/dt =0$). A system that starts off with only local $G_{ij}$ will require a time $t>t_{\min}$ to develop long range correlations \cite{sergey,jens2} consistent with our new results concerning (i) a required minimal time scale for changing the energy density of the system at a finite rate (\ref{LR_explain}) and (ii) the appearance of nontrivial correlations once the energy density varies (the central result of this paper). The above also applies to general intensive quantities $q$ different from the energy density. In Section \ref{2-Ham}, we will sharpen other considerations related to ${\cal{\tilde{U}}}(t)$ to arrive at exact inequalities. A brief summary of earlier known long range correlations is provided in \ref{rev:long-range}. In what follows, we first turn to product states where no broad distributions of intensive quantities arise. For product states undergoing an evolution with a locally separable Hamiltonian, the system degrees of freedom cannot couple to a common environment in Eq. (\ref{usfe}). In the sections thereafter, we will demonstrate that in general quantum systems (not constrained to a product state structure), broadening may be quite prevalent. Prevalent non-factorizable states generally allow for a coupling to a common environment.

\section{Product states and bounded separable Hamiltonians}
\label{sec:product}
Prior to demonstrating that energy density broadening may naturally accompany a cooling or heating of the system, we first 
discuss (within the framework of procedure (1) of 
Section \ref{sec:sketch} for which the detailed considerations of Eq. (\ref{usfe}) (Figure \ref{coupling.}) 
do not apply) states associated with individually decoupled local subsystems. Our focus is on systems with separable bounded local interactions.  
For a density matrix $\rho(t)$ that, at a time $t$, is a direct tensor product of local density matrices $\{ \rho_{l}(t) \}_{l=1}^{M}$ acting on disjoint spaces,
with $M= {\cal{O}}(N)$, 
\begin{eqnarray}
\label{rhofac}
\rho(t) = \rho_{1}(t) \otimes \rho_{2}(t)  \otimes \cdots \otimes \rho_{M}(t),
\end{eqnarray}
the standard deviation $\sigma_{H}$ of the Hamiltonian of Eq. (\ref{HsH'}) at this time will,
in accord with the central limit theorem, generally be ${\cal{O}}(\sqrt{N})$ even when the rate of change of the energy $dE/dt$ may be extensive (i.e., $\propto N$). This result applies to quantum and classical systems.
In classical theories, $\{\rho_{l}\}$ portray the probability distributions of independent decoupled local degrees of freedom. In the quantum arena, Eq. (\ref{rhofac}) also describes states in which no entanglement exists. 

As a case in point, we may consider the initial (spin $S=1/2$) state $| \psi_{Ising}^{0} \rangle = | s^{0}_{1} s^{0}_{2} \cdots s^{0}_{N} \rangle$ to be a low energy eigenstate of an Ising model $H_{I}=- \sum_{\langle ij \rangle} J_{ij} S_{i}^{z} S_{j}^{z}$ that is acted on during intermediate times by a transverse magnetic field Hamiltonian ($H_{tr}=-B_{y}(t) \sum_{i} S_{i}^{y}$) that causes a precession around the $S^{y}$ axis and thus alters the energy as measured by
$H_I$ (thereby heating or cooling the system). Here, $s_{i} = \pm 1$ denote the scaled eigenvalues of the local spin operators $S_{i}^{z}$. The transverse field Hamiltonian $H_{tr}$ may be explicitly written a sum of decoupled terms each of which acts on a separate local subspace, $H_{tr} \equiv \sum_{i=1}^{M} {\cal{H}}_{i}$, with $M=N$. The initial state $| \psi_{Ising}^{0} \rangle$ (and its associated density matrix) can be written as an outer product of $M=N$ single spin states (density matrices) defined on the same $M$ decoupled separate spaces. Thus an evolution, from an initial product state, with $H_{tr}$ will trivially lead to a final state which still is of the product state form. All product states $| \psi \rangle = |s_{1} s_{2} \cdots s_{N} \rangle$ are eigenstates of $H_{I}$.  A uniform rotation, between an initial time ($t=0$) and a final time $t_{f}$,
of all of the $N$ spins around the $y$ spin axis by the transverse field Hamiltonian $H_{tr}$ by an angle of $\pi/2$ will transform $| \psi_{Ising}^{0} \rangle$ to a final state $| \chi\rangle$ that is an equal modulus superposition of all Ising product states (all eigenstates of $H_{I}$),
viz., 
\begin{equation}
|\chi \rangle =  2^{-N/2} \sum_{s_{1} s_{2} \cdots s_{N}} (-1)^{\sum_{i=1}^{N} (\delta_{s_{i}^{0},-1} \delta_{s_{i},-1})}  | s_{1} s_{2} \cdots s_{N} \rangle \nonumber,
\end{equation}
with $\delta_{\sigma_i,\sigma_j}$ the Kronecker delta. We next discuss what occurs when the exchange constants $J_{ij}$ are of finite range but are otherwise arbitrary. The standard deviation of the energy (i.e., the standard deviation of $H_{I}$) associated with this final rotated state (and any other state during the evolution) of the initial 
Ising product state scales as ${\cal{O}}(\sqrt{N})$
while the energy change can be extensive \cite{exam1}. 
The state $|\chi \rangle$ corresponds to the infinite temperature limit
of the classical Ising model of $H_{I}$ (its energy density is equal to that of the system at infinite temperature and similarly all correlation functions vanish). A key point is that generic finite temperature states are {\it not} of the type of Eq. (\ref{rhofac}). In fact, general thermal states (i.e., eigenstates of either local or nonlocal Hamiltonians that are elevated by a finite energy density difference relative to the ground state) typically display volume law entanglement entropy \cite{volume1,volume2,volume3,volume4} in agreement with the Eigenstate Thermalization Hypothesis \cite{eth1,eth2,eth3,eth4,rigol,pol,polkovnikov1,polkovnikov2,von_neumann} while ground states
and many body localized states of arbitrarily high energy \cite{MBL1,MBL2,MBL3,MBL4,MBL5,MBL6,MBL7,noMBL,yesMBL} may exhibit area law entropies \cite{area}. The entanglement entropy of individual quantum ``thermalized" states imitates the conventional thermodynamic entropy of the macroscopic system that they describe \cite{plain}. 
In order to further elucidate these notions, in \ref{Ising_example}, we illustrate that correlations in finite energy density eigenstates of the Ising chain mirror those in equilibrated Ising chains at positive temperatures. In the one dimensional Ising model and other equilibrium systems at temperatures $T>0$, the high degree of entanglement and mixing between individual product states leads to contributions to the two point correlation functions that alternate in sign and ultimately lead to the usual decay of correlations with distance. Our central thesis is that an external driving Hamiltonian (such as that present in cooling/heating of a system) may lead to large extensive fluctuations. While the appearance of such extensive fluctuations may seem natural for non-local operators (such as (Heisenberg picture) time evolved local Hamiltonian terms in various examples), these generic fluctuations may also appear for local quantities (e.g., the local operators $\{{\cal{H}}_{i}\}$ in Eq. (\ref{central})). In Section \ref{sec:dual}, we will study systems for which the relevant $\{{\cal{H}}_{i}\}$ are, indeed, local. 

When all of the eigenvectors of the density matrix are trivial local product states that do not exhibit entanglement, the system described by $\rho$ is a classical system (with different classical realizations having disparate probabilities). In the next sections, we will demonstrate that large fluctuations of any observable may naturally arise for all system sizes (including systems in their thermodynamic limit). The calculations in the studied examples will be for single quantum mechanical states. Any density matrix (also that capturing a system having a mixed state in any region ${\cal S}$) may be expressed as $\rho = |\psi \rangle \langle \psi |$ with a pure state $| \psi \rangle$ that extends over a volume ${\cal{I}}' \supset {\cal{S}}$ \cite{lieb-pure,pure}.

As suggested in Section \ref{intuition}, our effect may be realized in both quantum and classical systems. 
Our analysis will allow for entangled states. These states describe general situations in which the evolution operator or the environment ${\cal{E}}$ in Eq. (\ref{usfe}) are non-factorizable and long-range coupling/correlations between the sites in ${\cal{S}}$ may result. 

\section{Dual examples with constant external driving fields}
\label{sec:dual}
The existence of finite connected correlations $|G_{ij}|$ (Eq. (\ref{central})) for far separated sites $|i-j| \to \infty$ is at odds with common lore. Before turning to more formal general aspects, we illustrate how this occurs in two classes of archetypical systems-
(i) any globally $SU(2)$ symmetric (arbitrary graph or lattice) spin $S=1/2$ model in an external magnetic field (discussed next in Section \ref{sec:gspin}) and (ii) dual hard core Bose systems on the same graphs or latices (Section \ref{gbose}).
In these models, the external fields/terms (magnetic fields in (i) or doping in (ii)) are constant. Although (i) and (ii) constitute two well known (and very general) intractable many-body theories, as we will demonstrate, the analysis of the fluctuations becomes identical to that associated with an integrable one body problem. In the context of example (i), this effective single body problem will be associated with the total system spin $\vec{S}_{tot}$. This simplification will enable us to arrive at exact results. 
Similar to Section \ref{sec:product}, the analysis below is within the framework of procedure (1) of 
Section \ref{sec:sketch}- that of an explicitly time varying Hamiltonian in a closed system with no environment. 
The initial system states that we will consider are eigenstates of the system Hamiltonian. Thus, these states match like the equilibrium states have a vanishing variance of the energy density $\sigma_{\epsilon} =0$. When the Eigenstate Thermalization Hypothesis \cite{eth1,eth2,eth3,eth4,rigol,pol,polkovnikov1,polkovnikov2,von_neumann} holds, an eigenstate may represent an equilibrium state. Repeating the calculations in this Section, one may verify that superposing eigenstates of nearly equal energy will not alter our finding of a finite $\sigma_{\epsilon}$ after the system couples to an external field such that it energy density varies at a finite rate $|d \epsilon/dt|$. These initial states will, nonetheless, display nontrivial correlations that are elaborated on in significant depth in the Appendices. In Section \ref{sec:CSM}, we will analyze other models with initial states that do not exhibit any nontrivial correlations. 

\subsection{Rotationally invariant spin models on all graphs (including lattices in general dimensions)}
\label{sec:gspin}

In what follows, we consider a general rotationally symmetric spin model ($H_{symm}$) of local spin-$S$ moments augmented by a uniform magnetic field.
\begin{eqnarray}
\label{hferro}
H_{spin} = H_{symm}- B_{z} \sum_{i} S^{z}_{i}.
\end{eqnarray}
Amongst many other possibilities, the general rotationally symmetric Hamiltonian $H_{symm}$ may be a typical spin interaction of the type 
\begin{eqnarray}
\label{HEIS*}
H_{Heisenberg} =  -\sum_{ij} J_{ij} \vec{S}_{i} \cdot \vec{S}_{j}  - \sum_{ijkl} W_{ijkl} (\vec{S}_{i} \cdot \vec{S}_{j})(\vec{S}_{k} \cdot \vec{S}_{l}) + \cdots,
\end{eqnarray}
with arbitrary Heisenberg spin exchange couplings $\{J_{ij}\}$ augmented by conventional higher order rotationally symmetric terms. We reiterate that the model of Eq. (\ref{hferro}) is defined on any graph (including lattices in any number of spatial dimensions). 

\subsubsection{Quantum Spin System}
\label{qss}

In the upcoming analysis, we will label the eigenstates of $H_{spin}$  (and their energies) by 
$\{|\phi_\alpha \rangle\}$ (having, respectively, energies $\{E_{\alpha}\}$). We will employ the total spin operator $\vec{S}_{tot} = \sum_{i=1}^{N} \vec{S}_{i}$. Since $[\vec{S}_{tot},H_{spin}]=0$, all eigenstates of $H_{spin}$ may be simultaneously diagonalized with $S_{tot}^{z}$ (with eigenvalue $m \hbar$) and $\vec{S}^{2}_{tot}$ (with eigenvalue $S_{tot}(S_{tot} +1) \hbar^{2}$). Thus, any eigenstate of Eq. (\ref{hferro}) may be written as $|\phi_\alpha \rangle = |\upsilon_{\alpha}; S_{tot} ,S_{tot}^{z} \rangle$ with 
$\upsilon_{\alpha}$ denoting all additional quantum numbers labeling the eigenstates of $H_{spin}$ in a given sector of $S_{tot}$ and $S_{tot}^{z}$ \cite{ssremark}. Although our results apply for local spins of any size $S$, in order to elucidate certain aspects, we will often allude to spin $S=1/2$ systems. For any eigenstate having a general $S_{tot}^{z} \neq \pm S_{\max} = \pm NS$, the associated density matrix is not of the local tensor product form of Eq. (\ref{rhofac}). Rather, any such eigenstate is a particular superposition of spin $S=1/2$ product states having a total fixed value of $S_{tot}^{z}$. The state of maximal total spin $S_{tot} = S_{\max}$ (which can be trivially shown to be a non-degenerate eigenstate for any value of $S_{tot}^{z}$, see \ref{trivial-maxS}) corresponds to a symmetric equal amplitude superposition of all such product states of a given $S_{tot}^{z}$
(i.e, such a sum of all product states of the type $| \uparrow_{1} \uparrow_{2} \downarrow_{3} \uparrow_{4} \downarrow_{5} \uparrow_{6}  \cdots \uparrow_{N-1} \downarrow_{N} \rangle$ in which there are a total of $(N/2 \pm S_{tot}^{z}/\hbar)$ single spin of up/down polarizations along the $z$ axis). We set an arbitrary eigenstate $| \phi_{\alpha} \rangle$ to be the initial state (at time $t=0$) of the system $| \psi^{0}_{Spin} \rangle$. The energy density (and the global energy itself) will have a vanishing standard deviation in any such initially chosen eigenstate, $\sigma_{\epsilon} =0$. 
We next evolve this initial ($t=0$) state via a ``cooling/heating process'' wherein the energy (as measured by $H_{spin}$) 
is varied by replacing, during the period of time in which the system is cooled or heated, the Hamiltonian of Eq. (\ref{hferro}) by a time dependent transverse field Hamiltonian (see Section \ref{cause}
for restrictions imposed by causality) 
\begin{eqnarray}
\label{htr}
H_{tr}(t') = -B_{y}(t') \sum_{i} S^{y}_{i}.
\end{eqnarray}
At $t=0$, the system Hamiltonian varies instantaneously (a particular realization of procedure (1) of Section \ref{sec:sketch}) from $H_{spin}$ to $H_{tr}$.
Once the ``cooling/heating process'' terminates at a final time ($t=t_{f}$), the system Hamiltonian becomes, once again, the original Hamiltonian of Eq. (\ref{hferro}). Once again, in this case, the change of the Hamiltonian at the final time $t_{f}$ is instantaneous. In accord with the discussion in Section \ref{intuition}, in Eq. (\ref{htr}), a finite fraction (in this case all) of the system degrees of freedom (i.e., the spins) couple to the external drive/bath (the external transverse field).  
Such a global coupling is necessary to achieve a finite $d \epsilon/dt$. During the evolution with $H_{tr}$, the spins globally precess about the $y$ axis. Thus, after a time $t$, the energy per lattice site is changed (relative to its initial value $\epsilon_{0}$) by an amount $\epsilon(t_{f})- \epsilon_{0} = B_{z} \frac{S^{z}_{tot}}{N} (1-\cos \theta(t_{f}))$. Here, $\theta(t) \equiv \int_{0}^{t}~ B_{y}(t') ~dt'$. In the terminology of \cite{chris',chris,campisi,Talkner,natana}, this energy density shift represents the work done per site. When $B_{z}S^{z}_{tot}>0$, the energy density of the system is generally increased relative to its initial value while for negative $B_{z}S^{z}_{tot}$, the system is ``cooled'' relative to its initial energy density. For all $S^{z}_{tot}$, the energy density $\epsilon(t)$ exhibits consecutive cooling and heating periods. Employing the shorthand $w \equiv S_{tot}^{z}/(\hbar S_{tot})$, the standard deviation of $\frac{H_{spin}}{N}$ is 
\begin{eqnarray}
\label{se}
\sigma_{\epsilon}(t_{f}) =  \frac{S_{tot} \hbar| B_{z} \sin \theta(t_{f})|}{N\sqrt{2}} 
\sqrt{1+ \frac{1}{S_{tot}} - w^2}.
\end{eqnarray}
We briefly elaborate on the physically transparent derivation of Eq. (\ref{se}). The applied transverse field generates a global Larmor precession of the spins about the 
$y-$axis. While the first term of Eq. (\ref{hferro}) is manifestly invariant under rotations, the second term (that of $( - B_{z} \sum_{i} S^{z}_{i}$)) will change. In the Heisenberg picture
after the evolution with the transverse field, each local $(B_{z} S^{z}_{i})$ transforms into $ B_{z}(  S^{z}_{i}  \cos (\int_{0}^{t_{f}}~ B_{y}(t') ~dt')+ S^{x}_{i} \sin (\int_{0}^{t_{f}}~ B_{y}(t') ~dt'))$. Since in any eigenstate of $S^{z}_{tot}$ (including $| \psi^{0}_{Spin} \rangle$), the expectation value $\langle S^{x}_{tot} S^{z}_{tot} \rangle = \langle S^{x}_{tot} \rangle \langle S^{z}_{tot} \rangle (=0)$, the only non-vanishing contributions to the variance of the Hamiltonian 
of Eq. (\ref{hferro}) will originate from the expectation value of the square of the second term of $H_{spin}$ and thus (up to a trivial prefactor of ($B_{z}^{2} 
\sin^{2} (\int_{0}^{t_{f}} dt' B_{y}(t'))$)) from
\begin{eqnarray}
\label{s2}
 && \sigma^{2}_{S ^{x}_{tot}} = \langle (S^{x}_{tot})^{2} \rangle= \frac{1}{2}  \langle (S^{x}_{tot})^{2} + (S^{y}_{tot})^{2} \rangle  
 =\frac{1}{2} \langle (\vec{S}_{tot})^{2} - (S^{z}_{tot})^{2} \rangle \nonumber
\\ && = \frac{1}{2} \Big( \hbar^{2} S_{tot}(S_{tot} + 1) - (S^{z}_{tot})^{2}  \Big).
\end{eqnarray}
Substituting $w \equiv S^{z}_{tot}/(\hbar S_{tot})$ (and rescaling by a factor of $N^{2}$ to
determine the variance of the energy density) leads to the square of Eq. (\ref{se}). A standard deviation comparable to that of Eq. (\ref{se}) appears not only for a single eigenstate of $H_{spin}$ but also for any other initial states having an uncertainty in the total energy that is not extensive. When $w = 1$ (or $-1$) with the total spin being maximal, $S_{tot}=S_{\max}$, the initial state $| \psi^{0}_{Spin} \rangle$ is a product state of all spins being maximally up (or all spins pointing maximally down). Even in the state of maximal spin $S_{tot} = S_{\max}$, so long as $|w|<1$, the standard deviation will generally be $\sigma_{\epsilon} = {\cal{O}}(1)$. 
Furthermore, although they are statistically preferable values for $S_{tot}$ when adding angular momenta in the large $N$ limit (e.g., \ref{trivial-maxS}), regardless of the form of $H_{symm}$ (for instance, irrespective of the specific couplings in Eq. (\ref{HEIS*})), in this $N \gg 1$ limit, states of vanishingly small $\frac{S_{tot}}{N}$ will not allow for a for a finite change of the energy density, $\Delta \epsilon= B_{z} \frac{S^{z}_{tot}}{N} (1-\cos (\int_{0}^{t_{f}}~ B_{y}(t') ~dt'))$, via the application of the transverse field (as embodied by the Hamiltonian $H_{tr}$). Indeed, the central point that we wish to emphasize and is evident in our example of Eq. (\ref {hferro}) is that, generally, when the energy density $\Delta \epsilon$ does change at a non-vanishing rate, a finite $\sigma_{\epsilon}>0$ is all but inevitable.

 Away from the singular $S_{tot}^{z} = \pm \hbar S_{\max}$ limit, spatial long range entanglement develops. When $(1-|w|) = {\cal{O}}(1)$, the scaled standard deviation of the energy density is, for general times, $(\frac{1}{\hbar B_{z}})  \sigma_{\epsilon}= {\cal{O}}(1)$ and, as we will elucidate in \ref{maxtotspin}, a macroscopic (logarithmic in system size) entanglement entropy appears. A comparable standard deviation $\sigma_{\epsilon}$
appears not only for the eigenstate but also for states initial having an energy uncertainty of order ${\cal{O}}(1)$ (in units of $B_{z} \hbar$) 
(e.g., $ c_{1} |S_{tot}, S_{tot}^{z} \rangle + 
c_{2} |S_{tot}, S_{tot}^{z}-\hbar \rangle$ with $c_{1,2} = {\cal{O}}(1)$).
In the following, we briefly remark on the simplest case of a constant (time indeodent) $B_{y}$.
Here, the time required to first achieve $ \frac{1}{B_{z} \hbar} \sigma_{\epsilon} = {\cal{O}}(1)$ starting from an eigenstate of $H_{spin}$ is ${\cal{O}}(1/B_{y})$.
This requisite waiting time is independent of the system size (as it must
be in this model where a finite $\sigma_{\epsilon}$ is brought about by the sum of local decoupled transverse magnetic field terms in $H_{tr}$). 
The large standard deviation implies (Eq. (\ref{central})) that long range connected correlations of $S_{i}^{z}$ emerge once the state is rotated 
under the evolution with $H_{tr}$. This large standard deviation of $\frac{1}{N} \sum_{i=1}^{N} S_{i}^{z}$ appears in the rotated state displaying (at all sites $i$) 
a uniform value of $\langle S_{i}^{z} \rangle$. Even though there are no connected correlations of the energy densities themselves
in the initial state, the non-local entanglement enables long range correlations of the local energy densities once the system is evolved with a transverse field. 
The variance $\sigma_{\epsilon}$ should not, of course, be confused with the spread of energy densities that the system assumes as it evolves (e.g., for the $S_{tot}^{z}=0$ state, $\sigma_{\epsilon} = {\cal{O}}(1)$ while 
the energy density $\epsilon(t)$ does not vary with time). We nonetheless remark that the standard deviation $\sigma_{\epsilon}$ vanishes at the discrete times $t_{k}= k \pi/B_{y}$ (with $k$ an integer)- the very same times where the rate of change of the energy density $\epsilon(t)$ is zero. 


We now turn to the higher order moments of the fluctuations of the $t>0$ states evolved with Eq. (\ref{htr}),
$\langle (\Delta \epsilon)^{{\sf p}} \rangle \equiv \frac{1}{N^{\sf p}} \langle (H^{H}_{spin} - \langle H^{H}_{spin} \rangle)^{\sf p} \rangle$ with ${\sf p} >2$. 
(The standard deviation of Eq. (\ref{se}) corresponds to ${\sf p}=2$.)
Here, $H^{H}_{spin}(t) = ({\cal{T}} e^{ \frac{i}{\hbar} \int_{0}^{t} iH_{tr}(t') dt'})  H_{spin} {\cal{T}} (e^{-\frac{i}{\hbar} \int_{0}^{t} H_{tr}(t') dt'})$ is the Heisenberg picture Hamiltonian and the expectation value is taken in the initial state 
$| \psi^{0}_{Spin} \rangle$. If $N \gg 1$ and 
$1>|w|$ then $S^{tot}_{\pm} | S_{tot}, S_{tot}^{z} \rangle =  \hbar \sqrt{S_{tot}(S_{tot}+1) - m (m \pm 1)} | S_{tot}, m \pm 1 \rangle \sim S_{tot} \hbar \sqrt{1- w^{2}} |S_{tot}, m \pm 1 \rangle$ where $S_{tot}^{z} = m \hbar$. Trivially, for all $m$ and $m'$, the matrix element of $ \delta S_{tot}^{z} \equiv S_{tot}^{z} - \langle S_{tot}^{z} \rangle$ between any two eigenstates, $\langle S_{tot}, m|  \delta S_{tot}^{z} |S_{tot}, m'  \rangle =0$. Thus, the only non-vanishing contributions to
$\langle (\Delta \epsilon)^{p} \rangle$ stem from $\langle (S_{tot}^{x})^{p} \rangle$. This expectation value may be finite only for even ${\sf p}$.
Thus, in what follows, we set ${\sf p}=2g$ with $g$ being a natural number. For $S_{tot} = {\cal{O}}(N)$, when expressing the expectation value of $\langle (\Delta \epsilon)^{2g} \rangle$ longhand in terms of spin raising and lowering operators, one notices that, in this large $N$ limit, each individual term containing an equal number of raising and lowering operators 
yields an identical contribution (proportional to $(S_{tot} \hbar  \sqrt{1- w^{2}})^{2g}$)
to the expectation value $\langle (\Delta \epsilon)^{2g} \rangle$.
Since there are ${2g \choose g}$ such contributions, for all $g \ll N$ in the thermodynamic ($N \to \infty$) limit, 
the expectation value $\langle (\Delta \epsilon)^{2g} \rangle =   {2g \choose g} (\frac{\sigma_{\epsilon}^{2}}{2})^{g}$. We write the final (Schrodinger picture) state at time $t=t_{f}$ as $| \psi_{Spin} \rangle = \sum_{\alpha} c_{\alpha} | \phi_{\alpha} \rangle$. The probability distribution of the energy density of Eq. (\ref{PER}) reads 
\begin{eqnarray}
\label{distribution}
P(\epsilon') = \sum_{\alpha} |c_{\alpha}|^{2} \delta(\epsilon'-\frac{E_{\alpha}}{N}).
\end{eqnarray}
In this example, the Heisenberg picture Hamiltonian $H^{H}_{spin}$ (and the associated operators ${\cal{H}}_{i}$) remains local for all times. In general systems, the time evolved Heisenberg picture Hamiltonian need not be spatially local. Eq. (\ref{distribution}) describes the probability distribution associated with the ``wave packet'' intuitively discussed in Section \ref{intuition} (a ``packet'' that is now given by the amplitudes $\{c_{\alpha}\}$
in our eigenvalue decomposition of the final state $| \psi_{Spin} \rangle$). The averaged moments of $\Delta \epsilon' \equiv (\epsilon'-\epsilon)$ are $\langle (\Delta \epsilon)^{2g} \rangle= \int d \epsilon' ~P(\epsilon') ~(\epsilon'- \epsilon)^{2g}$. Here, as throughout, $\epsilon = \frac{1}{N} \langle \psi_{Spin} | H_{spin} | \psi_{Spin} \rangle = - (\sum_{ij} J_{ij} + B_{z} S_{tot}^{z} \cos \theta(t_{f}))/N$
is the energy density in the final state (i.e., the average of the energy density $\epsilon'$ when weighted with $P(\epsilon')$). More generally, the expectation value of a general function $f(\frac{H_{spin}^{H}}{N})$ in the state $| \psi^{0}_{Spin} \rangle$ (or, equivalently, of $f(\frac{H_{spin}}{N})$ in the above defined final Schrodinger picture state $| \psi_{Spin} \rangle$) is given by $\langle f(\frac{H_{spin}^{H}}{N}) \rangle = \int d \epsilon' f(\epsilon') P(\epsilon')$.
The mean value of each Fourier component $e^{i \sf{q}  (\Delta \epsilon')}$ when 
evaluated with $P(\epsilon')$ is 
\begin{eqnarray}
\label{joq}
\langle e^{i \sf{q}  (\Delta \epsilon')} \rangle= \sum_{g=0}^{\infty} \frac{(i \sf{q})^{2g}}{2^{g}(2g)!} {2g \choose g} \sigma_{\epsilon}^{2g} =  \sum_{g=0}^{\infty} \frac{(-1)^{g}}{(g!)^{2}} (\frac{\sf{q} \sigma_\epsilon}{\sqrt{2}})^{2g} = J_{0}(\sf{q} \sigma_{\epsilon} \sqrt{2}),
\end{eqnarray}  
where $J_{0}$ is a Bessel function.
An inverse Fourier transformation then yields 
\begin{eqnarray}
\label{Heavy}
P( \epsilon') = \frac{\theta( \sigma_{\epsilon} \sqrt{2}-| \Delta \epsilon'|)}{\pi \sigma_{\epsilon}  \sqrt{2 - \frac{(\Delta \epsilon')^{2}}{  (\sigma_{\epsilon})^{2}}}}.
\end{eqnarray} 
Here, as earlier,  $\Delta \epsilon'$ denotes the difference between $\epsilon'$ and the value of the energy density $\epsilon(t)$.
The Heaviside function $\theta(z)$ in Eq. (\ref{Heavy}) captures the fact that the spectrum of $H_{spin}$ is bounded. Similar results apply to boundary couplings \cite{boundary}. 
The distribution of Eq. (\ref{Heavy}) may also be rationalized geometrically as we will shortly discuss (Eq. (\ref{pe-classical})).  
Comparing our result of Eq. (\ref{Heavy}) to known cases, we remark that, where it is non-vanishing, the distribution of Eq. (\ref{Heavy}) is the reciprocal of the Wigner's semi-circle law governing the eigenvalues of random Hamiltonians and the associated distributions of Eq. (\ref{distribution}), e.g., \cite{NJP14}. We stress that Eq. (\ref{Heavy}) is exact for
the general spin Hamiltonians of Eqs. (\ref{hferro},\ref{htr}) and does not hinge on assumed eigenvalue distributions of effective random matrices. 

Performing additional calculations, we find qualitatively similar results for analogous ``cooling/heating'' protocols. For instance, one may consider, at intermediate times $0 \le t \le t_{f}$, the Hamiltonian governing the system to be that of a time independent $H_{tr}$ (i.e., one with a constant $B_{y}(t)=B_{y}$) augmenting $H_{spin}$ instead of replacing it. 
That is, we may consider, at times $0 \le t \le t_{f}$, the total Hamiltonian to be 
\begin{eqnarray}
\label{eq:augment}
H_{a}=H_{spin} + H_{tr}.
\end{eqnarray}
For such an augmented total Hamiltonian $H_{a}$, the total spin $\vec{S}_{tot}$ precesses around direction of the applied external field $(B_{y} \hat{e}_{y} + B_{z} \hat{e}_{z}) \equiv B \hat{e}_{n}$. An elementary calculation analogous to that leading to Eq. (\ref{se}) then demonstrates that the corresponding standard deviation  $\sigma^{a}_{\epsilon}$ of the energy density at $t=t_{f}$,
 \begin{eqnarray}
 \label{sigmaat}
 \sigma^{a}_{\epsilon}(t=t_{f}) = \frac{ |B_{z} B_{y}| S_{tot} \hbar}{ N B \sqrt{2}}   \sqrt{1+ \frac{1}{S_{tot}} - w^{2}} \nonumber
 \\ \times \sqrt{ \sin^{2} (Bt_{f}) +  \frac{B_{z}^{2} (1- \cos  (Bt_{f}))^{2}}{B^{2}}}.
 \end{eqnarray}
 We wish to stress that if $S_{tot} = {\cal{O}}(N)$ and $|w| <1$ then, as in Eq. (\ref{se}), the standard deviation $\sigma^{a}_{\epsilon} = {\cal{O}}(N)$ for general times $t_{f}$.  The distribution of the energy density following an evolution with this augmented Hamiltonian will, once again, be given by Eq. (\ref{Heavy}) for macroscopic systems of size $N \to \infty$. The reader can readily see how such spin model calculations may be extended to many other exactly solvable cases. The central point that we wish to underscore is that {\it a broad distribution of the energy density}, $\sigma_{\epsilon} \neq 0$, is obtained in all of these exactly solvable spin models in general dimensions. 
 
\subsubsection{Semi-classical spin systems and a geometrical interpretation}
\label{poincare}
\begin{figure*}
	\centering
	\includegraphics[width=.85  \columnwidth, height=.135 \textheight, keepaspectratio]{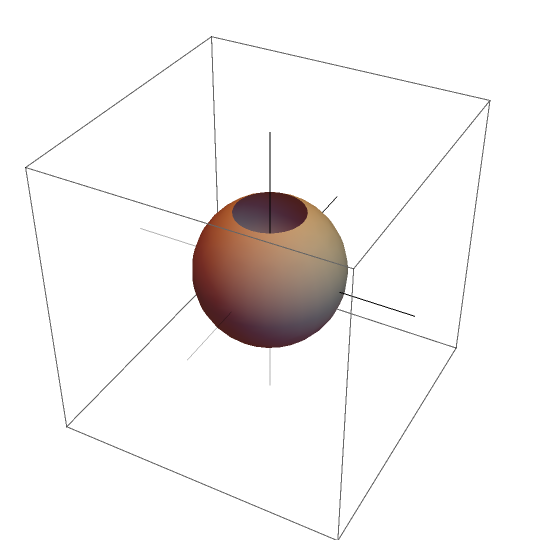}
	\caption{(Color Online.) Semi-classically, the total spin $\vec{S}_{tot}$ may, with equal probability, correspond to any vector connecting the origin of a sphere of radius $\hbar S_{tot}$ to a point along a ring forming ``a line of latitude''. In the figure above, this ``line of latitude'' ring is defined by boundary of the shaded spherical cap near the ``north pole''. All points along the line of latitude share the same value of $S^{z}_{tot}$. Here, in the initial state, the polar angle $\theta =0$.}
	\label{Cap1.}
\end{figure*}

\begin{figure*}
	\centering
	\includegraphics[width=.85  \columnwidth, height=.135 \textheight, keepaspectratio]{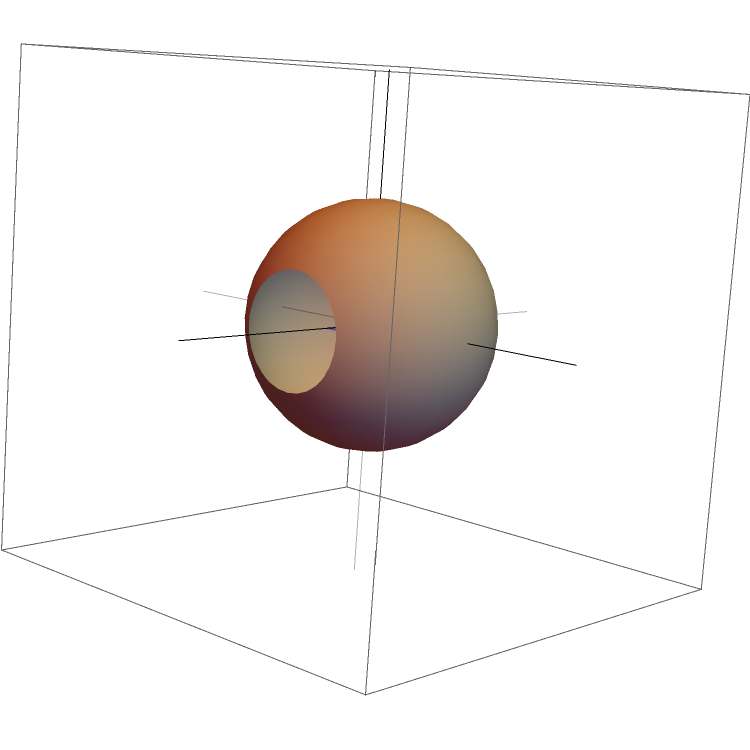}
	\caption{(Color Online.) Applying the transverse field of Eq. (\ref{htr}) to the ring of Figure \ref{Cap1.} leads to precession about the $S^{y}_{tot}$ axis. For the above displayed ring, $\theta(t) = \pi/2$. During the precession, the semi-classical total spin vectors $\vec{S}_{tot}$ on the ring acquire a range of possible $S^{z}_{tot}$ values leading to the standard deviation $\sigma_{\epsilon}$ of the energy density of Eq. (\ref{hferro}). The simple (semi-classical) calculation of Eq. (\ref{pe-classical}) for the distribution of $S^{z}_{tot}$ values for such a uniform ring leads anew to Eq. (\ref{Heavy}).}
	\label{Cap2.}
\end{figure*}

The results that we just derived are valid for any spin $S$ realization of the Hamiltonians of Eqs. (\ref{hferro}, \ref{htr}). The standard deviations of Eqs. (\ref{se}, \ref{sigmaat}) remain finite for all $S$ (with a scale set by the external magnetic field energies in these Hamiltonians).  As long known \cite{Lieb1973,Simon1980}, the $S \to \infty$ limit yields classical renditions of respective quantum spin models. Thus, the finite standard deviation of the energy density in individual eigenstates (Eqs. (\ref{se}, \ref{sigmaat})) and in thermal states formed by these eigenstates implies that the standard deviation of the energy density {\it remains finite in the classical limit} (as was suggested by the general arguments associated with Eq. (\ref{usfe})). More strongly, all that mattered in our earlier calculation of Section \ref{qss} were the $S_{tot}$ and $S_{tot}^{z}$ values. If $S_{tot}= {\cal{O}}(N)$ then even if the size of the spin $S$ at each lattice site is small, the total system spin $\vec{S}_{tot}$ is a macroscopic classical quantity and our results may be reproduced by a computation for semi-classical spins. Indeed, an explicit calculation for classical spin states trivially illustrates that a finite standard deviation $\sigma_{\epsilon} >0$ may arise in semi-classical systems \cite{explain_classical_spin}. To make this explicit, we now perform such a computation. This rather elementary calculation will link the geometry of the manifold of possible $S^{z}_{tot}$ values to the full distribution $P(\epsilon')$ of the possible energy densities. Towards this end, we parameterize the semi-classical total spin by a vector $\vec{S}_{tot}$ on a sphere of fixed radius $S_{tot}$ (the application of the transverse field Hamiltonian of Eq. (\ref{htr}) does not alter $(\vec{S}_{tot})^{2}$). Herein, at any time $t$, the vector $\vec{S}_{tot}$ may correspond, with equal probability, to any vector on a circular ring, see, e.g., Figures \ref{Cap1.} and \ref{Cap2.}.

In Eq. (\ref{Heavy}), $\Delta \epsilon'$ denotes the difference between $\epsilon'$ and the average energy density $\epsilon(t)$. At time $t$, along a ring (see, e.g., Figure \ref{Cap2.}), that is further parameterized by an azimuthal angle $\varphi'$,
the possible values of $S_{z}$ are given by $S_{tot}^{z}(\varphi',t) = \langle S_{tot}^{z}(t) \rangle+ S_{tot} \sqrt{1-w^{2}} \sin \theta(t) \cos \varphi'$. Here, $\theta(t)$ 
becomes the polar angle of the center of mass of the ring (i.e., $\theta(t)$ is the angle between (i) a vector connecting the origin to the center of the center of the ring (see, e.g., Figures \ref{Cap1.} and \ref{Cap2.}) and (ii) a vector along the positive $S^{z}_{tot}$ axis). The expectation value $\langle S_{tot}^{z} \rangle $ is that of $S_{tot}^{z}$ in the time evolved state (classically, it is the average of $S_{tot}^{z}$ around the full ring ($0 \le \varphi' < 2 \pi$) at time $t$), i.e., $S_{tot}^{z}(t) = S_{tot}^{z} \cos \theta(t)$. The possible values of $S_{z}(\varphi')$ appear symmetrically twice in the interval $0 \le \varphi' < 2 \pi$. We may thus consider only 
$0 \le \varphi' < \pi$. By the normalization of the probability distribution for $\varphi'$ and the corresponding probability distribution for the energy density, $P(\epsilon') d \epsilon' = \frac{d \varphi'}{\pi}$.
Thus, 
\begin{eqnarray}
\label{pe-classical}
P(\epsilon') = \frac{1}{\pi} \Big|\frac{d \varphi'}{d \epsilon'} \Big| =   \frac{N}{ \pi \Big| B_{z}\frac{\partial S_{tot}^{z}(\varphi')}{\partial \varphi'} \Big|}.
\end{eqnarray}
Combining Eq. (\ref{se}) (which may derived from a geometric analysis of Figure \ref{Cap2.} as we next explain) with Eq. (\ref{pe-classical}) then provides Eq. (\ref{Heavy}). We may indeed readily calculate the spread $\sigma_{S^{z}_{tot}}$ of $S^{z}_{tot}$ values and rationalize the finite standard deviation $\sigma_{\epsilon}$ of Eq. (\ref{se}) from simple geometric considerations, when $1/S_{tot}$ is set to zero (the semi-classical limit). Performing a geometric analysis, one finds that $\sigma_{S^{z}_{tot}} = \frac{R_{\sf ring}}{\sqrt{2}} |\sin \theta(t)| \equiv R_{g} |\sin \theta(t)|$ where $R_{\sf ring} = S_{tot}  \hbar \sqrt{1- w^{2}}$. Here, $R_{g}$ is the radius of gyration of the ring of Figure \ref{Cap2.} (corresponding to $\theta = \pi/2$) about an axis parallel to the $S^{z}_{tot}$ axis that passes through the center of mass of this ring. The finite radius of gyration $R_{g} \neq 0$ implies a spread of energy densities $\sigma_{\epsilon} =  \frac{|B_{z} \sin \theta(t)| R_{g}}{N} \neq 0$ at general times. This semi-classical 
result for $\sigma_{\epsilon}$ coincides with Eq. (\ref{se}). We will further comment on the $|w|=1$ states below and at the end of Section \ref{cause}. We now first briefly comment on another trivial limiting case.
When $w=0$, the initial state will correspond, in the description of Figure \ref{Cap1.}, to the equator.
Applying a transverse field will then lead to a rotation of the equator around the $S^{y}_{tot}$ axis; this so generated ring (another great circle on the sphere) will, generally, display a non-vanishing spread of $S^{z}_{tot}/N$ values (leading to $\sigma_{\epsilon} \neq 0$). However, when the initial state has $w=0$, such a rotation will not yield any change in the energy density, $d \epsilon/dt =0$. This trivial limiting case illustrates that, as a matter of principle, a finite rate of variation of the energy density is not mandatory in order to a finite have $\sigma_{\epsilon}$. As we demonstrate in the current work, the converse statement holds (a finite $d \epsilon/dt$ implies a finite $\sigma_{\epsilon}$). 

Although the Hamiltonian of Eq. (\ref{hferro}) is extremely general as are its eigenstates of high total spin $S_{tot} = {\cal{O}}(N)$ (e.g., states of large total spin in typical low temperature ferromagnets), characteristic equilibrium states of this Hamiltonian will correspond to a special subset having $|w|=1$ (that is, the total spin will be polarized along the externally applied field direction). As we discussed earlier, such equilibrium states will thus emulate product states (in which all individual spins assume the same polarization). Thus, as was indeed evident in Eqs. (\ref{se}, \ref{sigmaat}), when $w = \pm 1$, the broadening $\sigma_{\epsilon} =0$. In a related vein, the fully polarized state- a coherent spin state on a sphere of radius $S_{tot}$- is rotated ``en block'' without any other change of the wavefunction under the action of a transverse field. To see the effect for our exactly solvable system, we have to go away from the limit $|w|=1$. Away from this limit, the state of the system evolves non-trivially. In the parlance of Section \ref{intuition}, when evolving under the transverse field Hamiltonian of Eq. (\ref{htr}), the $|w| \neq 1$ spin state is not merely ``translated'' (rotated on a sphere of radius $S_{tot}$) with no other accompanying changes. \ref{appendix:preparation} discusses a gedanken experiment in which starting from an equilibrium state, one may apply transverse fields and let the closed system equilibrate anew so as to generate a state $| \psi_{Spin}^{0} \rangle$ of total spin $S_{tot} = {\cal{O}}(N)$ with $w \neq \pm 1$.

\subsubsection{Causality, correlations, and a finite $\frac{d \epsilon}{dt}$}
\label{cause}

We now return to the qualitative discussion of Section \ref{intuition} concerning the causal generation of long range correlations in real physical systems. Eq. (\ref{usfe}) suggests that long-range correlations emerge from the coupling between an external environment (which we have not explicitly included in the model system in this Section) to the system bulk (e.g., the global coupling of Eq. (\ref{htr})). As we will demonstrate in \ref{LR_explain}, compounding the lack of causal correlations in relativistic systems,
when the environment is included also in non-relativistic systems obeying Lieb-Robinson type bounds \cite{sergey,jens2,Lieb_Robinson,ali1,ali2}, a finite rate of variation of the energy density cannot appear at short times $t< t_{\min} = {\cal{O}}(L/v_{\sf  LR})$. Thus, generally, effective global couplings such as those of Eq. (\ref{htr}) cannot appear instantaneously. We wish to reiterate this particular point. Without the bulk coupling of Eq. (\ref{htr}) (and ensuing correlations), the system cannot exhibit a finite rate of change of its energy density (i.e., without such a global coupling, the latter rate of change $\frac{d \epsilon}{dt} =0$). It is only {\it after long enough times} (such as those implied by the Lieb-Robinson bounds of \ref{LR_explain}), at $t>t_{\min}$, that a global coupling such as that of Eq. (\ref{htr}) may appear in effective descriptions not explicitly involving an external environment. Only at these sufficiently long times, our obtained results for the correlations hold.  

Another point is also worth mentioning anew here. In an equilibrium state of the Hamiltonian of Eq. (\ref{hferro}), the total spin will be polarized along the applied field direction and $w=1$. In such a case, for the realization of various gednaken experiments (e.g., \ref{appendix:preparation}), long-range correlations (\ref{sxsx}) may indeed appear in the system after a time that scales with the system size. 

As noted after Eq. (\ref{se}), the calculation of the energy density and its standard deviation for a $|w|=1$ system evolving under Eq. (\ref{htr}) is identically the same as that for a product state of $S_{tot}/S$ spins. In the representation of Section \ref{poincare}, such an initial ferromagnetic state will correspond to a single point on the sphere (the north or south pole) instead of the ring in Figure \ref{Cap1.}; a rotation by a transverse field as depicted in Figure \ref{Cap2.} will then lead to this point rotated elsewhere- there will not any spread of the $S^{z}_{tot}$ values and $\sigma_{\epsilon} =0$. That such a ferromagnetic state (akin to the product states
discussed earlier) exhibits no spread of the energy density is consistent with Section \ref{sec:product}. Further, in tandem with our main thesis concerning a typical general trend between the energy changes and long range correlations, for $w \neq \pm 0, 1$ states, at those times at which the energy density changes at a vanishing rate $d \epsilon/dt =0$ (corresponding to $\theta(t) \equiv 0 (\sf mod \pi))$, the standard deviations of the energy density (and the associated long-range correlations that it implies) also vanishes, $\sigma_{\epsilon} =0$.

 \subsection{Itinerant hard core Bose systems}
 \label{gbose}
 
Our spin model of Section \ref{sec:gspin} can be defined for local spins of any size $S$. The function $P(\epsilon')$ of Eq. (\ref{Heavy}) characterizing our investigated states in this system is not a very typical probability distribution. However, the non-local entangled character of states having a finite energy density relative to the ground state 
is pervasive for thermal states. This model can be recast in different ways. In what follows we focus on the spin $S=1/2$ realization of Eq. (\ref{hferro}). The Matsubara-Matsuda transformation \cite{MM,MM-explain} maps the algebra of spin $S=1/2$ operators onto that of hard core bosons. Such hard core bosons may, e.g., emulate Cooper pairs in superconductors in the limit of short coherence length. Specifically, the hard core bosonic number operator at site $i$ is $n_{i}=b_{i}^{\dagger} b_{i} =0,1$ with $b_{i}$ and $b^{\dagger}_{i}$ the annihilation and creation operators of hard core bosons ($(b_i^{\dagger})^{2}=b_i^{2}=0$, $[b_{i},b^{\dagger}_{j}] = (1-2 n_{i}) \delta_{ij}$). Following this transformation, the spin Hamiltonian of Eq. (\ref{hferro}) is converted into its hard core bosonic dual, 
\begin{eqnarray}
\label{BoseH}
H_{Bose} = - \sum_{ij} J_{ij} ((b_{i}^{\dagger} b_{j} + h.c.) + n_{i} n_{j})  -  \sum_{i} (B_{z}- \sum_{j} J_{ij}) n_{i}. 
\end{eqnarray}
The above Hamiltonian describes hard core bosons hopping (with amplitudes $J_{ij}$) on the same $d-$dimensional lattice, featuring attractive interactions and a chemical potential set by $(B_{z}-\sum_{j} J_{ij})$. Here, the transverse field cooling/heating Hamiltonian $H_{tr}$ transforms into 
\begin{eqnarray}
H_{Bose-doping} = - \frac{iB_{y}(t)}{2} \sum_{i} (b_{i}^{\dagger} - b_{i}),
\end{eqnarray}
a Hamiltonian that alters the number of the bosons (thereby ``doping'' the system). 
In the context of Cooper pairs of short coherence length emulating hard core bosons, $H_{Bose-doping}$ may describe the effect of Cooper pairs injected/removed from the system from a surrounding environment comprised of a bulk superconductor.  The hard core Bose states are symmetric under all pairwise permutations $P_{ij}$ of the bosons at occupied sites. The bosonic dual of, e.g., the specific spin product state
$| \uparrow_{1} \uparrow_{2} \downarrow_{3} \uparrow_{4} \downarrow_{5} \uparrow_{6}  \cdots \uparrow_{N-1} \downarrow_{N} \rangle$
corresponds to the symmetrized state of a fixed total number of hard core bosons that are placed on the graph (or lattice) sites $(1,2, 4, 6, \cdots, (N-1))$. Thus, the bosonic dual of an initial spin state $| \psi^{0}_{Spin} \rangle$ with a total spin $S_{tot} = S_{\max} =N/2$ is an initial hard core Bose state $|\psi^0_{Bose} \rangle$ that is an equal amplitude superstition of all real space product states with the same total number of hard core bosons ($\sum_{i=1}^{N} n_{i}= m + \frac{N}{2}$) distributed over the $N$ lattice sites (an eigenstate of $H_{Bose}$ that adheres to the fully symmetric bosonic statistics). Evolving (during times $0 \le t \le t_{f}$) this initial state with $H_{doping}$, the standard deviation of Eq. (\ref{se}) and the distribution of Eq. (\ref{Heavy}) are left unchanged, apart from a trivial rescaling by $\hbar$ (e.g., $\sigma^{Bose}_{\epsilon} =  \frac{|B_{z} \sin \theta(t_{f})|}{2\sqrt{2}} \sqrt{1+ \frac{2}{N} - w^2}$ for $S_{tot} =N/2$). 
Similar to our discussion of the dual spin system of the previous subsection, the finite standard deviation in this energy density (and of 
the associated particle density $n= \frac{1}{N} \sum_{i} n_{i}$) does not imply that the ``doping'' is, explicitly, spatially inhomogeneous (indeed, at all times,
the expectation value of the particle number $\langle n_{i} \rangle$ stays uniform for all lattice sites $i$). 

We conclude this subsection with three weaker statements regarding viable extensions of the rigorous results that we derived thus far for hard core bosonic systems on general graphs (these graphs include lattices in general dimensions).

\noindent (a) We may relate the above lattice theory to {\it a continuum scalar field theory} in the usual way. Doing so, it is readily seen that for a continuous scaled $\varphi(x)$ field replacing $(b_{i}+b_{i}^{\dagger})$, the canonical Hamiltonian density
\begin{eqnarray}
\label{scalar}
{\cal{H}}[\varphi] = \frac{1}{2} (m^{2} \varphi^{2} + (\nabla \varphi)^{2}) + u \varphi^{4}
\end{eqnarray}
qualitatively constitutes a lowest order continuum rendition of the hard core Bose lattice model of Eq. (\ref{BoseH}) for a system with uniform nearest neighbor couplings $J_{ij}$.  A large value of the constant $u$ in generic bosonic $\varphi^4$ field theories of the type of Eq. (\ref{scalar}) yields a large local repulsion between the bosonic fields endowing them with hard core characteristics. The continuum analog of $H_{Bose-doping}$ is the volume integral of the momentum conjugate to 
$\varphi(x)$. Thus, during various continuous changes of the Hamiltonian, such generic scalar field theories (and myriad lattice system described by them) may exhibit the broad $\sigma_{\epsilon}$ that we derived for some of their lattice counterpart in this subsection. 

\noindent (b) The models of Eqs. (\ref{hferro},\ref{BoseH}) were defined on arbitrary graphs (including lattices in general spatial dimensions). Identical results apply for {\it spineless fermions} on one dimensional chains
with non-negative nearest neighbor hopping amplitudes/coupling constants $\{J_{ij} \}$ and analogs of $H_{Bose-doping}$
capturing a non-local coupling of the system to the external bath.  
These spinless Fermi systems may be trivially engineered by applying the Jordan-Wigner transformation \cite{JW} to 
Eq. (\ref{hferro}). 

\noindent (c) {\it Phonons in anharmonic solids}. One may apply the Holstein-Primakoff transformation, 
\begin{eqnarray}
\label{HP-WU}
S_{i}^{+} = \hbar \sqrt{2} \sqrt{ 1- \frac{a_i^{\dagger} a_i}{2S}} a_i, 
~S_{i}^{-} = \hbar \sqrt{2}a_i^{\dagger}  \sqrt{1- \frac{a_i^{\dagger} a_i}{2S}}, 
~S_{i}^{z} = \hbar (S- a_i^{\dagger} a_i),
\end{eqnarray}
to express the local spin operators in Eq. (\ref{hferro}) in terms of bosonic creation/annhilation operators ($a_{i}^{\dagger}$ and $a_{i}$). 
The resulting bosonic Hamiltonian may then be expanded in a series in $1/S$ (as in conventional $1/S$ expansions) \cite{Assa-book}. 
When Fourier transformed, the Hamiltonian describes coupled bosonic modes (involving the bosonic creation/annihilation operators $a^{\dagger}_{k}$ and $a_{k}$ at different Fourier modes $k$) 
such as those of phonons in anharmonic solids. Here, the heating/cooling protocol of Section \ref{sec:gspin} corresponds to the creation/annihilation of phonons and leads to identical results for $\sigma_{\epsilon}$. 
(Contrary to the anharmonic system, in harmonic theories, the eigenstates have a product state form and some of intuition underlying the product states of Section \ref{sec:product} comes to life. For completeness, we remark that for harmonic systems, the individual interactions terms in Eq. (\ref{HsH'}) are unbounded unlike those discussed in Section \ref{sec:product}.) 
A Schwinger boson representation may similarly express the spin system of Eqs. (\ref{hferro}, \ref{htr})  in terms of bosonic modes. 

\section{Long range correlations induced by a common environment- simple solvable limits}
\label{sec:CSM}

We now turn to systems akin to those of type (2) of Section \ref{sec:sketch} that illustrate the possible effect of an environment common to all the local degrees of freedom. As noted in Section \ref{intuition} (and schematically depicted in Figure \ref{coupling.}), in order to achieve a finite rate of change of the system energy density, there must be a coupling between the bulk of the system and its environment.  The models that we will study in this Section will explicitly include such a coupling. We will consider situations in which the driving environment will not initially be in an eigenstate of the full Hamiltonian, and thus exhibit fluctuations. Hence, some of the tractable models that we introduce in this Section may also be viewed as belonging to category (1) of Section \ref{sec:sketch} in which (unlike the models of Section \ref{sec:dual}), the driving parameters in the Hamiltonian (including any external fields) are replaced by operators that display a finite variance. 

In the general evolution operator of Eq. (\ref{usfe}), the coupling between the system and the environment $H_{{\cal S}-{\cal E}}$ may include both local stochastic effects of the environment coupling to the system (e.g., photon/phonon/\dots exchange coupling local degrees of freedom in the system ${\cal{S}}$ to local ones in the environment ${\cal E}$) as well coupling between collective degrees of freedom (if any) characterizing an external drive and the system bulk. For instance, in Joule's heating experiment
in which a large dropping mass heats a fluid by causing a paddle to stir, the height of the macroscopic dropping mass serves as a collective coordinate $\overline{q}$ associated with the environment that, at sufficiently long times $t>t_{\min}$ may couple to a finite fraction of the fluid (the system) that it heats a non-vanishing rate. Similarly, an external piston pressing on a gaseous system may couple and lead to bulk effects. In other instances, $\overline{q}$ may correspond to another collective degree of freedom (or ``switch'') that leads to a bulk coupling of the system to its environment. In these and other cases, the coupling between the environment and the individual system degrees of freedom is, on average, of uniform sign (see, e.g., \ref{app:triv1} for further discussion and simple proof concerning uniform sign correlations mandated by a finite rate of change of the energy density). Augmenting changes in such collective coordinates $\overline{q}$, there are many other local stochastic degrees of freedom of the environment that couple to those of the system. 

In this Section, we will compute correlation functions associated with exceptionally simple ``central spin model'' (CSM) type Hamiltonians capturing the caricature of Figure \ref{coupling.}; the ``central spin'' represents the common driving environment that couples to the bulk system spins or masses. These models are not introduced to portray real systems but rather as solvable examples. By comparison to Section \ref{sec:product}, the CSM type Hamiltonians studied in this Section are not separable. In Sections \ref{suchatrivialexampleatleastnowrefereeswillnotcomplain}, \ref{sec:Isingchainc}, and \ref{edwustl}, we consider the environment to be a single spin. By contrast, in Sections \ref{igtm} (in particular, Section \ref{oscenv}) and \ref{cosom}, we solve systems in which the environment ${\cal{E}}$ is of macroscopic (${\cal{O}}(N)$ or larger) size. The special solvable systems that we consider might be realized in, e.g., trapped ion systems in which spin-spin interactions are mediated by coupling to a common laser or other source \cite{porras2004,Kim2011} in which we will now allow for fluctuations. In all of the examples studied in this Section, we will illustrate the generic existence of connected local range correlations but assuming the converse--taking the initial state to be a simple product state with no such correlations-- and illustrate that the system evolves to a state with long range covariance. Thus, the initial product states of this Section will, unlike those in Section \ref{sec:dual}, be devoid of any non-trivial connected correlations. Similar to the models of Section \ref{sec:dual}, we consider these initial states are  eigenstates of the system Hamiltonian (trivially satisfying $\sigma_{\epsilon} =0$ as expected in equilibrium systems in the absence of an external environment driving the system). 


\subsection{Non-interacting Ising system}
\label{suchatrivialexampleatleastnowrefereeswillnotcomplain}

In the notation of Eq. (\ref{usfe}), we will first consider the spin $S=1/2$ time independent Hamiltonians
\begin{eqnarray}
H^{CSM}_{\cal{S}} = - B_{z} \sum_{i} S^{z}_{i}, \nonumber
\\ H^{CSM}_{{\cal S}-{\cal E}} =  - J_{\cal{E}} \sum_{i=1}^{N} S^{x}_{i} {\cal{P}}_{{\cal{E}}},  
\label{CSMT}
\end{eqnarray}
where the environment only Hamiltonian $H^{CSM}_{\cal{E}}$ may be any function ($f_{\cal{E}}$) of ${\cal{P}}_{{\cal{E}}}$. Here, ${\cal{P}}_{{\cal{E}}}$ is a projection operator on the $S=1/2$ ``central spin'' (the environment ${\cal{E}}$) that couples to each of the system spins in Figure \ref{coupling.}. Specifically, we choose $ {\cal{P}}_{{\cal{E}}} =( \frac{1}{2} - \frac{S_{\cal{E}}^{z}}{\hbar})$. Unlike the models of Section (\ref{sec:dual}) and Eq. (\ref{htr}) in particular, the effective transverse magnetic field $(J {\cal{P}}_{{\cal{E}}})$ is not a constant c-number but rather an operator that exhibits a finite standard deviation in general states of the system-environment hybrid. 
In the limit of dominant $ H^{CSM}_{{\cal S}-{\cal E}}$, 
the temporal evolution with the the full Hamiltonian of Eq. (\ref{usfe}), $\tilde{H}^{CSM} = (H^{CSM}_{\cal{S}} + H^{CSM}_{\cal E}+ H^{CSM}_{{\cal S}-{\cal E}})$ may be replaced by one with $H^{CSM}_{{\cal S}-{\cal E}}$. In the Heisenberg picture, as employed in Section \ref{sec:gspin}, the system spins will perform standard precessions yet now with a ``transverse field'' that is not a constant c-number but rather a bona fide (collective) degree of freedom $\overline{q}$ (that of the environment ${\cal{E}}$), i.e., the Heisenberg picture operator 
$S^{zH}_{i}(t)= e^{i\tilde{H} t /\hbar} S^{z}_{i} e^{-i \tilde{H} t/\hbar}  =  (S_{i}^{z} \cos ({\cal{P}}_{{\cal{E}}} J_{\cal E} t)- 
 S_{i}^{y} \sin({\cal{P}}_{{\cal{E}}} J_{\cal E} t))$. 
 To motivate the general emergence of long range connected correlations as the system evolves, we consider the initial ($t=0$) state to enjoy no such correlations. Specifically, we consider the initial state to be given, in the local $S^{z}$ product basis, by a simple spin $S=1/2$ ferromagnetic product state (an eigenstate (ground state) of $H^{CSM}_{\cal{S}}$) of the system multiplied by the $\hbar/2$ eigenstate of the environment $S=1/2$ spin $S^{x}_{\cal{E}}$ coupling to all system spins, 
 \begin{eqnarray}
 \label{psi0csm}
 |\psi^{0}_{CSM} \rangle = | \uparrow_{1} \uparrow_{2} \cdots \uparrow_{N} \rangle \otimes| \rightarrow_{\cal{E}} \rangle  \end{eqnarray}
 
 Evolving under $\tilde{H}^{CSM}$, at time $t$, 
 the rate of change of the system energy density 
 \begin{eqnarray}
 \label{ratee} 
\frac{d}{dt} \frac{1}{N} \langle H^{CSMH}_{\cal{S}} (t) \rangle  =
 \frac{\hbar B_{z} J_{\cal{E}}}{2} \sin(J_{\cal{E}}t),
 \end{eqnarray}
 with $H^{CSMH}_{\cal{S}}(t)$ denoting the Heisenberg picture system Hamiltonian. 
 Concurrently, the connected correlator  
 \begin{eqnarray}
 \label{connect2refereesbrains}
 \langle S_{i}^{zH} S_{j}^{zH} \rangle - \langle S_{i}^{zH} \rangle \langle S_{j}^{zH} \rangle =\frac{\hbar^{2}}{4} \sin^{4} (\frac{J_{\cal{E}} t}{2}).
 \end{eqnarray}
For general times $ t \not \equiv 0 ({\sf mod}  ~\pi /(2J_{\cal{E}}))$, this finite covariance is (by the very nature of this problem) the same for all pairs $ i \neq j$ and is thus trivially independent of the spatial distance $|i-j|$ between sites $i$ and $j$. This independence is not surprising since, in the absence of interactions in $H_{\cal{S}}$, the effective coupling between any two system spins at sites $i$ and $j$ to each other via the ``central''  spin that is afforded by the environment ${\cal{E}}$ is independent of the separation between the two sites (the graph of Figure \ref{coupling.} in the absence of intra-system couplings). The non-vanishing covariance between the spins in this example can be traced to the fluctuations of the environment (the variance of ${\cal{P}}_{{\cal{E}}}$ in the state $|\psi^{0}_{CSM} \rangle$). In \cite{comment-CSM*}, 
we briefly discuss the standard deviations of the energy density in this system and related aspects.

\subsection{Spin chain} 
\label{sec:Isingchainc}
We next consider a particular spin $S=1/2$ chain (IC) (with periodic boundary conditions) with nearest neighbor (n.n.) interactions
coupled to a central ($S=1/2$) spin ${\cal{E}}$, 
\begin{eqnarray}
H^{IC,CSM}_{\cal{S}} &&= - J_{n.n.} \sum_{i=1}^{N} S^{z}_{i} S^{x}_{i+1} \equiv - J_{n.n.} \sum_{i} b_{i}, \nonumber
\\ H^{IC,CSM}_{{\cal S}-{\cal E}} &&=  - J_{\cal{E}} \sum_{i=1}^{N} S^{x}_{i} {\cal{P}}_{{\cal{E}}}.
\label{CSMT'}
\end{eqnarray}
Similar to the example of the previous subsection, $H^{IC,CSM}$ may be a general function of ${\cal{P}}_{{\cal{E}}}$ where ${\cal{P}}_{{\cal{E}}} =( \frac{1}{2} - \frac{S_{\cal{E}}^{z}}{\hbar})$. Once again, for simplicity, we consider the limit, where the system evolves under the Hamiltonian $\tilde{H}^{IC,CSM} = H^{IC,CSM}_{{\cal S}-{\cal E}}$ and the initial state of the environment to be $|\rightarrow_{\cal{E}} \rangle$ (the eigenstate of $S^{x}_{\cal{E}}$ corresponding to an eigenvalue of $(\hbar/2)$). Longhand, the time evolved system ``bonds'' in the system Hamiltonian $H^{IC,CSM}_{\cal{S}}$ trivially become $b^{H}_{i}(t)=
(S_{i}^{z} S_{i+1}^{x} \cos (J_{\cal E} t {\cal{P}}_{{\cal{E}}}) - S^{y}_{i} S_{i+1}^{x} \sin (J_{\cal E} t {\cal{P}}_{{\cal{E}}}))$.
The sinusoidal time variation of $b^{H}_{i}(t)$ implies that for general states, the system energy density $\frac{\langle H^{IC,CSM}_{\cal{S}} \rangle}{N}$ may similarly vary at a finite rate.  We consider (in order to demonstrate, by contradiction, the existence of connected long range correlations at finite $t$) the initial system state to exhibit no long range covariance between the bilinears $(S^{a}_{i} S^{b}_{i+1})$ and $(S^{a'}_{j} S^{b'}_{j+1})$ for all spin polarizations (each of the spin components $a,b,a',b'$ may be $x,y,$ or $z$) for far separated sites $|i-j| = {\cal{O}}(N)$ (e.g., simple product form states of the form of Eq. (\ref{psi0csm}) and other generic states with short range correlations). In other words, in the initial state, $\langle S^{a}_{i} S^{b}_{i+1}) (S^{a'}_{j} S^{b'}_{j+1} \rangle = \langle S^{a}_{i} S^{b}_{i+1} \rangle \langle S^{a'}_{j} S^{b'}_{j+1} \rangle$ for $|i-j| = {\cal{O}}(N)$. Inserting $b^{H}_{i}(t)$, the covariance of Eq. (\ref{central})), for such distant sites 
\begin{eqnarray}
G_{ij} = 
 J^{2}_{n.n} \Big(\langle S^{z}_{i} S^{x}_{i+1} \rangle^2 \sin^{4} (\frac{J_{\cal E} t}{2}) &&+ \frac{1}{2} (\langle S^{z}_{i} S^{x}_{i+1} \rangle \langle S^{y}_{j} S^{x}_{j+1} \rangle +
\langle S^{z}_{j} S^{x}_{j+1} \rangle \langle S^{y}_{i} S^{x}_{i+1} \rangle
) \nonumber
\\ &&\times \sin (J_{\cal E}t) ~\sin^{2} (\frac{J_{\cal E} t}{2}) \Big).
\end{eqnarray} 
We reiterate that in computing the expectation value above, we took the environment state to be in the $(\hbar/2)$ eigenstate of $S^{x}_{\cal{E}}$. Now, the expectation value $\langle S^{z}_{i} S^{x}_{i+1} \rangle$ is, up to a constant multiplicative factor, the energy density, i.e., $\langle S^{z}_{i} S^{x}_{i+1} \rangle = -\frac{1}{NJ_{n.n.}} \langle H^{IC,CSM}_{\cal{S}}  \rangle$ which for general equilibrium states is finite. This, in turns, implies a finite $G_{ij}$ at general times for such distant sites $|i-j| = {\cal{O}}(N)$.

\subsection{Jaynes-Cummings type model}
\label{edwustl}

We next examine examine an analog of Eqs. (\ref{CSMT}) that, somewhat like the Jaynes-Cummings model \cite{Jaynes}, includes a coupling between local two state ($S=1/2$) degrees of freedom and a bosonic field (a ``central oscillator'' coordinate $\overline{q}$ in our case). Here, 
\begin{eqnarray}
&&H^{JCCM}_{\cal{S}} = - B_{z} \sum_{i=1}^{N} S^{z}_{i} \equiv \sum_{i=1}^{N} {\cal H}^{JCCM}_{i}, \nonumber
\\ &&H^{JCCM}_{{\cal S}-{\cal E}} =  - \lambda \overline{q} \sum_{i=1}^{N} S^{x}_{i}, 
\label{CSMT''}
\end{eqnarray}
and $H^{JCCM}_{\cal E}$ is any function of $\overline{q}$ (yet not containing the conjugate momentum $\overline{p}$- the environment does not evolve in time). 
In this model, $\overline{q}$ is a (bosonic displacement) degree of freedom that couples linearly to each of local two level degrees of freedom at site $i$. 
Similar to Section \ref{suchatrivialexampleatleastnowrefereeswillnotcomplain}, we take 
the initial state to be the product state of a ferromagnetic system completely polarized along the $z$ axis multiplied by the state of the environment, $|  \uparrow_1 \uparrow_2 \cdots \uparrow_N \rangle  \otimes | {\cal E} \rangle$. For concreteness, we set $\langle \overline{q} | {\cal E} \rangle$ to be in the a Gaussian in $\overline{q}$ 
of standard deviation $\sigma_{\overline{q}}$ and zero mean. 
Similar to Sections \ref{suchatrivialexampleatleastnowrefereeswillnotcomplain} and \ref{sec:Isingchainc}, we assume
$||H^{JCCM}_{{\cal S}-{\cal E}} || \gg ||H^{JCCM}_{\cal{S}}||$. Ignoring backaction effects of the system on the environment, the time evolved spin operators $S^{zH}_{i}(t)= e^{i\tilde{H} t /\hbar} S^{z}_{i} e^{-i \tilde{H} t/\hbar}  =  (S_{i}^{z} \cos (\lambda \overline{q} t)- 
 S_{i}^{y} \sin(\lambda \overline{q} t))$. As in the previous subsections, the sinusoidal variation of $S^{zH}_{i}(t)$ may lead, in general states, to a finite rate of variation of the expectation value of the time evolved energy density  $\frac{H^{JCCMH}_{\cal{S}}}{N}$. The covariance \cite{var-cos-Gau}
\begin{eqnarray}
 \label{connect2refereesbrains''}
&&\langle S_{i}^{z H}(t) S_{j}^{z H}(t) \rangle - \langle S_{i}^{z H}(t) \rangle \langle S_{j}^{z H}(t) \rangle \nonumber
\\ 
&=& \frac{\hbar^{2}}{4}( \langle {\cal E}| \cos^{2}  (\lambda \overline{q} t) | {\cal E} \rangle - \langle {\cal E} | \cos ( \lambda \overline{q} t) | {\cal E} \rangle^{2}) \nonumber
\\  &=& \frac{\hbar^{2}}{8} (1- e^{-  \lambda^{2} t^{2} \sigma^{2}_{\overline{q}}})^{2},
 \end{eqnarray}
implying connected correlations between ${\cal H}^{JCCM,H}_{i} (t)$ and ${\cal H}^{JCCM,H}_{j} (t)$ for all system sites $i$ and $j$ (including arbitrarily large $|i-j|$). 
For this model, the probability density of Eq. (\ref{PER}, \ref{distribution}) for ${\sf \varepsilon'} \equiv - 2 \epsilon'/(\hbar B_{z})$ \cite{pe'},
\begin{eqnarray}
\label{pe'-eq}
 P({\sf \varepsilon}') = \frac{\exp  \Big( -  \frac{(\cos^{-1} {\sf \varepsilon}')^{2}}{2 \alpha^2 \sigma_{\overline{q}}^{2}}\Big)  \vartheta_{3} \Big(    \frac{i\cos^{-1} {\sf \varepsilon}'}{2 \alpha \sigma_{ \overline{q}}^{2}}, e^{- \frac{ \pi^{2}}{2 \sigma^2_{\overline{q}}}} \Big)}{\alpha \sqrt{2 \pi \sigma_{\overline{q}}^{2} (1 - ({\sf \varepsilon}')^{2})}}, 
\end{eqnarray}
with $\vartheta_{3}$ the third Jacobi function and $\alpha \equiv  \lambda t$.  
If a kinetic term (involving the collective environment momentum $\overline{p}$) is included and backaction effects are not negligible, then the system will generally modify the environment. In such cases, with $\overline{q}^{H}(t)$ denoting the Heisenberg picture oscillator coordinate, instead of Eq. (\ref{connect2refereesbrains''}), there will be contributions to the covariance $\langle S_{i}^{z H}(t) S_{j}^{z H}(t) \rangle - \langle S_{i}^{z H}(t) \rangle \langle S_{j}^{z H}(t) \rangle$ that are of the form
\begin{eqnarray}
 \label{connect2refereesbrains'''}
\langle  \cos^{2}  (\lambda \int_{0}^{t} dt' \overline{q}^{H}(t') )  \rangle - \langle \cos (\lambda \int_{0}^{t} \overline{q}^{H}(t') dt')  \rangle^{2}.
\end{eqnarray}
These expectation values are, once again, generally non-vanishing (the initial state does not, generally, need to be an eigenstate of
$\cos (\lambda \int_{0}^{t} \overline{q}^{H}(t') dt')$) and long range connected long range correlations will appear in the system.

\subsection{Ideal gas type models}
\label{igtm}

\subsubsection{Static environment}
\label{sec:staticenvironment}
We next consider an ideal gas (IG) type system ${\it S}$ with a general bilinear mechanical coupling to a static external environment ${\cal{E}}$,
\begin{eqnarray}
\label{COIGM}
&&H^{IG}_{\cal{S}} = \sum_{i=1}^{N} \frac{p_{i}^{2}}{2m} \equiv \sum_{i=1}^{N} {\cal H}^{COIG}_{i} , \nonumber
\\&& H^{IG}_{{\cal S}-{\cal E}} =  - \lambda \overline{q} \sum_{i=1}^{N} x_i \equiv \sum_{i=1}^{N} {\cal H}^{IG}_{{\cal S}-{\cal E},i} ~~, 
\end{eqnarray} 
and $H^{IG}_{\cal E} $ a general function of $\overline{q}$. 
We take the Initial state to be a product state of each of the particle and the ground state of that of the environment\begin{eqnarray}
|\psi_{mech}^{0} \rangle = |\psi_{1}^{0} \rangle  \otimes |\psi_{2}^{0} \rangle \otimes \cdots \otimes |\psi_{N}^{0} \rangle \otimes  |\cal{E} \rangle.
\end{eqnarray}
The system accelerates under the external force $\lambda \overline{q}$, 
\begin{eqnarray}
\label{pxH1}
&&p^{H}_{i}(t)  = p_{i} + \lambda \overline{q} t, \nonumber
\\  && x^{H}_{i}(t)  = x_{i} + \frac{p_{i}}{m} t + \frac{\lambda \overline{q}}{2m} t^{2},
\end{eqnarray}
with $p_{i}$ and $x_{i}$ denoting the momentum and position operators at time $t=0$.  
Whenever Eqs. (\ref{COIGM}) with a static $H^{IG}_{\cal E}$ (the latter Hamiltonian is only a function of $\overline{q}$ and including no kinetic terms) are valid, then the rate of change of the system energy density 
$\frac{d \epsilon}{dt} = \frac{\lambda^{2} t \langle \overline{q}^{2} \rangle}{m}$,
and, trivially, 
\begin{eqnarray}
\langle p^{H}_{i}(t) p^{H}_{j}(t) \rangle - \langle p^{H}_{i}(t) \rangle \langle p^{H}_{j}(t)\rangle = \lambda^{2} \sigma^{2}_{\overline{q}} t^{2}. 
\end{eqnarray}
For all $i$ and $j$, the connected correaltor of Eq. (\ref{central}) between the time evolved local system energy densities ${\cal{H}}^{COIGH}_{i}(t)$ and ${\cal{H}}^{COIGH}_{j}(t)$ (i.e., the covariance between the kinetic terms   
$(p^{H}_{i}(t))^{2}/(2m)$ and $(p^{H}_{j}(t))^{2}/(2m)$) is 
\begin{eqnarray}
G_{ij} &= \frac{\lambda^{4} t^{4}}{4m^{2}} (\langle {\cal E}| \overline{q}^{4} | {\cal E} \rangle - \langle {\cal E}| \overline{q}^{2} | {\cal E} \rangle^2). 
\end{eqnarray} 
Fluctuations in $\overline{q}$ may thus trigger connected long range correlations. Classical fluctuations of the environment may yield a similar result if the density matrix of the system-environment hybrid is of the product form $\rho_{\cal S} \rho_{\cal E}$ and the variance of $\overline{q}$ is computed with the probability density matrix $\rho_{\cal E}$.

\subsubsection{Oscillatory environment}
\label{oscenv}
Similar to Eq. (\ref{connect2refereesbrains'''}), if the backaction effects of the system on its environment are not negligible then integrals over the Heisenberg picture operators $\overline{q}^{H}$ may more generally be written. 
For concreteness, instead of a static environment Hamiltonian $H^{IG}_{\cal E}$ having no kinetic terms, we consider the environment to be a central mechanical oscillator. The system-environment hybrid defined by Eq. (\ref{COIGM}) with $H^{COIG}_{\cal E}=  \frac{1}{2} \overline{M} \overline{\Omega}^{2} \overline{q}^{2} + \frac{\overline{p}^{2}}{2 \overline{M}}$ is exactly solvable since the full Hamiltonian $\tilde{H} = H^{IG}_{\cal{S}}  + H^{IG}_{{\cal S}-{\cal E}} + H^{COIG}_{\cal E}$ is quadratic \cite{RMP1980}. There are only two nontrivial mechanical eigenmodes appearing the Hamiltonian $\tilde{H}$ that involves the $N$ system particles and the single collective coordinate $\overline{q}$ of the environment ${\cal E}$. By virtue of the uniform coupling in $ H^{IG}_{{\cal S}-{\cal E}}$, all of the system degrees of freedom only appear through their center of mass coordinate and momentum. All other $(N-1)$ linearly independent combinations of the system coordinates that are orthogonal to center of mass displacements do not couple to the environment. The coupled Heisenberg (or classical) equations of motion for (I) center of mass of the system 
\begin{eqnarray}
\label{xcm}
x_{cm} \equiv \frac{\sum_{i=1}^{N} x_{i}}{N},
\end{eqnarray}
and (II) external environment $\overline{q}$ collective coordinate are
\begin{eqnarray}
\label{xht0}
\begin{split}
    \frac{d^{2}}{dt^{2}} \begin{pmatrix}
           x^{H}_{cm} (t) \\
           \overline{q}^{H} (t)
                    \end{pmatrix} 
                 = -   &\begin{pmatrix} 0 & \frac{\lambda}{m} \\
                 \frac{N \lambda}{\overline{M}}  & \overline{\Omega}^{2} 
                 \end{pmatrix}  
                 \begin{pmatrix}
           x^{H}_{cm} (t) \\
           \overline{q}^{H} (t) \end{pmatrix} 
            \equiv - \overline{D}    &\begin{pmatrix}
           x^{H}_{cm} (t) \\
           \overline{q}^{H} (t)
                    \end{pmatrix}.&
\end{split}
\end{eqnarray}
The eigenvalues of the dynamical matrix $\overline{D}$ trivially yield one oscillatory eigenmode 
($u(t) = A \sin \overline{\omega} t + B \sin \overline{\omega} t$ (with $A$ and $B$ constants)) of frequency 
\begin{eqnarray}
\label{omegaomega}
\overline{\omega} = \frac{1}{\sqrt{2}}  \sqrt{\sqrt{\overline{\Omega}^{2} + \frac{4 N \lambda^{2}}{m\overline{M}}}+ \overline{\Omega}^{2} },
\end{eqnarray}
and another eigenvector $v(t) = C e^{-\overline{\alpha} t} + D e^ {\overline{\alpha} t}$ (with constant $C$ and $D$) where  
\begin{eqnarray}
\label{alphaomega}
\overline{\alpha} = \frac{1}{\sqrt{2}}  \sqrt{\sqrt{\overline{\Omega}^{2} + \frac{4 N \lambda^{2}}{m\overline{M}}} - \overline{\Omega}^{2}}.
\end{eqnarray}
The results of Section \ref{sec:staticenvironment} correspond to the static environment operator $\overline{q}^{H}(t)$ arising in the limit $\overline{M} \to \infty$ of Eq. (\ref{alphaomega}); the momentum in the accelerating system (Eq. (\ref{pxH1})) is qualitatively similar in its unbounded linear increase to the exponential $e^ {\overline{\alpha} t}$ in the $M \to \infty$ limit (reminiscent of $\lim_{n \to 0}\frac{1}{n}(z^{n}-1) = \ln z$ with $z$ replaced by an exponential in $t$). Expressing $x^{H}_{cm}$, $\overline{q}^{H}$ as linear combinations of $u$ and $v$ and solving for $A,B,C,$ and $D$ by setting $x^{H}_{cm}(t=0)=x_{cm}$, $\overline{q}^{H}(t=0)= \overline{q}$, and total system and environment momenta $Nm\frac{d x_{cm}(t)}{dt}|_{t=0}= p_{tot}$ and $\overline{M} \frac{d \overline{q}^{H}(t)}{dt}|_{t=0}=\overline{p}$ yields
\begin{eqnarray}
\label{xcmt}
x_{cm}(t) =  \frac{1}{\overline{\omega}^{2} + \overline{\alpha}^{2}} 
\Big( \frac{m\overline{ \omega}^2 \overline{\alpha}^{2} }{\lambda} 
 \big( \overline{q} (\cos \overline{\omega} t - \cosh \overline{\alpha} t) 
+ \frac{\overline{p}}{\overline{M}} (\frac{\sin \overline{\omega} t}{\overline{\omega}} - \frac{\sinh \overline{\alpha} t}{\overline{\alpha}}) \big) \nonumber
\\   + x_{cm} ( \overline{\alpha}^{2} \cos \overline{\omega} t + \overline{\omega}^{2} \cosh \overline{\alpha} t) +\frac{p_{tot}}{Nm}  (\frac{\overline{\alpha}^{2}}{\overline{\omega}} \sin \overline{\omega} t + \frac{\overline{\omega}^{2}}{\overline{\alpha}} \sinh \overline{\alpha} t) \Big).
\end{eqnarray}
For an initial product state of the system $| {\cal S} \rangle$ (that does not display long range connected correlations between the system degrees of freedom) and its environment $| {\cal E} \rangle$, in the large system size ($N$) limit, the corresponding initial variances $\sigma^{2}_{x_{cm}}=0$ and $\frac{1}{N^{2}} \sigma^{2}_{p_{tot}}  =0$. Evaluating, using Eq. (\ref{xcmt}), the variance of $x_{cm}(t)$ at time $t$ when the environment $| {\cal  E} \rangle$ is the $\overline{n}$-th eigenstate of
the Harmonic oscillator Hamiltonian $H^{COIG}_{\cal E}$, 
\begin{eqnarray}
\label{s2xh}
\sigma^{2}_{x^{H}_{cm}(t)} =\frac{m^{2}\overline{\omega}^4 \overline{\alpha}^{4} (\overline{n} + \frac{1}{2})^{2} \hbar^{2} }{\lambda^2 \overline{M}^{2} ( \overline{\omega}^{2} + \overline{\alpha}^{2})^2} \Big( 
\frac{(\cos \overline{\omega} t - \cosh \overline{\alpha} t)^{2}}{\overline{\Omega}}   + \overline{\Omega} 
( \frac{\sin \overline{\omega} t}{\overline{\omega}} - \frac{\sinh \overline{\alpha} t}{\overline{\alpha}})^{2} \Big).
\end{eqnarray}
  
When present, a finite standard deviation of $\sigma_{\overline{q}}$ at time $t=0$ implies a finite $\sigma^{2}_{x_{cm}} = {\cal{O}}(1)$ at positive times. From Eq. (\ref{xcm}), this implies (similar to Eq. (\ref{central})) long range covariance between the local oscillator displacements, 
\begin{eqnarray}
 \frac{1}{N^{2}} \sum_{i=1}^{N} \sum_{j=1}^{N} (\langle x^H_{i}(t)  x^{H}_{j}(t) \rangle - \langle x^H_{i}(t) \rangle \langle  x^H_{j}(t) \rangle) = {\cal{O}}(1). \nonumber
 \end{eqnarray} 
 
 The results of Eqs. (\ref{xcmt}, \ref{s2xh}) undergo only a trivial change if the uniform
 coupling between $\overline{q}$ and the environment in $H^{IG}_{{\cal S}-{\cal E}}$ of Eq. (\ref{COIGM}) is generalized to any other bilinear coupling between the environment and system.
 For instance, we may replace $H^{IG}_{{\cal S}-{\cal E}}$ by   $(-N \lambda_{k} \overline{q}_{k}  x_{k}))$
 with the (un-normalized) Fourier mode $x_{k} = \frac{1}{N} \sum_{i=1}^{N} x_i e^{ikx}$ and
 $H^{COIG}_{\cal E}=  (\frac{1}{2} \overline{M} \overline{\Omega}_{k}^{2} \overline{q}_{k}^{2} + \frac{\overline{p}_{k}^{2}}{2 \overline{M}})$. In such a case, Eqs. (\ref{xht0},  \ref{omegaomega}, \ref{alphaomega}, \ref{xcmt}, \ref{s2xh}) will triivally hold with the substitutions 
 $\overline{\Omega} \to \overline{\Omega}_{k}$, $\lambda \to \lambda_{k}$, $x_{cm} \to (x_{k}/N)$,
 and $\overline{q} \to \overline{q}_{k}$. 
   
 \subsection{Central Oscillator-system oscillators Model}
\label{cosom}
We next consider local Harmonic oscillators,
\begin{eqnarray}
{\cal H}_{i} = \frac{p^{2}_{i}}{2m} + \frac{1}{2} m \omega_{i}^{2} x^{2}_{i},
\end{eqnarray}
with a global coupling to the environment of the form
\begin{eqnarray}
H^{COM}_{{\cal S}-{\cal E}} = - \lambda \overline{q} \sum_{i=1}^{N} x_{i}
\end{eqnarray}
and where $H^{COM}_{\cal{E}}$ is any function of $\overline{q}$ alone (i.e., no kinetic term, an infinitely ``heavy'' environment). In such a system, an effect of $ H^{COM}_{{\cal S}-{\cal E}}$ is  
to trivially shift the equilibrium positions ($x_i$) of all $N$ system oscillators from their initial value (that in the absence of coupling to the environment) by $\frac{\lambda}{m \omega_{i}^{2}} \overline{q}$. 
As a consequence, at all times $t$, 
\begin{eqnarray}
\label{xHxH}
\langle x^{H}_{i}(t) x^{H}_{j}(t) \rangle - \langle  x^{H}_{i}(t) \rangle  \langle  x^{H}_{j}(t) \rangle = \frac{\lambda^{2}}{m^{2} \omega_i^{2} \omega_j^{2}} (\langle \overline{q}^{2} \rangle - \langle \overline{q} \rangle^{2}). 
\end{eqnarray}
Eq. (\ref{xHxH}) also trivially holds for a shift of $\overline{q}$ by an arbitrary constant, $H^{C'OM}_{{\cal S}-{\cal E}} = - \lambda ({\sf q_{0}}+  \overline{q}) \sum_{i=1}^{N} x_{i}$ with ${\sf q}_{0}$ a general constant. Such a constant shift amounts to a constant displacement of the location of the oscillator equilibrium, $x_i \to (x_{i} - \frac{\lambda {\sf q}_{0}}{m \omega^{2}})$. As in the earlier models solved in this Section, if the environment ${\cal E}$ exhibits a finite standard deviation of $\overline{q}$ then Eq. (\ref{xHxH}) implies long range connected correlations amongst the local displacements $x^{H}_{i}$. 

\subsection{External fluctuating fields}
\label{external_fluctuate}

In the examples of Sections \ref{igtm} and \ref{cosom}, if $\{x_i\}$ portray the heights of the masses then the effect of the environment may be viewed as that of a gravitational field coupling linearly to $\{x_{i}\}$. In these models, however, the latter effective ``gravitational field'' features fluctuations. Similarly, in Section \ref{suchatrivialexampleatleastnowrefereeswillnotcomplain}, the external central spin acts as a transverse field with fluctuations that led to a finite variance of $\sigma^{2}_{\epsilon}$. 
Thus, in general, as in the models that we studied in this Section, we may consider systems with a global external field that displays fluctuations. Such models may be viewed as hybrid of procedures (1) and (2) of Section \ref{sec:sketch}. In these systems, there is an external environment driving the system (as in procedure (1)) exhibiting fluctuations (a variance) resulting from the environment being a bona fide quantum (and/or thermal) degree of freedom. Along similar lines, one may examine theories with background gauge fields that (like the collective degrees of freedom $\overline{q}$ that we studied thus far) are linearly coupled to the matter degrees of freedom and exhibit global fluctuations. If such global background fields feature a non-vanishing standard deviation then, repeating the same reasoning in the earlier parts of this Section, long range correlations may arise. 

\section{Dyson type expansions for general evolutions} 
\label{sec:Magnus}
To make progress beyond intuitive arguments and specific tractable systems, we next compute the standard deviation of the energy density (and, by trivial extension, any other intensive quantity $q$). Towards this end, we return to 
procedure (1) of Section \ref{sec:sketch} involving no external environment and examine Dyson type expansions for a general non-adiabatic \cite{adiabatic} time dependent Hamiltonian $H(t)$ (of which the piecewise constant Hamiltonians $H_{spin}$ and $H_{tr}$ (or $H_{Bose}$ and $H_{doping}$) are 
particular instances). Our calculation will demonstrate that in general situations, a finite $\sigma_{\epsilon}$ will arise. Via a Magnus expansion, the general evolution operator, the time ordered exponential 
${\cal{U}}(t) = {\cal{T}} \exp( - \frac{i}{\hbar} \int_{0}^{t} H(t') dt')$,
may be written as 
${\cal{U}} = \exp(\Omega(t))$ with $\Omega(t) = \sum_{k=1}^{\infty} \Omega_{k}(t)$ where
\begin{eqnarray}
\label{Magnus}
&&\Omega_{1}(t) = - \frac{i}{\hbar} \int_{0}^{t} dt_{1} H(t_{1}), \nonumber
\\ && \Omega_{2}(t) = - \frac{1}{2 \hbar^{2}} \int_{0}^{t} dt_{1} \int_{0}^{t_{1}} dt_{2} [H(t_{1}), H(t_{2})], \nonumber
\\ &&\Omega_{3}(t) = \frac{i}{6 \hbar^{3}} \int_{0}^{t} dt_{1} \int_{0}^{t_{1}} dt_{2} \int_{0}^{t_{2}} dt_{3} \Big([H(t_{1}), [H(t_{2}), H(t_{3})]] \nonumber
\\ && ~~~~~~~~~~~~~~~~~~~~~~~~
~~~~
+[H(t_{3}),[H(t_{2}), H(t_{1})]] \Big), \nonumber
\\  && ~~~~~~\cdots . 
\end{eqnarray}
We may apply the above Magnus expansion to a Heisenberg picture operator $A^{H}(t) = {\cal{U}} ^{\dagger} A {\cal{U}}$, with $A$ an arbitrary fixed operator, with the above $\Omega(t)$ and subsequently 
invoke the Baker-Campbell-Hausdorff formula $e^{-\Omega} A e^{\Omega} = A -[\Omega,A] + \frac{1}{2!} [\Omega,[\Omega,A]] - \frac{1}{3!} [\Omega,[\Omega,[\Omega,A]]] + \cdots$. If no change occurs at intermediate times $t$ and the Hamiltonian is that of the initial system  (i.e., $H(t) = H$) then, of course, the standard deviation $\sigma_{\epsilon}(t)$ will remain unchanged when computed with the (time independent) equilibrium density matrix for which it trivially vanishes. Similarly, if the evolution of $H(t)$ is adiabatic at all times then no broadening of the distribution $P(\epsilon')$ will arise. Our interest, however, lies in the Hamiltonians $H(t) \neq H$ necessary to elicit a change of the energy density $d \epsilon/dt \neq 0$ in a macroscopic system. In particular, we wish to examine the variance of the total energy density, 
\begin{eqnarray}
\label{ss-referee}
\sigma^2_{\epsilon}(t) =  \frac{1}{N^{2}} \Big(Tr (\rho (H^{H}(t))^2 ) -
(Tr (\rho H^{H}(t)))^{2} \Big), 
\end{eqnarray}
with $\rho$ the initial density matrix the system (time $t=0$) when cooling or heating commences.  
(In the dual examples considered in Section \ref{sec:dual}, $\rho = | \psi^{0} \rangle \langle \psi^{0}|$ with
$|\psi^{0} \rangle$ the initial spin or bosonic wavefunction.)
If $\sigma_{\epsilon}$ is to vanish identically then the resulting series for Eq. (\ref{ss-referee}) must vanish, order by order, 
for any $H(t)$. Collecting terms to the first two nontrivial orders in $H(t>0)$,  
\begin{eqnarray}
\label{expand}
\sigma_{\epsilon}^{2}(t) = && \sigma_{\epsilon}^{2}(0)  +  \frac{1}{N^{2}}   \langle [(\Delta H)^{2}, \Omega_{1}] \rangle \nonumber
\\ &&+  \frac{1}{2N^{2}} \Big( \langle 
[\Omega_{1},[\Omega_{1},(\Delta H)^{2}]] 
+[(\Delta H)^{2},\Omega_{2}] \rangle \nonumber
\\ && - \langle [(\Delta H), \Omega_{1}] \rangle^{2} \Big)
+ {\cal{O}}( (H(t>0))^{3}) .
\end{eqnarray} 
Here, $\langle - \rangle$ denotes an average computed with $\rho$ and $\Delta H \equiv (H-E_{0})$ where $H \equiv H(t=0)$ and $E_{0}$ is the initial energy $\langle H \rangle$. 
We emphasize that if, at all times $t$, the standard deviation vanishes identically for the heated/cooled system with the time dependent Hamiltonian, then the sum of all terms of a given order in $H(t>0)$ in the expansion of Eq. (\ref{expand}) must vanish for a general $H(t)$. In the special case $\rho = | \phi_{n} \rangle \langle  \phi_{n} |$ with $| \phi_{n} \rangle$ an eigenstate of $H$, the expectation values $ \langle [\Delta H, \Omega_{1}] \rangle =  [(\Delta H)^{2},\Omega_{2}] \rangle =0$. For this density matrix $\rho$, to order ${\cal{O}}((H(t>0))^{2})$, the standard deviation is given by the norm $\sigma_{\epsilon} = \Big| \frac{\Delta H (i\Omega_{1}(t))}{N} | \phi_{n} \rangle \Big|$ or (recalling that $\Delta H|\phi_{n} \rangle =0$ and consequently  
$\sigma_{\epsilon} = \frac{1}{N} \Big| [\Delta H, \Omega_{1}(t)] | \phi_{n} \rangle \Big|$), 
\begin{eqnarray}
\label{norm-good}
\boxed{\sigma_{\epsilon}(t)  =  
%
 \frac{1}{N\hbar } \Big|  \int_{0}^{t}  dt_{1}[H,  H(t_{1})] | \phi_{n} \rangle  \Big|.}
\end{eqnarray}
Because the total energy of the system changes with time (at an ${\cal{O}}(N)$ rate), the commutator of Eq. (\ref{norm-good}) cannot identically vanish and is, typically, of order ${\cal{O}}(N)$. Nonetheless, it is possible that when acting on the eigenstate $| \phi_{n} \rangle$, this commutator will yield a vector of size $o(N)$ and thus a vanishing contribution to $\sigma_{\epsilon}$ in the $N \to \infty$ limit. 

Indeed, as is to be expected, in the special product state setting of Section \ref{sec:product}, we will obtain a vanishing $\sigma_{\epsilon}$. Specifically, if for all $t$, the Hamiltonian $H(t)= \sum_{i=1}^{N'} {\cal{H}}_{i}(t)$ is a sum of decoupled commuting local operators that, act on the same $M=N'= {\cal{O}}(N)$ disjoint subspaces (Eq. (\ref{rhofac})), then the eigenstates $| \phi_{n} \rangle$ of $H(t=0)$ will be a product of $N'$ decoupled
states. Under the further constraint that, for all $t$, the operator norm $||{\cal{H}}_{i}(t)|| \le Y = {\cal{O}}(1)$, one observes that $ (\Delta H ) \Omega_{1}(t) | \phi_{n} \rangle$ becomes the sum of $N'$ orthogonal local product state vectors, each of which is of length ${\cal{O}}(1)$. 
Then, from Eq. (\ref{norm-good}), to second order in $H(t>0)$, 
\begin{eqnarray}
\label{local-good}
\sigma^{\sf local}_{\epsilon}(t) \lesssim \frac{t\sqrt{N'}}{\hbar N} Y^{2}.
\end{eqnarray}
Hence, to this order in the expansion of Eq. (\ref{expand}), for such local product states $| \phi_{n} \rangle$, we have $\lim_{N \to \infty} \sigma^{\sf local}_{\epsilon}(t) =0$. 

Contrary to Eq. (\ref{local-good}), however, for {\it general}  non-product state density matrices $\rho$ and non-adiabatic evolution of $H(t)$ (for which the commutators appearing in 
the series for $\sigma_{\epsilon}$ tend to zero), the norm of Eq. (\ref{norm-good}) does not identically vanish as $N \to \infty$ for all functions $H(t)$ and initial density matrices $\rho$ (even if $\rho$ is a stationary under an evolution with the initial Hamiltonian $H$). We stress that the perturbative result of Eq. (\ref{norm-good}) may, generally, be valid only for short times. Our aim in this Section is to illustrate that, generally, the standard deviation of the energy density does not vanish at all times. That a resulting $\sigma_{\epsilon}=0$ in a closed driven system cannot appear identically at all times is also evident from our exactly solvable examples. As noted above, the non-vanishing series expansion result illustrates that when the system starts from an equilibrium state with a sharp energy density $\sigma_{\epsilon}(0) =0$, then notwithstanding any locality of the Hamiltonian, $\sigma_{\epsilon}$ may become finite (i.e., ${\cal{O}}(1)$) at later times $t$. Additional aspects and further connection with ``wave packet'' analogy of Section \ref{intuition} are discussed in \ref{app-boost}. 

The Dyson type expansion analysis is not limited to the energy density $\epsilon$ (similar results hold for any other intensive quantity $q$) nor to specific continuum or lattice systems. Thus, broad distributions may generally arise in systems displaying an evolution of their intensive quantities. Of course, constrained solutions to the equation $\sigma_{\epsilon}(t)=0$, at all times $t$, may be engineered. Indeed, particular solutions associated with operators that translate the system spectrum bring to life the intuitive analogy that we made with wave packets (Section \ref{intuition}) as well as the special character of product states 
(Section \ref{sec:product}). Similar results may also appear for classical systems; using Weyl quantization, the commutators in Eqs. (\ref{Magnus}, \ref{expand}) are replaced (to lowest order in powers of $\hbar$) by the corresponding Poisson brackets and all averages are evaluated with the classical probability density instead of the quantum probability density matrix $\rho$. Eqs. (\ref{Magnus}, \ref{expand}) are indeed solely a consequence of the ``canonical'' time evolution of the system given by Hamilton's equations in the classical arena (replacing the quantum Heisenberg equations of motion). We next study another situation in which a finite $\sigma_{\epsilon}>0$ arises rather trivially. 

\section{Short time averaged probability distribution}
\label{section:short}

The results of this Section are motivated by and will also apply to averages in classical systems. 
We examine a time averaged probability density matrix on ${\cal{S}}$,
\begin{eqnarray}
 \label{run_average}
 \rho_{\tilde{\tau}}(t) \equiv \frac{1}{\tilde{\tau}} \int_{t}^{t+\tilde{\tau}} \rho(t') dt'.
 \end{eqnarray} 
 Here, $\rho(t') =  {\cal{U}}(t) \rho {\cal{U}}^\dagger(t)$ is the (instantaneous) density matrix in the Schrodinger picture. 
 Ref. \cite{Fagotti} studied numerous aspects of the probability densities $\rho_{\tilde{\tau}}(t)$ for lattice spin systems. 
 We remark that, arguably, any real measurement of a macroscopic quantity $Q$ in large ``semi-classical'' systems is not instantaneous but rather requires a finite period of time $\tilde{\tau}$; thus the observed values correspond to $Tr(\rho_{\tilde{\tau}}(t) Q)$. Averaging with this probability distribution,
 \begin{eqnarray}
 \label{var-run}
&&\Big\langle \Big(\frac{H}{N} \Big)^{2}\Big\rangle_{\tilde{\tau}} \equiv \frac{1}{N^{2}} Tr( \rho_{\tilde{\tau}}(t) H^{2}) \nonumber
\\ = &&  \int_{t}^{t+\tilde{\tau}} \frac{Tr( \rho(t') H^{2})}{N^{2}\tilde{\tau}} dt'   \ge \int_{t}^{t+\tilde{\tau}} \frac{(Tr( \rho(t') H))^{2}}{N^{2}\tilde{\tau}} dt' \nonumber
\\ = &&\frac{1}{\tilde{\tau}} \int_{t}^{t+\tilde{\tau}} \epsilon^{2}(t') dt'.
 \end{eqnarray}
 Similarly, 
 \begin{eqnarray}
 \label{run--}
 \Big\langle \frac{H}{N} \Big\rangle_{\tilde{\tau}}  =  \frac{1}{\tilde{\tau}} \int_{t}^{t+\tilde{\tau}} \epsilon(t') dt'.
 \end{eqnarray}
 Hence,  
 $\sigma_{\epsilon,\tilde{\tau}}^{2} \equiv \Big\langle \Big(\frac{H}{N} \Big)^{2}\Big\rangle_{\tau}  -   \Big\langle \frac{H}{N} \Big\rangle_{\tau} ^2$ 
 will be finite for an energy density $\epsilon$ that varies at a finite rate in the interval $[t, t+ \tilde{\tau}]$. For a short time interval in which $\frac{d \epsilon}{dt'}$ is approximately constant, Taylor expanding 
 $\epsilon(t')$ to linear order in $(t'-(t+\frac{\tilde{\tau}}{2}))$ in the integrands of Eqs. (\ref{var-run}, \ref{run--}), 
 \begin{eqnarray}
 \label{twelve}
 \sigma_{\epsilon, \tilde{\tau}} \gtrsim \frac{\tilde{\tau}}{\sqrt{12}} \Big| \frac{d \epsilon}{dt} \Big|.
 \end{eqnarray}
 Putting all of the pieces together, we see, from Eq. (\ref{central}), that macroscopic range $\overline{G}>0$ will appear when all correlations evaluated with the time averaged density matrix $ \rho_{\tilde{\tau}}(t) $ of Eq. (\ref{run_average}). Albeit being trivial, this result is extremely general and applies to all density matrices and Hamiltonians whenever $\frac{d \epsilon}{dt} \neq 0$. Returning to the opening sentence of this Section, the inequalities of Eqs. (\ref{var-run}, \ref{twelve}) indeed also hold for classical systems (with the trace in Eq. (\ref{var-run}) replaced by phase space integrals or other sum over classical microstates and $\rho$ being a classical probability distribution). In classical ergodic systems, equilibrium (and various non-equilibrium) phase space probability distributions have their conceptual origin in long or finite time averages: an equilibrium ensemble average reproduces the long time expectation values. Although it is somewhat obvious, it is nonetheless important to emphasize that, in the quantum arena, having an instantaneous density matrix that is a product state does not imply a time averaged density matrix that is also a product state. This is much the same as the two spin $S=1/2$ density matrix $\frac{1}{2} (|\uparrow \uparrow \rangle \langle \uparrow \uparrow | + | \downarrow \downarrow \rangle \langle \downarrow \downarrow|) $; the latter is an average of the density matrices of 
 two product states yet it is not, of course, the density matrix of a product state. 
 
 In the next Section, we will demonstrate that under certain conditions, $\sigma_{\epsilon}$ must be finite when $d \epsilon/dt\neq 0$. The converse, however, does not follow: a static energy density $\epsilon$ does not imply that 
 $\sigma_{\epsilon,\tilde{\tau}}$ and $\sigma_{\epsilon}$ are zero. In Sections \ref{deviation} and \ref{2b2b}, we will further discuss what occurs once the system is no longer driven. Apart from its form as a time average expectation value, our result of Eq. (\ref{twelve}) implies that there are states for which the standard deviation $\sigma_{\epsilon}>0$ when the latter is evaluated for instantaneous expectation values in mixed and pure states evolving under a piecewise constant $H(t)$ (such as that of Section \ref{sec:dual}). To this end, we may equate $\rho_{\tau}$ to be the instantaneous density matrix $\rho^{new}(t)$ of a new mixed state or, alternatively, to be the partial trace of the density matrix of a pure state defined on an artificially constructed volume ${\cal{I'}}$ larger than the system volume 
 $({\cal{S}}$) on which the Hamiltonian $H$ acts (${\cal{I'}} =  {\cal S} \cup {\cal{E'}}$ with ${\cal{E'}}$ an artificially constructed ``environment'') following the ``purification'' procedure of \cite{lieb-pure,pure}. In the notation of \cite{pure}, the dimension ${\cal{D}}$ will correspond  to the number of time steps in a discretization of the integral of Eq. (\ref{run_average}). Herein, 
given original pure states $\{|\psi(t') \rangle\}$ (with $t'=t + {\sf k} \tau/{\cal{D}}$ with integer $1 \le {\sf k} \le {\cal{D}}$), the scaled density matrices $ \frac{| \psi(t') \rangle \langle \psi(t')|}{\tau}$ may be summed, as in Eq. (\ref{run_average}), to provide an instantaneous density matrix $\rho^{new}(t)$. The latter density matrix may, following \cite{pure}, be constructed such that its partial trace over the environment ${\cal{E'}}$ yields 
$\rho_{\tilde{\tau}}(t)$ (i.e., $\rho_{\tilde{\tau}}(t) =\rho^{new}(t)= Tr_{\cal{E'}}|\Psi(t) \rangle \langle \Psi(t) |$ with $| \Psi(t) \rangle$ a pure state in ${\cal{I'}}$). This demonstrates, once again, that the standard deviation $\sigma_{\epsilon}$ as evaluated with instantaneous probability density matrices or pure states can be trivially finite even for local Hamiltonians $H$.

\section{Generalized two-Hamiltonian uncertainty relations}
\label{2-Ham}

We next turn to a more specific demonstration that, in other settings, when evaluated with the instantaneous density matrix, the standard deviation $\sigma_{\epsilon}>0$ when the energy density exhibits a finite rate of change. In this Section, we consider non-relativistic systems ${\cal S}$ of arbitrary size $N$ (large or small) that satisfy certain conditions in the order of decreasing generality. \newline

We first derive exact inequalities for closed system-environment hybrids and discuss, once again, how our results relate to causality. We will then derive exact bounds for open system-environment hybrids. In this Section, we will formalize and study procedure (2) of Section \ref{sec:sketch}. We will explicitly include the effects of the environment. In Sections \ref{section:closed} and \ref{sec:patuach}, we will respectively analyze, situations in which the ensuing system-environment hybrids constitute larger closed or open hybrid systems. 
 
 \subsection{Closed system-environment hybrids}
 \label{section:closed}
 
 \subsubsection{Exact Inequalities for closed system-environment hybrids}


 In this subsection, we will derive inequalities when the following assumptions are satisfied: \newline

\noindent{\bf{ Assumption {(1)}}}: When combined with their physical environment (or ``heat bath'') ${\cal{E}}$, these systems constitute
a larger global {\it closed isolated} hybrid system ${\cal{I}} =  {\cal S} \cup {\cal{E}}$ (of ${\tilde{N}}$ sites) such that the sites in ${\cal S}$ do not interact with any sites that are not in ${\cal{I}}$. 
The number of particles or sites in both ${\cal S}$ and ${\cal{E}}$ is held fixed. $\diamond$ \newline

\noindent{\bf{ Assumption (2 - weak version)}}: The Hamiltonian $H$ describing ${\cal S}$ is time independent.  $\diamond$\newline

We stress that the Hamiltonian ${\tilde{H}}$ describing the full hybrid system ${\cal{I}}$ including interactions between ${\cal S}$ and ${\cal{E}}$ 
is, at this stage, kept general and may depend on time. \newline

Denoting the evolution operator (first discussed in Eq. (\ref{usfe})) of the full closed hybrid system ${\cal{I}}$ by
\begin{eqnarray}
\label{ute-trail}
{\cal{\tilde{U}}}(t) = {\cal T}  \exp (-\frac{i}{\hbar} \int_{0}^{t} \tilde{H}(t') dt'),
\end{eqnarray} 
the two Heisenberg picture Hamiltonians $H^{H}(t)= {\cal{\tilde{U}}}^{\dagger}(t) 
H {\cal{\tilde{U}}}(t) (t)$ and $\tilde{H}^{H}(t) ={\cal{\tilde{U}}}^{\dagger}(t)  \tilde{H} {\cal{\tilde{U}}}(t)$ describe, respectively, the open system ${\cal S}$
and the larger closed hybrid system ${\cal{I}}$ at time $t$. The energy of the system ${\cal S}$ is $E(t)  = Tr_{\cal{I}} (\tilde{\rho} H^{H}(t))$
where $\tilde{\rho}$ is the initial density matrix of ${\cal{I}}$. 
By the uncertainty relations \cite{mix1,mix2},
\begin{eqnarray}
\label{uncert}
\sigma_{\epsilon(t)} \sigma_{\tilde{H}^H(t)} \ge \frac{1}{2} \Big| Tr_{\cal{I}} (\tilde{\rho}[\frac{H^{H}(t)}{N},\tilde{H}^{H}(t)]) \Big|.
\end{eqnarray}
Here, $\sigma_{\epsilon(t)}$ and $\sigma_{\tilde{H}^H(t)}$ denote, respectively, the uncertainties associated with $H^{H}(t) /N$ and $\tilde{H}^{H}(t)$ (when these uncertainties are computed with the probability density matrix $\tilde{\rho}$). Combined with the Heisenberg equations of motion for the time independent $H$ (Assumption  (2- weak version)), we obtain an extension of the time-energy uncertainty relations for this two Hamiltonian realization, 
\begin{eqnarray}
\label{uncertain}
 \boxed{\sigma_{\epsilon(t)} \sigma_{\tilde{H}^{H}(t)} \ge \frac{\hbar}{2N} \Big|\frac{dE}{dt} \Big|.}
\end{eqnarray} 
Eq. (\ref{central}) then implies a lower bound on the average macroscopic range correlators in the subsystem ${\cal S}$, 
\begin{eqnarray}
\label{uncertain*}
{\overline{G}_{{\cal{S}}}}  \ge \frac{\hbar^{2}}{4 \sigma^{2}_{\tilde{H}^{H}(t)}} \Big(\frac{d\epsilon}{dt} \Big)^{2}.
\end{eqnarray}

The derivative in Eq. (\ref{uncertain*}) scales as ${\cal{O}}(N^{2})$ 
if the energy $E(t)$ of ${\cal S}$ changes at a rate proportional to the size of ${\cal S}$ (i.e., if the energy density changes at a finite rate). Eqs. (\ref{uncert},\ref{uncertain},\ref{uncertain*}) will remain valid if Assumption (1) is relaxed, i.e., if ${\cal{I}}$ is an open system with a Hamiltonian $\tilde{H}$ that, itself, is in contact with a yet larger system. We next examine what occurs if the local energy density correlators $G_{ij}$ decay with a correlation length $\xi$, i.e., with
\begin{eqnarray}
\label{gc+}
G_{ij} \sim A \frac{e^{-|i-j|/\xi_{H}}}{|i-j|^{p}},
\end{eqnarray}
with $A$ a finite constant. Transforming to hyperspherical coordinates, we see that on a $d$ dimensional hypercubic $L \times L \times ... \times L$ lattice with $L \gg \xi_{H}$, the average correlator of 
Eq. (\ref{central}) will, up to factors of order unity, be given by 
${\overline{G}_{{\cal{S}}}}  \sim 2 A  \frac{\pi^{d/2} \Gamma(d-p)}{\Gamma(\frac{d}{2})} (\frac{\xi_{H}}{L})^{d} \xi_{H}^{-p}$.
Combined with Eq. (\ref{uncertain*}), this implies a lower bound on the correlation length
\begin{eqnarray}
\label{xH}
\xi_{H} \gtrsim  L^{\frac{d}{d-p}}  \Big[\frac{\hbar^{2} \Gamma(\frac{d}{2})}{8 A  \pi^{d/2}  \Gamma(d-p) \sigma^{2}_{\tilde{H}^{H}}(t)}  \Big(\frac{d \epsilon}{dt} \Big)^{2}  \Big]^{1/(d-p)},
\end{eqnarray}
with $\epsilon(t) =E(t)/N$ the energy density of ${\cal{S}}$. Note that the lower bound of Eq. (\ref{xH}) on the correlation length is monotonic in the temporal variation of the energy density $\epsilon(t)$. That is, the larger the rate of change $|\frac{d \epsilon}{dt}|$ of the energy density, the larger the lower bound on the putative finite correlation length $\xi_{H}$. In particular, for finite $d \epsilon/dt$ and $\sigma_{\tilde{H}^{H}}$, such a lower bound will diverge as $L \to \infty$ (indicating that an assumption of small $\xi_{H}$ cannot be made self-consistently). Moreover, regardless of $p$, {\it if (in any dimension) $\sigma_{\tilde{H}^{H}} < {\cal{O}}(\sqrt{N})$ then Eq. (\ref{xH}) illustrates that $\xi_H$
cannot be finite in the 
$L \to \infty$ limit whenever $d \epsilon/dt$ is finite}. Thus, the reader can see how {\it divergent correlation lengths are mandated} whenever ${\cal{I}}$ exhibits fluctuations that are smaller than those of typical open systems
(i.e., when $\sigma_{\tilde{H}^{H}} =o(\sqrt{N}$)). The bound of Eq. (\ref{xH}) assumes Eq. (\ref{gc+}) and is only suggestive. In what follows, we will examine conditions that will enforce a finite $\sigma_{\tilde{H}^{H}}$ and thus divergent correlations when $d \epsilon/dt \neq 0$. Towards that end, we impose a more restrictive condition: \bigskip

\noindent{\bf{ Assumption {(2 - strong version)}} }: The {\it fundamental} interactions appearing in the global Hamiltonian $\tilde{H}$ describing ${\cal{I}}$ 
 are time independent. $\diamond$ \newline
 
This assumption (which, for brevity, we will henceforth simply refer to as {\bf{Assumption (2}})) implies Assumption (2 - weak version). This is so since the terms in $\tilde{H}$ include, as a subset, the interactions appearing in the Hamiltonian $H$ describing ${\cal S}$. When Assumption (2) holds, time dependence arises when the density matrix $\tilde{\rho}$ is not diagonal in the eigenbasis of the full Hamiltonian $\tilde{H}$. 

In what briefly follows, we make general colloquial remarks motivating our final result. We will then invoke a last assumption (either of Assumptions (3) or (3') to be introduced below), with the aid of which we will be able to rigorously derive our result. When Assumption (2) holds, the global Heisenberg and Schrodinger picture Hamiltonians coincide, $\tilde{H}^{H}(t)  ={\tilde{H}}$. If a time independent Hamiltonian ${\tilde{H}}$ governs the dynamics of the closed hybrid system ${\cal{I}}$, then the energy will not vary with time. Classically, there is no meaningful finite standard deviation $\sigma_{\tilde{H}}$: the energy of the closed system is conserved. By contrast, no quantum dynamics are possible unless $\sigma_{\tilde{H}} \neq 0$. That is, any eigenstate of $\tilde{H}$ (for which $\sigma_{\tilde{H}} =0$) is trivially stationary under an evolution with $\tilde{H}$. For a general initial state $| \tilde{\psi}^{0} \rangle$ of the closed hybrid system ${\cal{I}}$, the probability density, 
 \begin{eqnarray}
 \label{rhl}
 \tilde{\rho}(t) = \sum_{\tilde{n} \tilde{m}}e^{-i \frac{(\tilde{E}_{\tilde{n}} - \tilde{E}_{\tilde{m}})t}{\hbar}}  \langle \tilde{\phi}_{\tilde{n}} | \tilde{\psi}^{0} \rangle \langle  \tilde{\psi}^{0} |   \tilde{\phi}_{\tilde{m}} \rangle |\tilde{\phi}_{\tilde{n}} \rangle  \langle \tilde{\phi}_{\tilde{m}}|, 
 \end{eqnarray}
will typically vary on a time scale of order $\tau \equiv \frac{\hbar}{\sigma_{\tilde{H}}}$. In Eq. (\ref{rhl}), $\{|   \tilde{\phi}_{\tilde{n}} \rangle\}$ are the eigenstates of $\tilde{H}$. 
The off-diagonal spread of the density matrix (in the eigenbasis of $\tilde{H}$) determines the oscillation frequencies that it displays. For pure states $| \tilde{\psi}^{0} \rangle$ in the closed hybrid system ${\cal{I}}$, a large $\sigma_{\tilde{H}}$ implies large temporal fluctuations \cite{mix-stationary}. If, as in many closed energy conserving systems with a well-defined semi-classical limit, the representative frequencies governing the global dynamics (and probability density) do not scale with $N$, i.e., if ${\cal{O}}(\tau) = {\cal{O}}(1)$ \cite{example-oscillator} 
then $\sigma_{\tilde{H}}$ will, typically, also not 
 vary with $N$. Inserting $\sigma_{\tilde{H}} = \frac{\hbar}{\tau}$  in Eq. (\ref{uncertain}), 
 \begin{eqnarray}
 \label{average_t_tau}
 \sigma_{\epsilon(t)} \ge \frac{\tau}{2}  \Big| \frac{d\epsilon}{dt} \Big|.
 \end{eqnarray}
 This result is natural for a probability distribution that varies over time scales $\gtrsim \tau$. Along related lines, a time average of the form of Eq. (\ref{run_average}) applied to the density matrix $\tilde{\rho}$ on ${\cal{I}}$ (i.e., $\tilde{\rho}_{\tilde{\tau}}(t) \equiv \frac{1}{\tilde{\tau}} \int_{t}^{t+\tilde{\tau}} \tilde{\rho}(t') dt'$) will remove frequencies higher than a cutoff that scales as $\hbar/\tilde{\tau}$. That is, if $\tau<\tilde{\tau}$, then $\tilde{\rho}_{\tilde{\tau}}(t)$ will not exhibit the higher frequency oscillations present in $\tilde{\rho}(t) $. The removal of these high frequencies (associated with 
 short ``virtual events'') will render the system more ``semi-classical''; in a path integral representation, in the sum of the exponentiated classical action over all possible paths, fluctuations of phases generated by relative energy differences larger than ${\cal{O}}(\hbar/\tilde{\tau})$ will, for an evolution over a time of length $\tilde{\tau}$, lead to oscillatory phases that will cancel. The larger the waiting or averaging time $\tilde{\tau}$ is, the more narrow the range of eigenstates that are relevant to the system evolution will be (i.e., only those 
 with energies in a small window about the average system energy may be considered) on time scales $\ge \tilde{\tau}$.  
  
The above intuition can be made more accurate to bolster the considerations of Section \ref{section:short}. The bound of Eq. (\ref{uncert}) is an algebraic identity that may be extended to arbitrary probability density matrices. In particular, in Eq. (\ref{uncert}), we may replace $\tilde{\rho} \to \tilde{\rho}_{\tilde{\tau}}$ for general averaging times $\tilde{\tau}$ (Eq. (\ref{run_average})). 
This implies the inequality
\begin{eqnarray}
\label{uncertain"*}
 \sigma^{\tilde{\tau}}_{\epsilon(t)} \sigma^{\tilde{\tau}}_{\tilde{H}(t)} \ge \frac{\hbar}{2N} \Big|\frac{dE_{\tilde{\tau}}}{dt} \Big|,
\end{eqnarray} 
where $E_{\tilde{\tau}}(t)  \equiv Tr_{\cal{I}} (\tilde{\rho}_{\tilde{\tau}} H^{H}(t))$. In Eq. (\ref{uncertain"*}), $\sigma^{\tilde{\tau}}_{\epsilon(t)}$ and  $\sigma^{\tilde{\tau}}_{\tilde{H}(t)}$ denote, respectively, the standard deviations of $(H/N)$ and $\tilde{H}$ as computed with the time averaged probability distribution $\tilde{\rho}_{\tilde{\tau}}$. Thus, we can qualitatively relate the uncertainty relations to the trivial general bounds of Eqs. (\ref{var-run}, \ref{twelve}). 
That is, for any finite (system size independent) averaging time $\tilde{\tau}$, the density matrix $\tilde{\rho}_{\tilde{\tau}}(t) $ will display $\sigma_{\tilde{H}} \le \hbar/\tilde{\tau}$. Eq. (\ref{uncertain"*}) will (in agreement with 
Eqs. (\ref{var-run}, \ref{twelve})) then imply a finite 
$ \sigma^{\tilde{\tau}}_{\epsilon(t)} $ whenever $\frac{dE_{\tilde{\tau}}}{dt}$ is extensive. As emphasized earlier, of physical relevance are finite time ($\tilde{\tau}>0$) window measurements. 

While bounded system size independent frequencies are natural in quasi-classical and ``typical'' closed (energy conserving) quantum systems, that is certainly not the case for all constructible model states \cite{special1}. With this in mind, we consider the consequences of any one of two additional conditions 
(labelled Assumption (3) and Assumption (3') in the below). Either of these conditions will lead to a system size independent standard deviation for the energy density (when the latter is evaluated with the instantaneous density matrix $\tilde{\rho}$). 

\bigskip
\noindent{\bf{  Assumption {(3)}}}: The closed hybrid system ${\cal{I}}$ equilibrates at long times. Stated more precisely (and automatically accounting for Poincare recurrence type events), the asymptotic long time average of the probability density $\rho_{\cal{I}}$ in the larger {\it closed} hybrid system ${\cal{I}}$ veers towards the {\it microcanonical}  ({\sf mc}) density matrix applicable for closed energy conserving systems in equilibrium \cite{Huang-book}. That is, 
\begin{eqnarray}
\label{rhomicro}
\tilde{\rho}_{\sf mc;  {\cal{I}}} = \lim_{{\tilde{{\cal{T}}}} \to \infty} \frac{1}{{\tilde{{\cal{T}}}}} \int_{0}^{{\tilde{{\cal{T}}}}} \tilde{\rho}_{\cal{I}} (t') dt', 
\end{eqnarray}
with $\tilde{\rho}_{\sf mc; {\cal{I}}}$ the microcanonical ensemble density matrix for the closed hybrid system ${\cal{I}}$.
$\diamond$
\bigskip

In systems obeying Eq. (\ref{rhomicro}), the uncertainty in the energy of ${\cal{I}}$ at asymptotically long times (i.e., as computed with $\tilde{\rho}_{\sf mc; {\cal{I}}}$) will be system size independent, 
\begin{eqnarray}
\label{sH1}
\sigma_{\tilde{H}} = {\cal{O}}(1).
\end{eqnarray}
Eq. (\ref{sH1}) constitutes the defining textbook property of the microcanonical ensemble \cite{Huang-book}.
Since the closed system-environment hybrid ${\cal{I}}$ is governed by the time independent Hamiltonian $\tilde{H}$, the standard deviation $\sigma_{\tilde{H}}$ is time independent and Eq. (\ref{sH1}) trivially holds at all times $t$ when the variance $\sigma_{\tilde{H}}$ is computed with the density matrix $\tilde{\rho}(t)$. Assumption (3) and the preceding discussion may seem abstract. The semiclassical intuition underlying the somewhat axiomatic standard definition of the microcanonical ensemble is rather trivial. We repeat anew some elements below.

 For a classical ergodic hybrid system (e.g., that assumed for ${\cal{I}}$ governed by the time independent ${\tilde{H}}$), the probability density is that associated with the long time average. For a {\it closed} conservative system,
the total energy is conserved and the probability density defined in this way exhibits zero variance of the total energy. In the quantum arena, if the closed ergodic system exhibits non-trivial dynamics then the standard deviation of its Hamiltonian cannot vanish (since the eigenstates of ${\tilde{H}}$ are trivially stationary). Thus, the common assumption underlying the microcanonical ensemble is that the standard deviation of ${\tilde{H}}$ is finite (in order to allow for non-vanishing frequencies) yet, for {\it a closed system} does not diverge as the size increases. This intuition rationalizes the standard use of Eq. (\ref{sH1}) defining the microcanonical ensemble. 
 \newline
 \newline
 In the spirit of the above maxim, we next introduce an alternate assumption that does not rely on the closed hybrid system ${\cal{I}}$ being ergodic (nor the use of 
 the microcanonical ensemble):
 \newline
 \newline
\noindent{\bf{ Assumption {(3')}}}: A finite time step discretization ($t=t_{\sf k} = {\sf k} \Delta t$ with integer ${\sf k}$ and $\Delta t$ a sufficiently small system size independent time step) may effectively simulate the evolution of ${\cal{I}}$. Here, as before, the (pure) state of the {\it closed} hybrid system ${\cal{I}}$ may be described by a wavefunction. The uniform discretization of $t$ implies that any function $f(t)$ (including the associated density matrix $\tilde{\rho}(t)$ of Eq. (\ref{rhl})) may be expressed as a Fourier sum $f(t) = \sum_{p'} \hat{f}(\omega_{p'}) e^{-i \omega_{p'} t}$ with $\omega_{p'}$ lying in the ``first Brillouin zone'' ($|\omega_{p'}| \le \pi/\Delta t$). Thus, the uncertainty in the energy of the closed hybrid system ${\cal{I}}$ satisfies $\sigma_{\tilde{H}} \le   \pi \hbar  /\Delta t$- a realization of Eq. (\ref{sH1}); gauge invariant \cite{explain_gauge} expectation values of finite time gradients in ${\cal{I}}$ (including the standard deviation of the discrete time gradient approximation of the Hamiltonian $\tilde{H}= i \hbar \frac{\partial} {\partial t}$) are bounded from above by ${\cal{O}}(1/\Delta t$). $\diamond$
\newline
\newline
 Assumptions (1-3) (as well as Assumptions (1,2,3')) \cite{product33} imply that when the energy density varies at a finite rate ($dE/dt = {\cal{O}}(N)$) then, from
Eqs. (\ref{uncertain},\ref{sH1}), the standard deviation of the energy density of ${\cal{S}}$,  
\begin{eqnarray}
\label{central'}
\boxed{\sigma_{\epsilon(t)} = {\cal{O}}(1).}
\end{eqnarray} 
Thus, we discern from Eqs. (\ref{central}, \ref{uncertain*}) that long range correlations must appear during the cooling or heating period at which the energy density of the system (${\cal S}$) is varied at a finite rate. Analogs of Eq. (\ref{central'}) are also valid for any other intensive quantity $q$ (different from the energy density $\epsilon$) whenever $\frac{dq}{dt} \neq 0$. Analogs of Eq. (\ref{central'}) are also valid for any other intensive quantity $q$ (different from the energy density $\epsilon$) whenever $\frac{dq}{dt} \neq 0$. When the environment ${\cal{E}}$ is included for (as we do now), the evolution 
of the system itself (Figure \ref{dyn_map.}) is, generally, non unitary; this non unitary evolution lies in strong contrast to the earlier examples of Section \ref{sec:dual} in which the system evolved unitarily. One may, nonetheless, still make some non-rigorous pedagogical contact with the spin models of Section \ref{sec:dual}
for a special case exhibiting unitary time evolution \cite{example1-3} for which all of the above three assumptions hold. 
These assumptions are not met (in particular Assumption (3) does not hold) for the rather artificial (yet exactly solvable) models of Section \ref{sec:CSM} \cite{comment-CSM*}. Assumptions (1-3) are often employed in standard textbook derivations of the canonical ensemble for open systems ${\cal S}$ by applying the microcanonical ensemble averages for the larger equilibrated closed systems ${\cal{I}}$ that include the relevant environments ${\cal{E}}$ that are in contact (or ``entangled'') with ${\cal S}$. If, as evinced by measurements in prototypical states in the composite hybrid system ${\cal{I}}$ at asymptotically long times, ergodicity and equilibrium set in, then the microcanonical ensemble may be invoked. 
We next turn to the scales of the righthand sides of Eqs. (\ref{uncert},\ref{uncertain},\ref{uncertain*}) and their consequence for systems that are cooled/heat at finite rate. By Heisenberg's equation for the time independent Hamiltonian $H$, 
\begin{eqnarray}
\label{HHtH}
\frac{dH^{H}}{dt} = \frac{i}{\hbar} [\tilde{H}, H^{H}].
\end{eqnarray}
Therefore, in order to obtain a finite rate of change of the system energy density $d \epsilon/dt$ (or an extensive rate $dE/dt$), the total Hamiltonian $\tilde{H}$ of the large hybrid system ${\cal{I}}$ must have a commutator with the Hamiltonian $H$ of ${\cal S}$ that is of order $N$, i.e., $Tr_{\cal{I}} (\tilde{\rho} [\tilde{H}, H^H]) = {\cal{O}}(N)$. 
Hence, to achieve a finite global rate of cooling/heating, $\tilde{H}$ must couple to an {\it extensive number of sites in the volume of ${\cal S}$}-
it is not possible to obtain an extensive cooling/heating rate by a bounded strength coupling that extends over an infinitesimal fraction of the system size 
(see also the discussion at the end of Section \ref{intuition} and that appearing after Eq. (\ref{htr}) in Section \ref{sec:gspin}). Effectively, a finite fraction of the sites lying in the volume of ${\cal S}$ must couple to $\tilde{H}$ whenever $\frac{d \epsilon}{dt} = {\cal{O}}(1)$. The initial state of the system ${\cal{S}}$ prior to its cooling/heating (or variation in its other parameters) may have a well defined energy density $\epsilon$ and other state variables yet nonetheless still be far from a typical equilibrium state. One may introduce various probes, clocks, etc., that start the cooling/heating process in a particular way; the initial state need not be in equilibrium but may rather be specially crafted. We further wish to further underscore that the value of the (nearly) constant energy of the closed hybrid ${\cal{I}}$ (up to corrections that do not increase with the system size) as captured by Assumption (3) (as well as Assumption (3')) imply constraints between the environment ${\cal{E}}$ and the system ${\cal{S}}$. Thus, qualitatively, the resulting picture (literally and figuratively) is in accord with the schematic of Figure \ref{coupling.} with the same environment ${\cal{E}}$ coupling to a finite fraction of all sites in the system. Indeed, if this is not the case and the environment ${\cal{E}}$ is composed of ${\cal{O}}(N)$ microscopic decoupled reservoirs with each of these reservoirs independently, coupling to another local region of ${\cal{S}}$ (such that ${\cal{O}}(N)$ independent local system-environment hybrids appear each having a conserved energy up to ${\cal{O}}(1)$ fluctuations) then the total energy of ${\cal{I}}$ will exhibit ${\cal{O}}(N^{1/2})$ fluctuations (a sum of the $N$ independent random errors with each of these errors being of order unity). In such instances, the energy of the closed hybrid ${\cal{I}}$ would not be remain constant, up to system size independent errors, in time in the thermodynamic limit. 

If Assumptions (1-3) are met then at asymptotically long times, memory of the initial state will be lost and all observables may be computed via the microcanonical ensemble with its few thermodynamic state variables. In particular, the defining feature of the microcanonical probability distribution of closed equilibrated systems holds, Eq. (\ref{sH1}). 
For completeness, we note that the Dyson type expansion of Section \ref{sec:Magnus} may also be reproduced in the setting of the current subsection with a time independent $\tilde{H}$ (for which the evolution operator is $e^{-i \tilde{H} t/\hbar}$ and the global density matrix is given by $\tilde{\rho}$). 

\subsubsection{Remarks on causality}
In the earlier part of this subsection, the effect of the environment ${\cal{E}}$ driving the system was explicitly included and, as in basic theories, the form of the terms in the system-environment hybrid (i.e., those in $\tilde{H}$) was time independent. While the form of the fundamental interactions in 
$\tilde{H}$ is time independent, tracing over the environment (Figure \ref{dyn_map.}) may lead to complex dynamical maps. We now revisit, yet again, the constraints implied by causality.  As noted in Section \ref{intuition}, in models with local interactions, compounding relativistic bounds, Lieb-Robinson inequalities \cite{Lieb_Robinson} generally provide upper bounds on commutators in non-relativistic systems (such as those appearing in Eq. (\ref{uncert})). These relations lead to bounds on correlations \cite{sergey,jens2}. However, as we explained above, in driven systems for which the energy density is made to vary at a finite rate, commutators such as those of Eq. (\ref{uncert}) must be extensive; such commutators may only appear at sufficiently long times (we refer the reader, once againm to \ref{LR_explain} for an explicit proof of this assertion). In diverse physical situations (i.e., when cooling/heating leads to a finite rate of change of the system energy density or measured temperature), photons and/or other particles/quasiparticles emitted/absorbed by an extensive volume of the surrounding heat bath effectively couple to the system bulk (see \ref{extensive}). In the spin model of Section \ref{sec:gspin} (in which the system evolution was unitary), the time independent (for all times $t>0$) transverse field  ($B_{y}$) Hamiltonian of Eq. (\ref{htr}) played the role of $\tilde{H}$ acting on all $N$ sites (so as to have $[\tilde{H}, H^H]={\cal{O}}(N)$).

\subsection{Open system-environment hybrids}
\label{sec:patuach}

\begin{figure*}
	\centering
	\includegraphics[width=.8  \columnwidth, height=.17 \textheight, keepaspectratio]{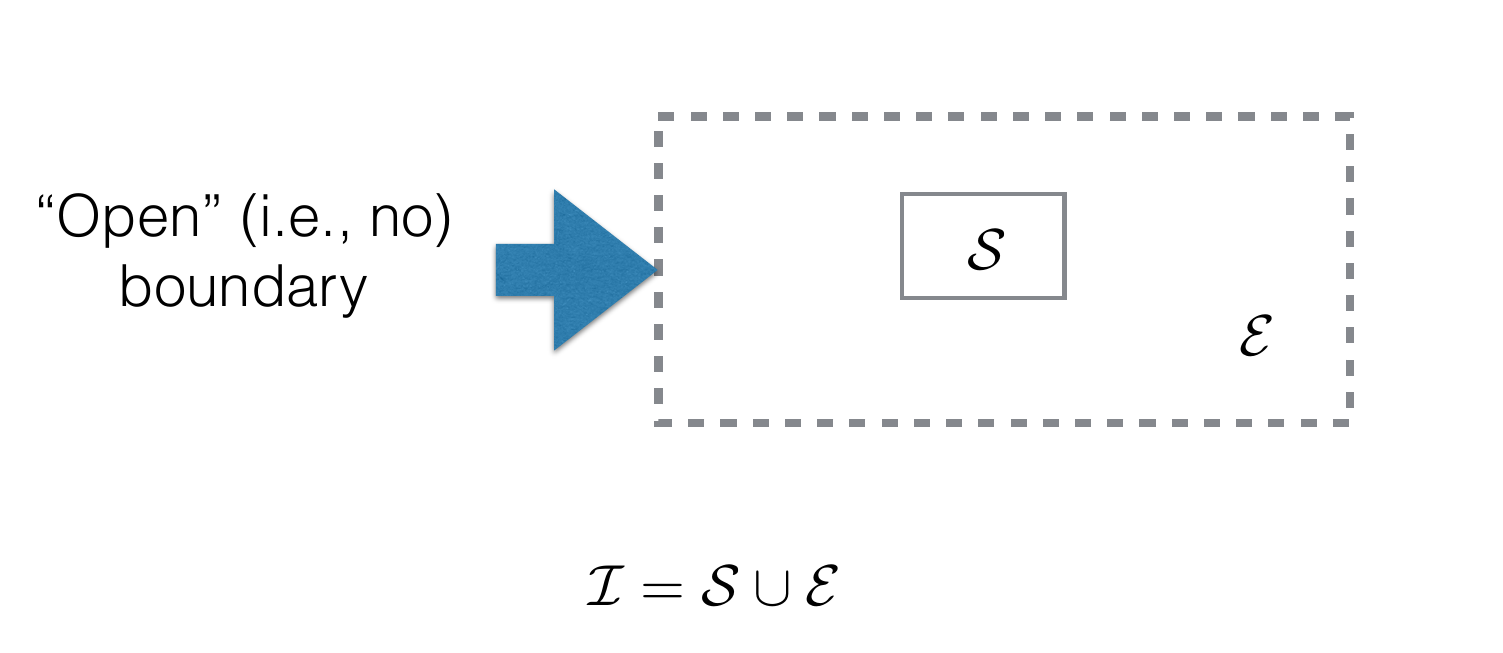}
	\caption{An open system-environment hybrid ${\cal{I}}$. The degrees of freedom in ${\cal{S}}$ may only interact with others in ${\cal{S}}$ and/or the environment ${\cal{E}}$. Unlike the analysis of \ref{section:closed}, however, the constituents of ${\cal{E}}$ may now interact also with others not in ${\cal{I}}$. (The shown open hybrid ${\cal{I}}$ lies in a larger (possibly infinite) volume $\Lambda$).}
	\label{opensys.}
\end{figure*}

As we noted above, for a closed system described by a wavefunction, a large $\sigma_{\tilde{H}}$ implies rapid temporal fluctuations. By contrast, the density matrix describing an open system can be time independent yet exhibit large $\sigma_{\tilde{H}}$ \cite{mix-stationary}. ``Canonical'' open systems ${\cal{I}}$ feature a large (by comparison to the energy uncertainties of the closed systems that we discussed earlier) $\sigma_{\tilde{H}} \sim {\tilde{N}}^{1/2}$ scaling. This larger value of $\sigma_{\tilde{H}}$ renders the corollaries of Eq. (\ref{uncertain}) weaker for open systems. Nonetheless, as we will next demonstrate by a simple ``proof by contradiction'' argument, if we consider an initial {\it open} thermal system composite ${\cal{I}}$ at an assumed temperature $T$ (instead of Assumption (1) for the closed systems of Section \ref{section:closed}), then there exists a limiting cooling/heating rate beyond which equilibrium is impossible. The bound that we will present encompasses the physical situation of a general uniform medium that is heated or cooled via contacts with an external environment. Our result pertains to what transpires if the subsystem ${\cal S}$ and the larger open hybrid system ${\cal{I}}$ containing it are in equilibrium with one another at a temperature $T$ (see Figure \ref{opensys.}). Specifically, we will invoke the following assumptions for open ($^{o}$) systems: \newline

\noindent{\bf{ Assumption {(1$^{o}$)}}}: When combined with their environment (or ``heat bath'') ${\cal{E}}$, these systems constitute
a larger {\it open} hybrid system ${\cal{I}} =  {\cal S} \cup {\cal{E}}$ (of ${\tilde{N}}$ sites) in which the sites in ${\cal S}$ {\it do not interact with any sites that are not in} ${\cal{I}}$. The open system-environment hybrid ${\cal{I}}$ is embedded in a larger volume $\Lambda$ (of size $N_{\Lambda}$). 
$\diamond$
\newline

We comment that with this assumption and definition of ${\cal{I}}$ as a volume containing all sites that ${\cal{S}}$ interacts with, when only short range interactions are present, we may choose, in the thermodynamic limit, ${\cal{I}}$ such that $\lim_{N \to \infty}\frac{\tilde{N}}{N} =1$ with $N$ denoting the number of sites in ${\cal{S}}$.
\newline

 
 \noindent{\bf{ Assumption {(2$^{o}$)}}}: The open hybrid system ${\cal{I}}$ is in thermal equilibrium with its environment at a fixed temperature $T$. In particular, the fluctuations (as computed with initial probability density matrix $\tilde{\rho}$) of extensive quantities are those of an equilibrated system at a temperature $T$. $\diamond$
\newline

\noindent{\bf{ Assumption {(3$^{o}$)}}}: The subsystem ${\it{S}} \subset {\cal{I}}$ is in thermal equilibrium with ${\cal{I}}$. 
$\diamond$
\newline 

This last assumption might be regarded as a consequence of Assumption (2$^{o}$) for the equilibrated
hybrid system ${\cal{I}}$ that includes ${\it{S}}$. Nonetheless, we wish to make Assumption (3$^{o}$)
explicit. 
\newline

The open hybrid system ${\cal{I}}$ (including ${\cal S}$) may be taken to lie deep in a uniform medium so that it is far from any external contacts that change its temperature. The Heisenberg picture Hamiltonian $\tilde{H}^{H}$ evolves with an operator different from Eq. (\ref{ute-trail})- one that involves also the sites exterior to ${\cal{I}}$. The latter coupling allows for a non-trivial time dependence. 
Equivalently, the Schrodinger picture probability density matrix $\tilde{\rho}(t)$ is generally a function of time \cite{QevolveQ}.
Macroscopic expectation values computed with $\tilde{\rho}(t)$ are those of equilibrated thermal systems yet measurable dynamics also appear (as in, e.g., an equilibrated gas with mobile molecules having correlations set by the diffusion equation). For a static $\tilde{\rho}(t)$, all expectation values will be trivially stationary. Since, by Assumption (2$^{o}$), the full hybrid system ${\cal{I}} = {\cal{S}} \cup {\cal{E}}$ is in equilibrium, the system ${\cal S}$ must be in equilibrium with its environment ${\cal{E}}$. From the zeroth law of thermodynamics, it then follows that ${\cal S}$ is also described by a (canonical) probability density matrix at the same inverse temperature $\beta$.

Because the sites in ${\it S}$ only interact with those in ${\cal I}$ (Assumption (1$^{o}$)), Eqs. (\ref{uncert},\ref{uncertain},\ref{uncertain*}) (as well as the bound of Eq. (\ref{xH}) for correlators of the form of Eq. (\ref{gc+})) remain valid. In what follows, following Assumption (2$^{o}$), we will set, in Eq. (\ref{uncertain}), the equilibrium values of standard deviations of the respective Hamiltonians in the appropriate (canonical) ensembles describing the open systems ${\cal{I}}$ and ${\it S}$. That is, 
\begin{eqnarray}
\label{sigmaHt}
\sigma_{\tilde{H}} = \sqrt{k_{B} T^{2} C_{v, \cal{I}}(T)}
\end{eqnarray}
(with $k_{B}$ the Boltzmann constant and $C_{v, \cal{I}}(T)$ the constant volume heat capacity of the large system composite ${\cal{I}}$) to be the standard deviation of the large open hybrid system ${\cal{I}}$, and equate 
\begin{eqnarray}
\label{sigmaHtt}
\sigma_{H} = \sqrt{k_{B} T^{2} C_{v,{\it S}}(T) },
\end{eqnarray}
where $C_{v,{\it S}}(T)$ is the heat capacity of the small system at temperature $T$, to be the standard deviation of the smaller subsystem ${\it S}$. We may repeat, {\it mutatis mutandis}, the steps that led to Eq. (\ref{central'}) when ${\cal{I}}$ was a closed system. Doing so and employing Eq. (\ref{uncertain}), we discover that
if the cooling/heating rate exceeds a threshold value for an equilibrated open hybrid system ${\cal{I}}$ (and any subsystem ${\it{S}} \subset {\cal{I}}$ that is in equilibrium with it (Assumption (3$^{o}$))),
\begin{eqnarray}
\label{canonical}
\boxed{\Big| \frac{dE}{dt} \Big| > \frac{2}{\hbar} k_{B} T^{2} \sqrt{  C_{v, \cal{I}}(T)  C_{v,{\it S}}(T)},}
\end{eqnarray}
then a simple contradiction will be obtained. That is, an assumption of having a sharp equilibrium energy density state variable 
(by coupling ${\cal{I}}$ to a larger external bath at a well defined temperature) \cite{phase_transition} becomes inconsistent once Eq. (\ref{canonical}) is satisfied. At sufficiently fast cooling or heating rates (given by Eq. (\ref{canonical})), the inequality of Eq. (\ref{uncertain}) will be violated when we substitute
the equilibrium open system values of $\sigma_{H/N}$ and $\sigma_{\tilde{H}}$. 

Using the exact inequality of Eq. (\ref{canonical}), it is illuminating to estimate the rate of the temperature variation beyond which equilibration of an open system is rigorously impossible to arrive at a typical thermalization bound. Towards that end, we assume that ${\cal I}$ and ${\cal S}$ are of comparable size (see also the comment following Assumption {(1$^{o}$)}), i.e., of size ${\cal{O}}(N)$, and that the heat capacity of both is, up to factors of order unity, given by $dNk_{B}$ and that the energy density is the order of $(dk_{B} T)$. Hence, if the energy variations fulfill a 
``Planckian rate'' inequality,
\begin{eqnarray}
\label{PlanckR}
\frac{\Big| \frac{d \epsilon}{dt} \Big|}{\epsilon} \gtrsim {\cal{O}} \Big(\frac{2k_{B} T}{\hbar} \Big),
\end{eqnarray}
then, in any dimension $d$, it might be impossible to satisfy all of our assumptions in unison. 
Interestingly, earlier work established that the thermalization rates for typical random states are given by $\frac{k_{B} T}{h}$ \cite{typical0}. The rigorous inequality of Eq. (\ref{canonical}) and 
its common realization of Eq. (\ref{PlanckR}) augment these relations to rigorously demonstrate that in typical situations (when all energy densities and heat capacities are set by the Botlzmann constant, the number of particles, and the energy), whenever {\it{the heating/cooling rate is larger than ${\cal{O}}(k_{B} T/\hbar)$ then no thermalization of the open system is possible.}} We arrived at this inequality by combining exact inequalities associated with the system dynamics
(Eqs. (\ref{uncert}, \ref{uncertain})) with the standard deviations (Eqs. (\ref{sigmaHt}, \ref{sigmaHtt})) of open thermal systems. 
 
 It is important to stress that the variations in the energy need not arise only as a result of an external drive.
Eq. (\ref{canonical}) also holds true for any system in equilibrated open systems for which the variations in the energy are {\it thermally self-generated fluctuations typical to the equilibrium state}. The bound of Eq. (\ref{PlanckR}) is similar that suggested in \cite{juan-martin} as a bound on Lyapunov exponents ($\lambda_{L} \le 2 \pi  k_{B} T/\hbar$) in thermal systems. At room temperature, $2k_{B} T/\hbar \sim 10^{14}$ Hz. Thus, at low temperatures, pulsed picosecond laser cooling/heating may, in principle, achieve these rates beyond which, as we just demonstrated, quantum uncertainty relations forbid thermalization (even for open systems). Our inequality of Eq. (\ref{canonical}) is rigorous. By contrast, Eq. (\ref{PlanckR}) only arises as an order of magnitude estimate. 

Our two results of Eqs. (\ref{uncertain}, \ref{canonical}) for, respectively, the closed and open composites ${\cal{I}}$ apply for any rate of the energy change $dE/dt$. These include situations in which $dE/dt$ scales as the surface area of the system (${\cal{O}}(N^{(d-1)/d})$) for which an extension of Eq. (\ref{central}) will, in turn, imply that $\overline{G} \ge {\cal{O}}(N^{-2/d})$. Eqs. (\ref{central'}, \ref{canonical}) further apply to any function $f(q)$ of an intensive quantity $q$ that is varied at a finite rate. In particular, setting $f(q) =q^{n}$, we find that the uncertainties in all moments of $q$ are, typically, finite if the rate $dq/dt$ is finite. With a formal proof at our disposal, we now briefly reflect back on the arguments of Section \ref{intuition} in which we explained why a varying quantity energy density (or any other intensive quantity $q$) with a finite rate of change $d \epsilon/dt$ (or general $dq/dt$) naturally suggests an uncertainty. The arguments of Section \ref{intuition} provide an intuitive basis for the time-energy uncertainty type relations that we derived and employed in this section for our two Hamiltonian system and, more generally for other intensive quantities. 

We next discuss inequalities that may also be derived when 
Assumption (3$^{o}$) is {\it not invoked}. Replacing the energy density $\epsilon$ in Eqs. (\ref{uncert}, \ref{uncertain}, \ref{uncertain*}) by a general self-adjoint quantity $Q$ having its support on a region {\it of arbitrary size} $N$,  
we discover that thermal fluctuations evaluated with the equilibrium many body density matrix $\rho_{\Lambda}$ must always satisfy
 \begin{eqnarray}
 \label{Qsee}
 \tau^{-1}_{Q} \equiv \frac{|\langle \frac{dQ}{dt} \rangle|}{\sigma_{Q}} \le \Big(\frac{2 \sqrt{k_{B} C_{v,{\cal I}}}}{\hbar}\Big) T.
 \end{eqnarray}
 This inequality is exact at all temperatures and times. The standard deviation $\sigma_{Q}$ and rate $\langle \frac{dQ}{dt} \rangle$ are computed with the reduced density matrix $\tilde{\rho}$ after a partial trace over all degrees of freedom not in ${\cal{I}}$ of the full density matrix describing the equilibrated system $\Lambda$ to which ${\cal{I}}$ generally belongs. If ${\cal{I}}$ is any subvolume of a system ($\Lambda$) that is in thermal equilibrium, then the Hamiltonian $\tilde{H}$ exhibits its own variance given by Eq. (\ref{sigmaHt}).
Similar to Eq. (\ref{PlanckR}), we find that if (i)  ${\cal{I}}$ is of comparable size to ${\cal S}$ (having ${\cal{O}}(N)$ sites) and (ii) if $C_{v,{\cal{I}}} \lesssim d N k_{B}$, then $\tau_{Q}$ cannot be shorter than ${\cal{O}}(\frac{\hbar}{2 k_{B} T \sqrt{dN}})$. Barring critical points/transition regions and/or strong anharmonoticities, in most substances, heat capacities are typically bounded by their (Dulong-Petit type) high temperature value of ${\cal{O}}(dN k_{B})$ making this order of magnitude inequality more stringent than might be suspected otherwise. As remarked above, in Eq. (\ref{Qsee}), $N$ may be of arbitrary size. Indeed, what matters is that in the uncertainty relations we may still approximate the equilibrium energy fluctuations in the larger hybrid system ${\cal{I}}$ by Eq. (\ref{sigmaHt})
and that ${\cal{S}}$ only interacts with sites in ${\cal{I}}$.
The environment ${\cal{E}}$ may be chosen to be the smallest volume such that all sites in ${\cal{S}}$ interact amongst themselves or with sites in ${\cal{S}}$ and as long as all observables in ${\cal{I}}$ (including fluctuations) are equal to those in thermal equilibrium. Generally, the upper bound of Eq. (\ref{Qsee}) becomes more stringent as ${\cal I}$ decreases (scaling with $\tilde{N}^{-1/2}$). Eq. (\ref{Qsee}) also provides a lower bound on the average long distance correlators,
 \begin{eqnarray}
 \label{GQeq}
\overline{G}_{Q} \equiv  \sigma^{2}_{q} = \frac{1}{N^{2}} \sum_{i,j} \Big(\langle {\cal{Q}}_{i} {\cal{Q}}_{j} \rangle - 
\langle {\cal{Q}}_{i} \rangle \langle {\cal{Q}}_{j} \rangle \Big)  \ge \frac{\hbar^{2}}{4 k_{B} T^{2} C_{v,{\cal I}}} \Big| \frac{dq}{dt}  \Big|^{2}.
 \end{eqnarray}
By the equilibrium fluctuation-response theorem, this inequality implies a lower bound on the uniform susceptibility $\chi_{Q}$ associated with a general order parameter or field $Q$ for an equilibrated open thermal system in which $Q$ fluctuates at a rate $(dQ/dt)$,
\begin{eqnarray}
\label{fluct-eqq}
\chi_{Q} \ge \frac{\hbar^{2}}{4 k_{B}^{2} T^{3} C_{v,{\cal I}}} \Big| \frac{dQ}{dt}  \Big|^{2}.
\end{eqnarray}

\subsection{Bounds on the rate of change of general local operators in translationally invariant  thermal systems}
\label{sec:lti}

To elucidate the meaning of our inequalities and illustrate how (a) the time-energy uncertainty inequalities arising from dynamics and (b) equalities in thermal equilibrium intertwine with one another, we next explicitly consider what occurs for the expectation values of local quantities in translationally invariant systems (LTI).
\newline

 \noindent {\bf{ Assumption{(1$^{LTI}$)}}}: We consider local quantities $Q$ defined on a spatial region ${\cal S} \subset {\cal{I}}$. The operators $Q$ commute with all terms in the Hamiltonian that do not involve sites in ${\cal{I}}$. The open system-environment hybrid ${\cal{I}}$ is embedded in a larger volume $\Lambda$ (of size $N_{\Lambda}$). 
Now, ${\cal{I}}$ itself does not need to constitute a sufficiently large region displaying typical thermal expectation values (i.e., Assumption {(2$^{o}$)} no longer holds). Instead of Assumption {(2$^{o}$)}, we impose an even weaker condition: \newline

\noindent{\bf{ Assumption {(2$^{LTI}$)}}}: Global expectation values in $\Lambda$ are given by equilibrium thermal averages. $\diamond$
\newline

\noindent{\bf{ Assumption {(3$^{LTI}$)}}}: The time independent Hamiltonian $H_{\Lambda}$ governing the dynamics of $\Lambda$ is translationally invariant. 
\newline

With these assumptions, we discuss general (possibly infinite volume) theories on a spatial region $\Lambda$ that evolve according to a fixed translationally Hamiltonian $H_{\Lambda}$ and ask what occurs when the initial probability density matrix to be $\rho_{\Lambda}$ describes the large volume $\Lambda$ in equilibrium. We will derive bounds only slightly weaker than those of Eq. (\ref{Qsee}) when only Assumptions (1$^{LTI}$), (2$^{LTI}$), and (3$^{LTI}$) are invoked instead of Assumptions (1$^{o}$), (2$^{o}$), and (3-$^{o}$). Towards that end, we explicitly define $\tilde{H}^{H} \subset H_{\Lambda}$ to be the set of all terms in $H_{\Lambda}$ that do not commute with the quantity $Q^{H} = e^{iH_{\Lambda} t/\hbar} Q e^{-iH_{\Lambda} t/\hbar}$ and thus (by Heisenberg's equation of motion) contribute to its time derivative, $\frac{dQ^{H}}{dt} = \frac{i}{\hbar}[\tilde{H}^{H}, Q^{H}]$. As emphasized above, $\tilde{H}$ may be the sum of all terms in the global Hamiltonian $H_{\Lambda}$ that do not commute with $Q$ and thus endow $Q$ with dynamics. To make the above explicit, we derive Eq. (\ref{Qsee}) in a general setting, longhand,
 \begin{eqnarray}
 \label{longlong}
\Big| \langle  \frac{dQ^{H}}{dt} \rangle \Big|^{2} \equiv  \Big|Tr(\rho_{\Lambda} \frac{dQ^{H}}{dt}) \Big|^{2}  = \Big|Tr (\rho_{\Lambda}( \frac{i}{\hbar} [H_{\Lambda}, Q^{H}(t)])) \Big|^{2}
\nonumber 
\\ =  \Big|Tr (\rho_{\Lambda}( \frac{i}{\hbar} [\tilde{H}^{H}(t), Q^{H}(t)])) \Big|^{2}
\le \frac{4}{\hbar^2} \sigma^{2}_{\tilde{H}^{H}(t)} \sigma^{2}_{Q^{H}(t)}.
 \end{eqnarray}
 Eq. (\ref{longlong}) is valid at all times and constitutes a minor twist of the standard time-uncertainty relations. In the third equality of Eq. (\ref{longlong}), we picked out of the full many body Hamiltonian $H^{H}_{\Lambda}(t) \equiv e^{iH_{\Lambda} t/\hbar} H_{\Lambda} e^{-iH_{\Lambda} t/\hbar} = H_{\Lambda}$,
 the sum $(\tilde{H}^{H}(t))$ of all terms in $H_{\Lambda}$ that do not commute with $Q^{H}(t)$. That is, $ [H_{\Lambda}, Q^{H}(t)] = [\tilde{H}^{H}(t), Q^{H}(t)]$. The last inequality in Eq. (\ref{longlong}) is, once again, the standard uncertainty identity, now applied to the two self-adjoint operators $\Delta \tilde{H}^{H}(t) \equiv (\tilde{H}^{H}(t) - Tr (\rho_{\Lambda} \tilde{H}^{H}(t)))$ and $\Delta Q^{H}(t) \equiv (Q^{H}(t) - Tr(\rho_{\Lambda} Q^{H}(t)))$ (the deviation, at time $t$, of $Q$ from its equilibrium value as computed with the density matrix $\rho_{\Lambda}$), viz., 
 \begin{eqnarray}
 \label{rhodelta1}
 \Big(Tr(\rho_{\Lambda} (\Delta \tilde{H}^{H}(t))^{2}) \times  Tr(\rho_{\Lambda}(\Delta Q^{H}(t))^{2}) \Big) \ge \frac{1}{4} \Big|Tr(\rho_{\Lambda} [\tilde{H}^{H}(t) , Q^{H}(t)])\Big|^{2}. 
 \end{eqnarray}
 As we remarked earlier (Section \ref{section:closed}), such an uncertainty inequality applies both to pure states (the typical case) as well as mixed states with general density matrices \cite{pure}. The above equations were a consequence of the system dynamics. We next discuss what occurs in thermal equilibrium. If $\Lambda$ is in thermal equilibrium, the variance $\sigma^{2}_{\tilde{H}^{H}(t)}$ will be that of the operator $\tilde{H}^{H}(t)$ computed in the thermal state $\rho_{\Lambda}$.  Inserting Eq. (\ref{sigmaHt}) in Eq. (\ref{longlong}) leads to Eq. (\ref{Qsee}) anew.
 The heat capacity $C_{v, \cal{I}} (T)$ in Eq. (\ref{Qsee}) is that associated with the fluctuations $\sigma_{\tilde{H}^{H}(t)}$ when computed with the density matrix $\rho_{\Lambda}$ (leading to Eq. (\ref{sigmaHt})). It is important to explain the physical content of $\langle \frac{dQ}{dt} \rangle$. For all quantities $Q$ if the density matrix $\rho_{\Lambda}$ depends solely on the time independent Hamiltonian $H_{\Lambda}$ (e.g., the density matrix associated with the canonical ensemble) the evolution operator ${\cal{U}}(t) = \exp(- iH_{\Lambda} t/\hbar)$ will then commute with $\rho_{\Lambda}$ and all expectation values will be stationary. This identical stationarity does not capture the local dynamics in thermal systems. In, e.g., an equilibrated gas, the atomic positions of the particles are not stationary (i.e., the average computed with the exact density matrix describing the gas will be time dependent). However, the expectation value of the velocity of any given particle when computing this average with the equilibrium canonical density matrices is identically zero; there is a finite probability density for the particles to assume any velocity and only the mean velocity vanishes. By the ``mean'', we may refer to an (i) ensemble average or one over (ii) long times or as we will focus when Assumption (2-$^{o}$ weak version) holds, for systems with translationally invariant Hamiltonians $H_{\Lambda}$, (iii) a global average over all of space. Indeed, while the global velocity average in an equilibrated ideal gas is zero, the local velocities are finite. The density matrix $\rho^{\sf canonical}_{\Lambda}= \frac{e^{-\beta H_{\Lambda}}}{Z_{\Lambda}}$ with $Z_{\Lambda}= Tr(e^{-\beta H_{\Lambda}})$ the partition function yields the correct average over all local observables yet does not describe the dynamics in the equilibrium system. That is, the global average 
 \begin{eqnarray}
 \label{O1N}
 \overline{{\cal{O}}} \equiv \frac{1}{N_{\Lambda}} \sum_{i=1}^{N_{\Lambda}} \langle {\cal{O}}^{H}_{i} (t) \rangle
 \end{eqnarray}
 of any Heisenberg picture expectation value of ${\cal{O}}$ evaluated with $\rho_{\Lambda}$ is stationary if $\Lambda$ is an equilibrium system with a value given by 
 \begin{eqnarray}
 \label{O1N1}
 \overline{{\cal{O}}} = \frac{1}{N} \sum_{i=1}^{N} Tr( \rho^{\sf canonical}_{\Lambda} {\cal{O}}^{H}_{i}).
 \end{eqnarray}
From Eqs. (\ref{longlong},\ref{rhodelta1}), for identical local operators $\{Q_{i} \}_{i=1}^{N_{\Lambda}}$, each with a corresponding $\tilde{H}_{Hi}$ defined on a region ${\cal{I}}_{i}$, we have
\begin{eqnarray}
\label{longQsee}
\sum_{i=1}^{N_{\Lambda}} Tr(\rho_{\Lambda} (\Delta \tilde{H}_{i}^{H}(t))^{2}) \ge  \frac{\hbar^{2}}{4} \sum_{i=1}^{N_{\Lambda}} \frac{Tr(\rho_{\Lambda} \frac{dQ_i^{H}}{dt})}{Tr(\rho_{\Lambda}\Delta Q_{i}^{H}(t))^{2})}.
\end{eqnarray} 
Invoking Eqs. (\ref{O1N}, \ref{O1N1}), the righthand side of Eq. (\ref{longQsee}) can be rewritten as a canonical thermal average of the fluctuations $ (\Delta \tilde{H}_{i}^{H}(t))^{2})$ whose value is set by the heat capacity and temperature, 
\begin{eqnarray}
\label{longQsee1}
&& Tr(\rho^{\sf canonical}_{\Lambda} (\Delta \tilde{H}_{i}^{H}(t))^{2})   \ge  \frac{\hbar^{2}}{4N_{\Lambda}} \sum_{i=1}^{N_{\Lambda}} \frac{Tr(\rho_{\Lambda} \frac{dQ_i^{H}}{dt})}{Tr(\rho_{\Lambda}\Delta Q_{i}^{H}(t))^{2})} \nonumber
\\  && \Rightarrow   k_{B} T^{2} C_{v,{\cal{I}}_{i}}  \ge \frac{\hbar^{2}}{4}  \overline{\cal{O}}.
\end{eqnarray}
Eq. (\ref{longQsee1}) constitutes a bound on general local measures of the dynamics in thermal systems. 
In the top line of Eq. (\ref{longQsee1}), we invoked translational invariance to the lefthand side of Eq. (\ref{longQsee}). The expectation value $ Tr(\rho^{\sf canonical}_{\Lambda} (\Delta \tilde{H}_{i}^{H}(t))^{2})$ is the same for all $1 \le i \le N_{\Lambda}$ (and is thus equal to the global average) and, for the assumed time independent $H_{\Lambda}$, is also the same at all times $t$.
In the bottom line of Eq. (\ref{longQsee1}), on the righthand side, 
the global average of Eqs. (\ref{O1N}, \ref{O1N1}) is applied to the operator 
\begin{eqnarray}
\label{OQi}
{\cal{O}} \equiv {\frac{\langle  \frac{dQ_i^{H}}{dt} \rangle}{ (\sigma_{Q_{i}}^{H}(t))^{2}}}.
\end{eqnarray}
On the lefthand side of the last equality of Eq. (\ref{longQsee}), we apply, analogous to Eq. (\ref{sigmaHt}), the identity 
$Tr(\rho^{\sf canonical}_{\Lambda} (\Delta \tilde{H}_{i}^{H}(t))^{2})  = k_{B} T^{2} C_{v,{\cal{I}}_{i}}$ where the heat capacity $C_{v,{\cal{I}}_{i}} \equiv \frac{d}{dT} Tr(\rho^{\sf canonical}_{\Lambda} \tilde{H}_{Hi})$. In Eq. (\ref{OQi}), both the expectation value of the local temporal derivative $\langle  \frac{dQ_i^{H}}{dt} \rangle$ and
 the local variance of $Q_{i}^{H}$ are calculated with $\rho_{\Lambda}$. Eqs. (\ref{longQsee1},\ref{OQi}) constitute an explicit local weaker rendition of Eq. (\ref{Qsee}) that require the use of the global average of ${\cal{O}}$ as defined in Eq. (\ref{O1N}). We stress that $\sigma_{Q_{i}}^{H}(t)$ is not an uncertainty due to purely quantum effects. Rather, $\sigma_{Q_{i}}^{H}(t)$ is the standard deviation of $Q_{i}^{H}$ in the system thermal system $\Lambda$ (i.e., $\sigma_{Q_{i}}^{H}(t)$ depicts fluctuations of $Q_{i}^{H}(t)$ from its average value as computed with the thermal density matrix $\rho_{\Lambda}$). The global average of the local variances $\sigma_{Q_{i}}^{H}(t)$ is that given by the canonical density matrix $\rho^{\sf canonical}_{\Lambda}$. 
 
 To elucidate the physical content of Eq. (\ref{longQsee1},\ref{OQi}), we may consider $Q$ to be a single position coordinate $Q=r_{i \ell}$ of particle $i$ of mass $m$ in a general many body thermal system $\Lambda$. The index $\ell=1,2, \cdots, d$ labels the Cartesian component of the particle  location in $d$ spatial dimensions. The full many body Hamiltonian $H_{\Lambda} = T + V$ contains kinetic energies ($T$) and {\it any} position dependent interactions $V(\{\vec{r}_{i}\})$. At any time $t$, the Heisenberg picture Hamiltonian $\tilde{H}_{i}^{H}(t)$ may be chosen to be the kinetic term $(p^{H}_{i \ell }(t))^{2}/(2m)$ since only this term in the Hamiltonian $H_{\Lambda}$ does not commute with $r^{H}_{i \ell}(t)$. 
 From Eq. (\ref{longQsee1}), we observe that in an equilibrated system at temperature $T$, the fluctuations of any Cartesian component of an individual particle displacement must obey a simple inequality,
 \begin{eqnarray}
 \label{rHi}
 \overline{ (\sigma_{r^{H}_{i \ell}})^{-1} 
 |\langle dr^{H}_{i \ell}/{dt} \rangle|}
 \le  \frac{k_{B} T \sqrt{2}}{\hbar},
 \end{eqnarray}
if the heat capacity associated with $\tilde{H}_{i}^{H}$ in the exact quantum system is lower than that computed in the classical limit $C_{v, \cal{I}} (T) \le C^{\sf classical}_{v, \cal{I}} = \frac{k_{B}}{2}$. By the equipartition theorem, for any interactions $V(\{\vec{r}_{i}\})$, the classical thermal average $\langle \frac{{p}^{2}}{2m} \rangle_{\sf classical} = \frac{k_{B} T}{2}$ and the associated heat capacity $C^{\sf classical}_{v, \cal{I}_{i}} = \frac{k_{B}}{2}$. Stated equivalently, Eq. (\ref{rHi}) will hold if the fluctuations $\sigma_{\tilde{H}^{H}}$ in the quantum system given by the exact $\rho_{\Lambda}$ are bounded from above by those of in the classical limit, $\sigma^{2}_{\tilde{H}^{H}_{i}} \le (\sigma^{2}_{\tilde{H}^{H}_{i}})_{\sf classical} = \frac{(k_{B} T)^{2}}{2}$. Here, $ (\sigma^{2}_{\tilde{H}_{i}})_{\sf classical}$ denotes the variance as computed with $\rho_{\Lambda}^{\sf classical} = \frac{e^{- \beta H_{\Lambda}}}{Z^{\sf classical}_{\Lambda}}$ (with $Z^{\sf classical}_{\Lambda}$ the classical partition function) instead of computing the standard deviation of $\tilde{H}$ with the exact density matrix $\rho_{\Lambda}$. For a classical thermal system, irrespective of the spatial dependence of the interaction $V$, the phase space integrals for computing $ (\sigma^{2}_{\tilde{H}_{i}})_{\sf classical}$ decouple into those over the position coordinates and those for the individual momentum components; the single momentum integral that does not cancel identically when averaging with $\rho_{\Lambda}^{\sf classical}$ involves only a Gaussian distribution $\frac{1}{\sqrt{2 \pi m k_{B} T}} e^{-p_{i \ell}^{2}/(2m k_{B} T)}$ similarly leading, as it must, to Eq. (\ref{rHi}). With a simple substitution, Eq. (\ref{Qsee}) is similarly realized with $C_{v,{\cal I}} = \frac{k_{B}}{2}$ if the self-adjoint $Q$ is any periodic function of an angle $\theta_{b}$ that, amongst all terms in $H_{\Lambda}$, does not commute only with a single kinetic term $\tilde{H} = \frac{p^{2}_{\theta_{b}}}{2I_{b}}$ (as happens when $[\theta_{b},H_{\Lambda}] =
[\theta_{b}, \tilde{H}]$). A physical realization is that of a molecular system with $\theta_{b}$ denoting, for any single molecule, the angle around a principal axis of rotation and $ \frac{p^{2}_{\theta_{b}}}{2I_{b}}$ an angular contribution to the kinetic energy with $I_{b}$ the associated moment of inertia. Here, $p_{\theta_{b}}$ is the orbital angular momentum conjugate to $\theta_{b}$ (i.e., $p_{\theta_{b}} = - i \hbar \frac{\partial}{\partial \theta_{b}}$).  
For such molecular systems, Eq. (\ref{rHi}) will hold anew when interchanging the Heisenberg picture $r^{H}_{i \ell} \to Q(\theta^{H}_{b})$ with $Q$ any $2 \pi$ periodic function of the angle $\theta_{b}$, if, as earlier, $\sigma^{2}_{\tilde{H}^{H}} \le (\sigma^{2}_{\tilde{H}^{H}})_{\sf classical} = \frac{(k_{B} T)^{2}}{2}$. The analogous inequalities for the fluctuations $\sigma_{p^{H}_{i \ell}}$ of the canonically conjugate momentum component of an individual particle are typically more involved since, generally, the momentum of an individual particle does not commute with multiple interaction terms (this becomes more acute in systems with long range interactions) that include the said individual particle coordinate $r_{i \ell}$. That is, the Hamiltonian $H_{\Lambda}$ giving rise to the dynamics of $p_{i \ell}$ (i.e., all potential energy terms whose sum is the associated force component $\frac{dp^{H}_{i \ell}}{dt}$) includes all interaction terms in $V(\{\vec{r}_{i}\})$ containing $r_{i \ell}$. 

The exact Planckian rate inequalities derived in this Section might possibly be extended to further establish suggested inequalities, e.g., \cite{damson2007,planck4,us2014,connie}. We conclude this Section by connecting our results concerning uncertainties in intensive quantities to conventional (non-weak \cite{weak,weakA,weakB,weakC}) quantum measurements. Qualitatively, interactions with the environment might be expected to mimic rapid repeated measurements that collapse the wavefunction and not allow Schrodinger type mixing states of significantly different energies to exist. Such a colloquial ``paradox''  is somewhat ill formed as we now explain. Continuous measurements by an environment will indeed not enable large uncertainties to appear. However, the putative existence of continuous collapses will also not allow for any change in the energy density or other intensive quantities. This situation is reminiscent to the well-known ``Quantum Zeno Effect'' \cite{zeno}. 
Progressively weaker  continuous measurements  \cite{weak,weakA,weakB,weakC} may allow for a more rapid evolution of various quantities hand in hand with larger uncertainties. We will discuss adiabatic process, quantum measurements, and thermalization in Section \ref{adiabatic-sec}.

\section{Deviation from equilibrium averages}
\label{deviation}
In the earlier Sections, we demonstrated that forcefully varying the set of intensive (typical state variable) parameters $\{q'\}$ characterizing the eigenstates of $H$ (such as the energy and particle number densities) at a finite rate generally leads to a widening of the distributions $P(\{q'\})$ of these quantities. This was investigated for systems both in the presence and absence of an explicitly included external environment with similar conclusions. Indeed, the causal constraints on the effective interactions associated with the environment was the greatest physical distinction of interest. In this Section, we wish to underscore that such a widening of the distributions $P$ allows for a natural departure from equilibrium behaviors.
That is, even if the expectation values of general observables in individual eigenstates coincide with equilibrium averages \cite{eth1,eth2,eth3,eth4,rigol,pol,polkovnikov1,polkovnikov2,von_neumann} and $H$ has no special many body localized eigenstates  
\cite{MBL1,MBL2,MBL3,MBL4,MBL5,MBL6,MBL7,noMBL,yesMBL}, once a broad distribution $P(\{q'\})$ is present, all averages differ from those in true equilibrium ensembles. 
This will occur since the broad probability distribution $P(\{q'\})$ describing the driven system is different from the corresponding probability distribution in equilibrium 
systems (where all intensive quantities have vanishingly small fluctuations); thus the broad distribution $P(\{q'\})$ will give rise to expectation values of typical observables that are different from those found in equilibrium. We write the equilibrium averages of quantities ${\cal{O}}_{c}$ that commute with the Hamiltonian $([{\cal{O}}_{c}, H]=0$) \cite{diag_common} in a general equilibrium ensemble ${\cal{W}}$ for large systems of arbitrary finite size,
\begin{eqnarray}
\label{equil_av}
\langle {\cal{O}}_{c} \rangle_{eq;\{q\};{\cal{W}}} = \int dq'  P_{eq; \{q\}}(\{q'\};{\cal{W}}) {\cal{O}}_{c}(\{q'\};{\cal{W}}).
\end{eqnarray}
Here, the integration is performed over the full set of intensive variables $\{q'\}$ and the function $P_{eq; \{q\}}(\{q'\};{\cal{W}})$ denotes the probability distribution in an equilibrium ensemble ${\cal{W}}$ for which the average of the various quantities $q = \int d q'  (q' P_{eq; \{q\}}(\{q'\};{\cal{W}}))$. Lastly, ${\cal{O}}_{c}(\{q'\};{\cal{W}}) \equiv \langle 
\phi(\{q'\}; 
{\cal{W}})| 
{\cal{O}}_{c}| \phi(\{q' \}; {\cal{W}}) \rangle$. 
Augmenting the set of intensive quantities $\{q'\}$ defining any of the standard equilibrium ensemble probability distributions, the index ${\cal{W}}$ may specify any additional quantum numbers. These quantum numbers may be associated with symmetries in which case ${\cal{W}}$ can label the orthogonal degenerate eigenstates $\{| \phi(\{q' \}; {\cal{W}}) \rangle\}$
of fixed energy or particle number or other global observables giving rise to the intensive quantities $q$. For instance, in Ising spin systems, the probability distribution $P_{eq; \{q'\}}(\{q'\}';{\cal{W}})$ may be finite only for states with a positive magnetization $ \frac{1}{N}  \sum_{i=1}^{N} \langle S_{i}^{z}\rangle$ as it is in these systems at temperatures below the ordering temperatures once time reversal symmetry is spontaneously broken. An essential feature of all systems in equilibrium is that they exhibit well defined thermodynamic state variables $\{q'\}$. For instance, as we alluded to in earlier Sections, the energy density exhibits 
${\cal{O}}(N^{-1/2})$ fluctuations in the open systems described by the canonical ensemble while it displays ${\cal{O}}(N^{-1})$ fluctuations in closed systems described by the microcanonical ensemble. In all equilibrium ensembles, the width $\sigma_{q}$ of any intensive quantity $q$ vanishes as $N \to \infty$. This sharp delta-function like characteristic of the probability distribution $P_{eq; \{q'\}}(\{q'\}';{\cal{W}})$ is diametrically opposite of $P(\{q'\})$ for which $\sigma_{q}$ is finite. Consequently, the expectation value in the driven system
$\langle {\cal{O}}_{c} \rangle_{driven}$ during the period in which $\{q'\}$ are made to vary with time (that will be given by Eq. (\ref{equil_av}) with the replacement of the equilibrium probability distribution $P_{eq}$ by its non-equilibrium counterpart with $P(\{q'\})$) will generally 
differ from the equilibrium average $\langle {\cal{O}}_{c} \rangle_{eq;\{q\};{\cal{W}}}$.

We now relate the equilibrium and non-equilibrium expectation values. Because the equilibrium distribution $P_{eq; \{q'\}}(\{q'\}';{\cal{W}})$ is, for large systems, essentially a delta-function in $\{q'\}$ (and all additional numbers ${\cal{W}}$), we may explicitly write the expectation values in the driven system as 
\begin{eqnarray}
\label{equil-neq}
\langle {\cal{O}}_{c} \rangle_{driven} = \int dq' P(q';{\cal{W}}) \langle {\cal{O}}_{c} \rangle_{eq;\{q'\};{\cal{W}}}. 
\end{eqnarray}
That is, {\it the expectation values of the observables ${\cal{O}}_{c}$ in the driven system may be expressed as weighted sums of the equilibrium averages} $\langle {\cal{O}}_{c} \rangle_{eq;\{q'\};{\cal{W}}}$ with the weights given by the finite width $\sigma_{q}$ distribution $P(q';{\cal{W}})$ that we focused on in the earlier Sections \cite{non-eq-comment}. The equilibrium expectation values $\langle {\cal{O}}_{c} \rangle_{eq;\{q'\};{\cal{W}}}$ of Eq. (\ref{equil_av}) are experimentally known in many cases. Thus, to predict the expectation values in the driven system, we need to know $P(q';{\cal{W}})$. In Eq. (\ref{equil-neq}), we allowed the probability distribution of the driven system to depend both on the general state variables characterizing the eigenstates of $H$ along with any additional quantum numbers ${\cal{W}}$ that might be selected to define various equilibrium ensembles  (e.g., the sectors of positive and negative magnetization in low temperature Ising systems or qualitatively similar sectors describing the broken translational and rotational symmetries of an equilibrium low temperature crystal). 

We next consider what occurs {\it if driven systems fail to equilibrate} at times $t'>t_f$ (when the parameters $\{q\}$ are no longer forcefully varied at a finite rate) and the system is effectively governed by the time independent Hamiltonian $H$ and the distribution $P(\epsilon')$ of energy densities as measured by the Hamiltonian $H$ will identically remain unchanged at all times $t'>t_{f}$. 
Towards this end, we remark that, for a system with any fixed time independent Hamiltonian $H$, the long time average of a general bounded operator ${\cal{O}}$ (that, unlike ${\cal{O}}_{c}$, need not commute with the Hamiltonian) is given by 
\begin{eqnarray}
\label{olta:eq}
{\cal{O}}_{l.t.a.} = Tr \Big( \frac{\rho(t_{f})}{\tilde{\cal{T}}} \int_{t_{f}}^{t_f+ \tilde{\cal{T}}} dt' {\cal{O}}^{H}(t') \Big) = Tr \Big( \frac{\rho_{\tau}(t_{f} + \tilde{\tau})}{\tilde{\cal{T}}} \int_{t_{f}+\tilde{\tau}}^{t_f+  \tilde{\tau}+ \tilde{\cal{T}}} dt' {\cal{O}}^{H}(t') \Big).
\end{eqnarray}
Here, $\rho(t_{f})$ the density matrix at the final time $t_{f}$ after which the Schorodinger picture density matrix no longer changes in time, the Heisenberg picture ${\cal{O}}^{H}(t') \equiv e^{iH(t'-t_{f})/\hbar} {\cal{O}} e^{-iH(t'-t_{f})/\hbar}$, and (as we have invoked it earlier) ${\tilde{{\cal{T}}}}$ is the said long averaging time. The instantaneous density matrix $\rho(t'>t_{f})$ is constant in time if and only if the density matrix $\rho_{\tilde{\tau}}(t')$ of Eq. (\ref{run_average}) is constant in time for $t'>t_{f} + \tilde{\tau}$. From the latter ``if and only if'' relation, the second line in Eq. (\ref{olta:eq}) follows.  

Now, by the Heisenberg equations of motion, for bounded operators ${\cal{O}}$, as ${\tilde{{\cal{T}}}} \to \infty$, the commutator $[H,\frac{1}{\tilde{\cal{T}}} \int_{t_{f}}^{t_f+ \tilde{\cal{T}}} dt' {\cal{O}}^{H}(t')]=  - \frac{i \hbar}{\tilde{\cal{T}}} \int_{t_{f}}^{t_f+ \tilde{\cal{T}}} dt' \frac{d {\cal{O}}^{H}(t')}{dt'} = - \frac{i \hbar}{ \tilde{\cal{T}}} \Big( {\cal{O}}^{H}(t_f+ \tilde{\cal{T}})  -  {\cal{O}}^{H}(t_f) \Big) = 0$. In other words, ${\cal{O}}_{l.t.a.}$ is trivially diagonal in the eigenbasis of the Hamiltonian \cite{diag_common}. (For classical systems, similar results are obtained when invoking Hamilton's equations with the commutators replaced by Poisson brackets.) For finite $\tilde{\cal{T}}$, there are corrections to the vanishing commutator that scale as $1/\tilde{\cal{T}}$.  Since, in the long time limit, ${\cal{O}}_{l.t.a.}$ commutes with the Hamiltonian, we may apply Eqs. (\ref{equil_av},\ref{equil-neq}). In particular, with the substitution ${\cal{O}}_{c} = {\cal{O}}_{l.t.a.}$, Eq. (\ref{equil-neq}) will provide the long time averages of arbitrary observables ${\cal{O}}$. 
For any ergodic system in equilibrium, the thermal average of the operator of any long time average ${\cal{O}}_{l.t.a}$ of Eq. (\ref{olta:eq}) is the equilibrium average. Substituting in Eq. (\ref{equil-neq}), one thus explicitly has
\begin{eqnarray}
\label{oolta}
{\cal{O}}_{l.t.a.}= \int dq' P(q';{\cal{W}}) \langle {\cal{O}} \rangle_{eq;\{q'\};{\cal{W}}}.
\end{eqnarray}
Along other lines, a similar conclusion was drawn in \cite{me}. Eq. (\ref{oolta}) is a general relation that holds true independent of the results derived in the earlier Sections and is true also in classical systems. From this equality we see that, excusing many body localized states \cite{MBL1,MBL2,MBL3,MBL4,MBL5,MBL6,MBL7,noMBL,yesMBL}, the only way that long time averages may be different from equilibrium averages is that the distributions $P(q';{\cal{W}})$ are not simple delta functions or simple combinations thereof (e.g., if $\langle {\cal{O}} \rangle_{eq;\{q'\};{\cal{W}}}$ has no dependence on some set of $q'$ values then these may be superposed) that reproduce equilibrium expectation values. 
Eq. (\ref{oolta}) holds for general local and global observables. In the special case in which ${\cal{O}} = \sum_{i} {\cal{O}}_{i}$ is a sum of local operators using Eq. (\ref{oolta}) to evaluate the long time average of ${\cal{O}}$ and ${\cal{O}}^{2}$ implies that the long time average of the pair correlators $\langle {\cal{O}}_{i} {\cal{O}}_{j} \rangle$ need not vanish for large spatial distances $|i-j|$ if the distribution $P(q';{\cal{W}})$ is associated with a broad distribution in ${\cal{O}}$ values (equilibrium averages for different  $q'$ and $W$ yield disparate values of ${\cal{O}}$). 
In what follows, we will ask whether an initially driven system may effectively saturate to a distribution $P(\epsilon')$ that relative to time independent Hamiltonian $H$ exhibits a vanishingly narrow ($\sigma_{\epsilon} =0$ as in equilibrium systems) or to a finite width ($\sigma_{\epsilon} \neq 0$) distribution. In Section \ref{2b2b}, we will consider a temperature ($T$) dependent $P(\epsilon')$. 

\section{Effective equilibrium in driven systems}
\label{sec:effective}

In this subsection, we will further explore a closed system sans an environment explicitly included in the calculations (procedure (1) of Section \ref{sec:sketch}). In this setting, given the general time ordered exponential ${\cal{U}}(t) = {\cal{T}} \exp( - \frac{i}{\hbar} \int_{0}^{t} H(t') dt')$, the Heisenberg picture Hamiltonian will evolve $H \to H^{H}(t) = {\cal{U}}^{\dagger}( t)  H {\cal{U}}(t)$. It thus follows that any initial (Schrodinger picture) conventional equilibrium probability distribution $\rho= f(H)$ with $f$ a function of the Hamiltonian will trivially evolve 
as 
\begin{eqnarray}
\label{rhorho}
\rho = f(H) \longrightarrow \rho(t) = f(H^{H}(t)).
\end{eqnarray} 

Thus, e.g., a general (canonical) Boltzmann distribution $f$ in $H$ will evolve into a corresponding one in $H^{H}(t)$. Eq. (\ref{rhorho}) may further enable the proof of other relations \cite{shivaji}. A corollary of Eq. (\ref{rhorho}) is that 
\newline

\fbox{\begin{minipage}{30em}
 $\bullet$ If the initial density matrix $\rho$ describes a system in thermal equilibrium then all of the usual thermodynamic relations may hold at later times with $H^{H}(t)$ being the Hamiltonian instead of $H$. 
 \end{minipage}}
 \newline
 
Specifically, in the equilibrium ensemble defined by $H^{H}(t)$, the system may exhibit equations of state. Thus, e.g., if the system started from a thermal state at inverse temperature $\beta$
(and, by Eq. (\ref{rhorho}), now has a (Schrodinger picture) canonical ensemble density matrix $\rho(t) = Z^{-1} e^{-\beta H^{H}(t)}$ with the time independent $Z= Tr[e^{-\beta H^{H}(t)}] = Tr[e^{-\beta H}]$) then, within this ensemble and the Heisenberg picture, all observables ${\cal{O}}$ of the driven system evolve according to 
\begin{eqnarray}
\label{dyndriv}
\boxed{\frac{d {\cal{O}}^{H}}{dt} = \frac{i}{ \beta \hbar}[{\cal{O}}^{H}(t), \ln \rho(t)].} 
\end{eqnarray}
If the dynamics in a given driven system obey local equations of motion (as they typically do) then the commutators of $H^{H}(t)$ 
with general observables must be as well. Equivalently, if at time $t=0$ the original Hamiltonian was local then {\it it will remain local}
in the time evolved Heisenberg observables of which it is a function. 
Eq. (\ref{dyndriv}) trivially holds also in the classical limit (with, as earlier, the commutator replaced by Poisson Brackets (PB) in the usual way,
$\frac{i}{\hbar} [A^{H},B^{H}] \to - \{A,B\}_{PB} \equiv - \sum_{\alpha} (\frac{\partial A}{\partial x_{\alpha}} \frac{\partial B}{\partial p_{\alpha}} - \frac{\partial A}{\partial p_{\alpha}} \frac{\partial B}{\partial x_{\alpha}})$, with the sum over all generalized coordinates $x_\alpha$ and their conjugate momenta $p_{\alpha}$). Thus, rather explicitly, for any driven classical system,  
\begin{eqnarray}
\label{dyndriv'}
\frac{d {\cal{O}}}{dt} = k_{B} T \{ \ln \rho(t), {\cal{O}}(t)\}_{PB}. 
\end{eqnarray}
The new general result of Eq. (\ref{dyndriv'}) implies the same equation also for overdamped dissipative systems systems so long as their microscopics are governed by an underlying Hamiltonian (as, indeed, all real physical systems are) and thus includes earlier analysis (e.g., \cite{Liverpool}) as particular limiting cases. Our general results of Eqs. (\ref{dyndriv}, \ref{dyndriv'}) further call into focus the important role of the modular Hamiltonian $(-\ln \rho)$ studied in previous works \cite{Louk1}. If the temperature varies with time then in Eqs. (\ref{dyndriv}, \ref{dyndriv'})
the relevant value of $T$ (and of the inverse temperature $\beta$) is that of the initial equilibrium state \cite{comment-T}.
We stress that the invariance of the partition function under the unitary temporal evolution does not imply that the states $\rho(t)$ do not change their character as the system evolves \cite{cite-duality}. 

If the system no longer varies (or varies weakly) in time (e.g., the system approaches a nearly stationary Heisenberg picture Hamiltonian $H^{H}(t)$) then the probability density matrix $\rho= f(H^{H}(t))$ will become (nearly) time independent. In particular, all expectation values computed with the probability density will be (nearly) time independent in much the same way that they were in the original equilibrium distribution. Note that if the evolution ${\cal{U}}(t)$ and initial Hamiltonian $H$ are both spatially uniform then the resulting $H^{H}(t)$ defining the effectively equilibrated system will also be translationally invariant. Thus, if $f$ represents the Boltzmann or any other distribution, then the standard deviation of $H^{H}(t)/N$ as computed with $f(H^{H}(t))$ will be zero. However, as we explained in the earlier sections, the variance of the original Hamiltonian $H$ (not the variance of $H^{H}(t)$) can generally scale as $N^2$. Having such a large variance ($\sigma_{H} = {\cal{O}}(N)$) allows for (yet, of course, does not mandate) rapid dynamics under $H$. In the special case of bounded local Hamiltonians $H^{H}$ and local operators ${\cal{O}}^{H}$, the commutators $[{\cal{O}}^{H},H^{H}]$ may only contain a finite number of bounded terms and thus the frequencies associated with the expectation value of any such local observable (and of its temporal derivatives) are bounded. Along more general lines, the von Neumann equation $\frac{\partial \rho(t)}{\partial t} = \frac{i}{\hbar}[\rho(t),H]$ allows for stationary $\rho(t)$ regardless of the magnitude of $\sigma_{H} = N \sigma_{\epsilon}$. A trivial example is afforded by a Schrodinger picture density matrix that is diagonal in the eigenbasis of $H$, and thus trivially stationary once the system evolves under $H$ in the absence of external driving terms. Similar to the discussion following Eq. (\ref{rhl}), the off-diagonal spread of $\rho$ determines its fluctuation frequencies. Indeed, some systems (e.g.,  glasses that we turn to next) do not adhere to the same equations of state as their conventional equilibrium (e.g., equilibrium solid and fluid) counterparts yet may, nonetheless, appear stationary on very long time scales. 

\section{``To thermalize or to not thermalize?''}
\label{2b2b}

\begin{figure*}
	\centering
	\includegraphics[width=1.  \columnwidth, height=.85 \textheight, keepaspectratio]{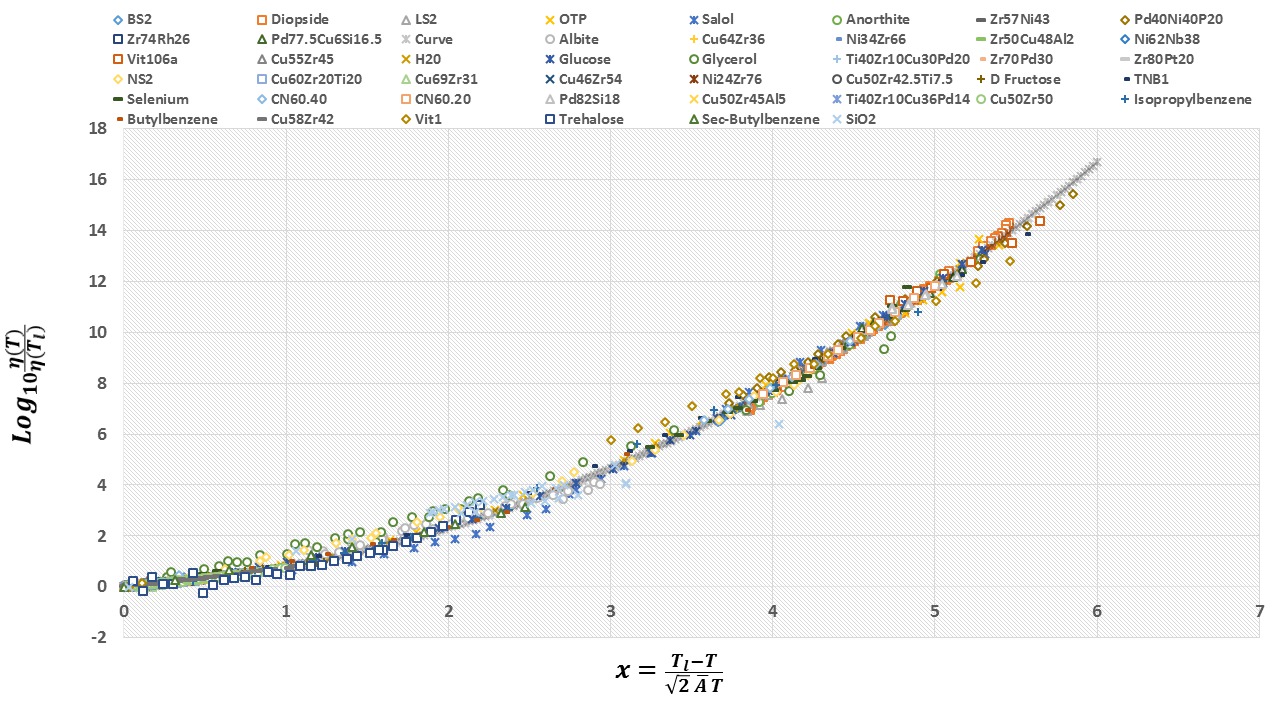}
	\caption{(Color Online.) Reproduced from \cite{us}. On the vertical axis, we plot the experimentally measured viscosity data divided by its value at the liquidus temperature ($\eta(T_{l})$) as a function of a dimensionless temperature ratio. 
	The viscosities of 45 liquids of diverse classes/bonding types (metallic, silicate, organic, and others) collapse on a single curve. The underlying continuous ``curve'' (more clearly visible at high viscosities where fewer data exist) is predicted by Eq. (\ref{vis}). Since $\overline{A}$ varies from fluid to fluid  (albeit weakly) \cite{us}, the shown collapse does not imply a corresponding collapse of the viscosity as a function of $T_{l}/T$ nor as a function of $T_{l}/(\overline{A} T)$ (due, relative to the latter, to an additional shift along the $x$ axis that is set by $-1/(\overline{A} \sqrt{2})$).}
	\label{Collapse.}
\end{figure*}

The above question alludes to possible differences between (i) an {\it effective equilibrium} 
density matrix associated with a density matrix $\rho(t)$ (including those for the systems discussed in Section \ref{sec:effective}) that becomes nearly stationary (and thus a nearly constant $\rho_{\tilde{\tau}}(t)$) at finite long times $t$ and (ii) the density matrix associated with the {\it truly asymptotic long time equilibrium} 
density matrix $\rho_{eq}$. Most of our focus thus far has been on intermediate times $0 \le t \le t_{f}$ during which the energy density (or any other intensive quantity $q$) varied. We showed that during these times,  the standard deviation of $q$ may be finite, $\sigma_{q} = {\cal{O}}(1)$. Thus, the variation of general quantities $q$ (including, notably, the energy density or temperature) may trigger long range correlations. As discussed in \ref{rev:long-range}, this effect may be further exacerbated by ``non-self-averaging'' \cite{derida,amnon,eitan,per} found in disordered systems. Our inequalities of Eqs. (\ref{central'}, \ref{canonical}) hold for general fluctuations (regardless of the magnitude of their  ``classical'' and ``quantum'' contributions \cite{class-quant+} to the variance). In most systems coupled to an external bath, after the temperature or field no longer changes (e.g., when $|d\epsilon/dt|$ vanishes at times $t>t$) thermalization rapidly ensues already at short times after $t_{f}$. Indeed, there are arguments (including certain rigorous results) that ``typical'' states \cite{typical0} 
might thermalize on times set by Planck's constant and the temperature, viz. the ``Planckian'' time scale ${\cal{O}}(\frac{h}{k_{B} T})$ encountered in Eq. (\ref{PlanckR}). Other, exceedingly short (as well as long), equilibration time scales may be present \cite{typical}. The Planckian rate of Eq. (\ref{PlanckR}) appears in a host of interacting systems, e.g., \cite{subir,planck1,planck3,musketeers,planck4,us2014}. Various reaction times are often given by such minimal Planckian time scales multiplied by $e^{\Delta G/(k_{B} T)}$ with $\Delta G$ the effective Gibbs free energy barrier for the reaction or relaxation to occur, e.g., \cite{us2014,eyring}. However, some systems such as glasses do not achieve true equilibrium: measurements on viable experimental time scales differ from the predictions of the microcanonical or canonical ensemble averages. (The difference between the microcanonical and canonical ensembles is irrelevant for all intensive quantities in the absence of long range interactions for which ``ensemble inequivalence'' is known to appear \cite{lr1,lr2,lr3,lr4}). In such cases (including, e.g., rapid supercooling of liquids that can lead to glass formation), the system may effectively exhibit self-generated disorder. Structural glasses are disordered relative to their truly thermalized crystalline counterparts. It is important to stress, however, that both structural glasses and crystalline solids are governed by the very same (disorder free) Hamiltonian. The effective disorder that glasses exhibit is not intrinsic but merely self-generated by the rapid supercooling protocol of non-disordered liquids. Thus, as hinted in Section \ref{deviation}, the question remains as to whether, once the energy density or other intensive quantity no longer varies (e.g., once the glass is formed and its temperature not lowered), the system will thermalize on experimental time scales (and display the rightmost distribution of Figure \ref{Idea.}) or not be able to do so. Similar to Assumption (3) of Section \ref{section:closed}, starting from a glassy state, supercooled liquids achieve their true equilibrium (crystalline) state only at asymptotically long times \cite{defn_glass}. In systems that do not thermalize on experimental time scales, the discrepancy between equilibrium ensemble averages and empirical observables hints that the width $\sigma_{\epsilon}$ of the energy density might become smaller than it was during the cooling process yet is not vanishingly small. Indeed, if $\sigma_{\epsilon} =0$ and no special ``many body localized'' states \cite{MBL1,MBL2,MBL3,MBL4,MBL5,MBL6,MBL7,noMBL,yesMBL} exist 
{\it then the long time averages of all observables must be equal their microcanonical expectation values}. Specifically, similar to Eq. (\ref{oolta}), the time average of a general quantity ${\cal O}$ over a long (finite) time ${\cal{T}}$ during which the probability distribution $P(q';{\cal{W}})$ is nearly stationary is identical to the equilibrium average, i.e., ${\cal{O}}_{l.t.a.} =  \langle {\cal{O}} \rangle_{eq;\{q'\};{\cal{W}}}$ when the distribution $P(q';{\cal{W}})$ is of a delta-function type nature in the energy density $\epsilon$ and all other intensive quantities $q$. If the expectation values of the thermodynamic equilibrium observables depend on the temperature or energy density (and are the same for all states related by symmetries of the Hamiltonian) then deviations of long time average values of observables ${\cal O}$ from their true equilibrium average values \cite{me}, 
\begin{eqnarray}
\label{more_is_different} 
{\cal{O}}_{l.t.a.} \neq  \langle {\cal{O}} \rangle_{eq;\{q\};{\cal{W}}}
\end{eqnarray}
will imply that {\it the width of the energy density may remain finite even after the system is no longer driven}, $\sigma_{\epsilon} >0$. In glassy systems that, by their defining character, cannot achieve true equilibrium 
(and thus satisfy Eq. (\ref{more_is_different})) on relevant experimental time scales, the link to the external bath is effectively excised since the dynamics are so slow that little flow may appear. Here, the finite long time averages of 
Eq. (\ref{olta:eq}) may be employed. If, in such instances, the probability density becomes time independent on measurable time scales then only an effective equilibrium (different from the true equilibrium defined by an equilibrium ensemble for the Hamiltonian defining the system) may be reached. That is, in systems with an effective equilibrium at sufficiently long times, see Section \ref{sec:effective}, the probability density $P(\epsilon')$ may be history independent and be a function of only a few global state variables yet differ from the conventional equilibrium statistical mechanics probability density in which the standard deviations of all intensive quantities vanish, e.g., $\sigma_{\epsilon} =0$. Since the probability density determines all observable properties of the system, interdependences between the state variables (i.e., equations of state) may result \cite{me}. Such a nearly static effective long time equilibrium distribution bears some resemblance to ``prethermalization'' in perturbed, nearly-integrable, models and other systems, e.g., \cite{prethermal1, prethermal2,prethermal3,prethermal4,prethermal5}. Indeed, if local observables do not vary rapidly in time then, by the Heisenberg equations of motion, these observables nearly commute with the Hamiltonian (and constitute nearly integrable constants of motion). We remark that by applying the Mastubara-Matsuda transformation \cite{MM}
(similar to that invoked in Section \ref{gbose}), we may map the prethermalized three-dimensional spiral spin states of \cite{prethermal5} to establish the existence of long-lived effective equilibrium crystals of hard-core bosons. As stated above (see also Section \ref{sec:effective}), at asymptotically long times, systems such as glasses finally truly thermalize to true equilibrium solids \cite{defn_glass}. However, prior to reaching  the true equilibrium defined by the any of the canonical ensembles for the full system Hamiltonian, over very long finite times, the supercooled liquid/glass may display a nearly static distribution $P$ and thus obeys equations of state, absence of memory effects and other hallmarks of effective equilibrium. Even when the equilibrium averages $\langle {\cal{O}} \rangle_{eq;\{q'\};{\cal{W}}}$ feature non-analyticities at specific $q'$, the smeared average of Eq. (\ref{oolta}) can be analytic (e.g., no measurable phase transitions might appear as $T$ is varied). In \cite{me}, we introduced this notion of an effective long time distribution $P$ of finite $\sigma_{\epsilon}$ and employed it to predict the viscosity of glass formers. This prediction was later tested \cite{us,us1} for the measured viscosity data of all known glass formers when these are supercooled below their melting temperature. Figure \ref{Collapse.} reproduces the result. Here, the probability density $P_{T}(\epsilon)$ at temperature $T$ is a normal distribution with the (finite) energy density width 
\begin{eqnarray}
\label{epsT}
\sigma_{\epsilon} = \overline{A} \frac{T (\epsilon_{melt}- \epsilon)}{T_{melt}- T}.
\end{eqnarray}
In Eq. (\ref{epsT}), $\overline{A}>0$ is a liquid dependent constant ($0.05 \lesssim \overline{A} \lesssim 0.12$ for all liquids with published viscosity data \cite{us,us1}). 
In equilibrium, such values of $\overline{A} \sim 0.1$ would be typically anticipated for effective classical harmonic solids/clusters (displaying a Gaussian distribution of the energy density with $\sigma_{\epsilon} = \sqrt{k_{B} T^{2} C_{v}}/N_{eff}$ where the heat capacity $C_{v} = dN_{eff}k_{B}$ and $\epsilon = dk_{B}T$) of $N_{eff} \sim 30$ atoms in $d=3$ dimensions. Albeit emulating such effective finite size equilibrium clusters, the energy densities $\epsilon$ and $\epsilon_{melt}$ in Eq. (\ref{epsT}) are, respectively, those of the genuinely {\it macroscopic} supercooled liquid or glass at temperature $T< T_{melt}$ and at the melting (or ``liquidus'') temperature $T_{melt}$. The wide distribution of Eq. (\ref{epsT}) mirrors that present in non-self-averaging disordered classical systems with an approximately linear in $T$ standard deviation and energy density $\epsilon(T)$. (All eigenstates of the density matrix may share the same energy while displaying a finite standard deviation $\sigma_{\epsilon}$.) In the models of Section \ref{sec:dual} (with the distribution of Eq. (\ref{Heavy}) that was far from the canonical normal form of equilibrium systems), the systems were driven by an external source whose effect on general quantities was cyclic in time. The situation may be radically different when the system is no longer forcefully driven out of equilibrium yet, nonetheless, is still unable to fully equilibrate. If, as in equilibrium thermodynamics, the final state maximizes the Shannon entropy for a given energy then the probability distribution of the energy density will be a Gaussian of width $\sigma_{\epsilon}= \frac{T \sqrt{k_{B} C_{v}}}{N}$ and standard $\frac{1}{\sqrt{N}}$ fluctuations result (with $\sigma_{\epsilon} \propto T$ for a nearly constant $C_{v}$). For systems of temperature $T$ that have not fully equilibrated, we may (as illustrated in the earlier Sections) find finite width $P_{T}(\epsilon')$. If the distributions $P_{T}(\epsilon')$ minimally differ in form from those in equilibrium then they may still be Gaussian with $\sigma_{\epsilon} \propto T$. Indeed, the general distribution that maximizes the Shannon entropy given a {\it finite} standard deviation $\sigma_{\epsilon}$ (and average energy density) is a Gaussian. Adhering to Occam's razor, the sole difference between the distribution of the energy density in equilibrium systems and those that we assume here for systems that have not yet achieved equilibrium is that in the latter systems $\sigma_{\epsilon} = {\cal{O}}(1)$ (while $\sigma_{\epsilon} =0$ for equilibrium systems in their thermodynamic limit.) Non-rigorous considerations further suggest the appearance of a Gaussian distribution once the system is no longer further cooled (or heated), see \ref{prompt} \cite{LG,plastic_deformation}. Assuming a normal distribution $P_{T}(\epsilon')$ of width 
$\sigma_{\epsilon}$, the viscosity $\eta$ of supercooled liquids at temperatures $T \le T_{melt}$ was predicted (by an application of Eq. (\ref{oolta})) to be \cite{me},
\begin{eqnarray}
\label{vis}
\eta(T) = \frac{\eta_{s.c.}(T_{melt})}{erfc \Big( \frac{\epsilon_{melt} - \epsilon(T)}{\sigma_{\epsilon} \sqrt{2}} \Big)} =
\frac{\eta_{s.c.}(T_{melt})}{erfc \Big( \frac{T_{melt} - T}{\overline{A} T \sqrt{2}} \Big)}.   
\end{eqnarray}
Eq. (\ref{vis}) is a direct consequence of Eqs.(\ref{oolta}, \ref{epsT}) and the Stokes' law \cite{me}. This prediction is indicated by the continuous curve in Figure \ref{Collapse.}. The coincidence between this non-perturbative prediction and the experimental data extends 16 decades of the viscosity increase and is a compilation of the analysis of the data of 45 fluids \cite{us}; the corresponding dimensionless ratio in the argument of Eq. (\ref{vis}), $x \equiv \frac{T_{melt} - T}{\overline{A} T \sqrt{2}}$ (the abscissa of Figure \ref{Collapse.}), varies up to a value of six. Unlike well known data collapse forms in equilibrium transitions and conventional critical phenomena in particular, the agreement between Eq. (\ref{vis}) and the experimental data does not wane for the larger $x$ (and viscosity) values. In fact, beyond an intermediate temperature range at which some scatter is seen in Figure \ref{Collapse.}, the quality of the data collapse improves as one progresses to lower temperatures more removed from the equilibrium melting temperature $T_{melt}$ \ref{explain-deviation}. At the so-called ``glass transition temperature'' $T_{g}$, the viscosity $\eta(T_{g}) =10^{12}$ ~Pascal $\times$ second \cite{berthier}. At lower temperatures $T<T_{g}$, the viscosity is so large that it is hard to measure it on experimental time scales. We note that at sufficiently low temperatures (energy densities), the deviations of $P_{T}(\epsilon)$ from a putative normal distribution (assumed in deriving Eq. (\ref{vis})) will become more important (since the probability of having states of energies lower than the ground state is strictly zero); other distributions such as log-normal have a strict cutoff below which their value vanishes. Furthermore, in deriving Eq. (\ref{vis}), an assumption \cite{me} was made  that the viscosity of the equilibrium solid is infinite; any finite contributions (no matter how small) to hydrodynamic transport from the equilibrium solid eigenstates will lead to larger hydrodynamic flow rates and viscosities lower than those predicted by Eq. (\ref{vis}). These effects may be of larger relevance at very low temperatures where the viscosity as predicted by Eq. (\ref{vis}) (in which these effects were excluded) becomes exceedingly large. Replicating the derivation for the viscosity in \cite{me} when the equilibrium solid displays activated flow and thus, at very low temperatures, the net contributions to the long time velocities $v_{l.t.a.}$ of \cite{me} from the occupied solid states overwhelm those from the sparsely populated equilibrium fluid states replaces, at these low energy densities, Eq. (\ref{vis}) by an activated Arrhenius form. Apart from predictions for the viscosity, more general transition and relaxation rates may be investigated along similar lines \cite{me,mexplain}.

One may, of course, use other words to rationalize the same basic physics that we suggest here regarding the relevant distributions of eigenstates of the Hamiltonian/classical modes/ ... of different energies/frequencies, etc. The very same distribution $P_{T}(\epsilon')$ invoked in deriving Eq. (\ref{vis}) may relate other properties of supercooled liquids and glasses to those of equilibrium systems. For instance, the experimentally measured thermal emission from supercooled fluids may differ in a subtle manner from one that is typical of equilibrium fluids. This deviation may be found by replacing Planck's law for the spectral radiance $I$ for photons of frequency $\nu$ in a system with well defined equilibrium temperature $T$ by a weighted average of Planck's law over effective equilibrium temperatures $T'$ that are associated with internal energy densities of equilibrium systems that are equal to $\epsilon'$,
\begin{eqnarray}
\label{prediction2}
I(\nu, T) = \frac{2 h\nu}{c^{3}} \int d T' \frac{\tilde{P}_{T}(T')}{e^{h \nu/(k_{B} T')} -1} +  I_{\cal{PTEI}}(\nu,T). \nonumber
\end{eqnarray}
Here, $\tilde{P}_{T}(T') = P_{T}(\epsilon') c^{eq}_{v}$ (with the equilibrium specific heat capacity $c^{eq}_{v} \equiv \frac{d \epsilon'}{dT'}$) is the distribution of effective equilibrium temperatures $T'$ associated with the probability distribution $P_{T}(\epsilon')$ of the energy densities. The second term, $I_{\cal{PTEI}} \equiv  \frac{2 h\nu}{c^{3}} \frac{\int_{\cal{PTEI}} d \epsilon' P_{T}(\epsilon')}{e^{h \nu/(k_{B} T)} -1}$, captures viable contributions from any 
``Phase Transition Energy Interval'' \cite{me} (wherein the energy density $\epsilon'$ of an equilibrium system may vary by an amount set by the latent heat without concomitant changes in the corresponding equilibrium temperature $T'$). More accurately stated, in Eq. (\ref{prediction2}), we may replace $\frac{2 h\nu}{c^{3}(e^{h \nu/(k_{B} T')} -1)}$ by $u(\nu,T')$- the energy density carried by photons of frequency $\nu$ when the equilibrium system is at a temperature $T'$ \cite{me}. We highlight that this prediction for the emission spectrum $I(\nu, T)$ is determined by the same distribution predicting the viscosity collapses of Figure \ref{Collapse.} (the Gaussian $P_{T}(\epsilon')$ of the width given by Eq. (\ref{epsT})). As such, this prediction may, in principle, be experimentally tested. This prediction is akin to that having an effective locally varying temperature (similar to that associated with observed heterogeneities) whose distribution is determined by $P_{T}$. Similarly, the temperature dependence of other observables (including various response functions) may be expected to have the same increase in the time scale as that characterizing the viscosity. Indeed, the time dependent heat capacity response follows exhibits a dynamical time that increases with temperature in a manner similar to the viscosity, e.g., \cite{Dixon}. We suspect that this increase in the relaxation time scale as the temperature is dropped may account very naturally for the experimentally observed smooth specific heat peak\cite{angell} near the glass transition temperature $T_{g}$ when the system is heated from lower temperatures (consistent with $T_g$ marking a dynamical crossover rather than a bona fide thermodynamic transition \cite{us,us1}). This is so since, at temperatures $T \le T_{g}$, on the time scales of the experiment, the system is essentially static (e.g., the viscosity of the Eq. (\ref{vis}) and the associated measured relaxation times are large). Consequently, the relatively stable nearly static structures that appear once the glass is formed need not significantly respond to a small amount of external heat. The situation is somewhat reminiscent of the extensive latent heat that is required to melt equilibrium crystals. Pronounced thermodynamic changes appear at the transition between equilibrium fluids and crystals. Once the supercooled liquid or glass becomes effectively static on experimental time scales at $T_{g}$, it may weakly emulate the latent heat signature of the equilibrium liquid to solid transition sans having true latent heat required to elevate the temperature. Contrary to the weak peak in the heat capacity on heating, when the system is cooled from temperatures above $T_{g}$, the heat capacity typically drops monotonically near $T_{g}$ and does not exhibit a peak (this may reflect a memory of larger mobility at higher temperatures). A finite 
$\sigma_{\epsilon}$ may naturally allow for a finite width temperature interval about $T_{g}$ where the empirically observed crossover in the heat capacity and/or other quantities can appear on experimental time scales. In line with our earlier discussion concerning general properties stemming 
$P_{T}(\epsilon')$, Eqs. (\ref{equil-neq},\ref{oolta}) \cite{me} further suggest that similar features may appear at other temperature at which other crossovers appear (i.e., the ratio of the width of the temperature range where a crossover is observed to the crossover temperature itself may be set by the scale of dimensionless parameter $\overline{A}$ appearing in Eq. (\ref{vis}) for the viscosity). Indeed, simple estimates illustrate that experimentally observed heat capacity crossover region is of the same scale as $\overline{A} T_{g}$  \cite{ATG}. 
More generally, by simple dimensionless analysis if the dimensionless parameter $\overline{A}$ is the most important feature of the system, the temperature window over which crossovers occur may scale as $f(\overline{A}) T_{g}$ with $f$ the appropriate function. By dimensional arguments, as a function of the temperature $\sigma_{\epsilon} = k_{B} T 
F(\{\frac{T_{a}}{T}\})$ where $\{T_{a}\}$ are any relevant temperature or associated energy scales (e.g., the melting temperature $T_{m}$ and any other) and $F$ is a function of the dimensionless temperature ratios; far away special temperature scales, one anticipates a largely linear dependence of $\sigma_{\epsilon}$ on temperature. A broadening due to the finite $\sigma_{\epsilon}$ may supplant any existing features of the equilibrium system (having $\sigma_{\epsilon}=0$). More general than heat capacity measurements alone, we stress that, experimentally, supercooled liquids indeed exhibit effective smooth crossovers instead of true singularities associated with thermodynamic phase transitions that appear at well defined transition temperatures. Thus, our suggestion is that the size of the temperature interval over which these crossovers arise/are enhanced as a result of smearing by the finite width distribution $P_{T}(\epsilon')$ is set by the effective crossover temperature scale multiplied by $\overline{A}$. An energy density distribution of a finite width $\sigma_{\epsilon}$ allows for a superposition of low energy density solid type eigenstates (that may break continuous translational and rotational symmetries) and higher energy density liquid type eigenstates \cite{me}. Such a general combination of eigenstates does not imply experimentally discernible equilibrium solid (crystalline) order. Sharp Bragg peaks need {\it not appear} in states formed by superposing eigenstates that, individually, display order \cite{me,nobel}. This absence of ordering reflects the possible lack of
 clear structure when, e.g., randomly superposing different Fourier modes with each Fourier mode displaying its defining periodic order. In an interesting preprint \cite{mines} that appeared after an earlier version of the current paper \cite{this_work}, it was found that effectively superposing (periodically replicated) finite size states of $16$ or $24$ atoms (so as have these states as unit cells that are repeated to span all space) according to their Boltzmann weights accurately reproduces the structure factor of certain structural glasses. This latter result is in accord with our approach to glasses; the size of these 16 and 24 atom states is not too dissimilar from the order of magnitude estimate, provided earlier in this Section, of the requisite number of atoms $N_{eff} \sim 30$ in an effective equilibrium solid that would lead to a Gaussian distribution of a width consistent with our theory of glasses and the ensuing collapse of Figure \ref{Collapse.}. It will be interesting to examine in more quantitative detail whether distributions associated with states similar to those examined in \cite{mines} adhere to the normal form that we invoked for $P$.
  
The mixing of eigenstates of different energy densities over a range set by $\sigma_{\epsilon}$ further suggests the appearance of non-uniform dynamics both in space and in time. The superposition of different modes suggests non-uniform spatial dynamics. Interestingly, in accord with this consequence of our theory, {\it dynamical heterogeneities} are empirically ubiquitous in supercooled fluids  \cite{DH1,DH2,DH3,DH4,DH5}; these large fluctuations are still present even after the fluids remain in contact with an external bath for a long time. To examine temporal fluctuations, we may invoke Eq. (\ref{oolta}) when the operator ${\cal{O}}$ is set to be $v_{i}^{2}$ and $v_{i}^{4}$ (i.e., the scaled kinetic energy of particle $i$ and its square), one may anticipate the standard deviation of $v_{i}^{2}$ divided by square of the average of $v_{i}^{2}$ itself (i.e., the dimensionless ratio $\frac{\sqrt{(v_{i}^{4})_{l.t.a.} - ((v_{i}^{2})_{l.t.a})^{2}}}{(v_{i}^{2})_{l.t.a}}$ which emulates the fluctuations in the local kinetic energy divided by the average local kinetic energy instead of similar ratios for the global energy density $\epsilon$).
Such a ratio may be naturally determined by the width of the total energy density distribution $P_{T}(\epsilon)$ divided by an energy scale set by the squared velocity to vary with $\overline{A}$; indeed, in equilibrium systems at a temperature $T'$ having potential energies that are independent of the momentum, the average local kinetic energy is, by the equipartition theorem, linear in $T'$ and from Eq. (\ref{oolta}) this ratio will yield the corresponding ratios for the equilibrium result when smeared by the weight $P_{T}(\epsilon)$ (or a similar distribution in the effective equilibrium temperatures $T'$
where the equilibrium internal energy density $u$ matches the energy density, $\epsilon = u(T')$). (Of course, in experiment, one typically does not directly measure $v^{2}_{i}$ in a given system but rather $v_{i}$.) 
The presence of a spatially non-uniform energy density is very natural during general heating or cooling processes (e.g., the exterior parts of a system being supercooled may be colder than its interior, see also the discussion towards the end of Section \ref{intuition}). Once supercooling stops, heat may diffuse through the system yet heterogeneities (that are borne in our framework from a distribution of finite $\sigma_{\epsilon}$) may persist for a long time \cite{cor'}. 

Eq. (\ref{oolta}) that enabled the prediction of the viscosity of Eq. (\ref{vis}) 
and others quantities does not rely on quantum effects. An advantage of the quantum approach described in this Section is that it allows
for an accurate definition of the (eigen)states of the systems as opposed to the more loosely defined classical microstates in which Planck's constant needs to be introduced by hand in order to produce a dimensionless number of states from phase space volumes \cite{micro-h}. Furthermore, in standard classical treatments, one often needs to integrate the equations of motion numerically in order to obtain results for various particular systems (this is particularly time consuming for slow glassy systems). Alternatively, if numerics are to be avoided, assumptions may be made about the classical energy landscape and configurational entropies. The quantum treatment invoked in this Section is devoid of such assumptions. Nonetheless, one may still translate the more fundamental and precise quantum description into a corresponding classical one \cite{me,us}.  

\section{Possible extensions to electronic and lattice systems}
\label{sec:electron}

The spin and hard core Bose models of Section \ref{sec:dual} were defined on lattices. In this Section, we will speculate and further discuss possible extensions to other, experimentally relevant, theories and lattice systems. 
The electronic properties of many materials are well described by Landau Fermi Liquid Theory \cite{NFL,AGD,LL,coleman}. This theory is centered on the premise of
well defined quasiparticles leading to universal predictions. Recent decades have seen the discovery of various unconventional materials displaying rich phases \cite{rich1,rich2,rich3,rich4,rich5,rich6,rich7,rich8,rich9,rich11,rich12,rich13,rich14,rich15,rich16,rich17,rich18,rich20,NFL} that often defy Fermi liquid theory.
Given the results of the earlier Sections, it is natural to posit that as these systems are prepared by doping or the application of external pressure and fields (in which case the varied parameter $q$ may be 
the carrier density, specific volume, or magnetization), a widening $\sigma_{q}$ will appear during the process.
This wide distribution might persist also once the samples are no longer experimentally altered. In such cases, the density matrices (and
associated response functions) describing these systems may exhibit finite standard deviations
$\sigma_{q}>0$. The broad distribution may trigger deviations from the conventional behaviors found in systems
having sharp energy and number densities ($\sigma_{n} =0$) or, equivalently, sharp chemical potentials 
and other intensive quantities. Theoretically, non-Fermi liquid
behavior may be generated by effectively superposing different density Fermi liquids (with each Fermi liquid having a sharp carrier concentration $n$) in an entangled state. 
Systems harboring such an effective distribution $P(\mu')$ of chemical potentials 
may be described by a mixture of Fermi liquids of different particle densities. Any non-anomalous Green's function is manifestly diagonal in the total particle number. 
Thus, the value of any such Green's function may be computed in each sector of fixed particle number and then subsequently averaged over the distribution of total particle numbers in order to determine its expected value when $\sigma_{n} \neq 0$. In particular, this implies that the conventional jump (set by the quasiparticle weight $Z_{\vec{k}, \mu'}$) of the momentum space occupancy \cite{NFL,AGD,LL,coleman}, 
in the coherent part of the Green's function ($G=G_{coh} + G_{incoh}$) will be ``smeared out'' when $\sigma_{\mu} \neq 0$. Similar to Eq. (\ref{equil-neq}), a distribution of chemical potentials (in a Lehmann representation like sum) will lead to the replacement of the coherent Green's function of ordinary Fermi liquids
by
\begin{eqnarray}
\label{gcoh}
G_{coh}(\vec{k}, \omega) = \int  d \mu' P(\mu') \frac{Z_{\vec{k}, \mu'}}{\omega- \epsilon_{\vec{k}} + \mu' + i/\tau_{\vec{k},\mu'}}. 
\end{eqnarray}
Here, $\tau_{\vec{k},\mu'}$ is the quasi-particle lifetime in a system with sharp $\mu'$ at wave-vector $\vec{k}$. 
The denominator in Eq. (\ref{gcoh}) corresponds to the coherent part of the Green's function of a Fermi liquid of
a particular chemical potential $\mu'$ and quasi-particle weight $Z =1$ \cite{NFL,AGD,LL,coleman}. 
Qualitatively, Eq. (\ref{gcoh}) is consistent with indications of the very poor Fermi liquid
type behavior reported in \cite{Dirk}. The effective shift of the chemical potential in Eq. (\ref{gcoh}) is equivalent to a change in the 
frequency dependence while holding the chemical potential $\mu$ fixed; the resulting nontrivial dependence of the correlation function on the frequency
(with little corresponding additional change in the momentum) is,
qualitatively similar to that advanced by theories of ``local Fermi liquids'', e.g., \cite{NFL,Si}.  
Our considerations suggest a similar smearing with the distribution $P(\mu')$ will appear for any quantity 
(other than the Green's function of Eq. (\ref{gcoh})) that is diagonal in the particle number. 
Analogous results will appear for a distribution of other intensive quantities. 
The prediction of Eq. (\ref{gcoh})
(and similar others \cite{me} in different arenas) may be tested to see whether a single consistent probability distribution function $P$ accounts for multiple observables.
General identities relate expectation values in interacting Fermi systems to a weighted average of
the same expectation values in free fermionic systems \cite{similar}. These relations raise the possibility of further related smeared averages, 
akin to those in Eq. (\ref{gcoh}), in numerous systems. Numerically, in various models of electronic systems that display non-Fermi liquid type behaviors, the energy density differences between contending low energy states $\{| \psi^{\alpha} \rangle\}$ (not necessarily exact eigenstates) are often exceedingly small, e.g., \cite{corboz}.
Since these states globally appear to be very different from one another, the matrix element of any local Hamiltonian between any two such orthogonal states vanishes, $\langle \psi^{\alpha} | H| \psi^{\beta} \rangle = 0$ for $\alpha \neq \beta$.
We notice that, given these results, arbitrary superpositions of these nearly degenerate states,
$\sum_{\alpha} a_{\alpha} | \psi^{\alpha} \rangle$, will have similar energies. Thus, for many body Hamiltonians modeling these systems, a superposition
of different eigenstates may be natural from energetic considerations. Towards the end of Section \ref{2b2b}, we remarked on the viable disordered character of the states formed by superposing eigenstates that break continuous symmetries. We now briefly speculate on the corresponding situation for eigenstates in electronic lattice systems that break discrete point group symmetries on a fixed size unit cell. Here, due to the existence of a finite unit cell in reciprocal space, a superposition of eigenstates that are related to each other by a finite number of discrete symmetry operations may not eradicate all Bragg weights. In other words, order may partially persist when superposing states on the lattice that, individually, display different distinct structures. 

\section{Thermalization and Quantum Measurements}
\label{adiabatic-sec}
As we demonstrated in the current work, rapidly driven systems may exhibit uncertainties in their energy and/or other densities. We now close our circle of ideas and focus on the diametrically opposite case of unitary evolutions- slow adiabatic processes (for which, obviously, $d q/dt=0$); this discussion will complement that of Section \ref{sec:effective}. In this Section, we will further speculate on relations concerning thermalization that superficially emulate those of quantum measurements. In line with the focus of the current work, the latter purely hypothetical connections suggest that the absence of thermalization may allow for broad distributions. 

As well known, a basic tenet of quantum mechanics is that a measurement will project or ``collapse'' a measured system onto an eigenstate of the operator being measured. A natural question to ask is whether such effective projections may merely emerge as a consequence of an effective very rapid thermalization of microscopic systems. To motivate this query and more generally examine effectively adiabatic processes, we consider a Hamiltonian 
\begin{eqnarray}
\label{HME}
H_{A \cup B}(t) = H_{A} + H_{AB}(t) + H_{B}
\end{eqnarray}
describing the combined system of two systems and the coupling between them ($H_{AB}$). This Hamiltonian emulates $\tilde{H}$ of the subsystem-environment hybrid of Eq. (\ref{usfe}). We first examine what occurs when the coupling $H_{AB}(t)$ changes adiabatically from zero. Consider the situation wherein, initially, at times $t \le 0$, systems $A$ and $B$ were in respective eigenstates $| \phi_{n_{A}} \rangle $ and $|  \phi_{n_{B}} \rangle$ of $H_{A}$ and $H_{B}$. That is, at times $t \le 0$, the state of the combined system $A \cup B$ was described by the product state of these two eigenstates. We further assume that at times $t<0$, the coupling $H_{AB}(t) =0$ and for times $t \ge 0$ an adiabatic change of $H_{AB}(t)$ ensues. Under these conditions, by the adiabatic theorem, at any later time $t$, the initial state has evolved into a particular eigenstate $ |\phi_{n_{AB}} (t) \rangle$ of $H_{A \cup B}(t)$, we have 
$| \phi_{n_{A}} \rangle |  \phi_{n_{B}} \rangle  \to |\phi_{n_{AB}} (t) \rangle$.
We may expand the density matrices $\rho_{A,B}$ of the initial system $A$ and $B$ in terms of the eigenvectors of $H_{A}$ and $H_{B}$. Explicitly expressing the density matrix $\rho_{A \cup B}(t)$ of the combined system at time $t$ in the eigenbasis of $H_{measure}(t)$, i.e., in the evolution from $t=0$ to a time $t>0$, the density matrix trivially evolves as
\begin{eqnarray}
&& \sum_{n_{A} n_{B}} \rho_{n_{A} n_{B}} | \phi_{n_{A}} \rangle |  \phi_{n_{B}} \rangle  \langle \phi_{n_{A}} | \langle \phi_{n_{B}} |  \to \sum_{n_{A} n_{B}} \rho_{n_{A} n_{B}}  |\phi_{n_{AB}} (t) \rangle  \langle \phi_{n_{AB}} (t) |.
\end{eqnarray}
Hence, if both systems $A$ and $B$ start from equilibrium (and thus have sharp energy densities- i.e., if at $t=0$ the eigenstates of $H_{A}$ and $H_{B}$ of significant amplitude were clustered around a given energy density) then an adiabatic evolution of $H_{AB}(t)$ will yield a density matrix $\rho_{A \cup B}$ having a sharp energy density, $\sigma_{\epsilon_{A \cup B}}(t) =0$. Thus, the notion that {\cal sufficiently slow processes enable systems to remain in equilibrium} is indeed consistent with the adiabatic theorem of quantum mechanics. 

We next comment on how such adiabatic processes (and later briefly discuss more general thermalization events that need not be adiabatic) may superficially emulate certain features of a wavefunction collapse. Towards that end, we consider the extreme case of a microscopic system $A$ (``being measured'') and a macroscopic system $B$ that we may regard as an environment that includes a coupling to an experimental probe at the measurement time $t_{measure}$.  As earlier, for a general adiabatic evolution, $ |\phi_{n_{AB}} (0) \rangle \to  |\phi_{n_{AB}} (t_{measure}) \rangle$. We now allow the coupling $H_{AB}(t)$ to be non-vanishing at all times $t$ (i.e., also including times $t \le 0$) and, due to its ease, first briefly discuss the case when its evolution is adiabatic. 

Under these circumstances, by the adiabatic theorem, $ |\phi_{A \cup B}(t_{measure}) \rangle$ must be an eigenstate of $H_{A \cup B}(t_{measure})$. Thus, such an adiabatic evolution emulates an effective ``collapse'' onto an eigenstate of the Hamiltonian that measures the state of the microscopic system $A$. We emphasize that the state $ |\phi_{A \cup B}(t_{measure}) \rangle$, describing both the microscopic system $A$ and the large system $B$, will be in an eigenstate of $H_{A \cup B}(t_{measure})$- i.e., not only the small system $A$ will be altered by the measurement. While, at any time $t$, the state $ |\phi_{A \cup B}(t) \rangle$ is an eigenstate of $H_{A \cup B}(t)$, its highly entangled content largely remains unknown. Thus, unique predictions for the outcome of other future evolutions cannot be made in such a case. It may be noted that certain ``realistic'' setups involving quantum measurements often entailing higher energy ``thermal'' states of the measurement device (e.g., the reaction between silver ions and the screen that they strike in a Stern-Gerlach type experiment creating visible spots on a screen). The collapsed system is in an excited state. 

The effective ``collapse'' brought about by such an adiabatic process may be nearly immediate for microscopic systems $A$. Typical lower bounds on time scales for adiabatic processes defined by an energy difference $\Delta E$ are set by $\hbar/\Delta E$ (for precise bounds see, e.g., \cite{adiabatic_G}). Such scales are consistent with the uncertainty relations and our bounds of Section \ref{2-Ham}. For small energy splittings $\Delta E$, this adiabatic time scale may become large. The above discussion of a hypothetical adiabatic evolution is merely illustrative. A potentially more practical question concerning realistic $H_{AB}(t)$ is that of the thermalization of the full system. At room temperature, the ``Planckian time" scale for the equilibrium thermalization of random initial states \cite{typical0} (see also Section \ref{sec:patuach}) is $h/(k_{B} T) \sim 10^{-13}$ seconds (e.g., the typical period of a thermal photon). The latter time scale may be smaller than that required for an adiabatic evolution yet is still finite; one may attempt to probe for such an effective finite time collapse borne by thermalization (cf., any such deviations from the textbook ``instantaneous collapse'') only at extremely low temperatures. The very rapid thermalization evolution suggested here allows for multiple measurement outcomes with different probabilities. A measurement provides only partial information on the many body entangled state $| \phi_{A \cup B}(t) \rangle$ formed by $A$ and $B$- it does not specify it. Conditional probabilities may be assigned to the possible future evolutions of this entangled state (and thus of future measurement outcomes thereof). Thus, our suggestion concerning thermalization is somewhat similar ``Quantum Bayesianism'' \cite{QBISM} and other frameworks relying on entanglement, e.g., \cite{Zeh,Zurek}. 

Further parallels between equilibration and certain features of an effective collapse in quantum measurements are  motivated by the Eigenstate Thermalization Hypothesis
 \cite{eth1,eth2,eth3,eth4,rigol,pol,polkovnikov1,polkovnikov2,von_neumann}. When valid, this hypothesis equates the results of local measurements of general observables ${\cal{O}}$ 
 in pure eigenstates $\{| \phi_{n} \rangle\}$ (of energies $\{E_{n} \}$) of a general Hamiltonian (including Hamiltonians describing a coupling between a measurement device and a microscopic system) with expectation values in equilibrated thermal systems defined by the full system Hamiltonian,
 \begin{eqnarray}
 \label{nOn}
 \langle \phi_{n} | {\cal{O}} | \phi_{n} \rangle = Tr(\rho(E_{n}) {\cal O}).
 \end{eqnarray}
 Here, $\rho(E_{n})$ is an equilibrium density matrix associated with the energy $E=E_{n}$ (and, when applicable, any other conserved quantities defining the state $| \phi_{n} \rangle$ and the thermal system). Taken to the extreme, Eq. (\ref{nOn}) suggests that we may relate two seemingly very different concepts:
  
 ({\bf i}){\it An effective collapse to an eigenstate}. The lefthand side of Eq. (\ref{nOn}) yields the results of quantum expectation values associated with (projecting the system onto) eigenstates of the Hamiltonian (also describing, as in a realization of Eq. (\ref{HME}), the measurement process- the substantial coupling $||H_{AB}|| \gg ||H_{A}||$ of the environment ($B$) containing a measurement device to the measured quantity $(A$) and) providing the dynamics.
 
 ({\bf ii}) {\it Equilibration.} The righthand side of Eq. (\ref{nOn}) reflects the outcomes of equilibration (in which, inasmuch as any observable ${\cal{O}}$ can inform, the system effectively becomes indistinguishable from an eigenstate of the very same Hamiltonian associated with item ({\bf i})). As noted above, in a realization of Eq. (\ref{HME}) describing a typical measurement, this Hamiltonian displays a dominant coupling between the measurement device and the quantity being measured, $||H_{AB}|| \gg ||H_{A}||$. 
 
 That is, denoting by $\rho_{collapse}$ the probability density matrix following the collapse to an eigenstate of the measurement device and $\rho_{equilibration}$ that associated with equilibration of the small system with the measurement device, Eq. (\ref{nOn}) suggests a very qualitative relation,
 \begin{eqnarray}
 \rho_{collapse} ``='' \rho_{equilibration}.
 \end{eqnarray}
  
More general than adiabatic processes alone, thermalization indeed shares other commonalities with quantum measurements. Just as a quantum measurement (and ensuing collapse) is not a time reversal invariant operation \cite{cnt} so, too, is a typical finite $T$ thermalization process in a highly entangled many body system. The second law of thermodynamics is consistent with an evolution of the entangled $A \cup B$ system displaying a non-decreasing entropy upon performing consecutive measurements (compatible with indeterminate outcomes for other subsequent measurements thereafter). We underscore that what we are suggesting the/asking is whether the evolution of a collapse to an eigenstate following a quantum measurement (an eigenstate of $H_{AB}$) is no different than a particular case of a unitary evolution with a Hamiltonian (the latter having $||H_{AB}|| \gg ||H_{A}||$). A sufficiently long time evolution with $H_{A \cup B}$ leads, in systems that thermalize with this Hamiltonian, to an equilibrium state. Whenever the Eigenstate Thermalization Hypothesis applies, the resulting equilibrium thermal state (to which the system will evolve) will be an eigenstate of the Hamiltonian $H_{A \cup B}$. In other words, after a (possibly very short) thermalization time after which the system equilibrates, the system will indeed ``collapse'' to an eigenstate of $H_{A \cup B}$ describing the coupling between $A$ and $B$. If the system is already an eigenstate of $H_{AB}$ then it will remain stationary- a ``measurement'' will lead to a constant outcome. If the system $A$ is not an eigenstate of $H_{AB}$ then it may evolve and ``precess'' due to the applied  external ``field'' $H_{AB}$. In a simple setting, the long time average of general observables will be the average over the measurements associated with these ``precessions'' (mirroring the phase fluctuations of Eq. (\ref{rhl})). The long time average associated with $H_{AB}$ (or any other Hamiltonian) will be given by Eq. (\ref{oolta}). If the Eigenstate Thermalization Hypothesis holds then a far stronger result will appear: the longtime average over microscopic precessions will correspond to an expectation value of the said observable within a single eigenstate of the full Hamiltonian $H_{A \cup B}$.  

A notional link between ({\bf{i}}) and ({\bf ii}) is naturally compatible with the appearance of wide distributions of various measurable quantities
in non-equilibrium systems. 
Regardless of the validity of the Eigenstate Thermalization Hypothesis of Eq. (\ref{nOn}), any equilibrium expectation value is an ensemble average over states having a sharp value of intensive state variables $q$. Thus, as alluded to in Section \ref{deviation}, barring special eigenstates \cite{MBL1,MBL2,MBL3,MBL4,MBL5,MBL6,MBL7,noMBL,yesMBL}, the system may rather straightforwardly exhibit non-equilibrium behaviors if the distribution of its intensive thermodynamic state
variables $q$ is, quite simply, not a delta function. We conclude this Section by underscoring that (as we explained in several of the previous Sections) the central result of the current paper regarding the existence of wide distributions in non-equilibrium systems does not rely on quantum effects nor the character of quantum measurement (on which we speculated above). Similar behaviors may appear in classical systems. The use of the quantum language in the current article merely made our considerations more precise and also gave rise to the bounds of Section \ref{2-Ham}. 

\section{Conclusions} 
\label{sec:conclusions}
We illustrated that a finite rate variation of general intensive quantities may lead to long range correlations. In the simplest variant of this effect, in systems having varying intensive observables $q$ (such as the energy density $\epsilon$) for which $\frac{d q}{dt} = {\cal{O}}(1)$, 
an average connected two site correlation functions need not vanish even
for sites are arbitrarily far apart. Trivial extensions hold for weaker variations of intensive quantities. For instance, if only short range effects of the environment appear (e.g., fluids featuring local hydrodynamic coupling to their boundaries) and, consequently, for an $N$ site system residing in $d$ spatial dimensions, $\frac{dq}{dt} = {\cal{O}}(N^{-1/d})$ then the average value of the connected two point correlation function for an arbitrary pair $(i,j)$ of far separated sites may be asymptotically bounded as $\overline{G} \ge {\cal{O}}(N^{-2/d})$ \cite{triv2/d}.

In the quantum arena, the general non-local correlations that we found relate to the macroscopic entanglement present in typical thermal states. Our results highlight that,
even in seemingly trivial thermal systems, one cannot dismiss the existence of long range correlations. Our analysis of non-equilibrium systems does not appeal to conventional coarsening and spinodal decomposition phenomena (although the departure from a spatially uniform true equilibrium state in spinodal systems is very naturally consistent with a distribution of low energy solid like and higher energy fluid like states). Cold atom systems may provide a controlled testbed for our approach. We speculate that our results may also appear in naturally occurring non-equilibrium systems. As we explained (Section \ref{2b2b}), the peculiar effect that we find may rationalize the unconventional behaviors of glasses and supercooled fluids. Our effect might further appear in electronic systems that do not feature Fermi liquid behavior (Section \ref{sec:electron}). Here, a broad distribution of effective energy densities and/or chemical potentials may appear. The validity of weighted averages such as that of Eq. (\ref{gcoh}) may be assessed by examining whether a unique distribution $P$ simultaneously accounts for all measurable quantities. In Section \ref{adiabatic-sec}, we illustrated how adiabatic processes maintain sharp thermodynamic quantities and speculated that a nearly instantaneous equilibration of small systems with macroscopic ones may emulate certain features of quantum measurements. We hope that our suggested effect and analysis will be further pursued in light of their transparent mathematical generality and ability to suggest new experimental behaviors (e.g., the universal viscosity collapse of supercooled liquids that it predicted and is indeed empirically obeyed over sixteen decades (Figure \ref{Collapse.})).

While deriving the above, we arrived at other results. These include the finding of {\it universal bounds relating thermalization and time derivatives of general observables} (Section \ref{sec:patuach}), explaining how driven system may be described by an effective
equilibrium distribution in which {\it the dynamics are universally generated by the logarithm of the corresponding probability density matrix} (Section \ref{sec:effective}), and speculatively pointing to similarities between unitary dynamics, thermalization, and quantum measurements (Section \ref{adiabatic-sec}). Additional technical details have been relegated to the Appendices. In \ref{prompt}, we motivate {\it the appearance of long time Gaussian distributions in both equilibrium and non-equilibrium systems}. 
 
{\bf{Acknowledgments}} \newline
I am thankful to interest by and discussions with M. Alford, N. Andrei, R. Bruinsma, K. Dahmen, A. Dymarsky, S. Ganeshan, A. K. Gangopadhyay, S.  Gopalakrishan, E. Henriksen, K. F. Kelton, A. Kuklov, M. Lapa, A. J. Leggett, K. Murch, F. Nogueira, S. Nussinov, V. Oganesyan, G. Ortiz, A. Polkovnikov, L. Rademaker, S. Ryu, C. Sa de Melo, M. Schossler, A. Seidel, D. Sels, K. Slagle,  and N. B. Weingartner. I thank S. Hartnoll, S. Nussinov, L. Rademaker, and an anonymous referee for critical reading of the manuscript and questions. This research was principally supported by the National Science Foundation under grant NSF 1411229 and further supported by NSF PHY17-48958 (KITP) and NSF  PHY-160776 (the Aspen Center for Physics).

\appendix

\section{Order of magnitude time estimates} 
\label{extensive} 

For radiation traveling at a speed $c$, during a time interval $\Delta t$, an extensive (i.e., volume proportional) amount of radiative heat $\Delta Q_{rad}$ may flow into a $d$ dimensional system of linear scale $L$ if $L \lesssim c (\Delta t)$. Thus, bulk effects from radiative heat exchange may only be present after a sufficiently long time $t \gtrsim L/c$ after radiative heating or cooling begins. Similarly, if the effective radiative absorption lengths $\ell_{S}$ and $\ell_{B}$ of, respectively, the media comprising the system and the surrounding heat bath satisfy $\ell_{S,B} \gtrsim L$ then the total system radiative heat flow rate may be proportional to its volume, $\Delta Q_{rad}/ \Delta t = {\cal{O}}(V)$.  The existence of a minimal time scale in non-relativistic systems may be proven from the Lieb-Robinson bounds (see \ref{LR_explain}).  

We now briefly provide order of magnitude estimates.  If, e.g., $L$ is the order of 1cm for a sample of index of refraction $\sim 1 $ and the relevant velocity $v=c$ is a typical radiation speed (as in, radiative cooling or heating) then the requisite minimal time scale 
$t_{\min} = \frac{L}{c} \sim 3 \times 10^{-11} \text{sec.}$
Experiments on supercooled liquids typically involve cooling at a rapid finite rate (thus, the experimental time scale $t \ge t_{\min}$). In metallic liquids (that form glasses when supercooled), often in experiments one uses (radiative) laser beam heating. In typical metals, both the heat and charge effectively travel at a finite fraction (typically of the order of $10^{-2}$) of the speed of light $c$ the effective speed of electrons in a metal; both effective heat and charge transport velocities are possibly the same in conventional metals obeying the Wiedemann-Franz law). The heat transfer rate is bounded by the rates of any of the individual (radiative/conduction/convection) processes that contribute to it. Thus, if either the typical radiative or conductive processes occur at speeds associated with a finite fraction of the speed of light $c$ then so, too, is the total heat transfer. In metals as well as in systems where the radiative penetration depth is larger than or of the scale of the linear dimension of the material, the speed associated with heat transfer is rather large and, correspondingly, the minimal time scale can, in these instances, become very short. 

The continuity equation for the local energy density, $\partial_{t} \epsilon(\vec{x}) + \vec{\nabla} \cdot \vec{j}(\vec{x}) =0$ where $\vec{x}$ denotes a spatial location in the continuum limit. If the average current flowing through the system surface $|\overline{j}| \equiv |\epsilon| v_{Q}$ where $\epsilon$ is the global average of the local energy density with $v_{Q}$ a speed characterizing heat or energy flow through a boundary of $\tilde{A}$) and the volume $V = {\cal{O}}(\tilde{A}L)$ then the rate $ (dE/dt)/E= {\cal{O}}(v_{Q}/L)$. That is, the time required to change the system energy density is proportional to $L$.

\section{A finite rate of change of intensive quantities and the Lieb-Robinson light cone}
\label{LR_explain} 

In driven systems with $d\epsilon/dt = {\cal{O}}(1)$, the commutators with expectation values equal to $dE/dt$ must be extensive. Specifically, both in (1) closed systems with a time dependent Hamiltonian (as in, e.g., Section \ref{sec:Magnus}), the commutator $[H^{H}(t), H]$ (where $H^{H}(t)$ is the Heisenberg picture Hamiltonian) as well as in (2) settings similar to those in Sections \ref{intuition} (Eq. (\ref{usfe}) therein) and \ref{2-Ham}, namely a subsystem with Hamiltonian $H$ in contact with the full system of Hamiltonian $\tilde{H}$, where the relevant commutator is given by Eq. (\ref{uncert}), the above two-Hamiltonian commutators are of order ${\cal{O}}(N)$. In both (1) and (2), for local Hamiltonians, one may examine the constraints implied by causality as these appear via the Lieb-Robinson bound \cite{Lieb_Robinson} for commutators $[{\cal A}_{H} (t), {\cal B}(0)]$ of local Heisenberg picture operators ${\cal A}$ and ${\cal B}$ that have their support centered about sites $i$ and $j$. In particular, whenever
the Lieb-Robinson bound applies, the operator norm
($||\cdot||$) of commutators between any two local quantities ${\cal A}$ and ${\cal B}$ is bounded from above by 
\begin{eqnarray}
|| [{\cal A}_{H}(t), {\cal B}(0)] || \le c' e^{(-a (|i-j|  - v_{\sf LR} |t|))}.
\label{LR_EQ+}
\end{eqnarray}
 Here, $a$ and $c'$ are constants and $v_{\sf LR}$ is the Lieb-Robinson speed of Section \ref{intuition}. The Lieb-Robinson speed plays the role of the velocity of light in relativistic theories. Since, by the Heisenberg equations of motion, the commutators in both cases (1) and (2) have an average given by the derivative of the energy $dE/dt$ and since the latter is of order $N$, i.e., $dE/dt = {\cal{O}}(N)$ when the energy density varies at a finite rate, the upper bounds on the two Hamiltonian commutators must also be of order $N$. Equivalently, as we next detail, the Lieb-Robinson ``light cone'' \cite{Lieb_Robinson}, during the times at which the energy density as measured by $H/N$ varies at a non-zero rate, is of the scale of the entire system. 

The Schrodinger picture Hamiltonian $\tilde{H}$ of the combined system (${\cal S}$) + environment (${\cal E}$) hybrid may be expressed as $\tilde{H} = H + H_{{\cal S}-{\cal{E}}} + H_{{\cal{E}}}$ where $H$ is the system Hamiltonian, $H_{{\cal S}-{\cal E}}$ denotes the coupling of the system to its environment, and $H_{{\cal{E}}}$ is the Hamiltonian of the environment. When only bounded local interactions appear in the system-environment hybrid, we will write the Hamiltonians (in the form of Eq. (\ref{HsH'}) (explicitly rewritten below) and its generalization),
\begin{eqnarray}
H= \sum_{i} {\cal{H}}_{i}
\label{hhi}
\end{eqnarray}
and 
\begin{eqnarray}
\label{seeh}
H_{{\it S}-{\cal E}} + H_{\cal E} = \sum_{j'} {\cal{H}}_{j'}
\end{eqnarray}
as, respectively, sums of the bounded local operators $\{{\cal{H}}_{i}\}$ and $\{{\cal{H}}_{j'}\}$. In what follows, as throughout the main text of the paper, $\tilde{\rho}$ will denote the density matrix of the system-environment hybrid. By Heisenberg's equations of motion, with $\epsilon = \frac{1}{N}Tr[\tilde{\rho} H^{H}(t)]$ the energy density of the system (where $H^{H}(t)= {\cal{\tilde{U}}}^{\dagger}(t) 
H {\cal{\tilde{U}}}(t)$, with ${\cal{\tilde{U}}}(t) = e^{-i\tilde{H}t/\hbar}$, for a time independent ${\tilde{H}}$ in Eq. (\ref{ute-trail})), the derivative $ i \hbar \frac{d \epsilon}{dt}$ is given by 
\begin{eqnarray}
\label{trivtrivtriv}
&&\frac{1}{N} \sum_{i} Tr(\tilde{\rho} [{\cal{H}}^{H}_{i}(t), \tilde{H}]) \nonumber
\\ =  &&\frac{1}{N} \sum_{i} Tr(\tilde{\rho} [{\cal{H}}^{H}_{i}(t), H^{H}(t) + H^{H}_{{\cal S}-{\cal{E}}}(t) + H^{H}_{{\cal{E}}}(t)]) \nonumber
\\ =  && \frac{1}{N} \sum_{i} Tr(\tilde{\rho} [{\cal{H}}^{H}_{i}(t), H^{H}_{{\cal S}-{\cal{E}}}(t) + H^{H}_{{\cal{E}}}(t)]).
\end{eqnarray}
The first equality of Eq. (\ref{trivtrivtriv}) invoked the trivial invariance of $\tilde{H}$ under time evolution with ${\cal{\tilde{U}}}(t) = e^{-i\tilde{H} t/\hbar}$ (i.e., $\tilde{H} = {\cal{\tilde{U}}}^{\dagger}(t) \tilde{H} {\cal{\tilde{U}}}(t) =  \tilde{H}^{H}(t)$). The last equality in Eq. (\ref{trivtrivtriv}) follows since, in the second commutator, $H^{H}(t) = \sum_{i} {\cal{H}}^{H}_{i}(t)$ similarly commutes with itself. For $t>0$, the norm of the above commutator average 
\begin{eqnarray}
\label{thh}
&&\frac{1}{N} | \sum_{i} Tr(\tilde{\rho} [{\cal{H}}^{H}_{i}(t), H^{H}_{{\cal S}-{\cal{E}}}(t) + H^{H}_{{\cal{E}}}(t)]) | \nonumber
\\ &&= \frac{1}{N} | \sum_{i,j'} Tr(\tilde{\rho} [{\cal{H}}^{H}_{i}(t), {\cal{H}}^{H}_{j'}(t)])| \nonumber
\\  &&\le \frac{c'}{N} \sum_{i,j'} e^{-a(|i-j'|-v_{LR}t)}.
\end{eqnarray}
The decomposition of the system Hamiltonian $H = \sum_{i} {\cal{H}}_{i}$ into a sum over local regions spans $N' = {\cal{O}}(N)$ terms- the number of sites in the system. In the last inequality, $c'$ is a constant, and $a$ and $v_{LR}$ denote the Lieb-Robinson decay constant (inverse correlation length) and speed respectively of Eq. (\ref{LR_EQ+}) \cite{Lieb_Robinson}. Rather explicitly,  
\begin{eqnarray}
\label{thh1}
 |Tr(\tilde{\rho} [{\cal{H}}^{H}_{i}(t), {\cal{H}}^{H}_{j'}(t)])| \le || [{\cal{H}}^{H}_{i}(t), {\cal{H}}^{H}_{j'}(t)]||.
 \end{eqnarray}
 \newline
 In order to derive Eq. (\ref{thh}), we note that the Lieb-Robinson bounds of Eq. (\ref{LR_EQ+}) \cite{Lieb_Robinson} applied to the local operators appearing in the Hamiltonian, 
$|| [{\cal{H}}^{H}_{i}(t), {\cal{H}}^{H}_{j'}(t)]|| \le c' e^{-a(|i-j'|-v_{LR}t)}$,
imply Eq. (\ref{thh}). For each $i \in {\cal S}$, there is a minimum distance $D(i)$ between $i$ and the surrounding region where $H_{{\it S}-{\cal E}} + H_{\cal E}$ has its support. For any such $i$, we may bound (from above) the sum over all $j'$ of the exponential $e^{-a(|i-j'|-v_{LR} t)}$ by a sum of this exponential over the larger domain external to a sphere of radius $D(i)$ around $i$ (such a volume contains ${\cal{E}}$ as a subset). For sufficiently short times $t$, in Eq. (\ref{thh}), the sum  
$\frac{c'}{N} \sum_{i,j'} e^{-a(|i-j'|-v_{LR} t)}$
tends to zero for macroscopic systems (since the minimal distance $D(i)$ of a typical $i \in S$ to its surrounding environment is of the order of the system length); for vanishingly small times, the latter sum of $e^{-a(|i-j'|-v_{LR} t)}$ over such a larger domain of $j'$ values with $|i-j'| \ge D(i)$ decays exponentially in $D(i)$. Specifically, in $d$ spatial dimensions, $\frac{c'}{N} \sum_{i,j'} e^{-a(|i-j'|-v_{LR} t)}$ scales as $O(D^{d}e^{-aD})$ for
large $D$. Putting all of the pieces together, we see that the Lieb-Robinson bounds imply that at vanishingly short times, $\frac{1}{N} |\sum_{i} Tr(\tilde{\rho} [{\cal{H}}^{H}_{i}(t), \tilde{H}]|)$ is bounded from above by a function that is exponentially small in the length of the system. In other words, under the above specified locality conditions, the energy density of a macroscopic system cannot change at a finite rate at sufficiently short times. A corollary of these inequalities is that in a local theory in which the the Lieb-Robinson bounds hold,
a transient time Hamiltonian describing the effects of the environment cannot change instantaneously in such a way as to give rise to a finite change in the energy density of the system. Thus, generally, the environment may not truly instantaneously couple to (nor decouple from) a finite fraction of a macroscopic system (in the form of an effective instantaneously varying Hamiltonian $H(t')$ (as in Section \ref{sec:dual}) when procedure (1) of Section \ref{sec:sketch} is invoked). The influences of the environment (and variations in any Hamiltonian that emulate the effects of the environment) are limited those associated with ``light cone'' distances of size $(v_{LR} t)$. The above calculations may be replicated, nearly verbatim, for operators associated with other intensive quantities $q$ different from $\frac{H}{N}$ of Eq. (\ref{hhi}). 

\section{Relating Heisenberg's equation of motion to correlations}
\label{app:triv1}
In what follows, we explicitly demonstrate that (when all interactions (i.e., $\tilde{H}$) are time independent): 
\newline

$\bullet$ If the energy density of the system changes at a finite rate then there must be {\it system length spanning correlations between the external environment and the system itself}. 
\newline

A formal proof of this assertion is straightforward. Using the notation of \ref{LR_explain} and the main text, by the Heisenberg equations of motion, 
\begin{eqnarray}
\label{lesse}
0< \Big |\frac{d \epsilon}{dt} \Big| = \frac{1}{N \hbar}  \Big|Tr(\tilde{\rho} [\tilde{H}, H^{H}]) \Big| =
\frac{1}{N \hbar}  \Big|Tr(\tilde{\rho} [\tilde{\delta H}, \delta H^{H}]) \Big|\nonumber
\\ =  \frac{1}{N \hbar} \Big| \sum_{i} Tr(\tilde{\rho} [{\delta \cal{H}}^{H}_{i}(t), \delta H^{H}_{{\cal S}-{\cal{E}}}(t) + \delta H^{H}_{{\cal{E}}}(t)])\Big|. 
\end{eqnarray}
For any of the Hamiltonians appearing in Eq. (\ref{lesse}) which we now generally represent by $Q$, we define $\delta Q \equiv (Q - \langle Q \rangle) \equiv  (Q - Tr(\tilde{\rho} Q))$. Apart from trivial shifts by $(- \langle Q \rangle$), Eq. (\ref{lesse}) and its derivation are identical to those of Eq. (\ref{trivtrivtriv}). For all operators ${\hat{{\cal A}}}$ and ${\hat{{\cal B}}}$,  
$\Big| Tr (\tilde{\rho}[{\hat{{\cal A}}},{\hat{{\cal B}}}]) \Big| 
\le 2 \times \max \Big\{ \Big|Tr (\tilde{\rho}({{\hat{\cal A}}} {\hat{{\cal B})}}) \Big|, 
\Big|Tr(\tilde{\rho}({\hat{{\cal B}}}{\hat{{\cal A}}}))\Big| \Big \}$. 
Thus, from Eq. (\ref{lesse}),
\begin{eqnarray}
\label{0<corr}
0&<&  \frac{2}{N} \sum_{i} \max \Big\{ \Big|Tr \Big(\tilde{\rho}({\delta \cal{H}}^{H}_{i}(t) (\delta H^{H}_{{\cal S}-{\cal{E}}}(t) + \delta H^{H}_{{\cal{E}}}(t))) \Big) \Big|, \nonumber
\\ &&\Big|Tr \Big(\tilde{\rho}(\delta H^{H}_{{\cal S}-{\cal{E}}}(t) + \delta H^{H}_{{\cal{E}}}(t)) {\delta \cal{H}}^{H}_{i}(t) 
\Big)\Big| \Big \}.
\end{eqnarray}
Since the number ($N'$) of system sites $i$ associated with the bounded local operators ${\cal{H}}^{H}_{i}$ (Eq. (\ref{HsH'})) is $N' = {\cal{O}}(N)$, from Eq. (\ref{0<corr}), we see that the average correlator between the local
  ${\delta \cal{H}}^{H}_{i}$ (that, apart from a set of vanishing measure, all lie in the system bulk at a distance $D = {\cal{O}}(L)$ from the surrounding environment) and the fluctuations $(\delta H^{H}_{{\cal S}-{\cal{E}}}(t) + \delta H^{H}_{{\cal{E}}}(t))$ must be finite. In other words (as is expected), the correlator between the bulk and the Hamiltonian coupling it to the surrounding environment is of order unity. Given Eq. (\ref{0<corr}),
this typical order unity correlator (the average over all sites $i$) between $(\delta H^{H}_{{\cal S}-{\cal{E}}}(t) + \delta H^{H}_{{\cal{E}}}(t))$ and ${\delta \cal{H}}^{H}_{i}(t)$ is of a uniform sign. (This uniform sign character is emulated by the uniform sign coupling between the collective (environment) coordinate $\tilde{q}$ in the solvable examples of Section \ref{sec:CSM}  to each of system degrees of freedom $i$ (corresponding to a uniform sign coupling for each of the spokes 
between the environment ${\cal E}$ and the system sites ${\cal S}$ in the cartoon of Figure \ref{coupling.}). For random sign couplings of uniform strength between the system and its environment, the energy density might vary at an ${\cal{O}}(N^{-1/2})$ rate.) The above holds irrespective of how large $N$ may be so long as $\Big(\frac{d \epsilon}{dt} \Big)$ is non-vanishing. We next consider what occurs when, similar to \ref{LR_explain}, we invoke Eq. (\ref{seeh}) and express the Hamiltonian of the environment and its coupling to the system as a sum of local terms ($\{ {\cal{H}}_{j} \}$ with $j \not \in {\cal S}$). In such a case, Eq. (\ref{0<corr}) will imply that if there is an exponential decay length $\xi$ associated with the larger of the two connected correlations functions $G_{ij'}(t) \equiv  \langle {\delta \cal{H}}_{i} (t) {\delta \cal{H}}_{j'} (t) \rangle \equiv Tr (\tilde{\rho} ({\delta \cal{H}}_{i} (t) {\delta \cal{H}}_{j'} (t)))$ and $G_{j'i}(t) \equiv  \langle {\delta \cal{H}}_{j'} (t) {\delta \cal{H}}_{i} (t) \rangle \equiv Tr (\tilde{\rho} ({\delta \cal{H}}_{j'} (t) {\delta \cal{H}}_{i}) (t))$ then $\xi \gtrsim {\cal{O}}(L)$. 
 Similarly, if the correlator decays algebraically, $|G_{ij'}| \sim |i-j'|^{-p}$, then Eq. (\ref{0<corr}) implies a finite rate of change of the energy density for large systems sizes $L$ only if $p<d$ with $d$ the spatial dimensionality of the system and the environment. It is noteworthy that the commutator of Eq. (\ref{lesse}) has (when evaluated with $\tilde{\rho}$) an imaginary expectation value for the Hermitian Hamiltonian operators. For semiclassical systems, the real component of the correlator $G_{ij'}$ is, typically, far larger than its imaginary part (which we bounded in the above). Stated equivalently, the expectation value of the anticommutator
 $\{ {\delta \cal{H}}_{i} (t),  {\delta \cal{H}}_{j'} (t) \}$ is, in semiclassical systems, normally far larger than
 the expectation value of the commutator $[{\delta \cal{H}}_{i} (t),  {\delta \cal{H}}_{j'} (t)]$. 

\section{Conditional probability arguments for long range correlations}
\label{semi-class-prob}

As we explained in \ref{app:triv1}, a driven system (one in which the intensive quantities change at a finite rate) must exhibit long range correlations between observables (${\cal{H}}_{i}$) at sites $i$ in the bulk to the environment (${\cal{E}}$). We now apply ``classical'' probability arguments to demonstrate that
when these long range correlations between different sites in the system and its environment are present, 
then the local Hamiltonian terms ${\cal{H}}_{i}$ at different sites in the system bulk may exhibit long range correlations. Towards this end, we write the classical joint probability distribution $P(E_{{\cal{E}}}, E_{i}, E_{j})$ associated with the values ($E_{i,j})$ of the energies ${\cal{H}}_{i}$ and ${\cal{H}}_{j}$ at the two sites $i, j$ in the bulk (in the system ${\cal S}$) and  the energy
 $(H^{H}_{{\cal S}-{\cal{E}}}(t) +  H^{H}_{{\cal{E}}}(t))$ affiliated with the environment ${\cal{E}}$ (denoted by $E_{\cal{E}}$). 
 In the context of \ref{app:triv1}, the joint probability distribution
 \begin{eqnarray}
 P(E_{{\cal{E}}}, E_{i}, E_j) \equiv  Tr\Big[\tilde{\rho}~ \delta(H^{H}_{{\cal S}-{\cal{E}}}(t) +  H^{H}_{{\cal{E}}}(t)- E_{\cal{E}}(t)) 
 \delta({\cal{H}}_{i} - E_{i}(t))~ \delta({\cal{H}}_{j} - E_{j}(t))\Big]. \nonumber
 \end{eqnarray} 
 Other joint probabilities are defined similarly. By the chain rule of conditional probabilities, 
 \begin{eqnarray}
 \label{chainprob}
 P(E_{{\cal{E}}},E_i, E_j) = P(E_i| E_{{\cal{E}}}, E_j) P(E_{{\cal{E}}}|E_j) P(E_j).
 \end{eqnarray}
 Here, $P(E_i|E_{{\cal{E}}},E_j) = \frac{P(E_{i},E_{{\cal{E}}},E_j)}{P(E_{{\cal{E}}},E_j)}$ is the conditional probability of measuring a local energy (with a local ``thermometer'') of value $E_{i}$ given a value of the local energy ($E_{j}$) at site $j$ and the above defined energy $E_{\cal{E}}$ associated with the environment ${\cal{E}}$. 
 Now, if $i$ is independent of $j$ then 
 \begin{eqnarray}
 P(E_i| E_{{\cal{E}}}, E_j ) = P(E_i| E_{{\cal{E}}}).
 \end{eqnarray} 
 Subsequently, Eq. (\ref{chainprob}) reduces to 
 \begin{eqnarray}
 P(E_{{\cal{E}}},E_i,E_j) = P(E_i| E_{{\cal{E}}}) P (E_{{\cal{E}}}| E_j ) P(E_j).
 \end{eqnarray} 
 The classical joint probability $P(E_i,E_j)$ then reads
 \begin{eqnarray}
 \label{independence_pass**}
 P(E_i,E_j) =& \sum_{E_{\cal{E}}} P(E_{{\cal{E}}},E_i,E_j) \nonumber
 \\  =& \sum_{E_{\cal{E}}} P(E_i| E_{{\cal{E}}}) P(E_{{\cal{E}}}| E_j) P(E_j) . 
 \end{eqnarray}
 This, in turn, implies that the conditional probability between the values of $E_{i}$ and $E_{j}$ 
 at the two sites in the system bulk is given by
 \begin{eqnarray}
 \label{condij}
 P(E_i|E_j) = \sum_{E_{{\cal{E}}}} P(E_i| E_{{\cal{E}}}) P(E_{{\cal{E}}}| E_{j} ) \nonumber
 \\ = \frac{\sum_{E_{\cal{E}}} P(E_i | E_{{\cal{E}}}) P(E_j| E_{{\cal{E}}}) P(E_{{\cal{E}}})}{
 \sum_{E_{\cal{E}}} P(E_j| E_{{\cal{E}}}) P (E_{{\cal{E}}})}.
 \end{eqnarray}
 In the second (alternate form) line of Eq. (\ref{condij}), we invoked Bayes' theorem. 
 \ref{app:triv1} demonstrated that in a (quantum) system in which the energy density varies at a finite rate,
 there the energy fluctuations in $i$ and ${\cal{E}}$ are not independent of one another. Similarly, the energy fluctuations in $j$ and in ${\cal{E}}$ are correlated and not independent of one another. Thus, in general, the conditional probabilities 
 \begin{eqnarray}
 \label{condpp}
 P(E_i|E_{{\cal{E}}}) \neq P(E_i) \mbox{ and } P(E_{{\cal{E}}}| E_j) \neq P(E_{{\cal{E}}}).
 \end{eqnarray} 
(Analogously, for the conditional probabilities appearing in the second line of Eq. (\ref{condij}), a coupling between the driving environment and the bulk implies (as formalized in \ref{app:triv1}) that $P(E_j| E_{{\cal{E}}}) \neq P(E_{j})$.) These inequalities are expected to generally hold for both quantum as well as classical systems since, at their core, these relations indeed reflect the bulk coupling between the environment driving the system and the system itself necessary to induce a finite rate of change of the energy density. (See also the discussion in \ref{app:triv1} concerning semiclassical systems.) When the inequalities of Eq. (\ref{condpp}) are substituted in Eq. (\ref{condij}), we will generally have
 \begin{eqnarray}
 P(E_i|E_j) \neq P(E_i).
 \end{eqnarray}
 That is, the local energy fluctuations at (arbitrarily far separated) sites $i$ and $j$ in the system bulk are {\it not independent} of one another as assumed in deriving Eq. (\ref{independence_pass**}). Thus, there are non-trivial correlation between any sites $i$ and $j$ in the driven system ${\cal S}$. With reference to Eq. (\ref{central}), we now see that (even for large $|i-j|$) the covariance
 \begin{eqnarray}
 G_{ij} = \sum_{E_{i},E_{j}} \Big(P(E_i|E_j) - P(E_i) \Big) P(E_j) E_{i} E_{j},
 \end{eqnarray}
 need not vanish (and may be of order unity). If the coupling to the environment is the dominant contribution to the correlations $G_{ij}$ when $|i-j|$ is large then when the coupling between the environment and different sites $i$ in the bulk is (nearly) constant, then all connected pair correlators $G_{ij}$ appearing in Eq. (\ref{central}) will be of (almost) uniform magnitude (and sign). Under these conditions, $\sigma_{\epsilon} = {\cal{O}}(1)$. 
 The above conditional probability arguments may be extended verbatim to general situations when general observables in ${\cal{E}}, i,$ and $j$ may carry time and/or other indices in addition to spatial ones. Indeed, in the simple conditional probability computations above the specific physical content of these labels was irrelevant.

\section{Other long range correlations}
\label{rev:long-range}
It has long been known that algebraic power law correlations may appear in non-equilibrium steady states of fluids and other systems in which the energy density and other intensive quantities do not vary with time and in which a coupling to spatially non-uniform external bath was a local boundary effect \cite{classical*,classical**,derida'}. The existence of a spatially non-uniform profile of the local energy density may enhance the large fluctuations that we find in the current work. We will briefly touch on related aspects towards the end of Section \ref{2b2b}. In classical systems with local interactions, broad distributions of various observables may also occur in the thermodynamic limit when these systems are disordered. This phenomenon is known as ``non-self-averaging'', e.g., \cite{derida,amnon,eitan,per}. In these disordered systems, an ensemble average of a physical observable computed over different disorder realizations may differ significantly from the expectation value of the same quantity in any single member of the ensemble. The systems that we will focus on in the current work need not be disordered nor critical. However, given the absence of self-averaging in such disordered classical systems, we remark that the broadening that we find will also apply to various systems when the (``ensemble of'') eigenstates of the density matrix effectively describe these different disorder realizations of classical critical systems. This is so since, in such cases, an average computed with the probability density matrix $\rho$ will reproduce the average associated with an ensemble of disordered classical states. 

In the driven system, the correlators $G_{ij}$ of Eq. (\ref{central}) may be finite. By evolving (forward and backwards) in time, one can examine the correlations of general quantities in the driven system. Eqs. (\ref{central}, \ref{usfe}) allow for other non-local covariances to be finite. Specifically, whenever Eq. (\ref{central}) holds, regarded as a formal operator, the Heisenberg picture Hamiltonian $H^{H}(t) =  {\cal{\tilde{U}}}^{\dagger} (t) H  {\cal{\tilde{U}}} (t)$, evaluated for times $t$ at which Eq. (\ref{central}) applies, will trivially, exhibit a standard deviation that is ${\cal{O}}(N)$ when computed with the initial density matrix $\tilde{\rho}$ at (i.e., prior to driving the system). The proof of this assertion is straightforward. If $\langle H^{H}(t) \rangle = Tr(\tilde{\rho} H^{H}(t))$ then,
\begin{eqnarray}
Tr \Big[ \tilde{\rho}(H^{H}(t) - \langle H^{H}(t)\rangle)^{2} \Big] = 
Tr \Big[{\cal{\tilde{U}}} (t)
\tilde{\rho} {\cal{\tilde{U}}}^{\dagger} (t)
(H-\langle H^{H}(t) \rangle)^{2}  \Big]. 
\label{trivHeisenb_4_referee}
\end{eqnarray}
Whenever Eq. (\ref{central}) holds,
\begin{eqnarray}
Tr \Big[ \tilde{\rho}(H^{H}(t) - \langle H^{H}(t) \rangle )^{2} \Big] 
= {\cal{O}}(N^{2}).
\label{trHHH@}
\end{eqnarray}
Thus, rather trivially, when evaluated with the initial probability density matrix $\tilde{\rho}$, the operator $(H^{H}(t)- \langle H^{H}(t) \rangle )$ exhibits an ${\cal{O}}(N^2)$ variance. This allows for non-local correlations similar to those in Eq. (\ref{central}) for operators different from $H$ also at initial times before
the system is driven. In special cases, when $H^{H}(t)$ will remain a sum of local terms similar to those in Eqs. (\ref{HsH'}), the simple derivation of Eq. (\ref{central}) may imply non-local correlations for operators do not appear in the Hamiltonian $H$ at time $t=0$. We will indeed precisely encounter such correlations and further elaborate on viable preparation of non-product form type states with these correlations in the example of Section \ref{sec:dual} (discussed in some detail in \ref{sxsx} and \ref{appendix:preparation}) where the correlations in the initial state assume a particularly simple form. In certain other instances, the operator $H^{H}(t)$ may become non-local and thus the long range covariance might not be too surprising. 

It should be stressed that in the current work we explain how long range correlations of the particular form of 
Eq. (\ref{central}) for the local energetic terms $\{{\cal{H}}_{i}\}$ may arise when the corresponding energy density ($\frac{1}{N} \sum_{i} \langle {\cal{H}}_{i} \rangle $) changes at a finite rate (and also explain how similar correlations appear when other intensive quantities vary at a finite rate). With the exception of a brief discussion at the very end of Section \ref{deviation}, we will not discuss results concerning macroscopic range correlations that are different from Eq. (\ref{central}). That is, in this paper we will largely analyze only correlations between the driven observables. For completeness, however, we must remark that many other nontrivial correlations may appear between quantities that are not driven. Indeed, long range correlations may even appear in equilibrated systems. As has been long known, systems such as the celebrated AKLT spin chains \cite{AKLT,denijs,danarovas,hasaki,KenTas} as well as Hubbard \cite{OgataShiba,Hubbard_Kruis,Hubbard_Kruis'}, $t-J_{z}$ \cite{ZA'} the Kitaev honeycomb \cite{hadongzohar},
and lattice Bose models \cite{ErezBerg} may indeed display nontrivial long range correlations. For instance, the AKLT spin chains exhibit non-trivial long range string correlations in their ground states \cite{denijs,danarovas,hasaki,KenTas} in addition to more mundane conventional short range nematic type correlations \cite{short-overlooked,topo-our-always-overlooked-work}. In \cite{short-overlooked,topo-our-always-overlooked-work}, a general algorithm was provided for the construction of non-vanishing string type and other correlators for general entangled ground states.

\section{Entangled Ising chain eigenstate expectation values produce thermal averages}
\label{Ising_example}

In order to explicitly illustrate how macroscopic entanglement may naturally appear in typical thermal states (even those of closed systems that have no explicit contact with an external bath), we turn to a simple example-
that of the uniform coupling one dimensional Ising model (the Hamiltonian $H_{I}$ of Section \ref{sec:product} on an open chain with uniform nearest neighbor coupling- $J_{ij} = J$). In these appendices, we will dispense with factors of $\hbar/2$ and use the conventional definition of the Ising model Hamiltonian with the spin at any site $r$ being $S^{z}_{r} = \pm 1$ (i.e., the diagonal elements of the Pauli matrix $\sigma^{z}_{r}$). In each Ising state product state, the value of  $\langle S^{z}_{r} S^{z}_{r'} \rangle$ is either 1 or (-1). This single Ising product state expectation value differs from that of the equilibrium system at finite temperatures. It is only if we compute the expectation value within a state formed by a superposition of many such product states (i.e., an expectation value within such a highly entangled state) or if we average under uniform translations of the origin (i.e., entangle with equal weights all states related by translation) that we will obtain the equilibrium result. The Ising operators $S^{z}_{i}$ are diagonal in the product basis; different product states are orthogonal to each other. In a superposition of different product states, only the diagonal (i.e., weighted Ising product expectation values) terms are of importance when computing $\langle S^{z}_{r} S^{z}_{r'} \rangle$.

We consider a highly entangled eigenstate $|\Psi \rangle$ of the one-dimensional Ising model.
Such an entangled state emulates, in real space, entangled eigenstates 
 $|\upsilon_{\alpha}; S_{tot}, S_{tot}^{z} \rangle$ with (for systems in their thermodynamic limit) $|S^{tot}_{z}/S_{\max}|<1$ (i.e., not product states of all spins maximally polarized up or down along the field direction) of the spin models discussed in Section \ref{sec:dual}. 
For an Ising model $H_{I}$ on a one dimensional chain of length $L$, given an eigenstate of energy $E$,
the frequency of low energy nearest neighbor bonds (namely, $S^{z}_{r} = S^{z}_{r'} = \pm 1$ (``$\uparrow \uparrow$ ''or ``$\downarrow \downarrow$'')) is $p$ 
and that of having higher energy bonds (i.e., ``$\uparrow \downarrow$'' or ``$\downarrow \uparrow$'') is $q$. 
Clearly, $p+q =1$ and  $(q-p)=  E/(LJ)$ where $J$ is the Ising model
exchange constant and $E$ is the total energy.
In the one dimensional Ising model there is no constraint on the nearest neighbor bonds $S^{z}_{i} S^{z}_{i+1}$ 
(these products are all independent variables that are ``+1'' or ``-1''  that sum to the scaled total energy $E/J$). 
Consider a spin at site $r$ which is, say, ``$\uparrow$''. We may now ask what is the average value of a spin at another site $r'$. Evidently, if there is an even number of domain walls (or even number of energetic bonds) between sties $r$ and $r'$
then the spin at site $r'$ is ``$\uparrow$" while if there is an odd number of domain walls between the two sites then the spin at site $r'$ is ``$\downarrow$". The average $\langle S^{z}_{r} S^{z}_{r'} \rangle = (p-q)^{|r-r'|}$. That is, if we have an even number of bad  domain walls (corresponding to n even power of $q$)
then the contribution to the correlation function will be positive while if we have
an odd number of domain walls (odd power of $q$) then the contribution to the correlation function will be negative. The prefactors in the binomial expansion of 
$(p-q)^{|r-r'|}$ account for all of the ways in which domain walls may be placed in the interval $(r,r')$. However, $(p-q) = (- E)/(LJ)$. Thus, the correlator
$\langle S^{z}_{r} S^{z}_{r'} \rangle = [(- E)/(LJ)]^{|r-r'|}$. This single eigenstate result using the binomial theorem indeed matches with the known results for correlations in the Ising chain in the canonical ensemble at an inverse temperature $\beta= \frac{1}{k_{B} T}$ where  $E = -J (L-1) \tanh \beta J$ and  $\langle S^{z}_{r} S^{z}_{r'} \rangle = (\tanh \beta J)^{|r-r'|}$.  The agreement of the spatially long distance correlator result in one eigenstate with the prediction of the fixed energy microcanonical ensemble 
is obvious. The above probabilistic derivation for general sites $r$ and $r'$ will hold so long as the eigenstate $|\psi \rangle$ is a sum of numerous Ising product states
(all having the same energy or, equivalently, the same number of domain walls). If this result holds for all site pairs $(r,r')$ then the entanglement entropy is expected to scale monotonically in the size (or ``volume'') of this one dimensional system. Indeed, a rather simple calculation (outlined in \ref{ent-ent-ent}) illustrates that if the $L$ site system is partitioned into subregions $A$ and $B$ of ``volumes'' $L_{A}$ and $L_{B}$ (with $L=L_{A} + L_{B}$) then if, e.g., $| \Psi \rangle$ is an equal amplitude superposition $|\Psi_{+} \rangle$ of all Ising product states  (i.e., an equal amplitude superposition of the product states $|s_{1} s_{2} \cdots s_{N} \rangle$ of Section \ref{sec:product}) that all have a given fixed energy  then the entanglement entropy between regions $A$ and $B$ scales as $\min \{\ln L_{A}, \ln L_{B} \}$. 

Broader than the specific example of this Appendix, the coincidence between the single (entangled) eigenstate expectation values with the equilibrium ensemble averages is expected to hold for general classical systems in arbitrary dimensions. To see why this is so consider the expectation value of a general observable (including any correlation functions) that is diagonal in the basis of degenerate classical product states. When computed in a state formed by a uniform modulus superposition of degenerate states (e.g., the equal amplitude sum of all local product states of the same energy), the expectation value of such an observable  may naturally emulate the microcanonical ensemble average of this observable over all classical states of the same energy. Finite energy density states (i.e., states whose energy density is larger than that of the ground state) formed by a uniform amplitude superposition of all product states generally exhibit macroscopic entanglement. As we have elaborated on in this Appendix, this anticipation is realized for the classical Ising chain. For the classical Ising chains discussed above, the below two general quantities are the same for a general observable ${\cal{O}}$:  (i) the mean of the  expectation values of ${\cal{O}}$ in all local product states that are superposed to form general ({\it not necessarily} an exact uniform modulus superposition of degenerate states) highly entangled states and (ii) the average of ${\cal{O}}$ as computed by a classical microcanonical ensemble calculation. As we emphasized earlier, general thermal states may exhibit ``volume'' law entanglement entropies \cite{plain}. However, not all eigenstates that display the equilibrium value of the correlators $\langle S^{z}_{r} S^{z}_{r'} \rangle$ need to exhibit volume law entanglement. As alluded to above, in the next Appendix, we will compute the entanglement entropy associated with $| \Psi_{+} \rangle$ and show that it is macroscopic even in one dimensional systems albeit being logarithmic in the ``volume''.

\section{Entanglement entropies of a uniform amplitude superposition of classical product states} 
\label{ent-ent-ent}

We next discuss the reduced density matrices and entanglement entropies associated with (1) the eigenstates $|\phi_\alpha \rangle = |\upsilon_{\alpha}; S_{tot} ,S_{tot}^{z} \rangle$ of Section \ref{sec:dual} when $S_{tot}$ happens to be maximal ($S_{tot} = S_{\max}$), (2) the symmetric quantum states described of \ref{Ising_example}, and a generalization thereof that we now describe. Specifically, we will consider general Hamiltonians that may be expressed as a sum of decoupled commuting local terms, $H= \sum_{i=1}^{L} {\cal{H}}_{i}$ (i.e., $N'=L$ in the notation of the Introduction) on a Hilbert space endowed with a simple local tensor product structure. We denote the eigenstates (of energies ${\cal{\varepsilon}}_{n_{i}})$ of each of the local operators ${\cal{H}}_{i}$ by $\{|\nu_{i}^{n_{i}} \rangle\}$.
For such systems, any product state $| c \rangle  = |\nu_{1}^{(n_{1})} \rangle \otimes | \nu_{2}^{(n_{2})} \rangle \otimes \cdots \otimes | \nu_{L}^{(n_{L})} \rangle$ is, trivially, a eigenstate of $H$ (of total energy $E_{c}= \sum_{i=1}^{L} {\cal{\varepsilon}}_{n_{i}}$). Formally, one may think of ${\cal{H}}_{i}$ as decoupled independent commuting 
``quasi-particle'' operators (i.e., colloquially, $H$ describes ``an ideal gas'' of such quasi-particles). We now explicitly write the states that are equal amplitude superpositions of all such product states $|c \rangle$ of a given total energy,
\begin{eqnarray}
\label{equal-am}
| \Psi_{+} \rangle \equiv \frac{1}{\sqrt{{\cal{N}}(E)}} \sum_{E_{c} = E} | c \rangle.
\end{eqnarray}
Similar to the discussion of \ref{Ising_example}, for observables ${\cal{O}}_{d}$ that are diagonal in the $\{|c \rangle\}$ basis, the single eigenstate expectation values $\langle \Psi_{+} | {\cal{O}}_{d} | \Psi_{+} \rangle$ are equal to the microcaonical equilibrium averages of $\langle {\cal{O}}_{d} \rangle_{eq;{\sf mc}}$ in which the energy $E$ is held fixed. In Eq. (\ref{equal-am}), ${\cal{N}}(E) = e^{{\sf S}(E)/k_{B}}$ is the number of product states $|c \rangle$ that have a total energy $E$ (and ${\sf S}(E)$ is the associated Boltzmann entropy). The states of Eq. (\ref{equal-am}) describe those of the Ising spin states alluded to in \ref{Ising_example}. Such states rear their head also in other arenas. For instance, since, in a many body spin system, the state of maximal total spin $S_{tot} = S_{\max}$ is a uniform amplitude superposition of all product states having a given value of $S_{z}^{tot}$ (i.e., a uniform amplitude superposition of all states of decoupled spins in a uniform magnetic field that share the same energy), states of the type $|\Psi_{+} \rangle$ include the eigenstates that we analyzed in Section (\ref{sec:dual}) (when these states are those of maximal total spin). The entanglement entropy that we will compute for $| \Psi_{+} \rangle$ will thus have implications for these and other systems. We partition the $L$ site system into two disjoint regions $A$ and $B$ and examine the entanglement between these two subvolumes. To facilitate the calculation, we will employ the symmetric combinations 
\begin{eqnarray}
\label{EAQ}
| E_{A}  \rangle_{+} \equiv \frac{1}{\sqrt{{\cal{N}}_{A}(E_{A})}} \sum_{E(\{c_{A}\})= E_{A}}  |\{c_{A}\} \rangle, \nonumber
\\ | E_{B}  \rangle_{+} \equiv \frac{1}{\sqrt{{\cal{N}}_{B}(E_{B})}} \sum_{E(\{c_{B}\})= E_{B}}  |\{c_{B}\} \rangle.
\end{eqnarray}
In the first of Eqs. (\ref{EAQ}), the sum is over all product states $\{|c_{A} \rangle\}$ having their support on the sites $1 \le i \le L_{A}$ that are of fixed energy $E_{A}$.
Similarly, the symmetric state $ |E_{B}  \rangle_{+}$ extends over the sites $ L_{A}+1 \le i \le L$. 
With these definitions, we rewrite Eq. (\ref{equal-am}) as
\begin{eqnarray}
\label{EAQ'}
 | \Psi_{+} \rangle =  \sum_{E_{A}} && \sqrt{\frac{{\cal{N}}_{A}(E_{A}) {\cal{N}}_{B}(E-E_{A})}{{\cal{N}}(E)}} \nonumber
\\ &&\times  |E_{A} \rangle_{+} |E_{B} = E-E_{A} \rangle_{+}.
\end{eqnarray}
The density matrix associated with this state is $\rho_{+} \equiv | \Psi_{+} \rangle \langle \Psi_{+}|$. To compute the entanglement entropy, we next write the
reduced density matrix 
\begin{eqnarray}
\label{rb}
  \rho_{B,+} \equiv && Tr_{A} \rho_{+} =  \frac{1}{{\cal{N}}(E)} \sum_{E_{A}} ({\cal{N}}_{A}(E_{A}) {\cal{N}}_{B} (E_{B} = E-E_{A}) \nonumber
\\ && \times | E_{B}=E-E_{A}\rangle_{+} \langle E_{B}=E-E_{A}|_{+}).  
\end{eqnarray}
If a given system is partitioned into two non-interacting subsystems $A$ and $B$ then the sole relation linking the two subsystems will be the constraint of total energy $E=E_{A} + E_{B}$. Of all possible ways of partitioning the total energy $E=E_{A} + E_{B}$, one pair of energies $\overline{E}_{A}$ and $\overline{E}_{B}$ will yield the highest value of ${\sf S}_{A}(\overline{E}_{A}) + {\sf S}_{B}(\overline{E}_{B})$. The ratios appearing in Eq. (\ref{rb}), 
\begin{eqnarray}
\label{NNS}
&& \frac{{\cal{N}}_{A}(E_{A}) {\cal{N}}_{B}(E-E_{A})}{{\cal{N}}(E)} \nonumber
\\ = &&e^{({\sf S}_{A}(E_{A})+ {\sf S}_{B}(E-E_A) - {\sf S}(E))/k_{B}},
\end{eqnarray}
 follow, upon Taylor expanding the ratio to quadratic order about its maximum at $\overline{E}_{A}$ and $\overline{E}_{B} = E - \overline{E}_{A}$, a Gaussian distribution with a standard deviation set by
\begin{eqnarray}
\label{sigmab}
\sigma_{B} = \sqrt{k_{B} T^{2} C_{v}^{eff}(T)}.
\end{eqnarray}
In Eq. (\ref{sigmab}), 
\begin{eqnarray}
\label{cvcvcv}
C_{v}^{eff}(T) \equiv \frac{C_{v}^{(A)} (T)  C_{v}^{(B)} (T)}{C_{v}^{(A)}(T) + C_{v}^{(B)}(T)}.
\end{eqnarray}
The latter Taylor expansion may be carried out for energy densities associated with finite temperatures. (In the vicinities of either the ground state value of the energy density or the highest energy density, the derivatives of the entropy relative to the energy diverge and the Taylor expansion becomes void.) The entropies ${\sf S}_{A}(E_{A})$ and ${\sf S}_{B}(E_{B})$ appearing in Eq (\ref{NNS}) are those of subsystems $A$ and $B$ that, as emphasized above, for non-interacting particles, are merely constrained by the condition that $E_{A} + E_{B} = E$. For this non-interacting system,
\begin{eqnarray}
e^{{\sf S}(E)/k_{B}} = \sum_{E_{A}} e^{{\sf S}_{A}(E_{A})/k_{B}} e^{{\sf S}_{B}(E_{B}=E-E_{A})/k_{B}}, 
\end{eqnarray} 
and thus, trivially, ${\sf S}(E) \ge {\sf S}_{A}(\overline{E}_{A}) + {\sf S}_{B}(\overline{E}_{B})$.
As throughout the current work, in Eqs. (\ref{sigmab}, \ref{cvcvcv}), $T$ denotes the temperature (set by the condition that the canonical ensemble equilibrium internal energy $Tr(He^{-\beta H})/Tr(e^{-\beta H})$ is equal to the total energy $E$).
The entropy of the Gaussian distribution scales as the logarithm of its width. Specifically, for the saddle point Gaussian approximation of Eqs. (\ref{NNS},\ref{sigmab},\ref{cvcvcv}), 
\begin{eqnarray}
\label{sent*}
{\sf S}_{ent,+} \equiv  - Tr (\rho_{B,+} \ln \rho_{B,+}) = \frac{1}{2} \ln (2 \pi \sigma_{B}^{2} + 1)  \sim  \ln \sigma_{B},
\end{eqnarray}
where in the last asymptotic form, we made manifest the assumed extensive $L_{A,B} \gg 1$ (and thus $\sigma_{B} \gg 1$). If $S_{A}(\overline{E}_{A}) = {\cal{O}}(L_{A})$ and $S_{B}(\overline{E}_{B}) = {\cal{O}}(L_{B})$ when $L_{A,B} \gg 1$ then, from Eqs. (\ref{sigmab}, \ref{cvcvcv}, \ref{sent*}), the entanglement entropy for states of finite temperature (i.e., states exhibiting a finite energy density above that of the ground state value),
\begin{eqnarray}
\label{sentl}
{\sf S}_{ent,+} = {\cal{O}}( \min \{ \ln L_{A}, \ln L_{B} \}).
\end{eqnarray}
We reiterate that generic states of fixed total energy will exhibit an entanglement entropy proportional to the system volume (see, e.g., the considerations of \cite{plain}). Even though a system of non-interacting particles is trivial and its properties may, generally, be exactly computed, its entanglement entropy may be macroscopic. 
We next discuss two specific realizations of Eqs. (\ref{sent*}, \ref{sentl}). 

\subsection{Maximal total spin eigenstates}
\label{maxtotspin}
As we noted above, for any fixed $S_{tot}^{z}$, the eigenstates of Eq. (\ref{hferro}), $| \Psi_{+} \rangle$ corresponds to a maximal total spin ($S_{tot} = S_{\max}$) state of the $L=N$ spins (with the given value of $S_{tot}^{z}$). 
In order to relate this to our general results of Eqs. (\ref{sent*}, \ref{sentl}) for the entanglement entropy, we consider the local Hamiltonians ${\cal{H}}_{i}$ forming the Hamiltonian $H = \sum_{i=1}^{N} {\cal{H}}_{i}$ to be given by ${\cal{H}}_{i} = - B_{z} S^{z}_{i}$. With this, $|\Psi_{+} \rangle$ of Eq. (\ref{equal-am}) is an eigenstate of $S_{tot}^{z}$ (with each product state $|c \rangle$
being an eigenstate of all $\{S^{z}_{i}\}$ operators). We consider what occurs if the $N$ spins are partitioned into the two groups $A$ and $B$ of approximately 
equal numbers $L_{A}$ and $L_{B}$, and $|w| \equiv |S_{tot}^{z}/(\hbar S_{tot})| <1$. In this case, 
the saddle point approximation of Eq. (\ref{NNS}) yields, as before, a Gaussian distribution and, a consequent logarithmic entanglement entropy,
\begin{eqnarray}
\label{sentsent}
{\sf S}_{ent,+} = {\cal{O}}(\ln N). 
\end{eqnarray}
Thus, as highlighted in Section \ref{sec:dual}, initial states $| \psi^{0}_{Spin} \rangle$ of maximal total spin and $|w| <1$ feature logarithmic in volume entanglement entropies. 

\subsection{Ising chains}
Returning to the considerations of \ref{Ising_example} and the notation introduced in Section \ref{sec:product}, we now consider the symmetric sum of all Ising product states that share the same energy (as measured by the Ising Hamiltonian $H_{I}$ of Section \ref{sec:product}).  As in Section \ref{maxtotspin}, we can transform the problem of computing the entanglement entropy of such symmetric states $| \Psi_{+} \rangle$ into that involving eigenstates of decoupled local Hamiltonians ${\cal{H}}_{i}$ that led to Eqs. (\ref{sentl}). Towards this end, we focus on the nearest neighbor spin products that were crucial to our analysis in \ref{Ising_example}, and define the operators 
 \begin{eqnarray}
1 \le i \le L-1: ~~~ && R_{i} \equiv S^{z}_{i} S^{z}_{i+1}, \nonumber
\\ && R_{L} \equiv S^{z}_{L}.
\end{eqnarray}
The Ising Hamiltonian now explicitly becomes a sum of the above defined decoupled commuting operators, $H_{I} = -J \sum_{i=1}^{L-1} R_{i}$. Using the vocabulary that we employed earlier, the ``quasi-particle'' operators $\{R_{i}\}_{i=1}^{L:-1}$ are associated with the existence ($R_{i} = -1$) or absence ($R_{i} = 1$) of domain walls between neighboring Ising spins.  On the two subregions $A$ and $B$, we define 
$H_{AI} = -J \sum_{i=1}^{L_{A}} R_{i}$ and $H_{BI} = - J \sum_{i=L_{A} +1}^{L-1} R_{i}$. The equal amplitude superposition of all Ising product states of fixed energy can be rewritten as 
\begin{eqnarray}
| \Psi_{I+} \rangle = \frac{1}{2^{L/2}} \sum_{r_{1}, r_{2}, \cdots, r_{L}} |r_1 r_2  \cdots r_{L} \rangle,
\end{eqnarray}
where $r_{i} = \pm 1$ denote the eigenvalues of $R_{i}$. When evaluating the reduced density matrix $\rho_{BI+}= Tr_{A} | \Psi_{I+} \rangle \langle \Psi_{I+}|$, the trace over all Ising spins $\{s_{i \le L_{A}} \}$ that lie in the spatial region $A$ is replaced by that over $\{r_{i \le L_{A}}\}$. Repeating the earlier calculations we find, once again, the entanglement entropy of Eqs. (\ref{sigmab},\ref{cvcvcv},\ref{sent*}) \cite{aln2}. Equating the internal energy of a system given by $H_{I}$ to $E$ we see that, when $L \gg 1$, the temperature appearing in Eqs.  (\ref{sigmab},\ref{cvcvcv},\ref{sent*}) is given by
\begin{eqnarray}
\frac{1}{k_{B} T} = - \tanh^{-1} \Big( \frac{E}{LJ} \Big).
\end{eqnarray}
In Eq. (\ref{cvcvcv}), the heat capacities of the Ising chain subsystems $A$ and $B$ (when $L_{A,B} \gg 1$) are
\begin{eqnarray}
&& C_{v}^{(A,B)}(T) = k_{B} L_{A,B} 
 \Big( (\beta J)^{2}-  \Big(\frac{\beta E_{A,B}}{L_{A,B}} \Big)^{2} \Big).
\end{eqnarray} 
Eq. (\ref{sentl}) provides the asymptotic scale of the entanglement entropy; similar to Eq. (\ref{sentsent}), if $L_{A}$ and $L_{B}$ are both of order of the system size, $ L_{A,B}={\cal{O}}(N)$ then the entanglement entropy ${\sf S}_{ent;+}$ of the symmetric state will scale logarithmically in $N$. General eigenstates may exhibit larger entanglement entropies (see \ref{Higher_entanglement_entropy_states}).

\section{The total spin of large systems}
\label{trivial-maxS}

We now discuss the total spin sectors that may appear in the spin model of Section \ref{sec:gspin}. Our aim is to highlight both statistical and physical aspects of the total spin and its scaling with the system size $N$.

All states with maximal total spin and definite eigenvalues of the total $S_{tot}^{z}$ operator are eigenstates of the general Hamiltonian $H_{spin}$ of Eq. (\ref{hferro}). (These eigenstates span the basis of all ferromagnetic spin states with spins uniformly polarized along different directions.) This assertion may be explicitly proven by the following simple observations:
(i) For any two spin $S=1/2$ operators, the scalar product $\vec{S}_{i} \cdot \vec{S}_{j} = \hbar^{2}(\frac{1}{2} P_{ij} - \frac{1}{4})$
where $P_{ij}$ is the operator permuting the two spins, (ii) Any state of maximum total spin 
($S_{tot} = S_{\max} = N \hbar/2$) is a symmetric state that is invariant under all permutations $\{P_{ij}\}$.
From properties (i) and (ii), it follows that any state $ |S_{tot} = S_{\max} =N \hbar/2, S_{tot}^{z} \rangle$
is an eigenstate of both the first and second terms of Eq. (\ref{hferro}) and therefore of the full Hamiltonian $H_{spin}$. 
Thus, any state of maximal total spin $S_{tot} = S_{\max}$ that is an eigenstate of $S_{tot}^{z}$ is automatically an eigenstate of $H_{spin}$ of Eq. (\ref{hferro}). In general, when $S_{tot} <S_{\max}$, only some linear combinations of the multiple states of given values of $S_{tot}$ and $S_{tot}^{z}$ are eigenstates of $H_{spin}$ 
(hence the appearance of additional quantum numbers $\upsilon_{\alpha}$ defining general eigenstates $|\phi_{\alpha} \rangle$).
To make this clear, we can explicitly write down the total spin for a system of $N$ spin $S=1/2$ particles. That is,
\begin{eqnarray}
\label{add+}
&& N=2: ~~~ \frac{1}{2} \otimes \frac{1}{2} = 1 \oplus 0, \nonumber
\\ && N=3: ~~~  \frac{1}{2} \otimes \frac{1}{2} \otimes \frac{1}{2} = \frac{3}{2} \oplus \frac{1}{2} \oplus \frac{1}{2}, \nonumber
\\&& N=4: ~~~ \frac{1}{2} \otimes \frac{1}{2} \otimes \frac{1}{2}  \otimes \frac{1}{2}  = 2 \oplus 1 \oplus 1 \oplus 1 \oplus 0 \oplus 0, \nonumber 
\\ &&  ~~~~~~~~~~~~~~ \cdots   ~~~~~~~~ .
\end{eqnarray}
The first (textbook type) equality of Eq. (\ref{add+}) states that singlet ($S=0$) and triplet $(S=1$) total spin combinations may be formed by adding two ($N=2$) spins of size $S=1/2$. Other well known relations are similarly tabulated for higher $N$. Since $H_{spin}$ is defined on a $(2S+1)^{N}$ dimensional Hilbert space, its eigenstates span all states in the direct product basis on the lefthand side of Eq. (\ref{add+}). For each $N$, the sector of maximal spin ($S_{tot} = S_{\max}=NS$) is unique. However, for $N>2$, all other total spin sectors in Eq. (\ref{add+}) exhibit a multiplicity ${\cal{M}}_{S_{tot}}$ larger than one. While it is, of course, possible to simultaneously diagonalize the Hamiltonian
of Eq. (\ref{hferro}) with the two operators $S_{tot}$ and $S_{tot}^{z}$, there are multiple states that share the same eigenvalues of
$S_{tot}$ and $S_{tot}^{z}$. Using the characters $\overline{\chi}_{S_{tot}}$ of spin $S_{tot}$ representations of SU(2), we find that there are 
 \begin{eqnarray}
 \label{MST}
{\cal{M}}^{N}_{S_{tot}} = \frac{N!(2S_{tot} +1)}{(\frac{N}{2} + S_{tot}+1)! (\frac{N}{2} - S_{tot})! }
\end{eqnarray}
sectors of total spin $0 \le S_{tot} \le \frac{N}{2}$
on the righthand side of Eq. (\ref{add+}). The decomposition into characters of $SU(2)$ has a transparent physical content. Consider a global rotation by of all spins an arbitrary angle $\theta'$ about the $z$ axis. The trace of the operator that implements this rotation is the same into the different basis appearing in Eq. (\ref{add+}): (1) the product basis (the lefthand side of Eq. (\ref{add+})) of  $N$ spins of size $S=1/2$ and (2) the basis comprised of the total spin sectors (the righthand side of Eq. (\ref{add+})). When expressing the basis invariant trace of the arbitrary rotation evolution operator in terms of the Laurent series in $e^{i \theta'/2}$ that arises when taking the trace of the rotation operator, the series must identically match in both of these 
bases of Eq. (\ref{add+}). Equating the trace as evaluated in (1) and (2) as discussed above, we explicitly have $(2 \cos \frac{\theta'}{2})^{N} = \sum_{S_{tot}} {\cal{M}}^{N}_{S_{tot}} \overline{\chi}_{S_{tot}}$ with $\overline{\chi}_{S_{tot}} = 
\sum_{{\sf{S}}=-S_{tot}}^{S_{tot}} e^{i {\sf S} \theta'} 
=\frac{\sin(2S_{tot}+1) \frac{\theta'}{2}}{\sin \frac{\theta'}{2}}$ from which Eq. (\ref{MST}) follows by Fourier transformation. Perusing Eq. (\ref{MST}), we see that for large $N$, the highest values of ${\cal{M}}^{N}_{S_{tot}}$ occur for small $S_{tot}$; in Eq. (\ref{add+}), a ``randomly'' (``infinite temperature'') chosen state of $N \gg 1$ spins is most likely to have $S_{tot} \le {\cal{O}}(\sqrt{N})$. Specifically, if we approximate, for fixed $N \gg S_{tot} \gg 1$, the distribution of binomial coefficients in Eq. (\ref{MST}) by a Gaussian, we trivially obtain 
\begin{eqnarray}
\label{MST10}
{\cal{M}}^{N}_{S_{tot}} \sim \frac{2^{N + \frac{5}{2}} e^{-\frac{2 S_{tot}^{2}}{N}}}{N^{\frac{3}{2}} \sqrt{\pi}} S_{tot}.
\end{eqnarray}
The binomial character of Eqs. (\ref{MST}) with the associated asymptotic Gaussian form of Eq. (\ref{MST10}) is not unexpected: a summation of $N \gg 1$ random classical spins (when these are viewed as uniform length displacement vectors) leads to a total spin that, similar to that appearing for the total displacement in random walks (sum of the uniform length displacements), follows a Gaussian distribution. As seen in our equations, the situation for quantum spins is qualitatively similar. Even though, when $N \gg 1$, states of low $S_{tot}/N \sim {\cal{O}}(N^{-1/2})$ are statistically preferred in the entries of Eq. (\ref{add+}), physically finite $S_{tot}/N$ ratios are naturally mandatory in numerous instances (including the ability to cool/heat the energy density of the system at a finite rate). For instance, sans symmetry breaking fields, in low temperature ferromagnetic states (having a finite magnetization density or, equivalently, an extensive total spin), the total spin value $S_{tot} = {\cal{O}}(N)$. In the presence of the applied symmetry breaking field in $H_{spin}$ of Eq. (\ref{hferro}), such a finite average of $(S_{tot}^{z}/N)$ arises at general finite temperatures. Furthermore, as noted above, in order to have a finite rate of change of the energy density by applying the transverse field $B_y$ of Eq. (\ref{htr}), we must have that the total spin $S_{tot} ={\cal{O}}(N)$.

\section{Correlations in rotationally invariant spin systems driven by a uniform field}
\label{sxsx}

\subsection{Long range correlations}
\label{lrra}
We will now briefly underscore that any eigenstate of $| \phi_{\alpha} \rangle$ of Eq. (\ref{hferro}) having $S_{tot} = {\cal{O}}(N)$ with $|w|<1$ displays long range correlations. As we will further explain, such macroscopic spin states with $|w|<1$ must appear if the application of a transverse field in the example 
of Section \ref{sec:gspin} leads to, e.g., either (1) finite second cumulants (i.e., variances) the change of the energy density (in addition to a finite rate of variation of the energy density as required for the systems that we analyze) or generally leads to (2) finite second derivatives of the energy density for  
time dependent external fields (such as those of Eq. (\ref{byt}) below).

First, we make the correlations in these states explicit by writing down the below two simple equalities,
\begin{eqnarray}
\langle (S_{tot}^{x})^{2} \rangle = \frac{1}{2} \Big\langle  \Big(\vec{S}^{2}_{tot} - (S_{tot}^{z})^{2} \Big) \Big\rangle = \frac{1}{2} \Big[S_{tot}(S_{tot} +1) \hbar^{2} - (S_{tot}^{z})^{2} \Big],
\label{totzz}
\end{eqnarray}
and 
\begin{eqnarray}
\langle (S_{tot}^{x})^{2} \rangle = \sum_{i \neq j} \langle S^{x}_{i} S^{x}_{j} \rangle + \sum_{i} \langle (S^{x}_{i})^{2} \rangle = \sum_{i \neq j} \langle S^{x}_{i} S^{x}_{j} \rangle
+ \frac{N \hbar^{2}}{4}.
\label{totxx}
\end{eqnarray}
Combining Eqs. (\ref{totzz}, \ref{totxx}), and noting that in any eigenstate $|\phi_{\alpha} \rangle$ of the $S^{z}_{tot}$ operator, the expectation value $\langle S^{x}_{i} \rangle = 0$, 
one finds that, on average, for all $i \neq j$,
the pair correlator
\begin{eqnarray}
\label{longxx}
\frac{1}{N(N-1)} \sum_{i \neq j} (\langle S^{x}_{i} S^{x}_{j} \rangle - \langle S^{x}_{i} \rangle \langle S^{x}_{j} \rangle) 
= \frac{(S_{tot}(S_{tot}+1) - \frac{N}{2}) \hbar^{2} - (S^{z}_{tot})^{2}}{2N(N-1)}.
\end{eqnarray}
For fully symmetric states $| \phi_{\alpha} \rangle$ (those associated with a maxima total spin, $S_{tot} = S_{\max} = NS$), all of the correlators when $i \neq j$ are equal to each other and given by the righthand side of Eq. (\ref{longxx}). The possibility of correlations in the initial state is consistent with our discussion following Eqs. (\ref{trivHeisenb_4_referee},\ref{trHHH@}). In the exactly solvable model system of Section \ref{sec:dual}, these correlations are of a particularly simple form of Eq. (\ref{longxx}). 

\subsection{Cumulants and higher order derivatives of the energy density for various fields}
\label{inevit.}

\subsubsection{Finite variances of the derivative of the energy density}
As noted earlier, in order for the system to display a finite rate of variation of its energy density (the focus of the systems discussed in our work), the spin system of Eq. \ref{hferro} must have macroscopic (${\cal{O}}(N)$) total $S_{tot}^{z}$ (as in a ferromagnet). While a finite average correlator for large $|i-j'|$ (such as that resulting when $S_{tot} = {\cal{O}}(N)$ and $|w| <1$) might appear paradoxical, one must recall that for these states $|\phi_{\alpha} \rangle$, the application of the transverse field of Eq. (\ref{htr}) led to a finite range of change of the energy density. That is, when evaluated in these states, the expectation value of the time derivative of the Heisenberg picture Hamiltonian $\frac{d \epsilon}{dt} = \frac{1}{N} \langle \frac{d H^{H}(t)}{d t} \rangle \neq 0$ for general times $t$. Indeed, the latter inequality defined our problem (that of a finite rate of change of the energy density). Given that, at most times $t$, 
the first moment of $\frac{d H^{H}(t)}{dt}$ in the state $| \phi_{\alpha} \rangle$ is finite, it is no surprise that its second cumulant (i.e., the variance) may also be finite at these or other times. Indeed, when $ \int_{0}^{t}~ B_{y}(t') ~dt' \equiv 0 ({\sf mod}~ \pi)$, 
\begin{eqnarray}
\label{2ndmoment}
\frac{1}{N^{2}}  \Big( \Big\langle \Big( \frac{d H^{H}(t)}{dt} \Big)^2 \Big\rangle -  \Big\langle  \frac{d H^{H}(t)}{dt} \Big\rangle ^{2} \Big) = \Big(\frac{B_{y}(t) B_{z}}{N}\Big)^{2} \Big\langle (S_{tot}^{x})^{2} \Big\rangle.
\end{eqnarray}
Thus, for those times $t$ at which $\theta(t) \equiv 0 ({\sf mod}~ \pi)$ (which, coincidently, for $w \neq 0, \pm 1$, are the only times at which $\frac{d \epsilon}{dt} = \sigma_{\epsilon} =0$) if the second cumulant of $\frac{1}{N} \frac{d H^{H}(t)}{dt}$ is finite then the initial state $|\psi^{0}_{Spin} \rangle
= | \phi_{\alpha} \rangle$ must display a finite 
$\langle \Big(\frac{S_{tot}^{x}}{N}\Big)^{2} \rangle$. From Eq. (\ref{totxx}), a non-vanishing $\langle \Big(\frac{S_{tot}^{x}}{N}\Big)^{2} \rangle$ implies a finite average value of 
$(\langle S^{x}_{i} S^{x}_{j} \rangle - \langle S^{x}_{i} \rangle \langle S^{x}_{j} \rangle)$ for far separated sites $i$ and $j$. Hence, the correlations of Eq. (\ref{longxx}) are not unexpected in systems generally exhibiting finite cumulants of  $\frac{1}{N} \frac{dH^{H}(t)}{dt}$. We must caution that, of course, the possibility of a finite first cumulant of $\frac{1}{N} \frac{dH^{H}(t)}{dt}$ at general times does not mandate the existence of a finite second cumulant (i.e., a variance). However, it certainly does not preclude it (as is indeed the case for our example of Section \ref{sec:gspin}). Generally, one anticipates a finite variance from the different local contributions to $\frac{d H^{H}(t)}{d t}$. These contributions are generally correlated due to the coupling between the local contributions (the local spins) to the external drive (the transverse field of Eq. (\ref{htr})) to all spins in the system so as to change the energy density at a finite rate (as motivated by the qualitative discussion of Eq. (\ref{usfe})). That the variance of $\frac{1}{N} \frac{dH^{H}(t)}{dt}$ is given by Eq. (\ref{2ndmoment}) may be explicitly seen as follows. In the Heisenberg picture, an evolution under the transverse field Hamiltonian of 
Eq. (\ref{htr}) leads to the precession
\begin{eqnarray}
S^{z}_{tot}(t) = S^{z}_{tot} \cos \theta(t) - S^{x}_{tot} \sin \theta(t),
\end{eqnarray}
where, as in the main text, $\theta (t) \equiv  \int_{0}^{t} B_{y}(t') dt'$. 
Invoking Eq. (\ref{hferro}), this yields
\begin{eqnarray}
\label{1st-der}
\frac{dH^{H}(t)}{dt} = B_{z} B_{y}(t) \Big( S^{z}_{tot} \sin \theta(t)- S^{x}_{tot} \cos  \theta(t) \Big).
\end{eqnarray}
This gives rise to Eq. (\ref{2ndmoment}) when $ \theta(t) \equiv 0 ({\sf mod}~ \pi)$.  

\subsubsection{Finite averages of the second order derivative of the energy density}
Higher order derivatives may be similarly examined. We next discuss the average of the second derivative of the energy density,  
\begin{eqnarray}
\label{diffS*}
\frac{d^{2}S^{z}_{tot}(t)}{dt^{2}} = -S^{z}_{tot} [B^{2}_{y}(t)  \cos \theta + \frac{dB_y}{dt} \sin \theta]  \nonumber
\\+ S^{x}_{tot}[B_{y}^{2} - \frac{dB_{y}}{dt} \cos \theta].
\end{eqnarray}
In the following, we will very briefly discuss two special simple cases: (1) a time dependent and (2) a constant external field. 
\newline
\newline

{\it Time dependent external field.}
\newline

From Eq. (\ref{diffS*}),  if $\frac{1}{N^{2}} \langle  ( \frac{d^{2} S^{z}_{tot} (t)}{dt^{2}})^2 \rangle = {\cal{O}}(1)$ then whenever $[B^{2}_{y}(t)  \cos \theta + \frac{dB_y}{dt} \sin \theta] =0$, the variance $\langle (S^{x}_{tot})^{2} \rangle = {\cal{O}}(N^{2})$. Since $B_{y}(t) = \frac{d \theta}{dt}$, this yields the ordinary differential equation
\begin{eqnarray}
\label{trivdiff}
\Big(\frac{d \theta}{dt} \Big)^{2} = -  \frac{d^{2} \theta}{dt^{2}} \tan \theta.
\end{eqnarray}
Explicitly integrating $(\frac{d^{2}\theta}{dt^{2}})/(\frac{d \theta}{dt}) = -  \frac{d \theta}{dt} \cot \theta$ once implies $\ln \frac{d \theta}{dt}  = -\ln (\sin \theta) + C_{1}$. An exponentiation and a second integration result in $\cos \theta = C_{2} - C t$ (with $C,C_{1,2,}$ arbitrary integration constants). Hence, if $\theta(0) =0$ then 
the solution to Eq. (\ref{trivdiff}) is, for $0 \le t \le \frac{2}{C}$, given by $\theta(t) = \cos^{-1} (1- Ct)$ for general $C>0$. Thus, if an applied field 
\begin{eqnarray}
\label{byt}
B_{y}(t) = \frac{d\theta}{dt}= \frac{1}{\sqrt{\frac{2t}{C}- t^{2}}},
\end{eqnarray}
not only trivially leads to a finite rate of change of the energy density but also to a finite  $\frac{1}{N^{2}} \langle \frac{d^{2} H^{H}}{dt^{2}} \rangle$ on a continuous time interval then $\langle (S^{x}_{tot})^{2} \rangle = {\cal{O}}(N^{2})$  (i.e., $|w| <1$) signaling, as discussed in \ref{lrra}, the existence of long range correlations.
\newline
\newline

{\it Constant external field.}
\newline

If apart from having a finite rate of change of the energy density, the square of the second order derivative $\frac{1}{N^{2}} \langle (\frac{d^{2} H^{H}}{dt^{2}})^2 \rangle > 0$ when $\theta(t) = \pi/2$ for a uniform time independent $B_{y}$ then, from Eq. (\ref{diffS*}),  
$\langle (S^{x}_{tot})^{2} \rangle = {\cal{O}}(N^{2})$, i.e., $|w| <1$ (implying long range correlations once again). 

\section{Preparation of the initial spin states of Section \ref{sec:gspin}}
\label{appendix:preparation}

The results of Section \ref{sec:gspin} hold for {\it any} initial state $| \psi^{0}_{Spin} \rangle$ that is an eigenstate of the Hamiltonian $H_{spin}$ of Eq. (\ref{hferro}) evolved under the transverse field 
Hamiltonian $H_{tr}$  of Eq. (\ref{htr}). We reiterate that a finite rate of cooling or heating can be achieved by $H_{tr}$ only when the initial state $| \psi^{0}_{Spin} \rangle$ is of a macroscopic total spin $S_{tot} = {\cal{O}}(N)$ (e.g., a ferromagnet) and the ratio $w \equiv S^{z}_{tot}/(\hbar S_{tot} ) \neq 0$. Furthermore, as noted earlier, the inequality $w  \neq \pm 1$ must be satisfied in order for the initial state to differ qualitatively from a product state in which all spins are polarized along the $z$ direction. Indeed, as we explained in Section \ref{sec:product}, for initial product states, no spreading is possible (i.e., $\sigma_{\epsilon}=0$). In a related manner, if $w = \pm 1$ then the transverse field Hamiltonian $H_{tr}$ will act as a pure displacement operator on the spin coherent state initially polarized along the $z-$ axis and lead to no spreading of the energy density as evaluated with Eq. (\ref{hferro}). It is only for the fully polarized states $w = \pm 1$ that no spreading occurs. 
The states $|\psi^{0}_{Spin} \rangle$ that we considered are, obviously, somewhat special (see also \ref{sxsx}). In this Appendix, we describe a purely gedanken experiment for preparing states (with either quantum or classical probability densities) of high spin $S_{tot} = {\cal{O}}(N)$ with $|w|<1$. Towards this end, we first consider the Hamiltonian of Eq. (\ref{hferro}) as that describing a typical ferromagnet ${\sf F}$ associated with the Hamiltonian $H_{Heisenberg}$ of  Eq. (\ref{HEIS*}) on a lattice of $N$ sites (having, e.g., all of the couplings in Eq. (\ref{HEIS*}) non-negative) that is subjected to, at low temperatures, to a longitudinal external field ($B_{z}$). The latter external field is created by a large permanent magnet ${\sf M}$ of size $N_{\sf M} = {\cal{O}}(N)$. The global magnetic field $B_{z}$ generated by ${\sf M}$ has small $\delta B_{z} = {\cal{O}}(N^{-1/2})$ fluctuations in its magnitude. We consider the ``${\sf F}- {\sf M}$'' hybrid to be, initially, in contact with a thermal bath. In equilibrium, at low temperatures, the spins in ${\sf F}$ become polarized with the resulting total magnetization being parallel to the applied external field $B_{z}$ (viz., $S_{tot}^{z} = S_{tot}  = {\cal{O}}(N)$). Next, we introduce a transverse field $B_{y}$ (captured by Eq. (\ref{htr}) or Eq. (\ref{eq:augment})) that acts on ${\sf F}$. Following the application of the transverse field, the total spin will precess about the $y$ axis (see Figures \ref{Cap1.} and  \ref{Cap2.}). Next, we turn off the transverse field and let the system evolve under Eq. (\ref{hferro}). As earlier, we do so by considering the ${\sf F}- {\sf M}$ hybrid which is now closed (i.e., with no connection to an external heat bath). Now that the total spin is no longer polarized along the $z$ axis, the fluctuations in the values of $B_{z}$ will lead to a spread of precession of the total spin  about the $z$ axis. After a time $\tau_{cover} \sim 2 \pi/\delta B_{z}$ (assuming that this time is larger
than the Lieb-Robinson time of Section \ref{intuition}, $\tau_{cover} > t_{\sf LR}$), the probability distribution for the total spin covers uniformly a ``line of latitude'' of fixed $S_{tot}^{z}$ (see Figure \ref{Cap1.}). 
This resulting probability distribution for the total spin emulates that associated with $| \psi^{0}_{Spin} \rangle$ of Section \ref{sec:gspin} or that affiliated with the semi-classical distribution of Section \ref{poincare}. Once a strong transverse field ($||H_{tr}|| \gg ||H_{spin}||$) is applied anew to this state, the results Eqs. (\ref{se}, \ref{Heavy}) will follow (the ring of Figure 
\ref{Cap1.} will rotate to that of Figure \ref{Cap2.}). Similarly, Eq. (\ref{sigmaat}) will yield the standard deviation of the energy density for the more general situation of Eq. (\ref{eq:augment}) for an arbitrary size $H_{tr}$ augmenting $H_{spin}$. The existence of a minimal time (beyond that required for the field to couple to all sites in the system) is reminiscent of 
lower bounds that we found in other model systems (e.g., the time scale required to have a fluctuation in the effective external field set by $\overline{q}$ satisfy $\sigma_{\overline{q}} = {\cal{O}}(1)$ in Section \ref{oscenv}). 

The simple example in this Appendix is a particular realization of the schematic of Figure \ref{coupling.}. All spins in the system couple uniformly to the external magnet (the ``environment'') ${\sf M}$. Fluctuations in 
the ``collective coordinate'' of the external drive, the global magnetization of ${\sf M}$, lead to a distribution of precession frequencies of the total spin in the ``system'' ${\sf F}$ and to the ensuing long range fluctuations of Eq. (\ref{longxx}) within it \cite{2unitary}. If the environment acts like a uniform stationary field with no fluctuations on a product state form then correlations will not arise as was seen in the example of Section \ref{sec:product}. 

\section{Finite rate shifts of the energy density $\epsilon$} 
\label{app-boost}
 
 For Hamiltonians $H(t)$, the single condition $\frac{d \epsilon}{dt}=0$ at all times may be satisfied by an infinite number of special Hamiltonians and/or density matrices $\rho$. A solution is afforded by Hamiltonians $H(t)$ that during the cooling/heating time have a commutator with $H$ that is a non-vanishing c-number. Under these circumstances, the evolution operator ${\cal{U}}$ has the form of a shift operator for the energy (shifting, during each interval of time $[t,t+dt]$, the Heisenberg picture Hamiltonian so as to have $H^{H}(t) \longrightarrow (H^{H}(t)-\frac{i dt}{\hbar} [H(t),H]))$ that brings to life the intuitive analogy 
with wave packets (Section \ref{intuition}). In order to obtain a general shift of the energy without widening the width of the energy density distribution, one may apply a shift operator (an evolution with a ``momentum'' conjugate to $H$). Indeed, the above evolution leads to a shift of the energy density with no additional changes. More comprehensive solutions to the equation $\frac{d}{dt} \sigma_{\epsilon} =0$ at all times $t$ (and thus solutions to $\sigma_{\epsilon(t)} =0$ at all $t$) given that $\sigma_{\epsilon}=0$ at time $t=0$ are afforded by combining multiple ``shifts'' of the above type with 
the product states of Section \ref{sec:product}. That is, we may set the initial state to be a general product of decoupled density matrices afforded by Eq. (\ref{rhofac}) with general values of $1 \le M \le N$. If all of the probability density matrices are local (have their support on regions of size ${\cal{O}}(1)$) 
then any Hamiltonian evolution is possible. Conversely, if the density matrices cannot be factorized beyond 
a region of size ${\cal{O}}(N)$ then only an innocuous shift with a constant $[H(t),H]$ will be possible. General hybrids where for (1) all non-local density matrices such innocuous shift appears while (2) the evolution of any local density matrices is arbitrary further satisfy $\sigma_{\epsilon(t)}=0$ at all $t$. Generally, as Eq. (\ref{norm-good}) illustrates, as the system is cooled or heated, an evolution from an initial sharp energy density will not only shift the initial delta function distribution of the energy density but will also lead to a  (non-vanishing) widening $\sigma_{\epsilon}$.

\section{Aspects of the viscosity fits for supercooled liquids}
\label{explain-deviation}
One may work backwards to extract an effective $\sigma_{\epsilon}$ needed to fit the experimental viscosity data when using the first equality in Eq. (\ref{vis}). This leads to $\sigma_{T} \equiv \frac{T_{melt} - T}{\epsilon_{melt} - \epsilon} \sigma_{\epsilon}$ (which according to Eq. (\ref{epsT}) is equal to $\overline{A} T$) exhibits larger deviations from a linear in $T$ near $T_{melt}$ than at temperatures far below $T_{melt}$ where
a nearly perfect linear behavior appears. Such deviations from a nearly perfect linear increase of the effective $\sigma_{T}$ at lower temperatures are seen in, e.g., Figure 3 in \cite{us} and Figures 16, 17, and S1 in \cite{us1}. Indeed, above melting, non-supercooled equilibrated high temperature fluids have (by virtue of being in equilibrium) a sharp energy density ($\sigma_{T}=0$) implying a breakdown of any putative increase of $\sigma_{T}$ with temperature and suggesting a possible departure from a perfect linear increase of $\sigma_{T}$ also before $T_{melt}$. In Figure \ref{Collapse.} of the current work, the logarithm of the scaled dimensionless viscosities of all liquids must collapse onto the single ordinate $\log(\eta(T)/\eta(T_{melt}))=0$ at $T=T_{melt}$. The smaller deviation from linear in $T$ behavior of the effective extracted $\sigma_T$ at lower temperatures is consistent with the better collapse at lower temperatures seen in Figure \ref{Collapse.}. Indeed, at temperatures far below $T_{melt}$, the only natural temperature scale is $T$ itself (suggesting by dimensional analysis a possibly better linear in $T$ behavior). Indeed, it is also possible that the crossover to an even more precise form similar to Eq. (\ref{vis}) includes temperatures that do not precisely coincide with $T_{melt}$. In such a case, at temperatures much lower than $T_{melt}$, such corrections will be irrelevant but close to $T_{melt}$ such deviations may become more important. Furthermore, in the derivation of \cite{me} for Eq. (\ref{vis}) an approximation was made that may become more accurate at the temperatures become lower. Specifically, the long time average $v_{l.t.a}$ of the speed of a dropped spherical object into a viscous fluid, given (similar to Eq. (\ref{oolta})) by the integral 
$\int_{\epsilon_{melt}^{+}}^{\infty} P_{T}(\epsilon') v_{eq}(\epsilon') d\epsilon'$
was approximated by the product
$(v_{eq}(\epsilon_{melt}^{+}) \int_{\epsilon_{melt}^{+}}^{\infty} P_{T}(\epsilon')  d \epsilon')$.
This approximation becomes more accurate if the distribution $P_{T}(\epsilon')$ drops sharply as the energy density is increased for
$\epsilon'> \epsilon_{melt}^{+}$; in such instances, most of the weight in the first integral above occurs in a narrow
region just above $\epsilon_{melt}^{+}$ so that the above replacement is justified. This is certainly the case for a Gaussian distribution $P_{T}(\epsilon')$ centered about energy densities far below $\epsilon_{melt}$;
in such cases, most of the weight of the integral $\int_{\epsilon_{melt}^{+}}^{\infty} P_{T}(\epsilon')  d \epsilon'$
arises from a very narrow interval just above $\epsilon_{melt}$. However, if the distribution $P_{T}(\epsilon')$
is centered about energy densities close to $\epsilon_{melt}$ (especially if the standard deviation of $P_{T}(\epsilon')$ is not too small) then the above approximation will become more inaccurate. The above integrals were used in \cite{me} (see also \ref{explain-deviation}) to compute the viscosity as given in Eq. (\ref{vis}) using Stokes' law, viz., $\eta = {\sf{const.}}/v_{l.t.a}$ (with ${\sf{const.}}$ denoting a temperature independent constant).

\section{Intuitive arguments for the appearance of long time Gaussian distributions}
\label{prompt}

The prediction of Eq. (\ref{vis}) for the viscosity of quintessential non-equilibrium liquids (supercooled liquids and glasses) that yielded the 16 decade collapse of Figure \ref{Collapse.} was first derived \cite{me} by computing long time averages and invoking a Gaussian distribution of finite width $\sigma_{\epsilon}$. At the other extreme, equilibrium systems also display a Gaussian distribution of their energy density $P(\epsilon')$. In \cite{me}, we motivated the presence of a Gaussian distribution by maximizing the Shannon entropy for a given $\sigma_{\epsilon}$. We now suggest that long time normal distributions (both in systems that exhibit long time equilibrium and those that do not such as glasses) might also be natural from other considerations. In general, the probability distribution $P(\epsilon')$ may be calculated along lines similar to those that led to Eq. (\ref{Heavy}) in our toy example of Eq. (\ref{hferro}) where the system was continuously driven by an external transverse field. However, unlike the models studied in 
Section \ref{sec:dual}, at long times, supercooled liquids and glasses are no longer driven by an external bath $H_{tr}$ that continuously cools/heats them in a predetermined fashion. Instead, for supercooled liquids and glasses, at long times, the external heat bath (similar to the situation in equilibrium thermodynamics), becomes a source of stochastic noise (whose strength is set by its temperature $T$). Thus, the initially driven (i.e., continuously cooled) supercooled fluids or glasses will, at these long times, be effectively exposed to random noise. Following
the reasoning that led to Eq. (\ref{Heavy}), we examine general moments of the Heisenberg picture Hamiltonian
\begin{eqnarray}
\label{aver}
\langle (\Delta \epsilon)^{p} \rangle \equiv \frac{1}{N^{p}} \langle (H^{H} - \langle H^{H} \rangle)^{p} \rangle \equiv \langle (\frac{\Delta H^{H}}{N})^{p} \rangle 
\end{eqnarray}
when evaluated in the initial equilibrium state prior to cooling $| \psi^{0} \rangle = \sum_{n} c_{n}^{0} | \phi_{n} \rangle$. 
Here, $\{c_{n}^{0}\}$ are the amplitudes of the initial state 
$| \psi^{0} \rangle$ in the eigenbasis of the system Hamiltonian $H$. 
 Writing
Eq. (\ref{aver}) longhand as a product of $p$ factors of $(\frac{\Delta H^{H}}{N})$, we have
\begin{eqnarray}
\label{aver1}
\langle (\Delta \epsilon)^{p} \rangle  = && \sum_{n_{1} n_{2} \cdots n_{p}}  c^{(0)^*}_{n_{1}}   c^{(0)}_{n_{p}} \Big( \frac{(\Delta H^{H})_{n_{1}n_{2}}}{N} \Big) \nonumber
\\ &&\times \Big( \frac{(\Delta H^{H})_{n_{2} n_{3}}}{N}\Big) \cdots \Big(\frac{(\Delta H^{H})_{n_{p-1} n_{p} }}{N}\Big),
\end{eqnarray}
where $(\Delta H^{H})_{ab}$ are the matrix elements of $\Delta H^{H}$ in the eigenbasis of $H$. 
If, at long times, the matrix elements of the scaled Heisenberg picture Hamiltonian $\frac{\Delta H^{H}}{N}$ 
(now evolved with the stochastic influence of the environment at long times) attain random phases relative to each other
then the only remaining contributions in Eq. (\ref{aver1}) will be those in which all matrix elements come in complex conjugate pairs of the type 
$\Big(\frac{(\Delta H^{H})_{ab}}{N} \Big) \Big( \frac{(\Delta H^{H})_{ba}}{N}\Big)$. More precisely, in the calculation of the long time average of Eq. (\ref{aver1}), only the temporal average of such complex conjugate pairs will not vanish. Thus, similar to the calculation that led to Eq. (\ref{Heavy}), only even moments $p=2g$ may be finite. 
Now, however, the number of non-vanishing contributions (the number of ways in which the elements of $H^{H}$ may be matched in complex
conjugate pairs) will scale as $\Big(\frac{(2g)!}{2^{g} g!}\Big)$. This, in turn, prompts us to consider the possibility that, approximately, 
\begin{eqnarray}
\label{random_phase}
\langle (\Delta \epsilon)^{2g} \rangle \sim \Big(\frac{(2g)!}{2^{g} g!}\Big) \sigma_{\epsilon}^{2g}.
\end{eqnarray}
 This is especially the case if the initial state $| \psi^{0} \rangle$ corresponds to a single eigenstate of the Hamiltonian $H$, i.e., $ c^{(0)}_{n_{1}}  = \delta_{n_{1},n}$ and $c^{(0)}_{n_{p}} = \delta_{n_{p},n}$. 
The appearance of phases ($c^{0}_{n} \to c^{0}_{n} e^{-i E_{n} t/\hbar}$) does not, of course, change the average energy $\langle E \rangle$ nor $\langle (H^{H})^{2} \rangle$ at any time (implying the invariance of the variance $\sigma^2_{\epsilon}$) due to identical phase cancellations for all $t$. However, higher order moments of $H^{H}$ will, generally, vary with time. For these higher order moments, the (essentially random) phases will only cancel only at large $t$ (not identically at all $t$) allowing for Eq. (\ref{random_phase}). If, for all $g$, these moments of $\Delta \epsilon$ are equal to those evaluated with a Gaussian distribution (as follows from Wick's theorem- the combinatorics of which essentially reappeared in the above), then the probability distribution $P(\epsilon')$ for obtaining different energy densities in the final state must, indeed, be a Gaussian. If the expectation value $\langle H^{H} \rangle = E$ is held constant, then similar results may still apply (the density matrix may now mix states \cite{me} each with the aforementioned Gaussian distribution in $\epsilon$ subject to such constraints of fixed $\langle H^{H} \rangle$). 
 The above simple (non-rigorous) derivation rationalizes the appearance of Gaussian distributions in systems that equilibrate at long times (standard thermal systems) as well as the conjectured Gaussian form of $P(\epsilon')$ for supercooled liquids (Section \ref{2b2b}) that led to Eq. (\ref{vis}). 

Long time steady states with constrained conserved quantities may enable memory loss of the initial conditions and the appearance of effective equations of state \cite{me}. For thermal fluctuations in standard (``canonical'') systems, the resulting Gaussian distribution in $\epsilon$ is defined by its average and a standard deviation linear in the temperature ($\sigma_{\epsilon} \propto T$) suggestive of Eq. (\ref{epsT}). In a somewhat qualitatively similar manner, the stochastic effects of the environment are often simulated by Gaussian distributed forces whose standard deviation depends on the temperature $T$. The assumption of random phases in the above derivation of the Gaussian form does not, of course, imply small variances; the standard deviation of the energy density $\sigma_{\epsilon}$ (possibly still linear in the temperature) may be finite. As emphasized in Sections \ref{2-Ham} and \ref{2b2b}, in thermal systems the (typically linear in $T$) standard deviation characterizing the distribution $P(\epsilon')$ is $\sigma_{\epsilon} = \sqrt{k_{B} T^{2} C_{v}}/N \sim {\cal{O}}( N^{-1/2})$. For the product states of Section \ref{sec:product}, the energy is be a sum of ${\cal{O}}(N)$ independent variables (associated with decoupled Hamiltonians acting on different states) for which, as discussed earlier, in accord with the central limit theorem
(applicable to the independent decoupled contributions to the total energy),
$\sigma_{\epsilon} = {\cal{O}}(N^{-1/2})$. For non-decoupled contributions on states that are not the simple product form, while the above considerations suggest a Gaussian distribution, the central limit theorem 
for decoupled variables cannot be applied and $\sigma_{\epsilon}$ may be finite. More complex multi-scale probability distributions are possible (e.g., Appendix 6 of \cite{me}.) The above arguments for long time Gaussian distributions may be replicated, by a change of variables, to general intensive quantities $q$ other than the energy density. For a general $q$, various distributions $P(q';{\cal{W}})$ may in some cases, lead to the same result in Eq. (\ref{oolta}) for certain measured observables ${\cal{O}}$. 

\section{High Entanglement Entropy States}
\label{Higher_entanglement_entropy_states}
As we underscored earlier, typical ``thermal states'' may exhibit an entanglement entropy that scales with the system volume \cite{plain}, not its logarithm. The eigenstates $| \Psi_{+} \rangle$ examined in 
\ref{ent-ent-ent} were special in two different ways: 
(i) The eigenstates were constructed as an equal weight symmetric combination of all local product states and 
(ii) The systems that we examined were endowed with a local ``quasi-particle'' structure embodied by the independent commuting operators $\{{\cal{H}}_{i}\}$ (and associated local product eigenstates).
In general, even when only property (i) is violated, larger entropies may arise. It is instructive to see why this is so and how the state $| \Psi_{+} \rangle$ is special inasmuch as the calculation of its entanglement entropy is concerned. 
In the space spanned by all product states $|c \rangle$ that given energies $E_{B}$ (instead of that performed in \ref{ent-ent-ent} in the 
basis of the symmetric basis of Eqs. (\ref{EAQ})), the reduced density matrix $\rho_{B,+}$ becomes block diagonal.
Repeating the calculation of \ref{ent-ent-ent} in this basis, we find that in each region of fixed energy $E_{B}$, the block matrix is equal to 

\begin{eqnarray}
\label{onem}
{\sf One} ={\left( \begin{array}{ccccc}
1 & 1 & 1 & \cdots & 1\\
1 & 1 & 1 & \cdots & 1 \\
. & . & . & \cdots & 1 \\
. & . & . & \cdots & 1 \\
1 & 1 & 1 & \cdots & 1
\end{array} \right)}
\end{eqnarray}  
\newline

multiplied by the factor $e^{({\sf S}_{A}(E-E_{B}) - {\sf S}(E))/k_{B}}$. The dimensions of the matrix ${\sf One}$ are determined by the number ${\cal{N}}_{B}(E_{B})= e^{{\sf S}_{B}(E_{B})/k_{B}}$ 
of degenerate states $\{|E_{B}, j  \rangle\}_{j=1}^{{\cal{N}}_{B}(E_{B})}$ that have an energy $E_{B}$ on the spatial region $B$. We may perform a unitary transformation to the discrete Fourier basis (spanned by the states $|E_{B}, k_{E_{B}}  \rangle  \equiv ({\cal{N}}_{B}(E_{B}))^{-1/2} \sum_{j=1}^{{\cal{N}}_{B}(E_{B})} e^{i  k_{E_{B}} j} |E_{B}, j  \rangle$ with the wavenumber $k = 2 \pi m/{\cal{N}}_{B}(E_{B})$ where  $m=0, 1, 2, \cdots, {{\cal{N}}_{B}(E_{B})}-1$). This transformation reduces the matrices of the form of Eq. (\ref{onem}) to a single non-vanishing entry. Indeed, up to a constant prefactor ($ {\cal{N}}_{B}(E_{B})$), the matrix ${\sf One} $ is the outer product
 $|E_{B}, k_{E_{B}}=0  \rangle \langle E _{B}, k'_{E_{B}}=0 |$.
To make the contact with \ref{ent-ent-ent} lucid, we remark that in the notation of Eqs. (\ref{EAQ}), the single non-vanishing Fourier mode $|E_{B}, k_{E_{B}}=0  \rangle = | E_{B} \rangle_{+}$. In each block of fixed energies $E_{B}$, all other discrete Fourier ($k_{E_{B}} \neq 0$) modes have a vanishing amplitude. Such a Fourier transformation yields the eigenvalue spectrum, 
\begin{eqnarray}
{\sf Spec} \{\sf One\} = \{ {\cal{N}}_{B}(E_{B}), \underbrace{0,0,0, \cdots, 0}_{{\cal{N}}_{B}(E_{B})-1}  \}.
\end{eqnarray}
Thus, upon performing a unitary transformation to the Fourier basis, the block diagonal matrix $\rho_{B,+}$ becomes sparser (each vanishing eigenvalue of the reduced density matrix $\rho_{B,+}$ does not contribute to the entropy) and only the completely symmetric states of Eq. (\ref{EAQ}) are of relevance. If the equal amplitude eigenstates $| \Psi_{+} \rangle$ are replaced by a general linear combination
$| \Psi_{ \{ a_{c} \}}\rangle = \sum_{E_{c} = E} a_{c} |c \rangle$ (with $\sum_{c} |a_{c}|^{2} =1$) then the associated reduced density matrix $\rho_{B, \{a_{c} \}} = Tr_{A}  | \Psi_{ \{ a_{c} \}}\rangle \langle  \Psi_{ \{ a_{c} \}}| $ will remain block diagonal. However, the block matrices that span each region of fixed energy $E_{B}$ will, generally, look very different from ${\sf One}$. Intuition concerning the larger entanglement entropies that may generally result can be gained by suggestive arguments. Towards this end, we may consider what occurs if each diagonal block of $\rho_{B,+}$ of the type ${\sf One}$ is replaced by other block diagonal matrices with a wider distribution of the eigenvalues such that, e.g., each of the non-vanishing eigenvalues of $\rho_{B,+}$ for energies $E_{B}$ (close to the energy $\overline{E}_{B}$ that maximizes the sum ${\sf S}_{A}(\overline{E}_{A} = E - \overline{E}_{B}) 
+ {\sf S}_{B}(\overline{E}_{B}))$ is, effectively, split into $K$ equal parts. In such a case, the entanglement entropy ${\sf S}_{ent,\{ a_{c} \}}$
will be larger than $S_{ent,+}$ by an additive contribution of $\ln K$. If, for $L_{B} < L_{A}$,
the logarithm $(\ln K) = {\cal{O}}({\sf S}_{B}(\overline{E}_{B})) = {\cal{O}}(L_{B})$ then this additive contribution to the entanglement entropy may be linear in the volume $L_{B}$ of subsystem $B$.


\begin{thebibliography}{28}
\bibitem{non-unitary} A trivial example illustrating this non-unitarity is that of two $S=1/2$ spins comprising, respectively, the system and the environment. Suppose that the initial state is $| \uparrow_{\cal{S}} \uparrow_{\cal{E}} \rangle$ (with the first entry denoting the state of the system and the second entry corresponding to that of the environment) and that at time $t$ is $\frac{1}{\sqrt{2}} (|\uparrow_{\cal{E}} \downarrow_{\cal{S}} \rangle + |\downarrow_{\cal{E}} \uparrow_{\cal{S}} \rangle)$. A unitary evolution connects these two states (e.g., a permutation between the initial and final states).  A trace over the ``environment'' yields the reduced density matrices $\rho_{\cal{S}} = diag(1,0)$ and $\rho_{\cal{S}}(t)= diag(\frac{1}{2}, \frac{1}{2})$. The eigenvalues of $\rho_{\cal{S}}$ and $\rho_{\cal{S}}(t)$ are different. Thus, there exists no unitary transformation connecting $\rho_{\cal{S}}$ and $\rho_{\cal{S}}(t)$.
\bibitem{adiabatic} If in case (1) of Section \ref{sec:sketch}, the Hamiltonian is varied adiabatically such that both its initial and final forms are equal to $\sum_{i} {\cal{H}}_{i}$ then, by the adiabatic theorem, no change in $P(\epsilon')$ (including the cooling sketched in Figure \ref{Idea.}) can occur between the final and initial states.
\bibitem{class-quant+} We denote the eigenvectors of a density matrix $\rho$ by $\{| \psi_{\ell} \rangle\}$ and their corresponding eigenvalues by $\{p_{\ell} \}$.
 The expectation values $E_{\ell} = \langle \psi_{\ell} | H | \psi_{\ell} \rangle$ mark the energies of given density matrix eigenstates.
The variance of the energy density $\sigma_{\epsilon}^{2} = \frac{Tr[\rho(H-E)^{2}]}{N^{2}}$ may then be expressed as a sum of 
``classical'' and ``quantum'' contributions,
\begin{eqnarray}
\sigma_{\epsilon}^{2} = (\sigma_{\epsilon}^{\sf{class.}})^{2} + (\sigma_{\epsilon}^{\sf{quant.}})^{2}.
\end{eqnarray}
Here,
\begin{eqnarray}
\label{class+++}
(\sigma_{\epsilon}^{\sf{class.}})^{2} \equiv \frac{1}{N^{2}} (\sum_{\ell} p_{\ell} E_{\ell}^{2} -  E^{2}),
\end{eqnarray}
(note that, generally, $\langle \psi_{k}| H^{2} | \psi_{\ell} \rangle \neq E_{\ell}^{2}$)
and 
\begin{eqnarray}
(\sigma_{\epsilon}^{\sf{quant.}})^{2} \equiv \frac{1}{N^{2}} (\sum_{\ell} p_{\ell} \sigma_{\ell}^{2}),
\end{eqnarray} 
where the variances
\begin{eqnarray}
\label{sl++}
\sigma_{\ell}^{2} \equiv \langle \psi_{\ell} | (H-E_{\ell})^{2} | \psi_{\ell} \rangle,
\end{eqnarray}
stem from the fluctuations of Hamiltonian in given eigenstates of the density matrix.
(Similar classical and quantum standard deviations may be defined for quantities $q$ different from the energy density $\epsilon$.) 
In classical systems where $H$ and $\rho$ commute, $(\sigma_{\epsilon}^{\sf{quant.}})^{2} =0$ and the sole contribution to the variance
is that of Eq. (\ref{class+++}) as it is in, e.g., equilibrium classical systems in the canonical ensemble. 
The standard deviation of $(H/N)$ (i.e., $\sigma_{\epsilon}$) should not, of course, be confused with 
$\sigma_{\epsilon}^{\sf{class.}}$. Indeed, in a general single pure state
(a situation for which $\sigma_{\epsilon}^{\sf{class.}} =0$), the standard deviation of the energy density may be finite, $\sigma_{\epsilon} = {\cal{O}}(1)$. Concrete realizations of this maxim will be provided in the examples of Section \ref{sec:dual}. In a quasi-static state, different eigenstates of the density matrix may exhibit similar energy densities, viz. $\sigma_{\epsilon}^{\sf{class.}}=0$.
\bibitem{explain-similar} Along nearly identical lines, for conventional thermal systems, the standard deviation scales as $\sigma_{\epsilon} \sim N^{-\alpha}$ with $\alpha = 1/2$ implying that the lattice average of the pair correlator $\overline{G_{q}} \ge {\cal{O}}(N^{-2 \alpha}) = {\cal{O}}(1/N)$. At temperatures $T \to \infty$, this $1/N$ scaling merely amounts to the existence of self-correlations (of typical magnitude ${\cal{O}}(1)$) $i=j$ and the absence of pair correlations when the two lattice sites are different, $i \neq j$. Analogously, for critical systems wherein $G_{ij} \sim r_{ij}^{-p}$ with $p=d-2+\eta^*$ with $\eta^*$ an anomalous exponent, the average correlator (over all pairs of lattice sites) scales as $\overline{G} \sim N^{-p/d}$.
\bibitem{2unitary}
Starting from an initial product state, a product of two unitary operators each of which acts on two neighboring sites at a time on the graph of Figure \ref{coupling.} may generate long range correlations between sites $i$ and $j$. For instance, in a spin $S=1/2$ type setting, the two step unitary evolution  involving neighboring sites on the graph of Figure \ref{coupling.}, the evolution
$|\uparrow_{i} \uparrow_{j} \uparrow_{\cal{E}} \rangle \to \frac{1}{\sqrt{2}} (|\uparrow_{i} \downarrow_{j} \uparrow_{\cal{E}} \rangle +  | \uparrow_{i} \uparrow_{j} \downarrow_{\cal{E}} \rangle ) \to \frac{1}{\sqrt{2}} 
(|\uparrow_{i} \downarrow_{j}  \downarrow_{\cal{E}} \rangle + | \downarrow_{i} \uparrow_{j} \uparrow_{\cal{E}} \rangle) \equiv | \psi_{u,2} \rangle$ generates a non-vanishing covariance $\langle \psi_{u,2} | \sigma^{z}_{i} \sigma^{z}_{j} | \psi_{u,2} \rangle - \langle \psi_{u,2} | \sigma^{z}_{i}  | \psi_{u,2} \rangle \langle \psi_{u,2} | \sigma^{z}_{j} | \psi_{u,2} \rangle = -1$.
In the simple limit wherein each of the $N$ spins (and/or other degrees of freedom) in the system uniformly couples to the environment alone (i.e., the system spins (and/or other degrees of freedom) do not interact amongst themselves), there is no notion of distance between the sites insofar as the interactions are concerned. If one starts from a spatially uniform state of the system then, at all future times, by symmetry, the covariance between any pair of sites will always be the same independent of their location (and thus spatial separation). In Section \ref{sec:CSM}, we explicitly solve several ``Central Spin'' models in which this intuition can be explicitly realized.  Another formal trivial exactly solvable system (of a similar type) is that of  \ref{appendix:preparation} in which the environment is, at intermediate times, a macroscopic magnet with a narrow distribution of its total magnetization that couples to all sites of the lattice leading to a precession of the total system spin with a distribution of frequencies that, after sufficiently long times, endows the system with long range correlations of the form of Eq. (\ref{longxx}).
\bibitem{exam1} As a concrete example if, on a hypercubic lattice of $N$ sites in $d$ dimensions that is endowed with periodic boundary conditions, the exchange couplings $J_{ij} =J$ for nearest neighbor pairs $(ij)$ with $J_{ij}=0$ otherwise, and the initial state was a ground state of uniform polarization (i.e., $s_{i} = s = \pm 1$) 
then the difference in energies between the final and initial states will be given by $E_{\chi} - E_{\psi^{(0)}} = NdJ$. In this example, the standard deviation of the energy in the final state $| \chi \rangle$ is equal to $J \sqrt{Nd}$.  
Note that for randomly chosen initial states, the difference in the energy densities of the final and initial states vanishes in the thermodynamic limit (i.e., the probability distribution of $(E_{\psi^{(0)}}/N)$ for all possible 
initial states $| \psi^{(0)} \rangle$ becomes a delta function as $N \to \infty$).
\bibitem{plain}
Regardless of the validity of the eigenstate thermalization hypothesis, we note that the volume law entanglement of states that describe thermal systems is extremely natural. Stated plainly, if $\rho$ describes the density matrix of a 
large region ${\cal{I}}$ formed by a thermal system ${\cal S}_{th}$ and its environment,  ${\cal{I}} = {\cal S}_{th} \cup {\cal{E}}$, then the reduced density matrix for region ${\cal S}_{th}$ formed by performing a trace of the full density matrix over the environment ${\cal{E}}$, must be of a form of an equilibrium (eq) density matrix $\rho_{eq;S_{th}}$. That is, $\rho_{S; th} \equiv Tr_{\cal{E}} \rho = \rho_{eq}(S_{th})$. It follows that the entropy of entanglement 
${\sf S}_{ent}$ between regions ${\cal S}_{th}$ and ${\cal{E}}$ will be identically equal to the standard thermodynamic entropy ${\sf S}_{eq}$ of a thermal system in ${\cal S}_{th}$, i.e., ${\sf S}_{ent} \equiv  -Tr[ \rho_{S;th} \ln \rho_{S;th}] = -Tr  [\rho_{eq}(S_{th}) \ln \rho_{eq}(S_{th})] \equiv {\sf S}_{eq}$. Since the entropy ${\sf S}_{eq}$ of an equilibrium system at temperatures $T>0$ is an extensive quantity, it follows that the entanglement entropy ${\sf S}_{ent}$ of an equilibrium system with its environment obeys a ``volume'' law (i.e., the entanglement entropy will be proportional to the spatial volume of ${\cal S}_{th}$).
\bibitem{pure} 
Using the initial ``purification'' procedure of \cite{lieb-pure} (a thermofield double type construct), the proof of this assertion can be made trivial as we now explain. Diagonalizing the density matrix of dimension ${\cal{D}}$ on a region ${\cal S}$ we have $\rho = \sum_{a=1}^{\cal{D}} p_{a}
 | \psi_{a} \rangle \langle \psi_{a} |$. If we associate to each state $| \psi_{a}\rangle$ that has its support in ${\cal S}$ a different orthogonal state $|\overline{\psi}_{a} \rangle$ (with $\langle \overline{\psi}_{a} | \overline{\psi}_{a'} \rangle = \delta_{aa'}$) having its support in the environment ${\cal{E'}}$ (i.e., ${\cal S} \cup {\cal{E'}} = {\cal{I'}}$ and ${\cal{S}} \cap {\cal{E'}} = \{\emptyset\}$). With these elements in tow, we will then have $\rho = Tr_{\cal{E'}}  | \Psi \rangle \langle \Psi |$ where $| \Psi \rangle \equiv \sum_{a=1}^{\cal{D}} \sqrt{p}_{a} | \psi_{a}  \rangle \otimes | \overline{\psi}_{a} \rangle$ is a pure state in ${\cal{I'}}$.
\bibitem{ssremark}  When adding the spin of $N$ particles, as in, e.g., Eq. (\ref{add+}), there is a unique sector of maximal total spin $S_{tot} = S_{\max}$. For lower values $S_{tot}$, the eigenstates $|\phi_{\alpha} \rangle$ of the Hamiltonian $H_{spin}$ need to be specified by an additional index ($\upsilon_{\alpha}$), see also \ref{trivial-maxS}.
\bibitem{boundary} We briefly comment on what transpires if the forms of the Hamiltonian $H_{spin}$ and, notably, the coupling of the spins to the transverse field as embodied by the Hamiltonian $H_{tr}$ apply only for spins on the surface of the system (the number of these spins $N_{surface}= {\cal{O}}(N^{(d-1)/d}$)). In such situations, the standard deviation in Eq. (\ref{se}) would naturally scale as $\sigma_{\epsilon} = {\cal{O}}(N^{-1/d})$. The distribution of Eq. (\ref{Heavy}) will retain its form. If this circumstance arises then, in the discussions and equations in the main text concerning $H_{spin}$ and $H_{tr}$, the total number of spins $N$ will be replaced by the number of surface spins $N_{surface}$.
\bibitem{explain_classical_spin} The calculation for any ensemble of initial microstates that is invariant under rotations about the $z$ axis that have fixed values of $S^{tot}_{z}$ and $\vec{S}_{tot}^{2}$ mirrors that leading to Eq. (\ref{se}) with the eigenstate expectation values replaced by ensemble averages. Eq. (\ref{s2}) will be trivially modified in the classical case. Specifically, in the last line of Eq. (\ref{s2}), the quantum $\hbar^{2} S_{tot}(S_{tot} +1)$ will be replaced by the classical $\vec{S}_{tot}^{2}$.
\bibitem{MM-explain}
Specifically, the Matsubara-Matsuda transformation \cite{MM} from spin $S=1/2$ raising and lowering operators to the creation and annihilation operators of hard core bosons is given by 
$\frac{S_{i}^{+}}{\hbar} \to b_{i}^{\dagger}$ and $\frac{S_{i}^{-}}{\hbar} \to b_{i}$ (from which it follows that $\frac{S_{i}^{z}}{\hbar} \to (n_{i} - \frac{1}{2})$ where $n_{i} = b_{i}^{\dagger} b_{i}$ is the hard core boson number operator).
\bibitem{JW} 
 With $c_{i}$ and $c_{i}^{\dagger}$ denoting respectively, the spineless Fermi annihilation and creation operators
(and with $n_{i} = c_{i}^{\dagger} c_{i}$), the Jordan-Wigner dual of the initial spin Hamiltonian of Eq. (\ref{hferro}) reads
\begin{eqnarray}
\label{FermiH}
H_{Fermi} =  \sum_{|i-j| = 1} J_{ij} ((c_{i}^{\dagger} c_{j} + h.c.) + n_{i} n_{j}) -  \sum_{i} (B_{z} - \sum_{|j-i|=1} J_{ij}) n_{i}.
\end{eqnarray}
We may consider the initial spineless fermion system of Eq. (\ref{FermiH}) that is governed, at intermediate times, by the (now non-local) Jordan-Winger dual of $H_{tr}$ or, equivalently, the fermonized dual of $H_{Bose-doping}$. Following a temporal evolution with the latter intermediate time Hamiltonian, an initial eigenstate of $H_{Fermi}$ formed by the equal amplitude superposition of all spineless Fermi states 
of a fixed total particle number (the dual of $| \psi^{0}_{Spin} \rangle$) will transform into a final state displaying the standard deviation and distribution of Eqs. (\ref{se}, \ref{Heavy}). (In order to make the transformation lucid, we remark that the spineless Fermi dual of the particular spin
$S=1/2$ state $| \uparrow_{1} \uparrow_{2} \downarrow_{3} \uparrow_{4} \downarrow_{5} \uparrow_{6}  \cdots \uparrow_{N-1} \downarrow_{N} \rangle$ describes an antisymmetrized superposition of all product states in which the fixed total number of spineless fermions $\sum_{i} n_{i}$ are distributed over the specific sites $(1,2, 4, 6, \cdots, (N-1))$.)
\bibitem{comment-CSM*}
In the state $| \psi^{0}_{CSM} \rangle$, the standard deviation $\sigma_{H^{CSM}_{{\cal S}-{\cal E}}} = \frac{\hbar}{2^{5/4}} \sqrt{N}  |J_{\cal{E}}| $ while $\sigma_{H^{CSM}_{\cal{S}}} = 0$. 
Eq. (\ref{connect2refereesbrains}) holds whenever  $H^{CSM}_{\cal{E}}$ is a function of ${\cal{P}}_{{\cal{E}}}$ alone.
For the particular case of $H^{CSM}_{\cal{E}}=0$, the standard deviation of the full Hamiltonian will scale with the system size as it does in typical (``canonical'') open thermal systems, $\sigma_{\tilde{H}^{CSM}} (= \sigma_{H^{CSM}_{{\cal S}-{\cal E}}}) = {\cal{O}}(\sqrt{N})$. Correlators such $\langle  \prod_{i=1}^{N} S_{i}^{zH}(t) \rangle = \frac{\hbar^{N}}{2^{N+1}}
(1+ \cos^{N}(J_{\cal{E}} t))$ contain Fourier components (frequencies) as large as $N |J_{\cal{E}}|$. However, the variance of the frequencies scales as $\sqrt{N}$. Since the cosine in the above form is an equal amplitude superposition of positive and negative frequencies $\pm J_{\cal{E}}$, the distribution of frequencies in $\cos^{N}(J_{\cal{E}} t)$ is indeed identical to that associated with a ``random walk'' of length $N$ with equal amplitude steps $\pm J_{\cal{E}}$ leading to a binomial distribution with a standard deviation scaling as $\sqrt{N}$. The distribution of frequencies in the expectation values of global products such as $ \prod_{i=1}^{N} S_{i}^{zH}(t)$ reflects that  
in the total Hamiltonian $\tilde{H}^{CSM}$. All expectation values in this system may only have Fourier components associated with differences between the energies of eigenstates appearing in the decomposition of the initial state into the eigenstates of $\tilde{H}^{CSM}$.
\bibitem{var-cos-Gau}
For $z' = \cos(\alpha \overline{q}')$ with $\overline{q}'$ drawn from the normal distribution of standard deviation $\sigma_{\overline{q}}$ and zero mean, one immediately obtains all moments computed with $P(z')$, i.e,
 $\langle (z')^{\sf p} \rangle = \frac{1}{2^{\sf p}} \sum_{\ell=0}^{\sf p} {{\sf p} \choose \ell} e^{-(\alpha (2 \ell - {\sf p}) \sigma_{\tilde{q}})^{2}/2}$. (The average of $e^{i c \overline{q}'}$ computed with the centered Gaussian distribution $\overline{P}(\overline{q}')$ is $e^{-c^{2} \sigma^{2}_{\overline{q}}/2}$ and thus the expectation value of $\cos^{\sf p} (\alpha \overline{q}')$ is, after expressing the cosine in terms of exponentials and using the binomial theorem, trivially given by the above expression for $\langle  (z')^{\sf p} \rangle$.) The moments computed for ${\sf p}=1,2$ immediately yield the variance of Eq. (\ref{connect2refereesbrains''}))
\bibitem{pe'}
Following Eq. (\ref{central}), the standard deviation of the energy density of the system $\sigma_{\epsilon}$ will, once again, be finite for macrosopic $N$. Similar to Eqs. (\ref{distribution},\ref{joq},\ref{Heavy}), one may readily calculate $\langle (\Delta \epsilon)^{\sf p} \rangle$ for all ${\sf p} \ll N$ for, e.g., a harmonic oscillator ground state $|{\cal E} \rangle$.
The calculation will amount to the expectation value of these moments with the a distribution formed by replacing $\epsilon'$ by $(- B_{z} \cos (\overline{q}' t))$ (or, equivalently, with the definitions of the main text, when replacing ${\sf \varepsilon}'$ by $ \cos(\alpha \overline{q}')$) when $\overline{q}'$ itself is drawn from a Gaussian distribution $\overline{P}(\overline{q}') = \frac{1}{\sqrt{2 \pi \sigma_{\overline{q}}^{2}}} e^{-\overline{q}'^{2}/(2 \sigma^{2}_{\overline{q}})}.$ In each interval of width $\pi$ in $\alpha \overline{q}'$ (where we may identify a unique $\overline{q}'$ with $\cos (\alpha \overline{q}') = {\sf \varepsilon'}$) 
such a distribution for $\overline{q}'$ would imply a corresponding distribution $P({\sf \varepsilon'})$ for the scaled energy density ${\sf \varepsilon'}$ such that $P({\sf \varepsilon'}) d{\sf \varepsilon'}= \overline{P}(\overline{q}') d \overline{q}' $. Thus, if the range of arccosine extended over all real ${\sf \varepsilon'}$ then $\overline{P}(\overline{q}')$ would equal $\frac{\overline{P}( (\cos^{-1}{\sf \varepsilon'})/\alpha)}{\alpha \sqrt{1 - {\sf \varepsilon'}^{2}}}.$ ``Folding'' ${\sf \varepsilon'}$ back to $[0, \pi]$ when using the principal arccosine branch (to physically capture the bounded energy density of the system ($-1 \le {\sf \varepsilon'} \le 1$) comprised of two-level states at each site), we sum over all consecutive intervals of width $\pi$, i.e., $P({\sf \varepsilon'})=\frac{1}{\sqrt{2 \pi \sigma_{\overline{q}}^{2} \alpha^{2} (1 - {\sf \varepsilon'}^{2}) } } \sum_{n=-\infty}^{\infty} e^{-( \pi n+ (\cos^{-1}{\sf \varepsilon'})/\alpha))^{2}/(2 \sigma^{2}_{\overline{q}})}$. Performing the sum (and employing $\vartheta_{3}(x,y)  = \sum_{n=-\infty}^{\infty} y^{n^{2}} e^{2nix}$) leads to Eq. (\ref{pe'-eq}). Similar to Eq. (\ref{Heavy}) for the spin and related models of Section \ref{sec:dual}, the energy density of the system comprised of two-level states at each site is trivially bounded ($-1 \le {\sf \varepsilon'} \le 1$ in the above distribution $P({\sf \varepsilon'})$).
\bibitem{mix2} We remark that a shorter, one line, proof of the uncertainty relations for mixed states follows from the fact that any density matrix may be expressed in terms of pure states \cite{lieb-pure,pure}. 
Thus if, in the notation of \cite{pure}, we take ${\cal{I}} \subset {\cal{I'}}$, then the uncertainty relation for pure states implies the corresponding uncertainty relations for general mixed states with arbitrary density matrices.
\bibitem{mix-stationary} In mixed states, all density matrices that are diagonal in the eigenbasis of $\tilde{H}$ are trivially stationary under the evolution with 
$\tilde{H}$. Such diagonal matrices may, however, display a standard deviation $\sigma_{\tilde{H}}$ that is large as $(\tilde{E}_{\max} - \tilde{E}_{\min})/2$ (where $\tilde{E}_{\max}$ and $\tilde{E}_{\min}$ denote the maximal and minimal eigenvalues of $\tilde{H}$). By contrast, if the pure state density matrix $ \tilde{\rho}(t)$ of Eq. (\ref{rhl}) exhibits a large standard deviation $\sigma_{\tilde{H}}$ then it also has high weight components fluctuating rapidly in time (of frequency scale $\sigma_{\tilde{H}}/\hbar$).
\bibitem{example-oscillator}  A simple example of a large closed (semi-classical) system with system size independent frequencies governing the global evolution is that of nearly harmonic elastic solids. The frequencies of the eigenmodes do not scale with the system size. (Note that in this closed system example, the different eigenmodes cannot couple independently to different baths- the total energy of the (semi-classical) system is conserved.)
\bibitem{special1} For instance, one may concoct states $| \tilde{\psi} \rangle$ (or density matrices $\tilde{\rho}$) in ${\cal{I}}$ that exhibit large variances $\sigma_{\tilde{H}}^{2} =( \langle \tilde{\psi} | \tilde{H}^{2} | \tilde{\psi} \rangle - (\langle \tilde{\psi} |\tilde{H} | \tilde{\psi}  \rangle)^{2})$ when the system size increases. This may be achieved by superposing eigenstates that span an extensive (${\cal{O}}(E) = {\cal{O}}(N)$) range of $\tilde{H}$ eigenvalues. As discussed in Section \ref{sec:product}, on their own, product states with separable Hamiltonians will lead to $\sigma_{\tilde{H}} \sim {\cal{O}}(\sqrt{N})$. The central spin models of Section \ref{sec:CSM} exhibit extensive $\sigma_{\tilde{H}}$. The microcaonical ensemble (for which $\sigma_{\tilde{H}}$ is bounded) will be rendered incompatible in states like these for which the standard deviation scales with the system size.
\bibitem{explain_gauge} The uncertainty relation of Eq. (\ref{uncertain}) was derived from Eq. (\ref{uncert}) by expressing $\tilde{H} = i \hbar \frac{\partial}{\partial t}$. The discrete time gradients of bounded quantities such as $(H/N)$ 
and the standard deviation of the time discretized version of $\tilde{H}$ cannot exceed ${\cal{O}}(1/\Delta t)$. However, the expectation values of gauge non-invariant quantities can be made arbitrarily large or small. In non-relativistic systems studied in the current paper, the energy can be shifted by a constant without altering the expectation values of all measurable observables 
$Q$ so long as these observables are not explicitly time dependent ($\frac{\partial Q}{\partial t}=0$). A shift of the energy by a constant $\Delta E$ is tantamount to a trivial ``gauge transformation''- a multiplication of the wavefunction by a linear in time phase factor, $| \psi \rangle \to e^{-i (\Delta E) t/\hbar} | \psi \rangle$. We emphasize that the measurable frequencies of these oscillations and their standard deviation are bounded in time discretized systems. The standard deviation of the energy density is a physical gauge invariant quantity (as are general relative energy differences) and all quantities evaluated with the density matrix $\tilde{\rho}(t)$ of Eq. (\ref{rhl}).
\bibitem{example1-3} Although Assumptions (1-3) of Section \ref{section:closed} are not always satisfied (see, e.g.,\cite{special1}), rather general systems do obey them and further exhibit a finite rate of change of the energy density $|\frac{d \epsilon}{dt}|$. To make contact with our earlier examples of Section \ref{sec:dual}, as a case in point, we may consider a simple example. We take $\tilde{H}$ to be $H_{spin}$ of Eq. (\ref{hferro}) with a short range $H_{symm}$ and in which the magnetic field $B_{z}$ is along the $z$ direction over the entire volume ${\cal{I}}$; in a similar spirit, we consider the Hamiltonian $H$ in the region ${\cal{S}}$ to be given by Eq. (\ref{hferro}) with an internal applied field $B_{y}$ parallel to the $y$ direction. In this example (apart from ``surface terms'' arising from the short range interactions in $H_{symm}$), the interaction between the system ${\cal{S}}$ and its external environment ${\cal{E}}$ is dominated by an external field ($\vec{B} = (0, -B_{y}, B_{z})$) applied from sites exterior to ${\cal{S}}$. This external field couples to all spins in ${\cal{S}}$; when added together with $H$, this field will reproduce the terms of ${\tilde{H}}$ appearing in 
${\cal{S}}$. We further set the density matrix to be $\rho = | \psi \rangle \langle  \psi |$, with $| \psi \rangle$ corresponding to an equal amplitude superposition of two eigenstates of maximal total spin $S_{tot}= {\cal{O}}(N)$ that differ by one quantum of $\hbar$ in their $S_{z}^{tot}$ eigenvalues. In the convention of Section \ref{sec:dual}, $| \psi \rangle = \frac{1}{\sqrt{2}} (|\upsilon_{\alpha}; S_{tot}, S^{z}_{tot} \rangle + |\upsilon_{\alpha}; S_{tot}, S^{z}_{tot} - \hbar \rangle )$. It is readily verified that Assumptions (1-3) are satisfied. The state $| \psi \rangle$ displays a standard deviation $\sigma_{\tilde{H}} = \frac{\hbar|B_{z}|}{2} = {\cal{O}}(1)$
(consistent with the finite standard deviation of the total energy characterizing the microcanonical ensemble on the full volume ${\cal{I}}$). However, within the smaller subvolume ${\cal{S}}$, at all times $t$, the standard deviation of the energy density (i.e., the standard deviation of $H/N$) is $\sigma_{\epsilon} = \frac{|B_{y}| \hbar S_{tot}}{2N} \sqrt{2 + \frac{2}{S_{tot}} - (w^{2} + (w -\frac{1}{S_{tot}})^{2})} = {\cal{O}}(1)$;  the standard deviation of the total energy of ${\cal{S}}$ is generally extensive, $\sigma_{H} = {\cal{O}}(N)$. In the above, as in Section \ref{sec:dual}, $w \equiv S_{tot}^{z}/(\hbar S_{tot})$. The rate of change of the energy density is $|\frac{d \epsilon}{dt}| = \frac{|B_{y} B_{z}|  \hbar S_{tot} \cos(B_{z} t)}{2N} \sqrt{1+ \frac{1}{S_{tot}} - w(w- \frac{1}{S_{tot}})}$. Thus, in this example, generally 
$|\frac{d \epsilon}{dt}| ={\cal{O}}(1)$ for finite $B_{y}$ and $B_{z}$ when $S_{tot} = {\cal{O}}(N)$ and $|w|<1$.
\bibitem{QevolveQ}
Generally, sites exterior to ${\cal{I}}$ may give rise to an evolution as given by Eq. (\ref{rhorho}) applied to ${\cal{I}}$ (and ${\cal{S}}$). In systems that exhibit non-equilibrium behavior, the time evolved $\tilde{H}^{H}(t)$ (and $H^{H}(t)$) may not be similar to the initial $\tilde{H}$ (and $H$). Consequently, the resulting probability distribution $\tilde{\rho}(t)$
will give rise to averages that differ from those in equilibrium. That is, Eq. (\ref{uncert}) will always be satisfied. However, the time evolved $H^{H}(t)$ and $\tilde{H}^{H}(t)$ will,
when Eq. (\ref{canonical}) holds, deviate from the initial Hamiltonians $H$ and $\tilde{H}$ and thus have expectation values and fluctuations different from those in equilibrium.
\bibitem{phase_transition}
A well defined temperature does not always imply a sharp energy density. For instance, a finite interval of the internal energy densities corresponds to the single melting (and other coexistence) temperature(s). This width of this interval is set by the latent heat of fusion at the melting temperature. At other temperatures in which no phase coexistence appears, there is a unique internal energy density (a sharp thermodynamic state variable) $\epsilon(T)$ associated with every temperature $T$.
\bibitem{diag_common}
If the spectrum of $H$ exhibits degeneracies then we will explicitly choose the states $\{|\phi(\{q' \}; {\cal{W}}) \rangle\}$ to be common eigenstates of both $H$ and ${\cal{O}}$.
\bibitem{non-eq-comment}
For completeness, we note that our generally derived broadening of the distribution $P(\epsilon')$ in driven systems would imply relations similar to Eq. (\ref{equil-neq}) also for systems that do not exhibit equilibrium properties, e.g.,\cite{MBL1,MBL2,MBL3,MBL4,MBL5,MBL6,MBL7,noMBL,yesMBL}; averages of observables ${\cal O}$ with the broadened distribution $P$ differ from those with a delta function type distribution.
\bibitem{comment-T} When the system is closed, its evolution is unitary and in line with the invariance of the partition function, the spectrum of the density matrix cannot change. A system displaying a Boltzmann distribution at an initial temperature cannot re-equilibrate to a Boltzmann distribution at a different final temperature (since the set of eigenvalues of equilibrium density matrices at different temperatures is not the same (yet under a unitary evolution the spectrum cannot change)).
\bibitem{LG} Since the energy density is bounded from below by its ground state value $\epsilon_{g.s.}$,
the probability distribution $P(\epsilon')$ vanishes for $\epsilon'< \epsilon_{g.s.}$. The exact distribution $P(\epsilon')$
(whether a Gaussian or of another approximate form) is, of course, cut off at such low $\epsilon'$.  
The same holds for the Gaussian distribution describing the energy density of a large finite size system 
within the canonical ensemble.
\bibitem{plastic_deformation} While it is natural to expect a continuous Gaussian distribution for supercooled fluids and glasses, the distribution $P(\epsilon')$ for plastically deformed crystals might be different. Since different spatial regions of an equilibrium crystal that has been cracked, etc., are locally similar to small finite volume patches of an equilibrium solid, the spatial overlap integral between the wavefunction describing the plastically deformed solid and the state of the equilibrium solid is anticipated to be finite.
Thus, $P(\epsilon')$ may be a sum of delta-functions characterizing the different finite volume patches that emulate the finite equilibrium crystal.
\bibitem{mexplain} The transition rates from initial $| \psi_{i} \rangle = \sum_{s} c_{s}(i) | \phi_{s} \rangle$ to final $| \psi_f \rangle = \sum_{s'} c_{s'}(f) |\phi_{s'} \rangle$ states are ${\cal{R}}_{i \to f} = \frac{d}{dt} |\langle \psi_i | U_{pert}(t) | \psi_f \rangle|^{2} = \sum_{\ell,\ell',s,s'} (c_{\ell}(i) c_{s'}(f) c^{*}_{s}(i) c_{\ell'}^{*}(f) 
 \frac{d}{dt} ( \langle \ell | U_{pert}(t)| \ell' \rangle^{*}  \langle s| U_{pert}(t) | s' \rangle))$. The matrix elements of the evolution operator $U_{pert}$ containing the effects of general (arbitrary magnitude) perturbations are temperature independent and only determined by the Hamiltonian governing the system (a particular supercooled fluid) and the perturbation in question. By contrast, the amplitudes $\{c_{s}(i)\}$ and $\{c_{s'}(f)\}$ (that determine the distributions $P$ associated with the initial and final states) will generally vary with the energy densities of the initial and final states (or their associated temperatures). Indeed, from Eq. (\ref{epsT}), the width of $P$ increases with the temperature (if, as assumed, $\overline{A}$ is kept constant) thus smearing the equilibrium result more significantly at elevated temperatures. The sum of transition rates to all possible final states orthogonal to the initial state can, similar to Eq. (\ref{equil-neq}), be expressed as a weighted average of the equilibrium transition rates at different energy densities \cite{me}.
\bibitem{ATG} A comparison to experimental and earlier theoretical fits (see, e.g., Figure4 of  \cite{ATG'}) for the heat capacity data yields a crossover temperature window $\Delta T_{h}$ (the width along the temperature axis of the ``hysteresis'' where the heating and cooling heat capacity curves differ) for different systems. Empirically, different heating (and cooling) rates yield different heat capacity curves yet these, too, tend to almost exactly coincide outside of the crossover temperature window of width $\Delta T_{h}$.
For OTP, Salol (only experimental results for the heat capacity on heating are given for this liquid in \cite{ATG'}), and Glycerol, the corresponding ratios  $\frac{\Delta T_{h}}{T_{g}} \sim  0.08, 0.087,$ and $0.13$ (corresponding to $\Delta T_{h} \sim 20$K, $20$K, and $25$K). By comparison, the values of the scale factor $\overline{A} \sqrt{2}$ appearing in Eq. (\ref{vis}) for the fit of Figure \ref{Collapse.} are, respectively, 0.070, 0.087, and  0.109 \cite{us}. We caution that the general numerical coincidence of the scales between $\frac{\Delta T_{h}}{T_{g}}$ and $\overline{A} \sqrt{2}$ is only suggestive ($\overline{A}$ does not vary dramatically across different fluids). Furthermore, here a comparison was only made for the three liquids (fluids examined in \cite{ATG'} that have, unambiguously, an identical composition to those whose viscosity was studied in \cite{us}). The experimental heat capacity traces in \cite{ATG'} are from \cite{OTP+} (for OTP), \cite{GLY+} (for glycerol), and \cite{SAL+} (for Salol).
\bibitem{cor'} Our approach suggests other possible consequences. For instance, since smaller systems may generally exhibit (already in equilibrium) larger fluctuations of their energy density, a corollary of our approach is that supercooling of small fluid droplets might  be more easy to achieve than that of a macroscopic system.
\bibitem{micro-h}
We allude here to the standard rule of thumb in which quantization afforded by Planck's constant must be effectively introduced when counting microstates in classical integrals in phase space in order to obtain a dimensionless number. In basic classical statistical mechanics, the sum over microstates of a system of distinguishable $N$ particles in $d$ spatial dimensions leads to the phase space integral multiplied by a factor which involves a constant that (when compared to the exact quantum results for ideal gases and other systems) is set by Planck's constant $h$, i.e.,  
\begin{eqnarray}
\sum_{n} \cdots \to \frac{1}{h^{dN}} \int d^{dN}x~d^{dN}p \cdots . 
\end{eqnarray}
Similarly, for identical particles, there is an additional prefactor of $\frac{1}{N!}$ that needs to also be introduced by hand in the classical description (to avoid Gibbs' paradox and related behaviors). This celebrated fudge factor emulates, once again, quantum states of identical particles (reflecting their invariance under permutations).
\bibitem{aln2} Since $r_{L} = \pm 1$ does not appear in $H_{I}$, there is an additive $\ln 2$ contribution to the entanglement entropy vis a vis the application of the formulas to the partition of $(L_{A} + L_{B}-1)$ decoupled Ising spins $\{r_{i}\}_{i=1}^{L-1}$ of energy $E= - J \sum_{i=1}^{L-1} r_{i}$ into two groups of $L_{A}$ and $(L_{B}-1)$ spins whose energies are given by the spatially uniform Hamiltonians $H_{AI}$ and $H_{BI}$ respectively. This does not impact the asymptotic scaling of Eq. (\ref{sentl}).
\end{thebibliography}
\end{document}